\documentclass[fleqn,usenatbib]{mnras}

\usepackage[T1]{fontenc}

\DeclareRobustCommand{\VAN}[3]{#2}
\let\VANthebibliography\thebibliography
\def\thebibliography{\DeclareRobustCommand{\VAN}[3]{##3}\VANthebibliography}

\usepackage{graphicx}	
\usepackage{amsmath}	
\usepackage{amssymb}	

\DeclareRobustCommand{\ion}[2]{%
\relax\ifmmode
\ifx\testbx\f@series
{\mathbf{#1\,\mathsc{#2}}}\else
{\mathrm{#1\,\mathsc{#2}}}\fi
\else\textup{#1\,{\mdseries\textsc{#2}}}%
\fi}
\newcommand{\hi}{\ion{H}{i}}

\usepackage{newtxtext,newtxmath}

\title[DEVILS Blended Spectra]{Deep Extragalactic VIsible Legacy Survey (DEVILS): DR1 Blended Spectra Search for Candidate Strong Gravitational Lenses}

\author[B.W. Holwerda]{
B. W. Holwerda$^{1}$\thanks{Contact e-mail: \href{mailto:benne.holwerda@louisville.edu}{benne.holwerda@louisville.edu}},
S. Knabel$^{1}$, 
J. E Thorne$^{2}$,	
S. Bellstedt$^{2}$,	
M. Siudek$^{3,4}$,	
L. J. M. Davies$^{2}$	
\\
$^{1}$ University of Louisville, Department of Physics and Astronomy, 102 Natural Science Building, 40292 KY Louisville, USA.\\
$^{2}$ ICRAR, The University of Western Australia, 35 Stirling Highway, Crawley, WA 6009, Australia\\
$^{3}$ Institut de F\'{\i}sica d'Altes Energies (IFAE), The Barcelona Institute of Science and Technology, 08193 Bellaterra (Barcelona), Spain \\ 
$^{4}$ National Centre for Nuclear Research, ul. Pasteura 7, 02-093, Warsaw, Poland \\
}

\date{Accepted 2021 November 13. Received 2021 November 11; in original form 2021 April 7.}

\pubyear{---}

\begin{document}
\label{firstpage}
\pagerange{\pageref{firstpage}--\pageref{lastpage}}
\maketitle

\begin{abstract}
Here, we present a catalog of blended spectra in Data Release 1 of the Deep Extragalactic VIsible Legacy Survey (DEVILS) on the Anglo-Australian Telescope (AAT). Of the 23197 spectra, 181 showed signs of a blend of redshifts and spectral templates. We examine these blends in detail for signs of either a candidate strong lensing galaxy or a useful overlapping galaxy pair. 

One of the three DEVILS target fields, COSMOS (D10) is close to complete and it is fully imaged with Hubble Space Telescope Advanced Camera for Surveys (HST/ACS) and we visually examine the 57 blended spectra in this field in the F814W postage stamps. Nine are classical strong lensing candidates with an elliptical as the lens, out to higher redshifts than any previous search with spectroscopic surveys such as SDSS or GAMA. The gravitational lens candidate success rate similar to earlier such searches (0.1\%). 

Strong gravitational lenses identified with blended spectroscopy have typically shown a high success rate (>70\%) which make these interesting targets for future higher resolution lensing studies, monitoring for supernovae cosmography, or searches for magnified atomic hydrogen signal.
\end{abstract}




\section{Introduction}

Strong gravitational lenses occur when a massive, distant  galaxy --the lens-- aligns with an even more distant galaxy --the source-- and bends the light from the source into our line of sight. Strong gravitational lensing provides a highly accurate measurement of the {\em total} enclosed mass in these elliptical galaxies that is robust when compared against current galaxy evolution models and assumptions \citep[][and reference therein]{slacs12}.  Gravitational lensing provides the best estimate of the total projected mass of the lens galaxy enclosed within the Einstein radius, delineated by the arcs projected on the sky from the background source galaxy.

As such, lensing elliptical galaxies may provide an excellent test case for other theories of gravity, such as the Emergent Gravity recently proposed by \cite{Verlinde17}, through a combination of lensing and kinematic measurements \citep[see][]{Tortora18a}. 
To date, gravitational lenses have proven General Relativity to be correct with the highest accuracy over galaxy-wide scales \citep{Collett18}. 

Because strong gravitational lenses appear in every respect to be just like other (elliptical) galaxies, results can therefore be generalized to all spheroidal galaxies {\em in the observed mass range} \citep{slacs9}. Gravitational lensing has been a powerful technique to measure the masses of {\em massive} elliptical galaxies \citep{slacs4,slacs5}, constraining the Fundamental Plane ($r_e-M_*-\sigma$) of elliptical galaxies \citep{slacs7}, and the stellar population's mass-to-light ratio and thus their initial mass function \citep[IMF;][]{slacs9}.

The observational drive for classical galaxy-galaxy strong lenses --an elliptical galaxy as a lens with an emission line galaxy as the source-- is now to measure the mass content throughout the spheroidal galaxy mass function \citep{slacs12}, explore the Fundamental Plane in different environments \citep{slacs8}, constrain the stellar mass-to-light ratio in nearby ellipticals \citep{slacs2,Collier18a,Collier18b}, and to discover dark matter substructure in known strong galaxy lens cases \citep{Cyr-Racine19}. Thanks to Integral Field Unit (IFU surveys, lensing arcs no longer need to be well outside the lens galaxy's isophote 
\citep{Smith17a,Lee17h,Talbot18,Smith20b}.

Thus far, lensing arcs have predominantly been identified in {\em massive} ($>10^{11}M_\odot$) spheroidal lens systems thanks to selection biases: SDSS spectroscopic targets are volume-weighted to greater mass, and visual identification favors well-separated arc and lens \citep[][]{slacs12}. The spectroscopically selected sample from the Galaxy And Mass Assembly spectroscopic survey \citep[GAMA,][]{Driver09,Liske15} opened a much greater range in lens masses thanks to the higher completeness and depth of GAMA \citep[][]{Holwerda15}. Spectroscopic identification of lensing galaxies have had a high rate of confirmation, e.g. \cite{Chan16} reported a $>$70\% confirmation rate for a subset of 14 of the \cite{Holwerda15,Knabel20} lens candidates. Similar numbers have been reported for SLACS.

Since the first spectroscopy successes, the search for classical strong gravitational lenses has entered a rapid expansion with the use of machine learning or citizen science on deeper and high resolution ground-based surveys \citep{Petrillo17,Petrillo18,Petrillo19,Li20c,Jacobs19,Huang20a}. The trade-off is between high levels of confidence in individual sources (spectroscopy) or larger target samples that still need confirmation and redshifts (machine learning). The three identification methods are very much complementary in redshift and mass as a comparison between the GAMA spectroscopic and KiDS Machine learning and citizen science identifications in the three GAMA equatorial fields as shown in \citep{Knabel20}. 
One can confirm machine learning identified strong lenses in a spectroscopic survey using the crosscorrelation metrics discussed below (Knabel et al. \textit{in prep.}).

Candidate strong lenses, even presently unresolved, are valuable potential targets for follow-up monitoring for lensed supernovae lightcurves \citep{Oguri10,Shu18,Goldstein19,Wojtak19,Holwerda21} for use of supernova time delay cosmography, the independent measure of the Hubble constant ($H_0$) from the delay between lensed images \citep[see the reviews by][]{Treu16,Oguri19}. As long as redshifts for lens and source galaxy are accurately known, the time-delay primarily depends on the Hubble constant. 

The search, confirmation, and scientific use of strong gravitational lensing systems is expected to change once more with the advent of sensitive and high spatial resolution surveys for atomic hydrogen (\hi). 
\cite{Lipnicky18a} report the first strongly lensed HI detection using a nearby edge-on spiral galaxy as the lens. The new \hi{} surveys, such as MIGHTEE \citep{Jarvis16,Delhaize21,Maddox21} and LADUMA \citep{Holwerda11seeon,Blyth15a,Blyth16,Baker18} using the MeerKAT radio telescope and WALLABY \citep{Koribalski20} on ASKAP \citep{ASKAP}, promise a boon of \hi{} detections lensed by a closer galaxy \citep{Deane15,Blecher19}. 
A magnified \hi{} detection would allow these surveys to probe below their sensitivity limit to verify for example the results from stacking \hi{} spectra for a given galaxy population \citep[cf][]{Delhaize13,Healy21,Chowdhury20,Chowdhury21}. 

\cite{Deane15} made  predictions for the CHILES and LADUMA deep \hi{} fields (COSMOS and CDF-S respectively) are at best a single strong lens to dozens. Depending on spectroscopic coverage and volume, more strong \hi{} lenses can be found in MIGHTEE and WALLABY. Blended optical spectra would point to where to look in position and frequency in the \hi{} cubes. The benefit of \hi{} confirmation is that it need not rely on a classical galaxy-galaxy gravitational lens, i.e. an elliptical or spheroidal galaxy acting as the lens. Any galaxy along the line of sight would be sufficient to magnify the source \hi{} signal. With this in mind, we report the position and redshifts of all the blended spectra in this paper.

The successor of GAMA and the precursor of the WAVES survey\footnote{Wide Area Vista Extragalactic Survey \cite{Driver19}.} is underway on the AAT: the Deep Extragalactic VIsible Legacy Survey \citep[DEVILS,][]{Davies18}. The DEVILS survey goes to a greater depth and higher redshift range than GAMA with the same high level of completeness. We present the candidate search as a pathfinder for the near future strong lenses searches using spectroscopy. 
Here we report the candidate strong gravitational lenses identified in the three fields targeted in the DEVILS survey specifically to facilitate follow-up observations, e.g. searches for magnified \hi{} signal or doubly imaged supernovae with a time delay. 

This paper is organized as follows: 
section \ref{s:devils} describes the DEVILS data-products used, section \ref{s:selection} describes the selection of blended spectra, section \ref{s:results} presents the characteristics for the three DEVILS target fields, section \ref{s:hst} discusses the COSMOS (D10) field Hubble imaging of blended spectra, and section \ref{s:conluding} are our concluding remarks for this catalog result. 
Throughout this paper we adopt a \cite{Planck-Collaboration15} cosmology ($H_0 = 67.7$ km/s/Mpc, $\Omega_m = 0.307$).

\section{DEVILS}
\label{s:devils}

The Deep Extragalactic VIsible Legacy Survey \citep[DEVILS,][]{Davies18} is a large spectroscopic campaign at the Anglo-Australian Telescope \citep[AAT,][]{Sharp06} aimed at bridging the near and distant Universe by producing a high completeness survey of galaxies and groups at intermediate redshifts ($0.3 < z < 1.0$). The target sample consists of $\sim$60.000 galaxies selected down to $Y<21.2$ mag, over $\sim6$ deg$^2$ spread over three well-studied deep extragalactic fields: the Cosmic Origins Survey field, (COSMOS, DEVILS D10 field), the Extended Chandra Deep Field South (ECDFS, D03), and the X-ray Multi-Mirror Mission Large-Scale Structure region (XMM-LSS, D02)

These three fields are all ``deep drilling fields" for the Legacy Survey of Space and Time (LSST) on the Vera C. Rubin Telescope. Target selection for DEVILS is from Y-band imaging performed by the deep Visible and Infrared Survey Telescope for Astronomy (VISTA) extra-galactic observations: VIDEO \citep{Jarvis13} and UltraVISTA \citep{McCracken12}, with photometry measured using the {\sc ProFound} package \citep{Robotham18}. Photometric star/galaxy separation is done on the basis of NIR colours, and has been validated by visual inspection described in Davies et al. (\textit{in prep}). 
These fields are the prime targets for the MIGHTEE \citep{Jarvis16,Delhaize21,Maddox21}, LADUMA \citep{Holwerda11seeon,Blyth15a,Blyth16,Baker18}, and CHILES \citep{Dodson15,Jones16a,Fernandez16,Hess19,Blue-Bird20} deep \hi{} surveys.

We will focus on the COSMOS \citep{Scoville07b} field (DEVILS D10) in section \ref{s:hst} because it is the closest to full completion of the three DEVILS fields and it has complete complementary Hubble Space Telescope archival imaging data. Blended spectra candidates from the other two fields, XMM-LSS (D02) and ECDF-S (D03) are presented for future follow-up. 
The field represents an early opportunity where lensed \hi{} could be found in the CHILES survey \citep{Fernandez16}. 

\section{Strong Lens Selection}
\label{s:selection}

Strong lensing candidates are selected from a clear double signal in the spectrum's redshift crosscorrelation fits with spectral templates. We note that the searches for a double signal in an observed spectrum can have two scientific motivations. One can look for lensing systems such as the original SLACS project \citep[Strong Lenses with ACS,][]{slacs1}. SDSS blended spectra i.e. passive galaxy spectra with one or more emission lines obviously at a higher redshift, were highly successful identifying strong gravitational lenses (70\% efficacy) for follow-up and analysis with Hubble. A similar success rate was reported by \cite{Chan16} for the \cite{Holwerda15} selections. 
Alternatively, one searches for overlapping galaxies that can be used to estimate galaxy transparency based on differential photometry \citep{kw99a,kw99b,kw00a,kw00b,kw01a,kw01b,Holwerda07c,Holwerda09,Holwerda13a,Holwerda13b,Holwerda16a}. The latter was the original motivation for the search in SDSS \citep{Holwerda07c} and GAMA \cite{Holwerda15} for blended spectra. 

\subsection{AUTOZ}
\label{ss:autoz}

The redshifts in the DEVILS survey are found using the crosscorrelation software {\sc autoz} \citep{Baldry14}. This uses a series of templates and crosscorrelates them with the spectra, producing a ranking of redshift and template combinations with different strengths of correlation. This often leads to second crosscorrelation peak because a strong emission line being misidentified at the incorrect redshift (aliasing). These are not likely candidates for a gravitational lens but cannot entirely be discounted because a strong lens might just lie on such a  

The DEVILS DR1 {\sc autoz} catalog has 23197 entries based on AAT spectroscopy and any archival spectra of similar or better quality. This first science release DR1 of DEVILS is about a third towards its science goal of 60.000 redshifts complete with ancillary products in three science fields.

\subsection{Blended Spectra}
\label{ss:blends}

Strong lenses can be identified from a second crosscorrelation peak that is much stronger than the third or fourth one and consists of the first, highest crosscorrelation peak at lower redshift with a passive template (no or few emission lines) and a second crosscorrelation peak at a higher redshift with a strong emission line spectrum template.
We follow the same setup as in \cite{Holwerda15} for the {\sc autoz} catalog from the DEVILS fields. We define the R value as follows: 
\begin{equation}
    R = {\sigma_2 \over \sqrt{\sigma_3^2/2 + \sigma_4^2/2 }}
\end{equation} 
\noindent where $\sigma_n$ is the height of the crosscorrelation peak for the nth ranked redshift and template combination.

\begin{figure}
    \centering
    \includegraphics[width=0.5\textwidth]{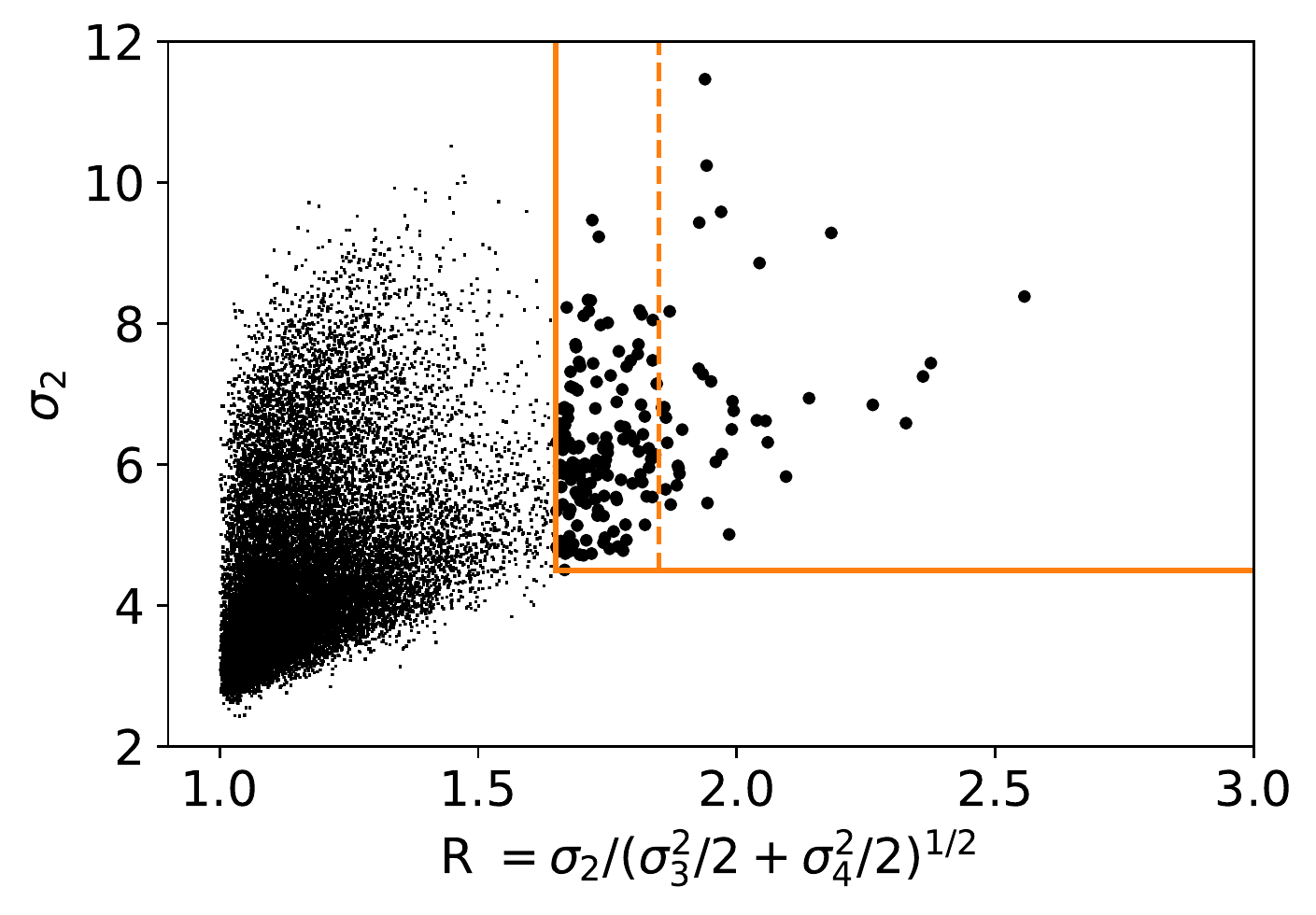}
    \caption{The value of R and the height of the second crosscorrelation peak ($\sigma_2$). The selection of strong candidates for blended spectra is shown as the solid orange line. The original R criterion from \protect\cite{Holwerda15} is shown as the dashed line. In the selected cases, there is a high second peak that is much higher relative to the subsequent ones.}
    \label{f:R-s2}
\end{figure}

Figure \ref{f:R-s2} shows the values for R and $\sigma_2$ (the height of the second crosscorrelation peak) in the DEVILS spectroscopic {\sc autoz} catalog. We opt for a slightly wider set of criteria (in R and $\sigma_2$) than \cite{Holwerda15} with a lower threshold for the R values. A comparison between {\sc autoz} of machine learning identified strong lenses (Knabel et al. \textit{in prep.}) has shown that the original criterion was very conservative in R. Spectra with a relatively high second crosscorrelation peak are our blended spectra targeted for further scrutiny. 

\begin{figure}
    \centering
    \includegraphics[width=0.5\textwidth]{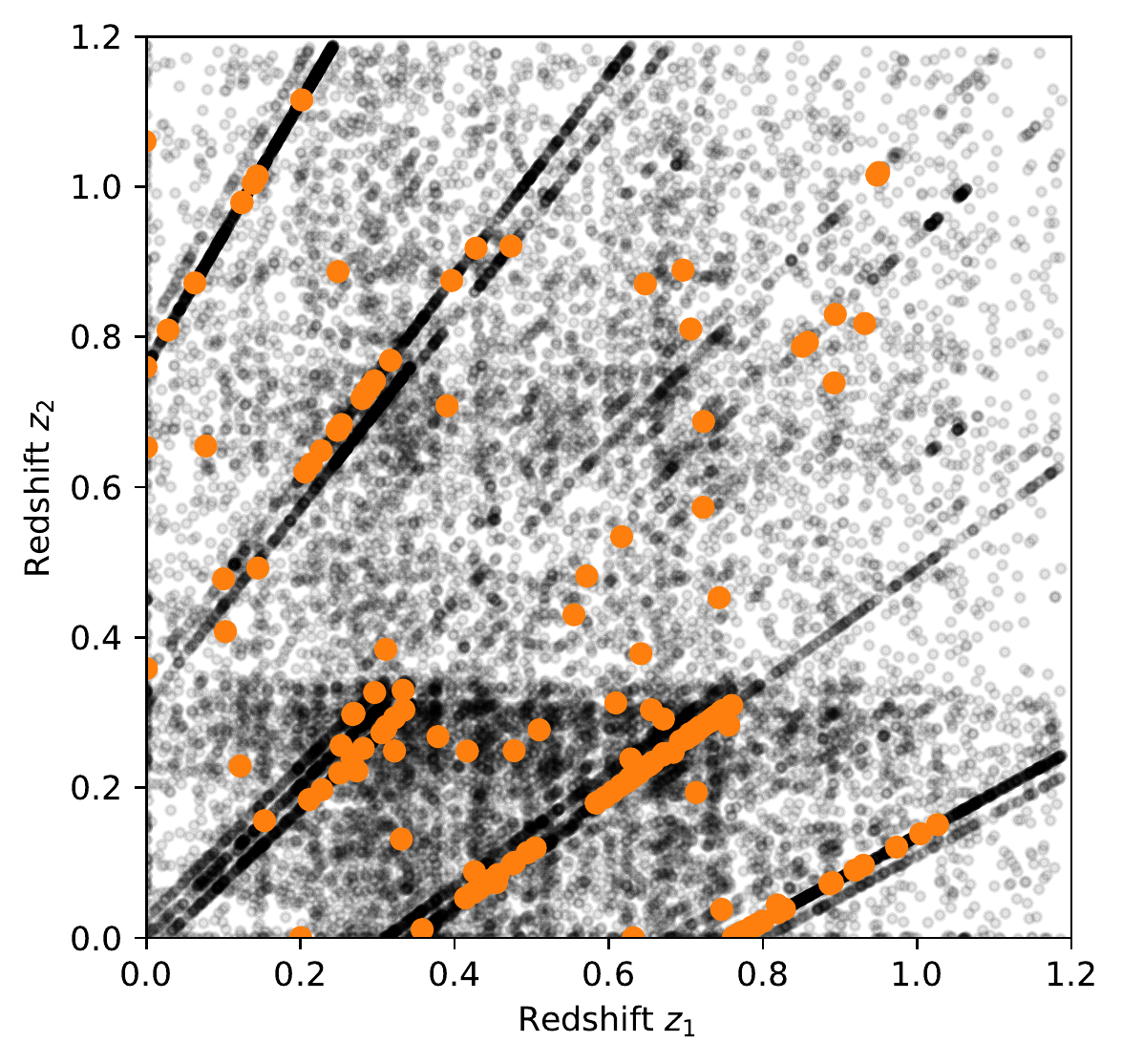}
    \caption{The first and second redshift as identified by {\sc autoz}. Alias lines are visible as diagonal lines of points. A number of the candidates selected in Figure \ref{f:R-s2} are on such alias lines.}
    \label{f:z1-z2}
\end{figure}

Figure \ref{f:z1-z2} shows the first and second peak redshift for all the DEVILS spectra and those selected as blended spectra systems. 
A large number of these are on aliasing lines and may well be artifacts of the {\sc autoz} fits. However, this does not preclude them from being a strong gravitational lens candidate. For example. a candidate on aliasing lines were later confirmed as strong lenses \cite{Knabel20}. For a search of magnified \hi, we argue a wide net of candidate positions and redshifts is most practical. 

\begin{figure}
    \centering
    \includegraphics[width=0.5\textwidth]{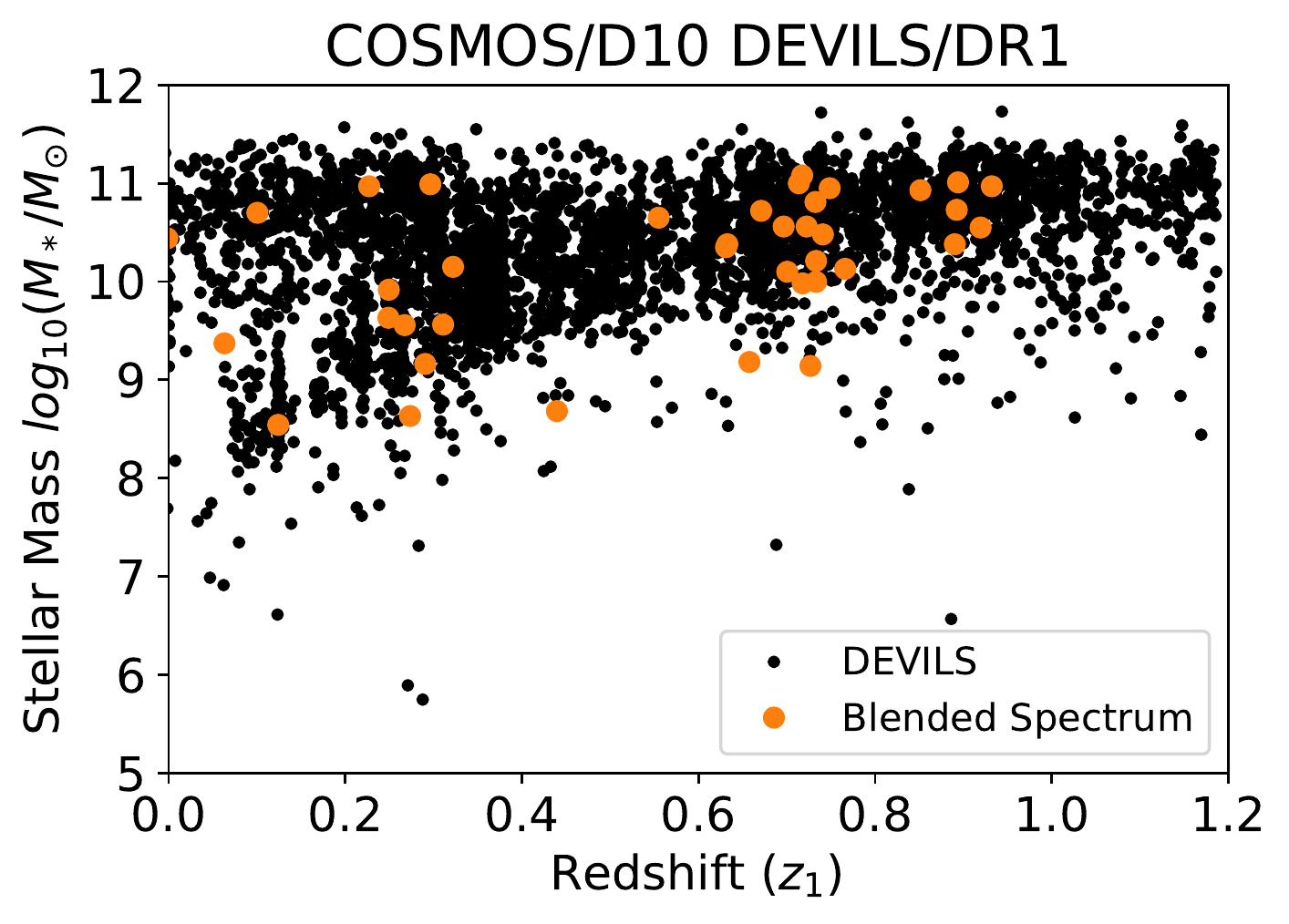}
    \caption{The redshift vs stellar mass plot for the DEVILS DR1 in COSMOS/D10 with blended spectra marked with orange markers. DEVILS DR1 is complete to $log(M^*) \sim 9.5$ out to $z\sim1$. Blended spectra appear to mostly within the completeness envelope. }
    \label{f:devils:mass-z}
\end{figure}

Figure \ref{f:devils:mass-z} shows the redshift of the primary cross-correlation and the stellar mass for COSMOS/D10 from \cite{Thorne21}, derived from the {\sc profound} photometry (Davies et al. \textit{in prep}) using {\sc ProSpect} tool \citep{Robotham20}. DEVILS DR1 is complete to $log(M_*/\rm{M}_{\odot}) \sim 9.5$ out to $z\sim1$. Blended spectra are mostly within this completeness envelope and follow redshift over-densities due to large scale structure because these increase the odds of two galaxies along the line-of-sight.

\begin{figure}
    \centering
    \includegraphics[width=0.5\textwidth]{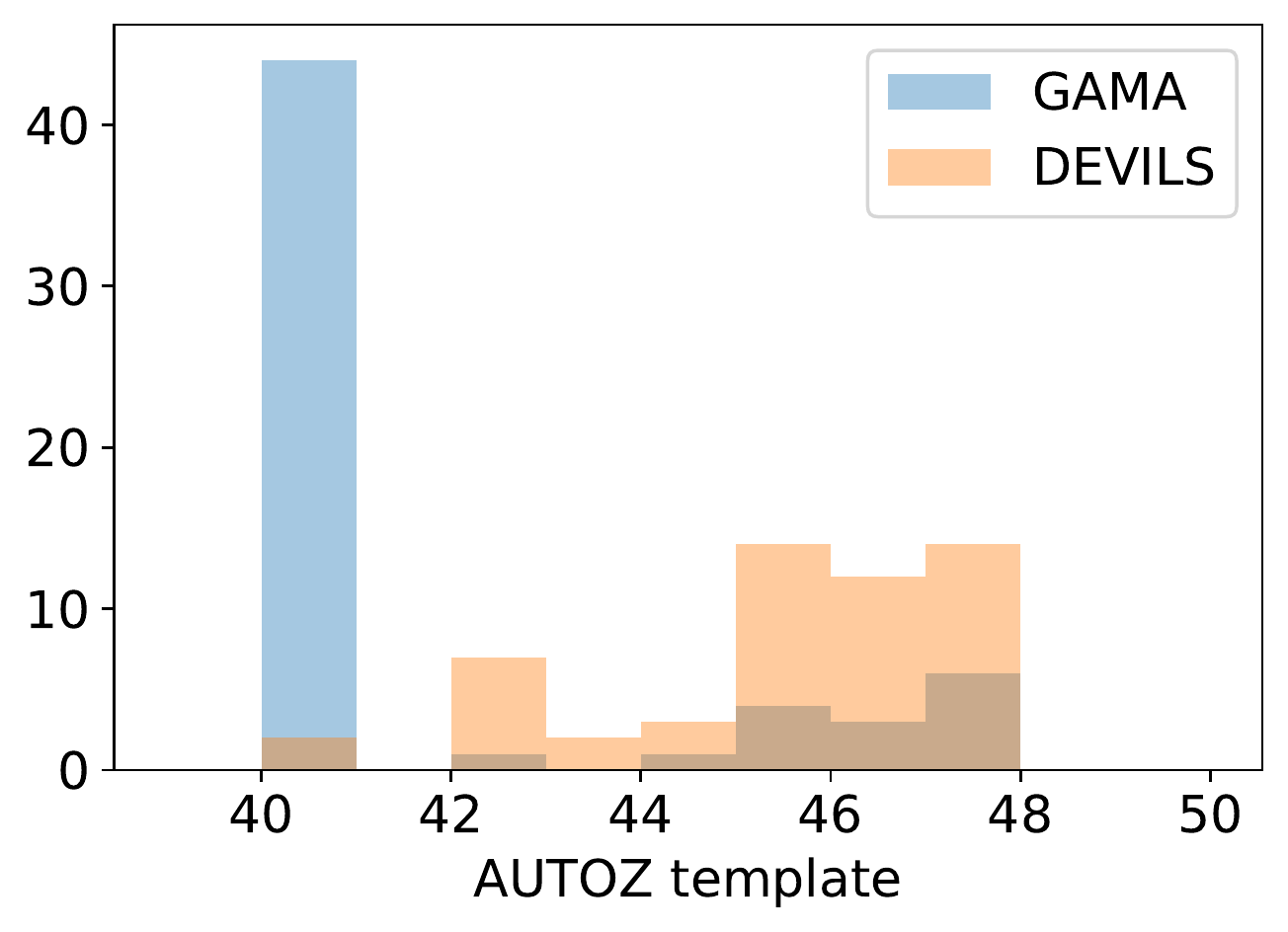}
    \caption{The histogram of template numbers (lower meaning more passive templates) for the GAMA blended spectra from \protect\cite{Holwerda15} and those from DEVILS DR1 catalog.}
    \label{f:templates:hist}
\end{figure}

\subsection{Templates}
\label{ss:templates}

Figure \ref{f:templates:hist} shows the template distribution from the {\sc autoz} for the lowest redshift galaxy for the new DEVILS blended spectra and those from \cite{Holwerda15}. To broadly classify the templates, \cite{Holwerda15} classify templates 40–42 as ``passive galaxies'' (PG) and 43–47 as ``emission-line galaxies'' (ELG). The GAMA lens candidates were selected to prefer passive templates while the DEVILS candidate selection is a mix of passive and emission line templates. The different mix of templates in DEVILS is because 
DEVILS probes to an earlier epoch and therefore is contains more emission line galaxies. Our lower selection threshold which may contain more contamination from aliasing strong emission lines.  DEVILS is still being completed and s/n is still being accumulated for a successful redshifts with stellar continuum dominated sources (PG) and a higher completeness in the target regions. 

\begin{figure*}
    \centering

    \includegraphics[width=0.49\textwidth]{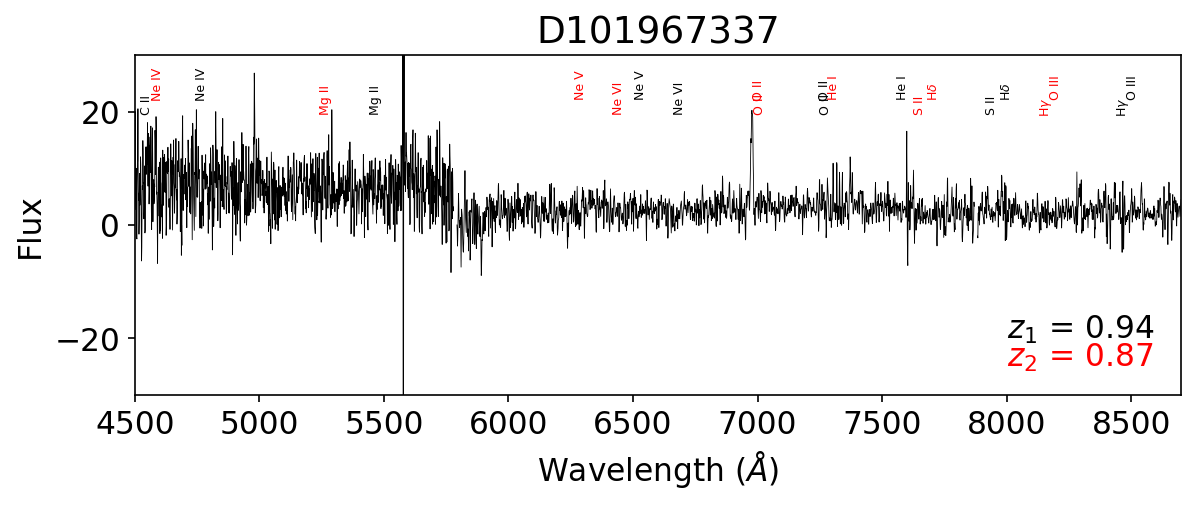}
    \includegraphics[width=0.49\textwidth]{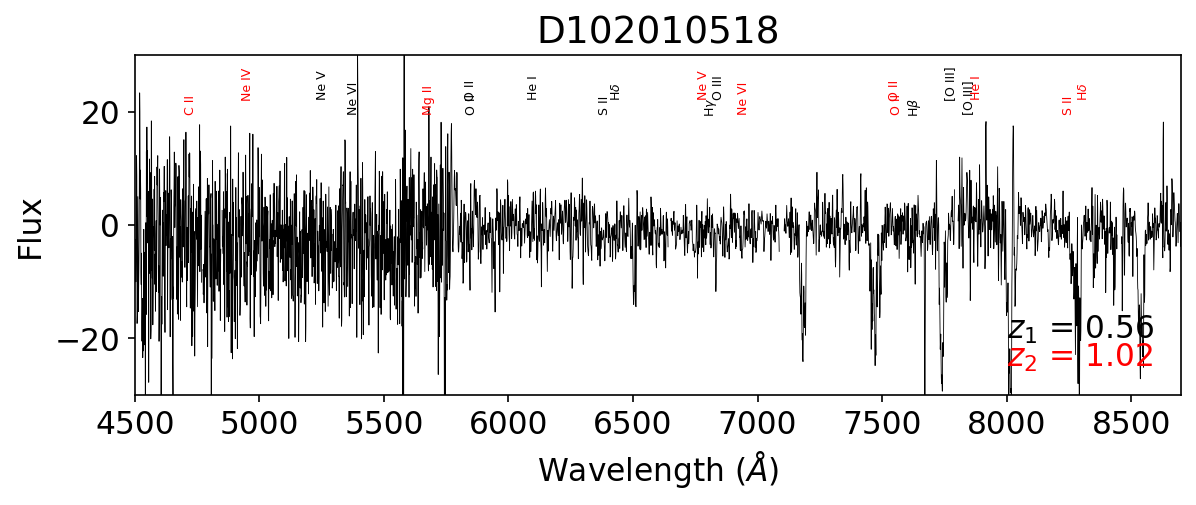}\\
    \includegraphics[width=0.49\textwidth]{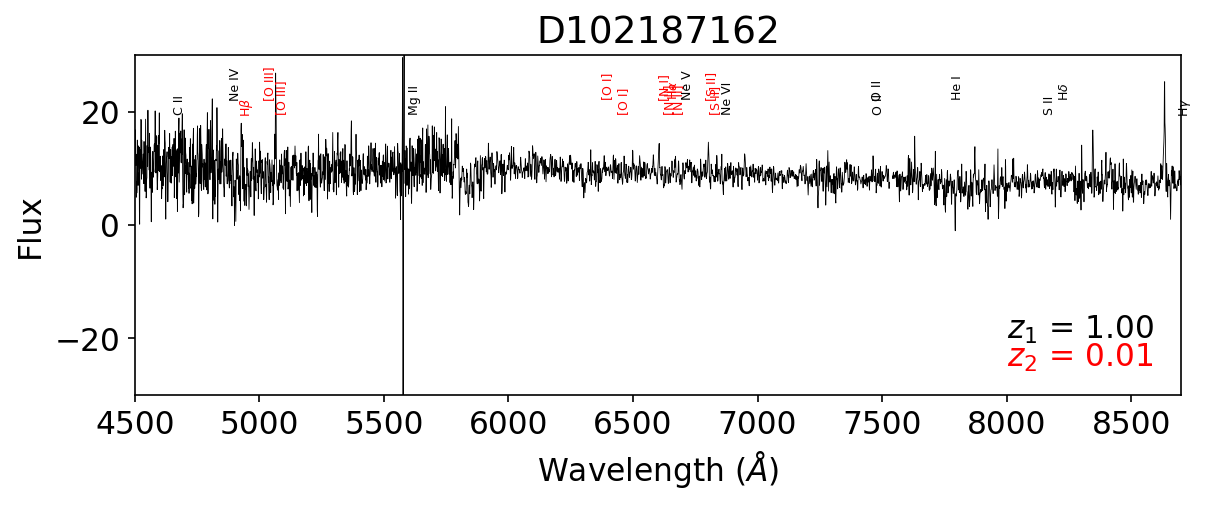}
    \includegraphics[width=0.49\textwidth]{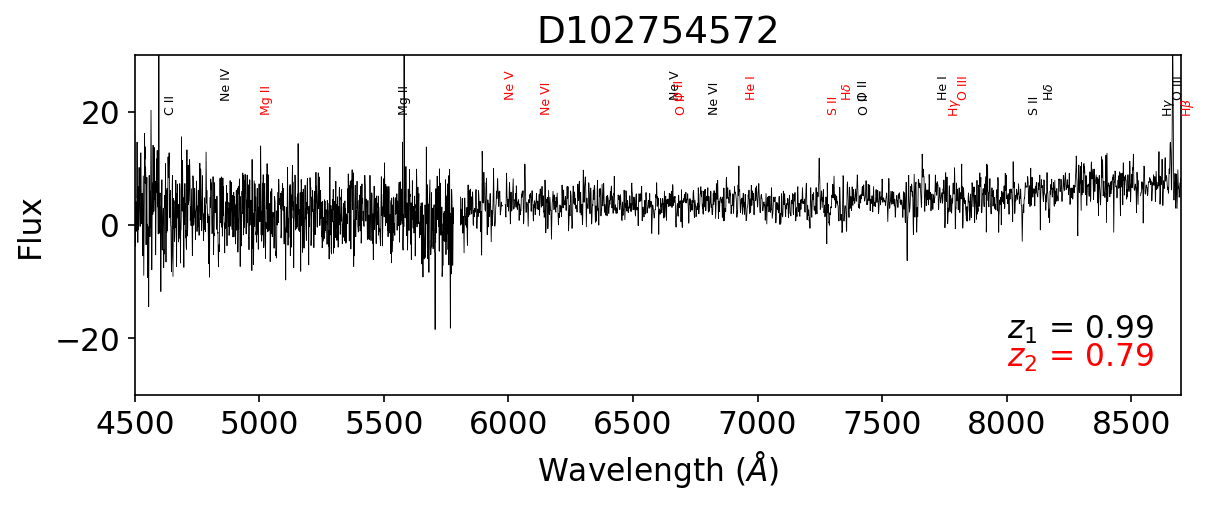}\\
    \includegraphics[width=0.49\textwidth]{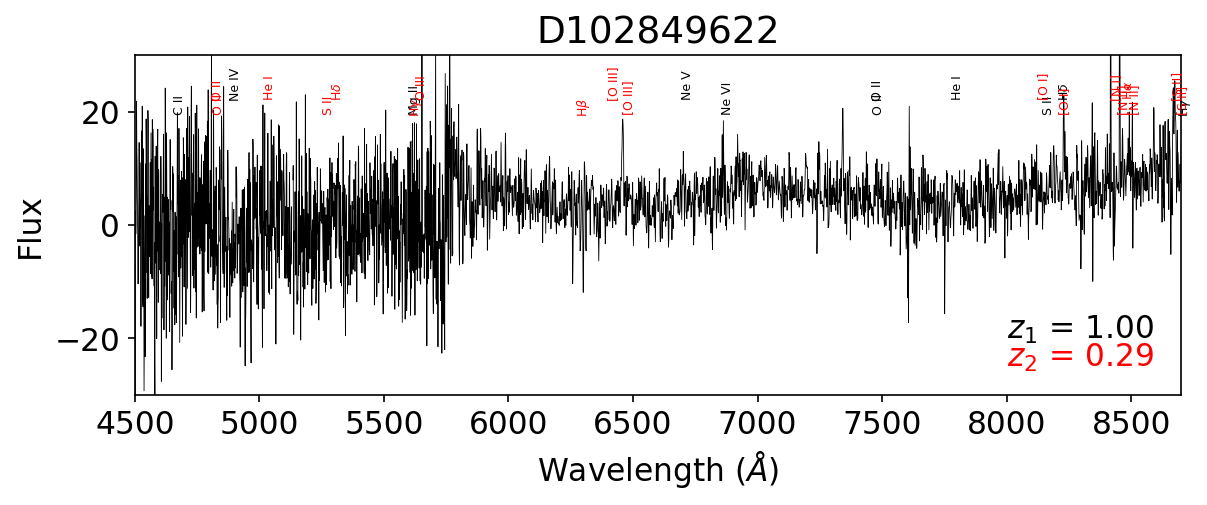}
    \includegraphics[width=0.49\textwidth]{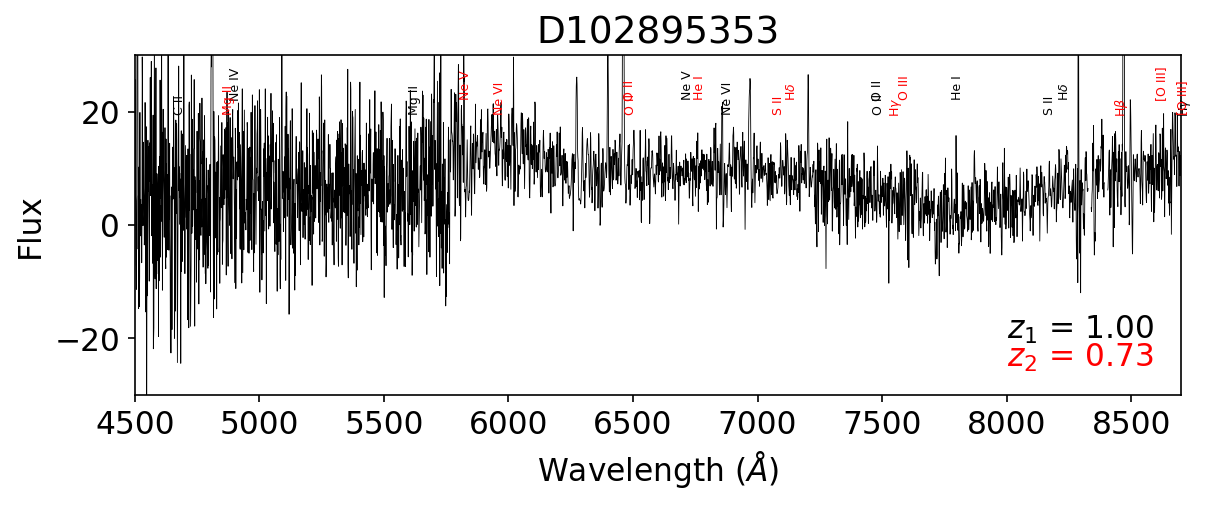}\\
    \includegraphics[width=0.49\textwidth]{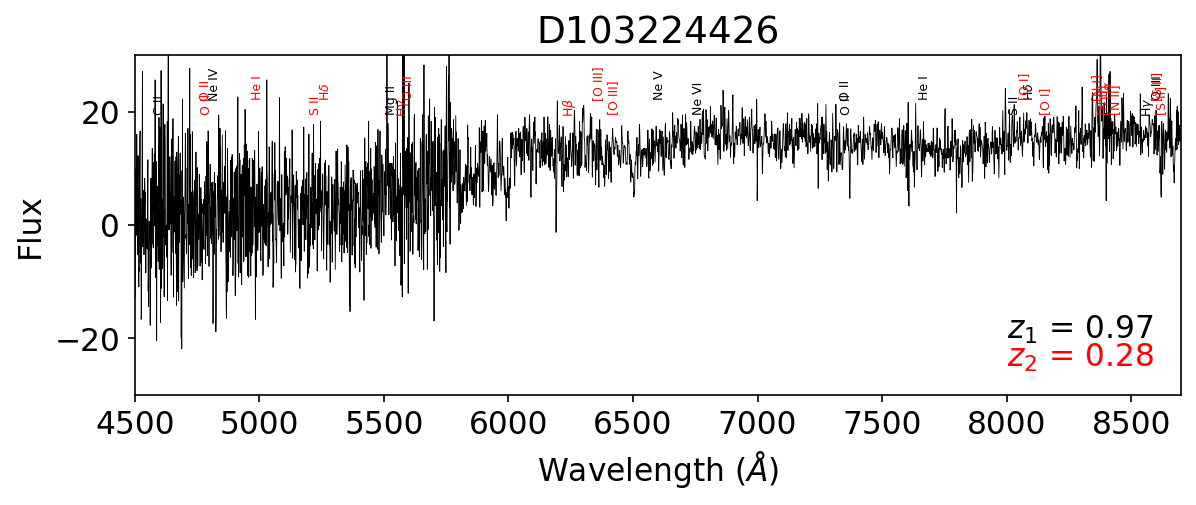}
    \includegraphics[width=0.49\textwidth]{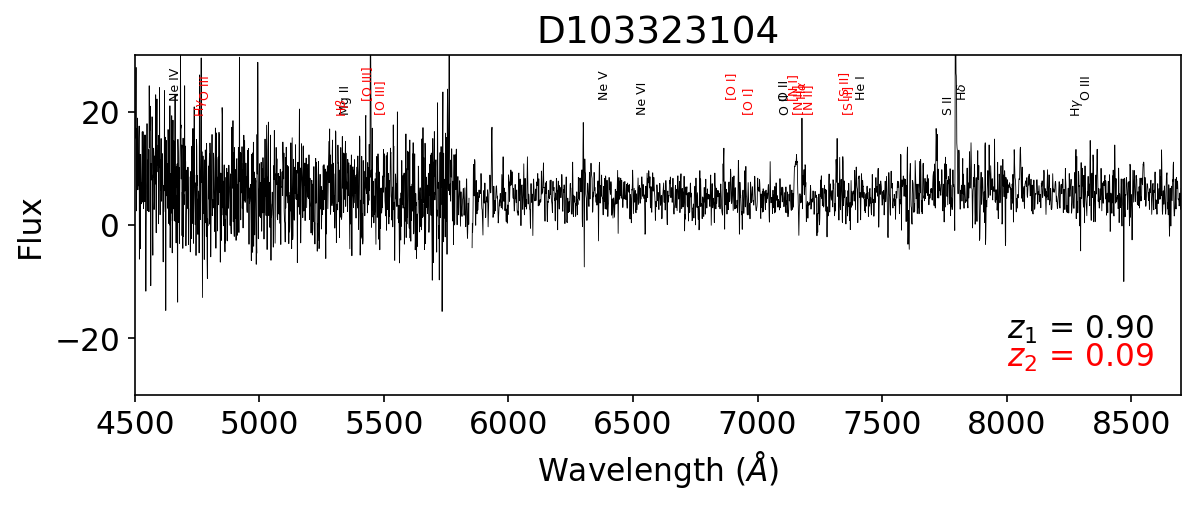}\\

   \caption{\label{f:spectra} Native DEVILS spectra in the COSMOS region with two redshift solutions that make our cross-correlation cut. Emission and absorption lines are marked for the first (black) and second (red) redshift solution. As an alternate view is shown in Figure \ref{f:spectra:appendix}.}
\end{figure*}

Figure \ref{f:spectra} shows a few example DEVILS spectra for the COSMOS fields. Expected emission and absorption features are marked for both of the crosscorrelation solutions ($z_1$ and $z_2$). It illustrates how certain spectral features drive a high crosscorelation with the {\sc autoz} templates. 

\section{Results}
\label{s:results}

\subsection{DEVILS Fields}
\label{ss:devils:fields}

\begin{figure*}
    \centering
    \includegraphics[width=0.32\textwidth]{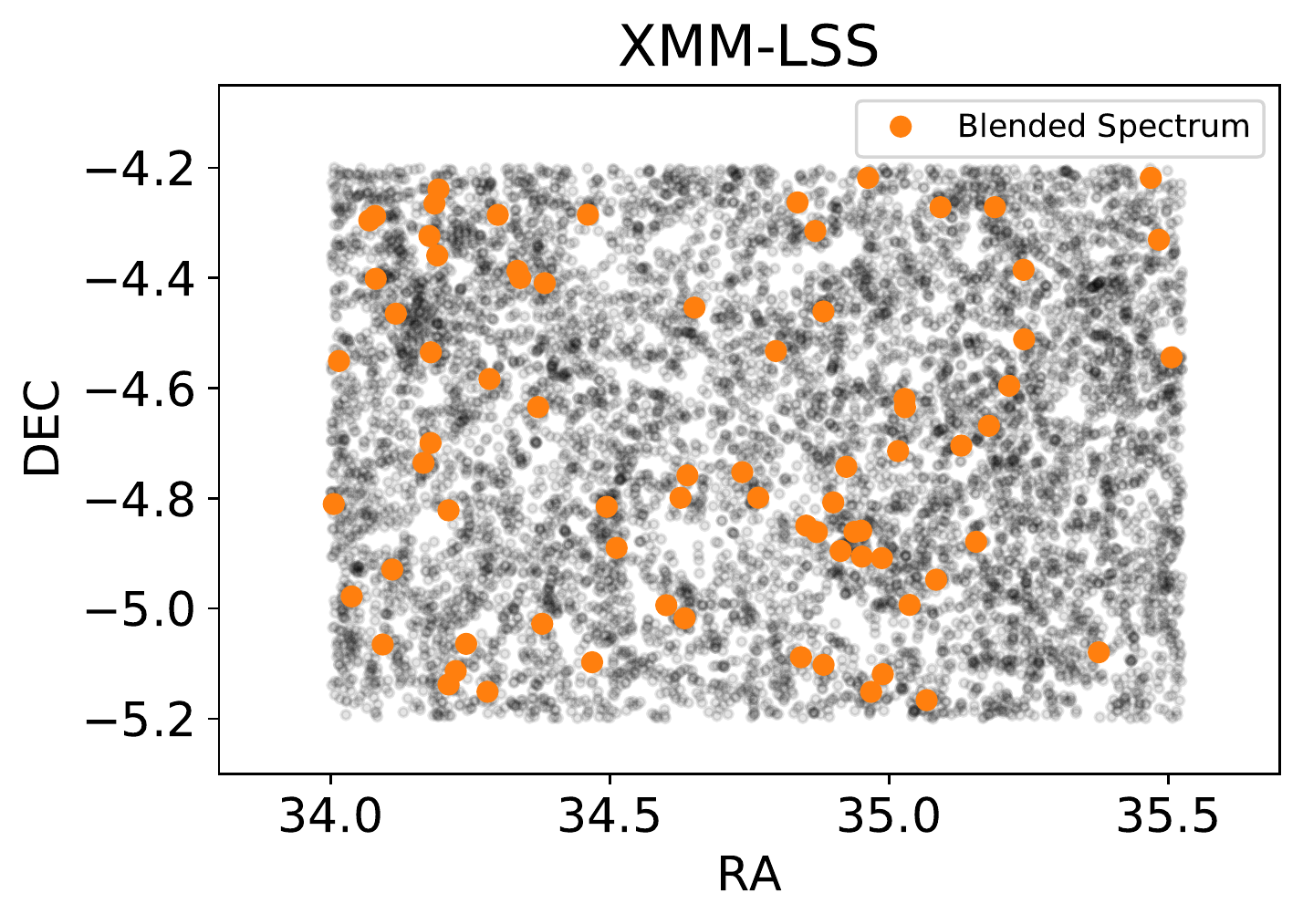}
    \includegraphics[width=0.32\textwidth]{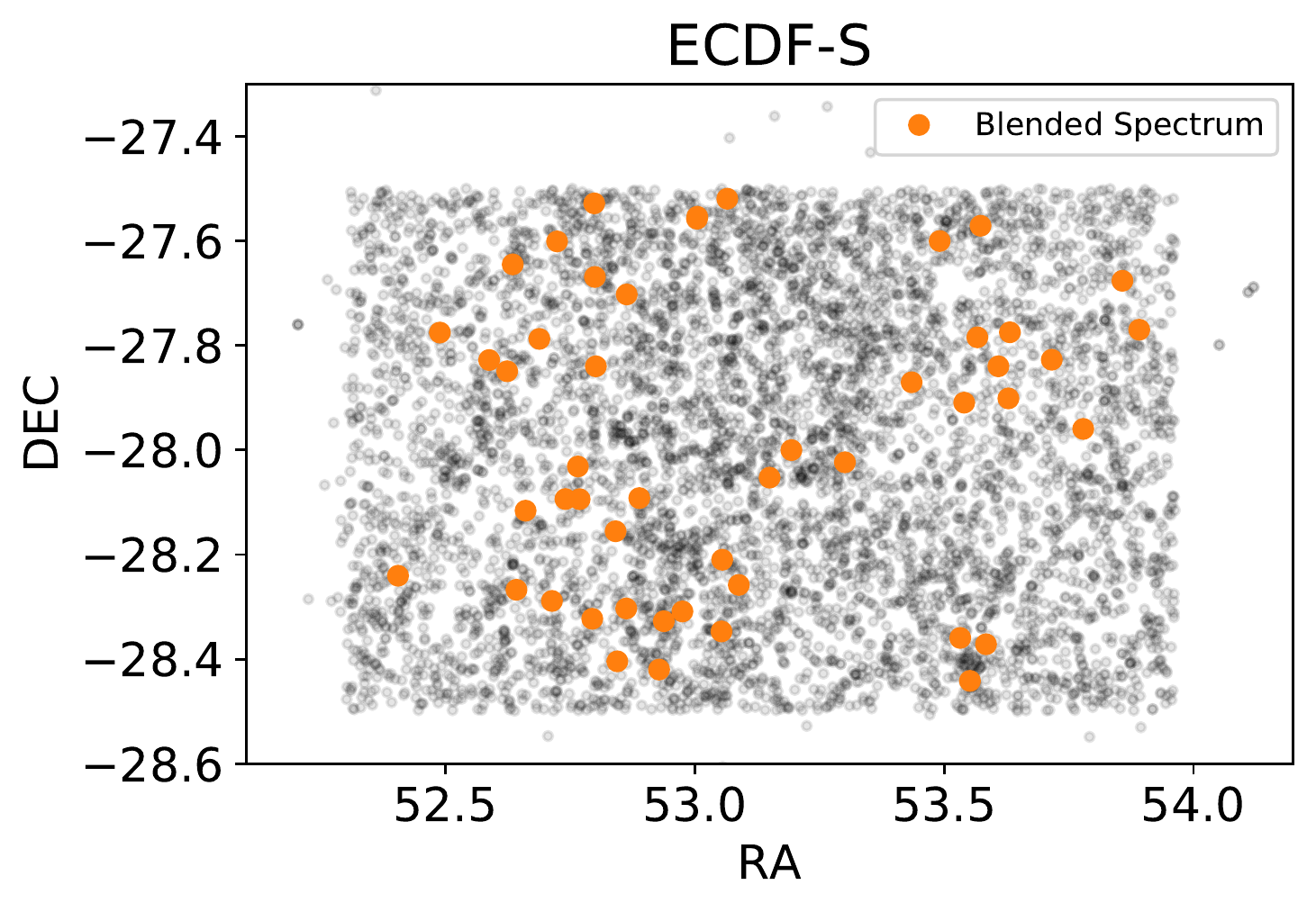}
    \includegraphics[width=0.32\textwidth]{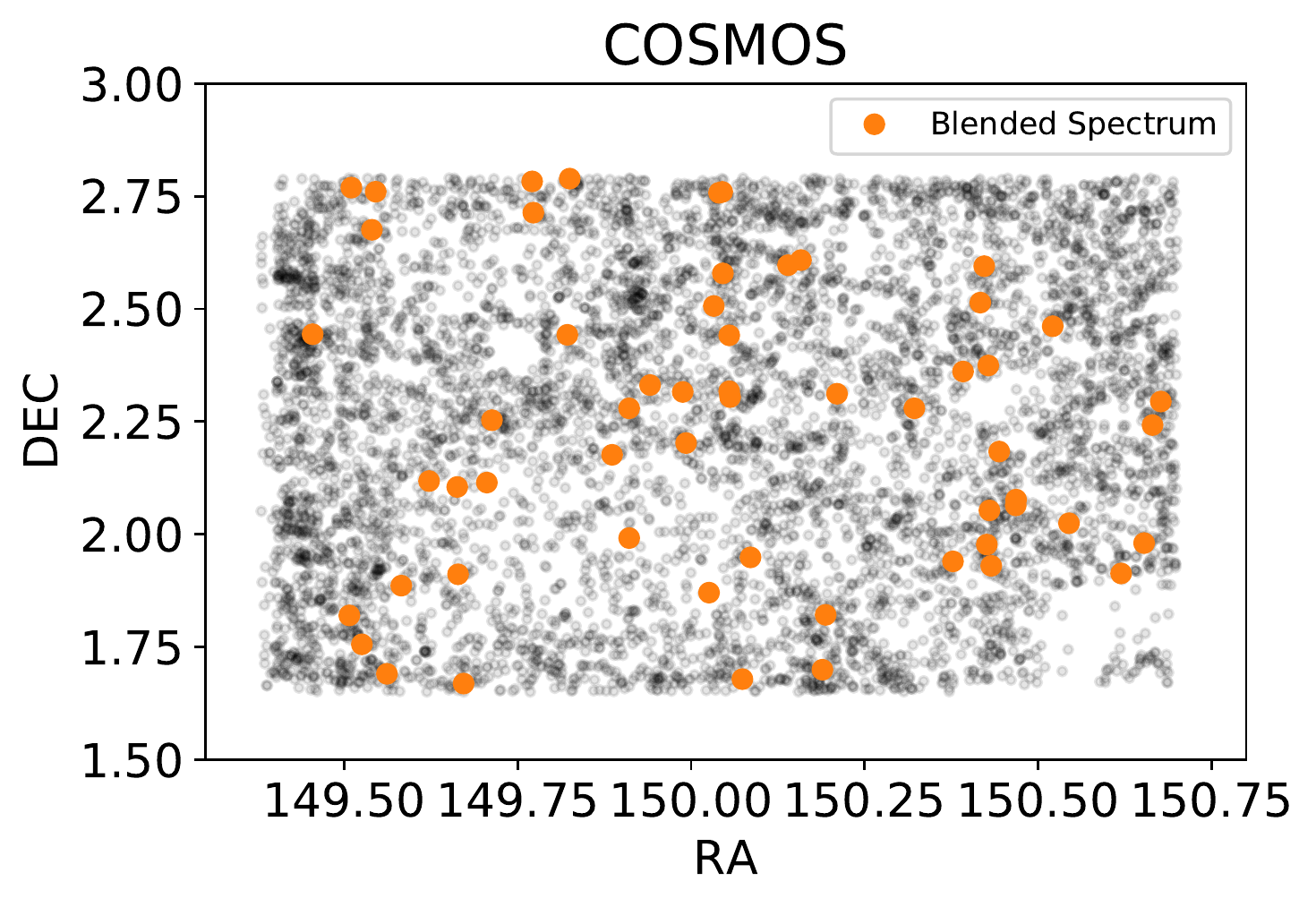}
    \caption{The positions of the DEVILS spectra in all three fields. }
    \label{f:devils:fields}
\end{figure*}

Figure \ref{f:devils:fields} shows the positions of DEVILS DR1 redshifts spread over the three fields. Of these three, COSMOS (D10) is the most complete. We identify 57 blended spectra in COSMOS, 75 in XMM-LSS, and 49 in ECDFS, which would be too high for the expected on-sky number density of strong lenses alone \citep[see the discussion in ][]{Knabel20} but an equal number of galaxy-galaxy alignments is expected \citep[e.g. the overlapping pairs of galaxies][]{Holwerda15}.

\begin{table*}
    \centering
    \begin{tabular}{l l l l l l l}
DEVILS  & Common    & RA    & Dec   & Input     & No. &  Blended \\
Name    & Name      & (Deg) & (Deg) & Survey    & Spectra & Spectra \\
\hline
D02 & XMM-LSS & 35.4975	& -4.825 & VIDEO & 9402 & 75 \\
D03	& ECDFS	 & 53.65 & -28.1	& VIDEO & 6574 & 49 \\
D10	& COSMOS &	150.1 &	2.1	& UltraVISTA & 7219 & 57 \\
\hline
    \end{tabular}
    \caption{DEVILS DR1 Fields and the number of blended spectra identified in each. }
    \label{t:devils:fields}
\end{table*}

Table \ref{t:devils:fields} lists the number of spectra and the number of blended spectra identified in DEVILS DR1. Typically 0.1\% of a large survey is a blended source \citep[][]{Holwerda15} but the rate is closer to 1\% in DEVILS. This is partly due to the lower threshold in R we set as well as the high completeness to a lower apparent magnitude of the survey, increasing the chances of alignments along the line-of-sight. We expect however that the majority of the blended spectra is chance alignments and not strong gravitational lenses. 

\begin{figure*}
    \centering
    \includegraphics[width=\textwidth]{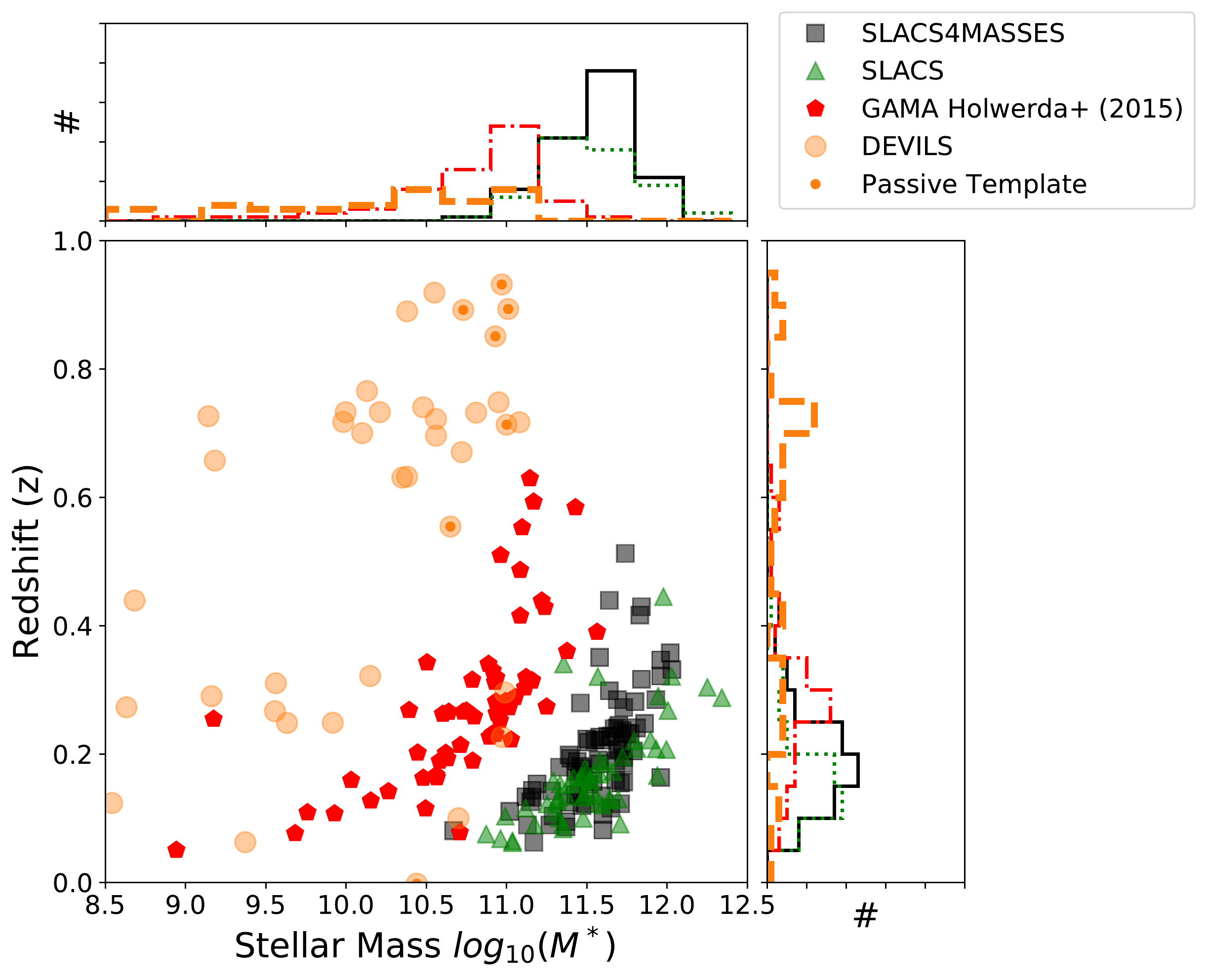}
    \caption{A comparison between spectroscopically identified strong lensing candidates. Stellar masses from \protect\cite{Thorne21} for the DEVILS/D10 candidates. }
    \label{f:3plot}
\end{figure*}

Figure \ref{f:3plot} shows the stellar mass from the {\sc ProSpect} \citep{Robotham20} spectral energy density fit to the DEVILS photometry and the first peak redshift in COSMOS (DEVILS/D10) field. 
We compare these to mass and redshifts from \cite{Holwerda15} and the SLACS \citep{slacs5} and SLACS for the Masses \citep[S4TM][]{Shu17} values. We note here that the values from \cite{Holwerda15} are also derived using {\sc magphys}. The SLACS values are reported for both a Salpeter and a Charbrier Initial Mass Function (IMF) and use that to infer a mass to light ratio (M/L) for the Johnson-I HST observation. The wider aperture for SDSS spectra (3\farcs) allows for the selection of more massive systems at lower redshift \citep[see][]{Knabel20}.

Figure \ref{f:3plot} shows how the SLACS and S4TM probe massive systems ($log(M_*/\rm{M}_{\odot}) > 11$) below $z\sim0.5$. The high completeness and depth of GAMA (together with narrower fiber aperture) resulted in a lens sample that is both at higher redshifts for the same mass as well as lower masses at low redshift. The expectation for DEVILS is that the final lens candidate sample from this survey probes both higher redshift again as well as lower masses at intermediate redshfits. The DEVILS markers in Figure \ref{f:3plot} are all the blended spectra without additional constraints. However, these include indeed lower masses at low redshift ($z<0.3$) and higher redshift candidates at the high mass end ($0.5<z<1$).

\subsection{Predicted Einstein Radii}
\label{ss:EinsteinR}

To estimate the Einstein radius of the candidate lenses, we approximate the galaxy lens as a thin-lens system with the lens and source positioned along the same line of sight, where the Einstein radius ($\theta_E$) is given by:
\begin{equation} \label{theta_e_pm_eqn}
\theta_E = \sqrt{\frac{M_E}{10^{8.09}M_\odot}} \sqrt{\frac{D_{LS}}{D_{L}D_{S}} \rm{Mpc}} \ \  \textnormal{arcsec},\end{equation}
where $M_E$ is the total \textit{enclosed} mass by the Einstein radius, $D_{LS}$ is the distance from lens to source, $D_{L}$ is the distance from observer to lens, and $D_{S}$ is the distance from observer to source. All distances are angular diameter distances calculated from redshift assuming flat $\Lambda$CDM. 

\cite{Auger10} modeled and analyzed 73 SLACS lenses and presented a power law relation between the total (lensing) mass enclosed within half the effective radius (close to the typical Einstein radius), and the stellar mass: 
\begin{equation} \label{enclosed_mass_eqn}
M_E = 0.0011 \times  \left(\frac{M_*}{M_{\odot}}\right)^{1.25}
\end{equation}

\begin{figure}
    \centering
    \includegraphics[width=0.5\textwidth]{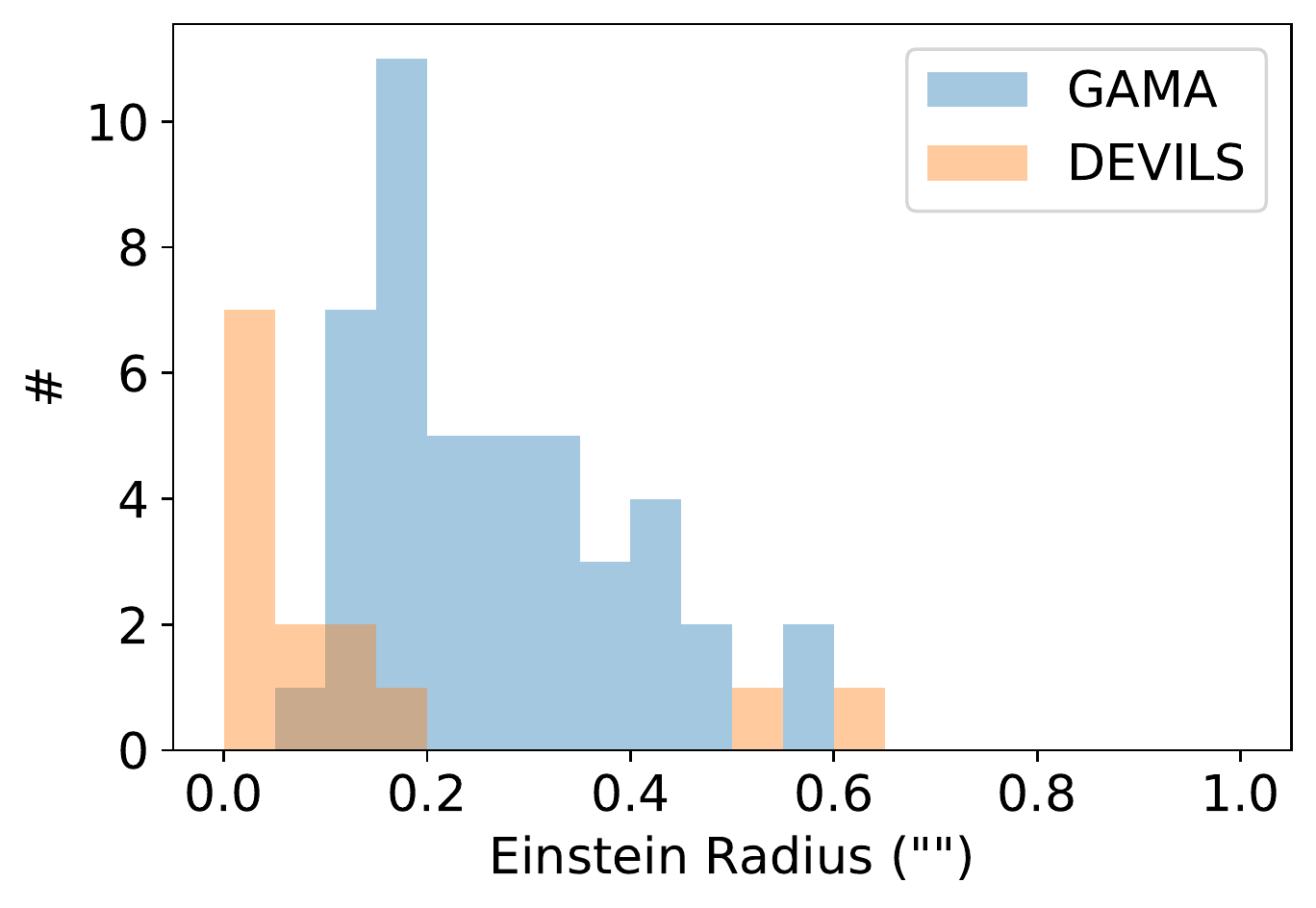}
    \caption{The histogram of predicted Einstein radii for the GAMA blended spectra from \protect\cite{Holwerda15} and those from DEVILS DR1 catalog (this paper).}
    \label{f:RE:hist}
\end{figure}

Figure \ref{f:RE:hist} shows a histogram for the GAMA blended spectra from \cite{Holwerda15} and the DEVILS DR1 COSMOS/DR1 stellar masses. A majority of DEVILS blended spectra predict Einstein radii of 0\farcs1 or below. 
In practice, we have found that the above estimate under-predicts the Einstein radii fractionally (Knabel et al. \textit{in prep.}).
Nevertheless, such compact lenses are incredibly difficult to meaningfully resolve, even with the Hubble Space Telescope or a ground-based AO system with an IFU but they may prove worthwhile targets for searches for magnified \hi{} lines or a next generation of AO-enabled telescopes.  
 
The odds that no magnified stellar emission can be observed go up with a smaller Einstein ring. The magnification of the lens goes as $\mu= {|\theta| \over |\theta| - \theta_{E} }$, where $|\theta|$ is the angular distance between source and lens in the image plane and $\theta_E$ is the Einstein radius \citep[cf][]{Mason15b}. With small Einstein radii, the sources have to overlap in the image plane to produce high magnifications i.e. there has to be a meaningful cross-section between the extent of the star-formation nebular line and stellar emission of the background galaxy and the caustic for there to be a detectable signal and ring respectively. 

The selection of sources as lenses depends on the survey considered and what level of magnification is considered a lens. \cite{Dobler08} modelled Sloan (SDSS) spectra to estimate the selection function of the SLACS survey. They had to account for the size of the source galaxies in emission line regions (e.g. OII emitting parts of star-forming galaxies at higher redshift). Most recent authors examined the selection functions of photometric imaging surveys for strong lenses \citep{Chang15a, Collett15,Collett17a}.
\cite{Dobler08} estimates high probabilities for false-positives (80\% and above), with higher false-positive rates for high ($z\sim 0.4$) redshift lenses. Extrapolating the trends seen in \cite{Dobler08}, can expect higher false positive rates for DEVILS (higher redshift lens galaxies) or lower magnifications. For a pair of galaxies along the line of sight to be selected, they need not be lensing \citep[defined as a magnification of at least $\mu=1.5$ by][]{Dobler08}. Compared to their figure 5, the GAMA lens rate of 70\% confirmed \citep{Chan16} of 0.05\% of the survey is typical for $\mu=2$ magnifications. 

The verification of these lenses though magnified \hi{} emission is a different proposition however. \hi{} disks are much more extended than the star-forming counterpart \citep[cf.][]{de-Blok20,Wang21} and closer to uniform in surface brightness. While the Einstein radius may not shrink, the angular separation of source and lens will. The rate of unassociated emission lines becomes commensurately lower and the lensing effect greater. With a position and redshift to look for these lenses in an \hi{} cube, together with likely a factor of several magnification, makes these DEVILS targets more likely to be galaxy-galaxy \hi lenses.

\section{HST Imaging}
\label{s:hst}

For one of the fields, COSMOS (D10), there is uniform HST/ACS imaging from the COSMOS collaboration \citep{Scoville07b} and it is straightforward to obtain image cutouts\footnote{ \url{https://irsa.ipac.caltech.edu/data/COSMOS/index_cutouts.html}}. Figure \ref{f:ACS:cutouts} shows all the 56 blended spectra for which a cutout is available. These cutouts are not available for the other two DEVILS fields.
The majority appear to be chance alignments in the aperture of the AAT spectrograph, i.e. a genuine occulting galaxy pair. 

Despite the high resolution of these images, they are not ideal to visually identify strong lenses. The COSMOS field was observed in the F814W filter (broad Johnson I) and this is good to map the stellar mass of higher redshift galaxies, a blue filter (F606W of F555W, i.e. Johnson V) is typically used to identify the source galaxy in a strong gravitational lens. 

Given the limited information in a single filter, shallow HST image, we limit our visual classification to Spiral-Elliptical, Spiral-Spiral or Elliptical-Elliptical superpositions or ``lens candidate'' (elliptical with possible circular structure), same as in \cite{Holwerda15}. This is a visual estimate of which galaxy looks in front as well the broad morphological type. In \cite{Holwerda15}, we found that the visual classification (based on SDSS imaging) and the template matches were relatively poor, and deeper and higher resolution imaging confirmed strong lens candidates \citep{Chan16}. We note which systems may useful for a future study of dust transparency using the occulting galaxy method \citep[cf][]{Holwerda16,Holwerda16a}; a partial overlap of two galaxies with good apparent symmetry. 

The total of blended spectra that are marked as potential classical lenses (the lensing galaxy is a foreground elliptical) after visual inspection is 9 ($\sim$0.1\% of DEVILS DR1 on D10, Figure \ref{f:ACS:lenscandidates}) which is more in line with expectations from previous searches for lenses and overlapping galaxies in spectroscopic surveys \citep{Holwerda07c,Holwerda15}. One should consider these confirmed actual lenses if multicolor HST imaging, high-resolution AO-enabled IFU, or magnified \hi{} can verify their nature.

\begin{figure*}
    \centering
\includegraphics[width=0.2\textwidth]{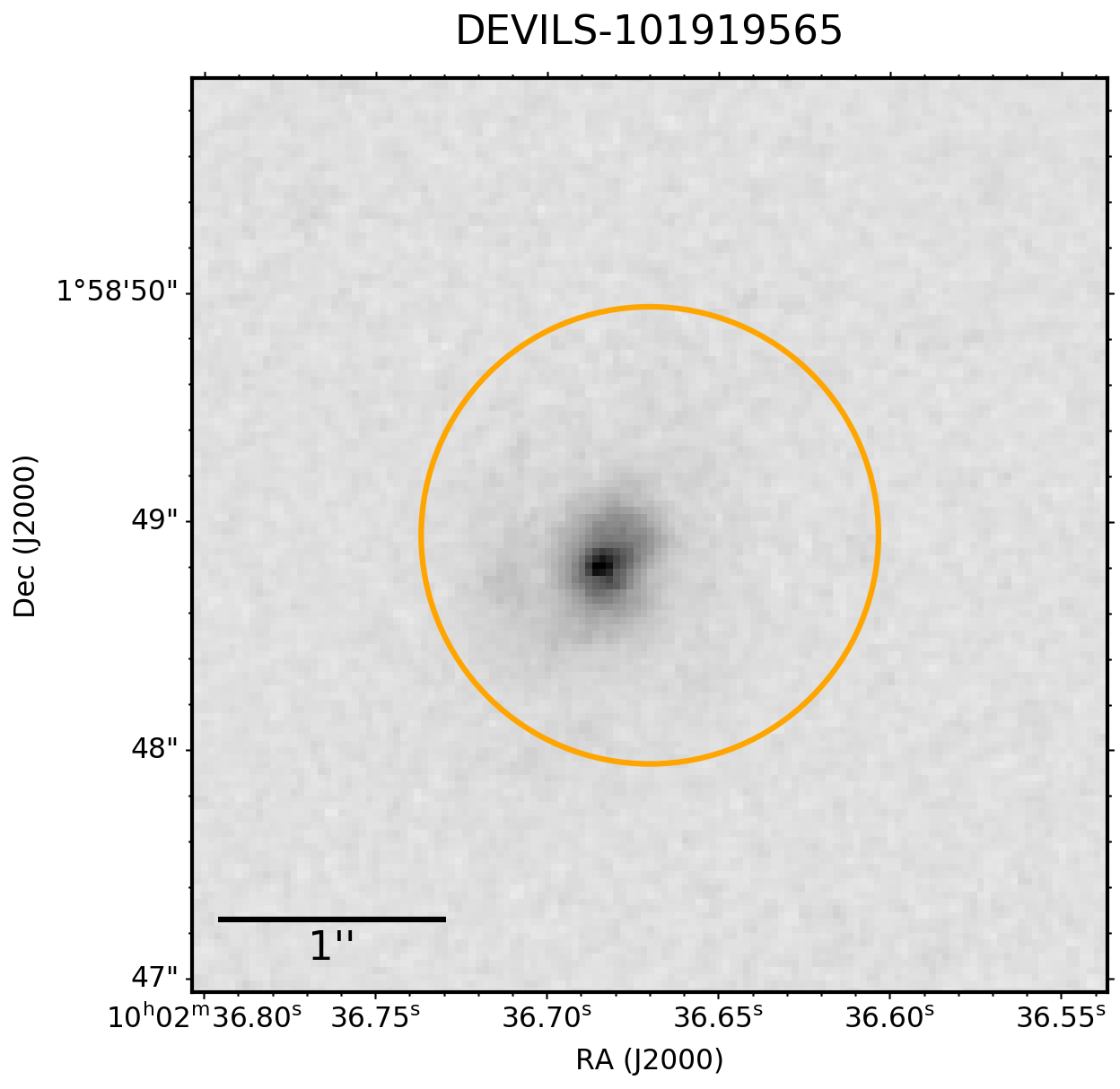}
\includegraphics[width=0.2\textwidth]{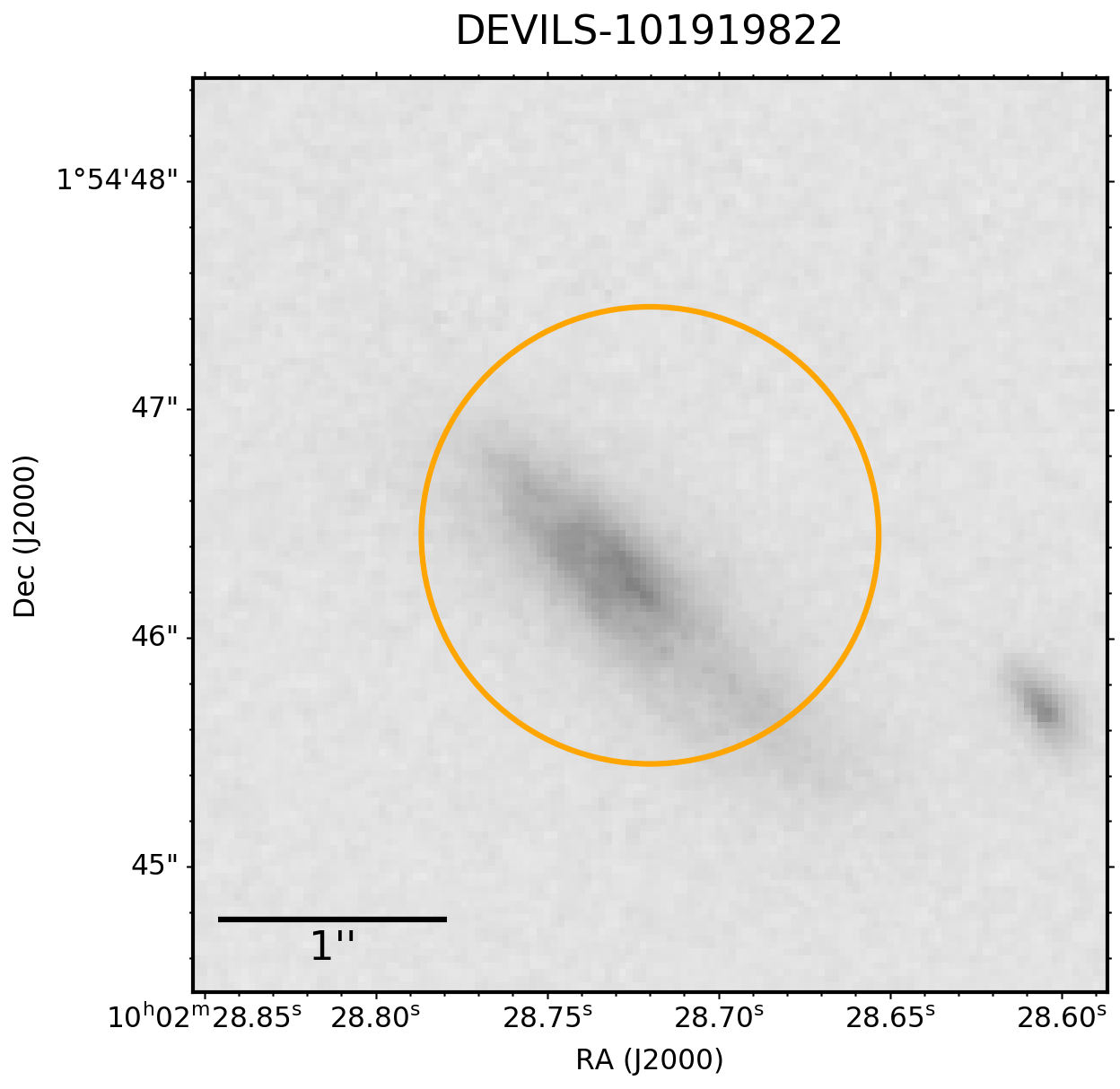}
\includegraphics[width=0.2\textwidth]{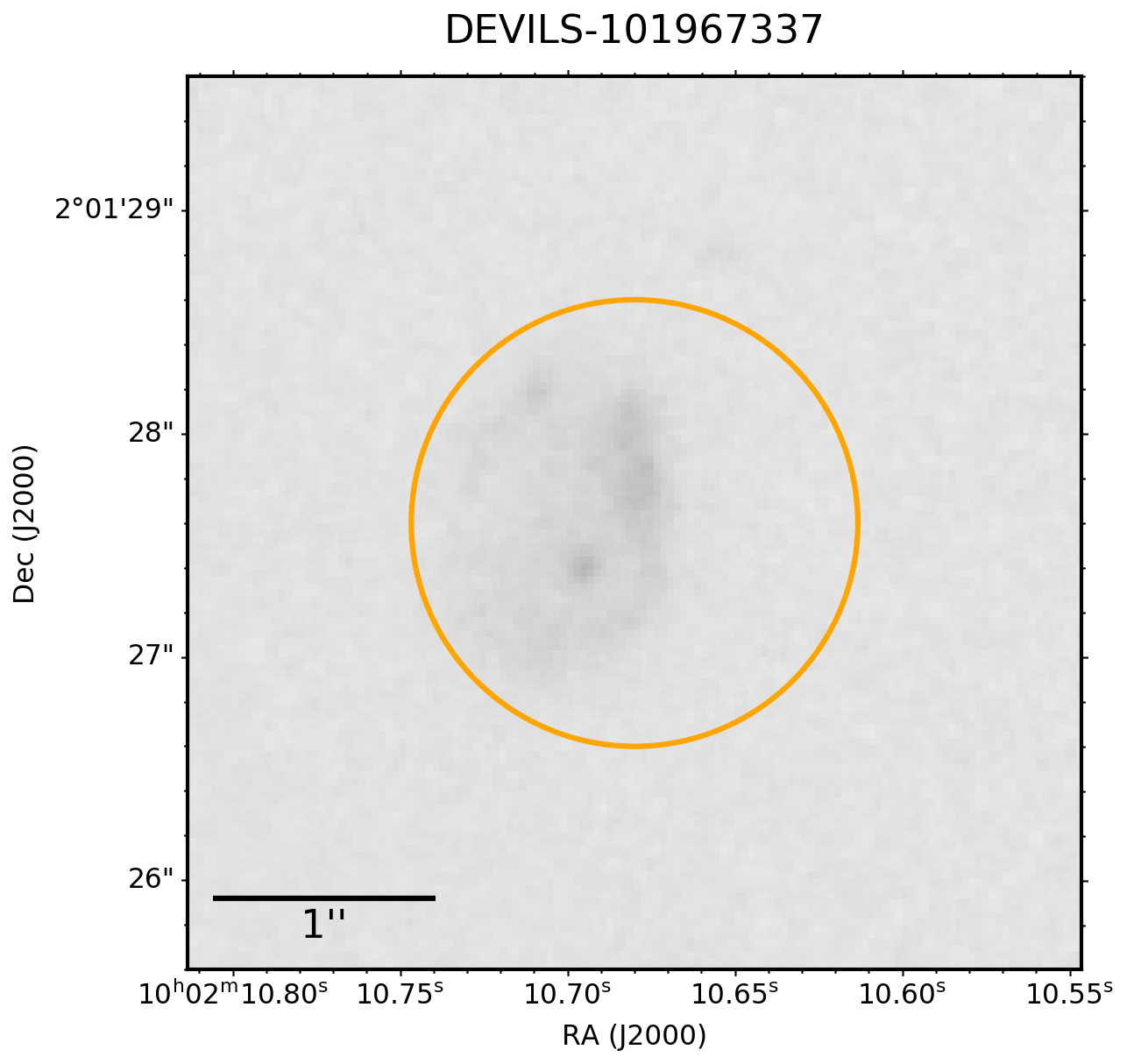}
\includegraphics[width=0.2\textwidth]{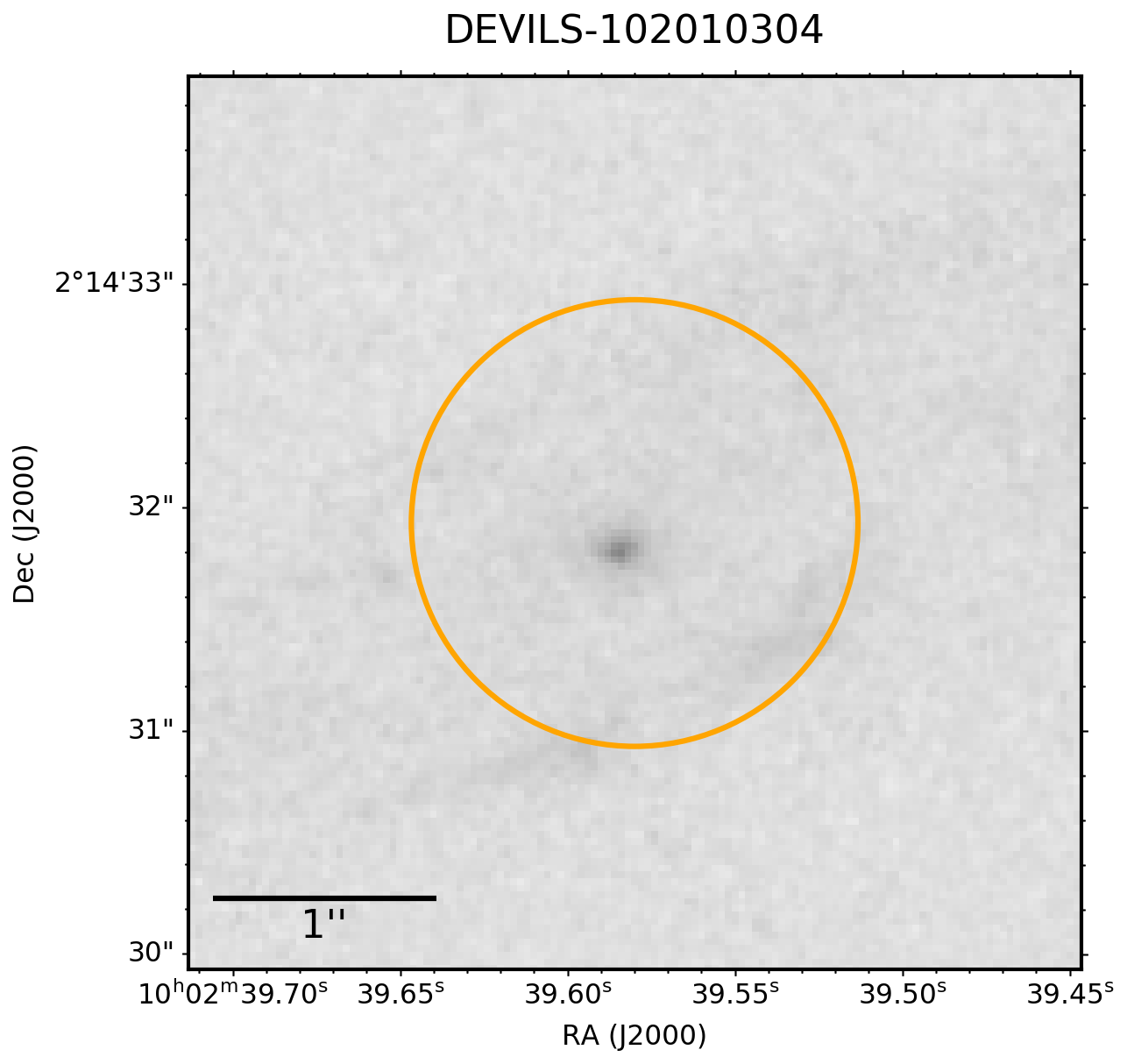}
\includegraphics[width=0.2\textwidth]{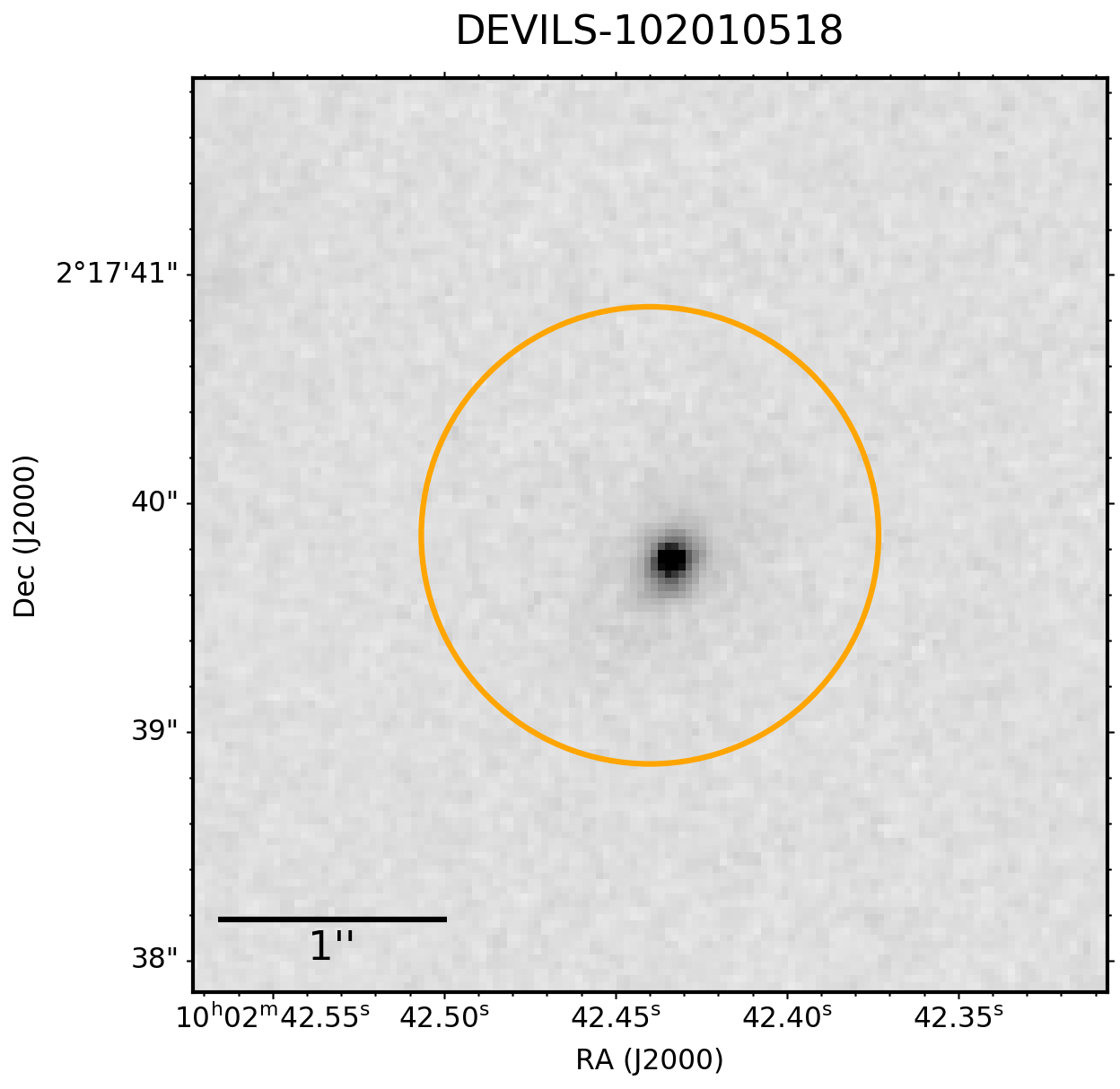}
\includegraphics[width=0.2\textwidth]{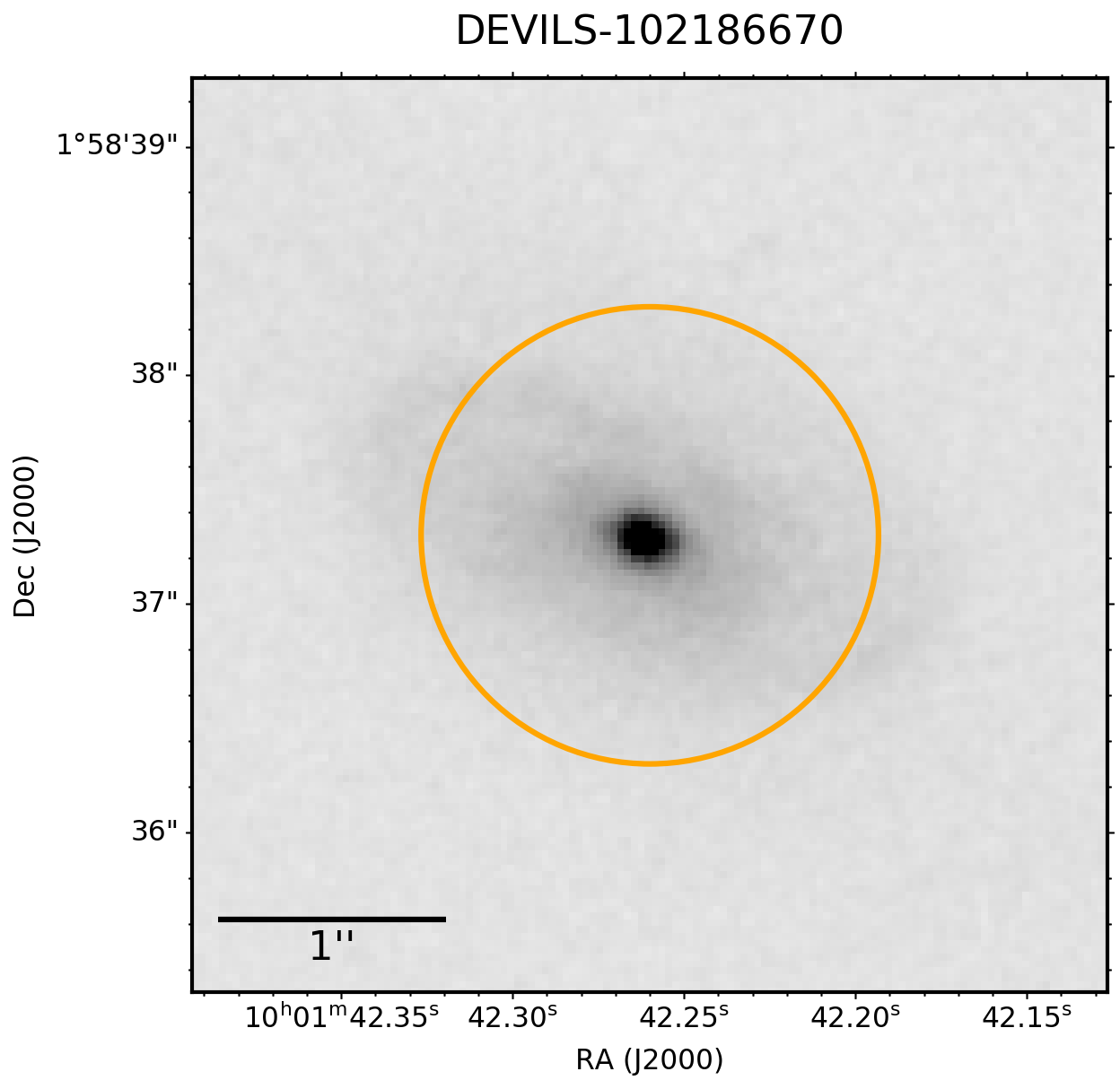}
\includegraphics[width=0.2\textwidth]{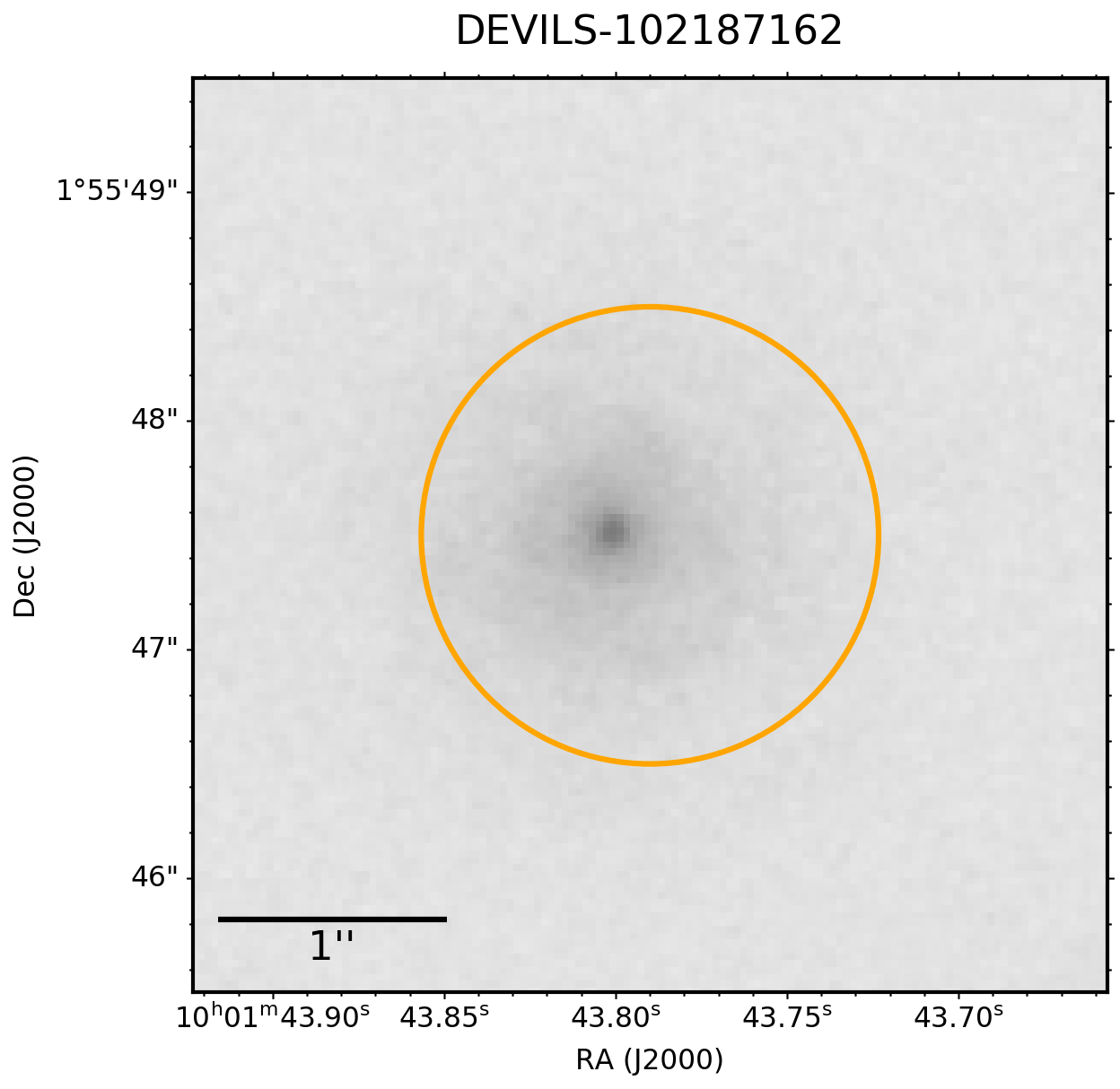}
\includegraphics[width=0.2\textwidth]{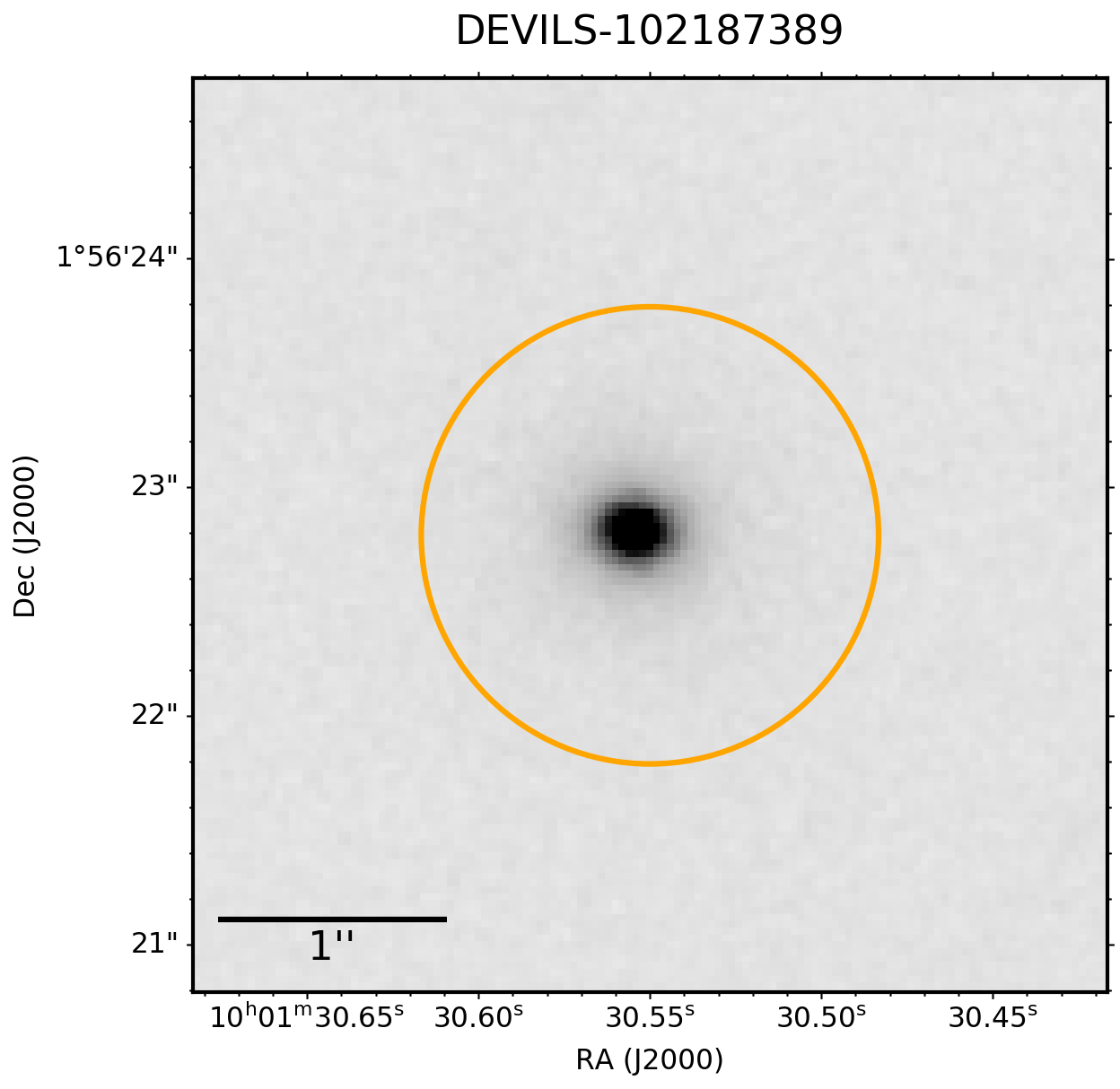}
\includegraphics[width=0.2\textwidth]{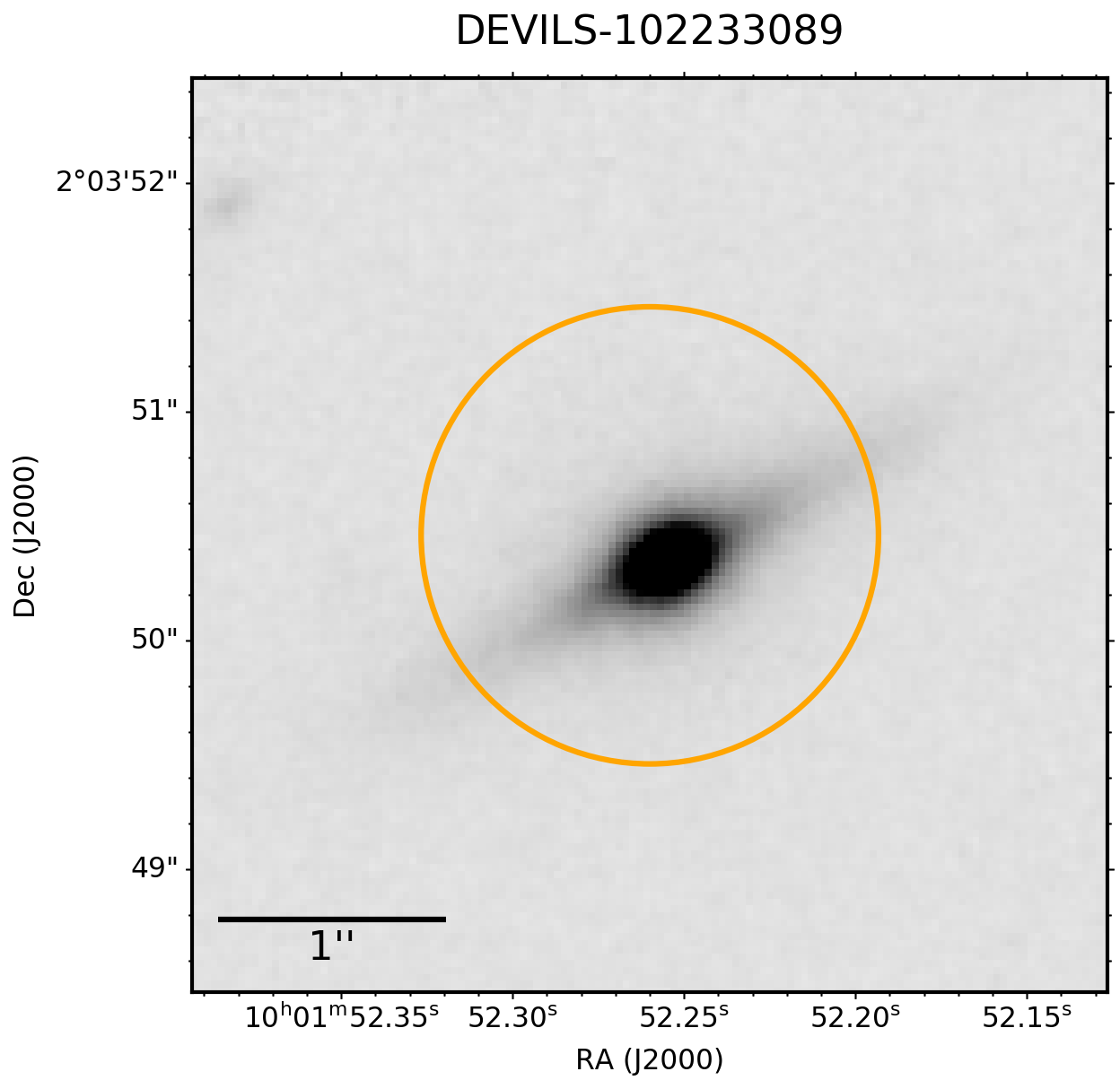}
\includegraphics[width=0.2\textwidth]{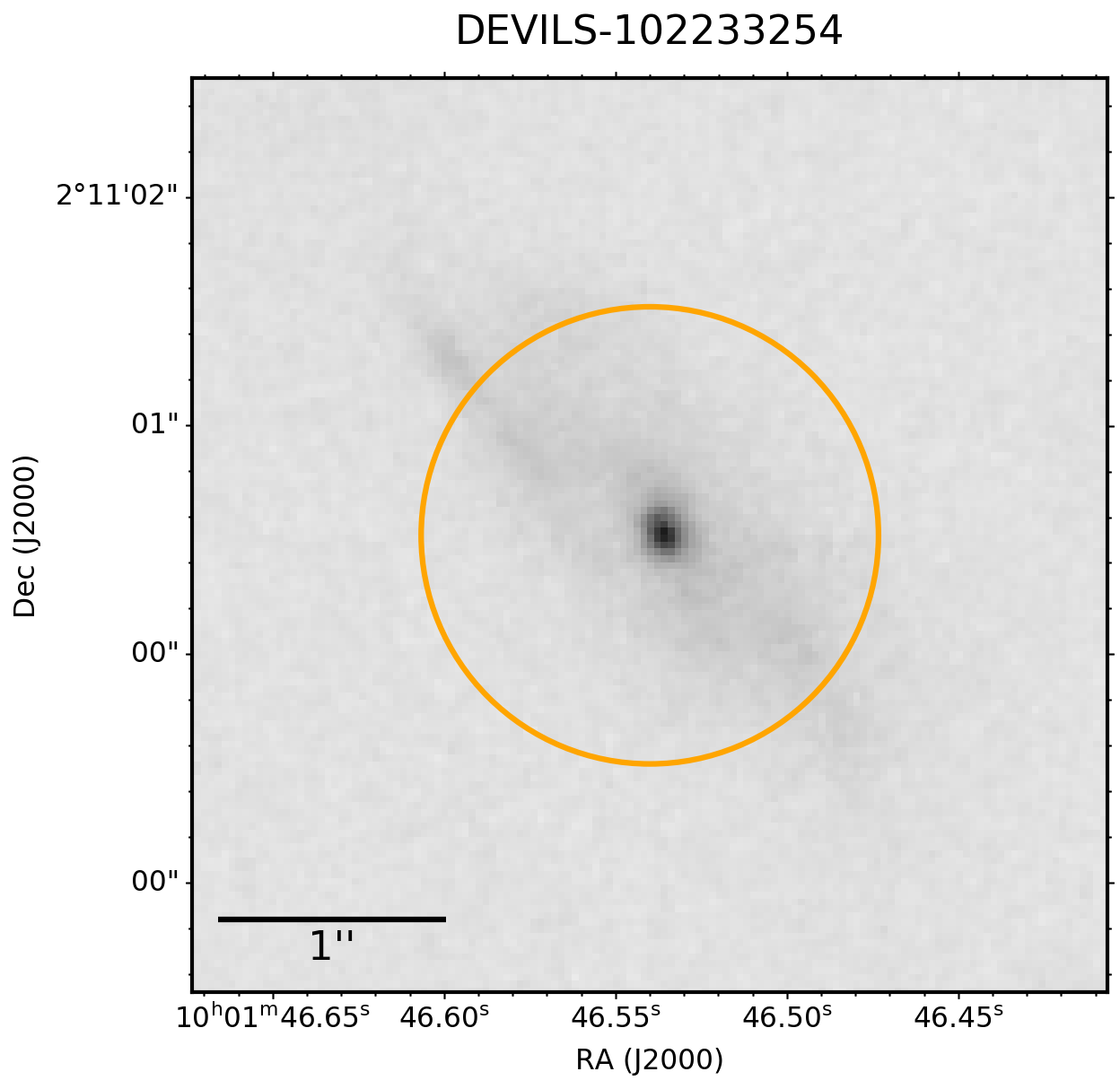}
\includegraphics[width=0.2\textwidth]{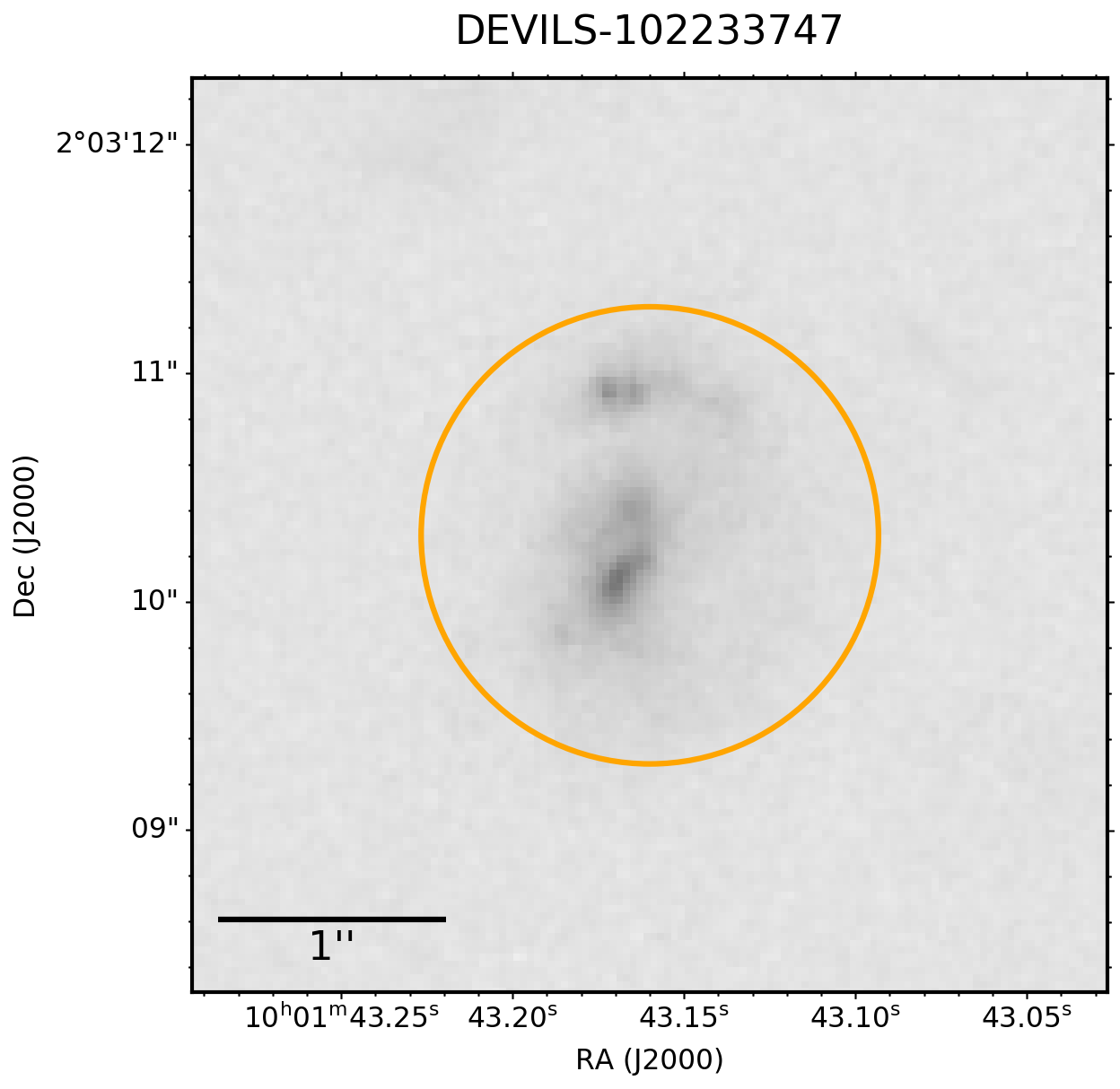}
\includegraphics[width=0.2\textwidth]{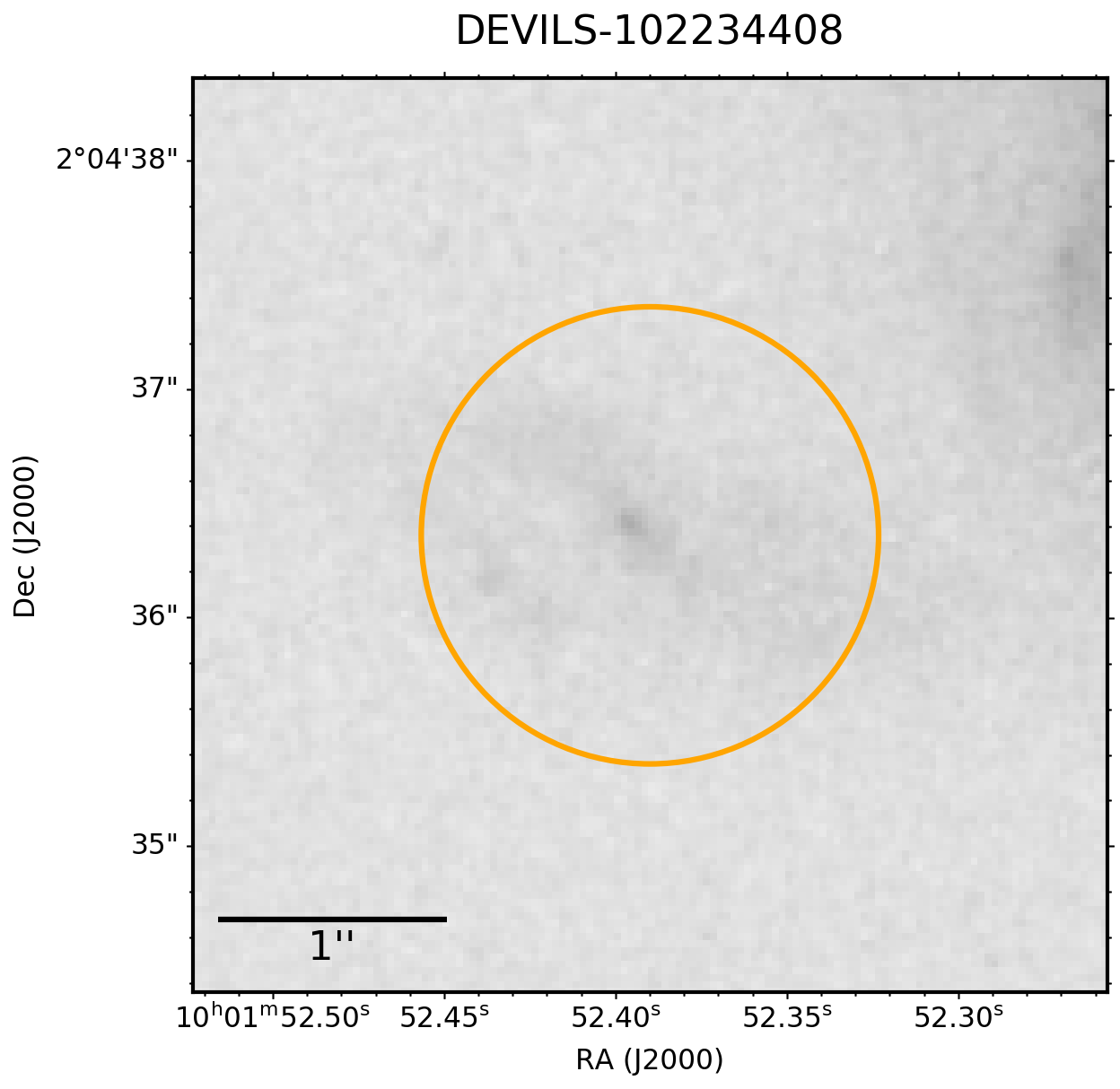}
\includegraphics[width=0.2\textwidth]{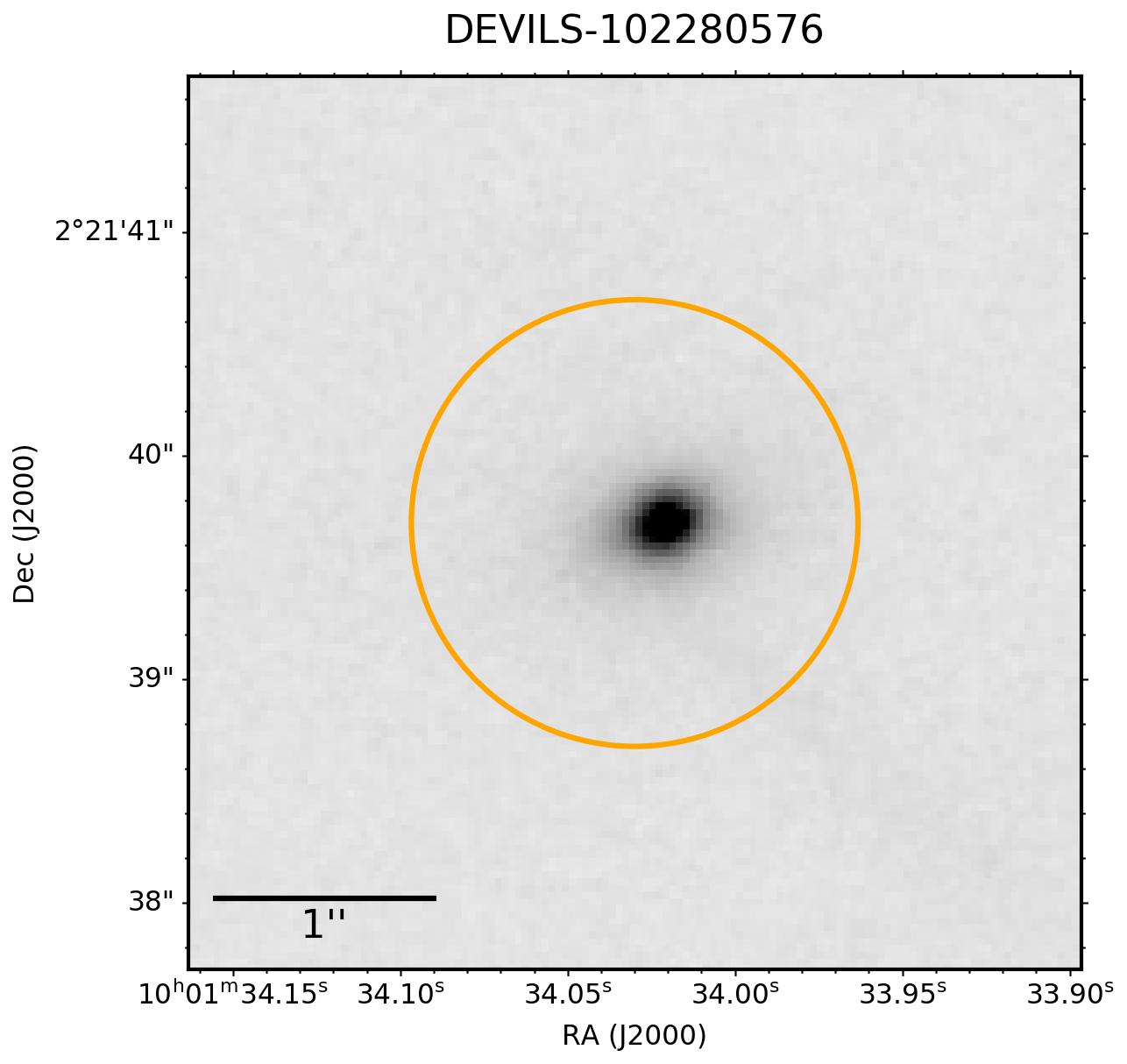}
\includegraphics[width=0.2\textwidth]{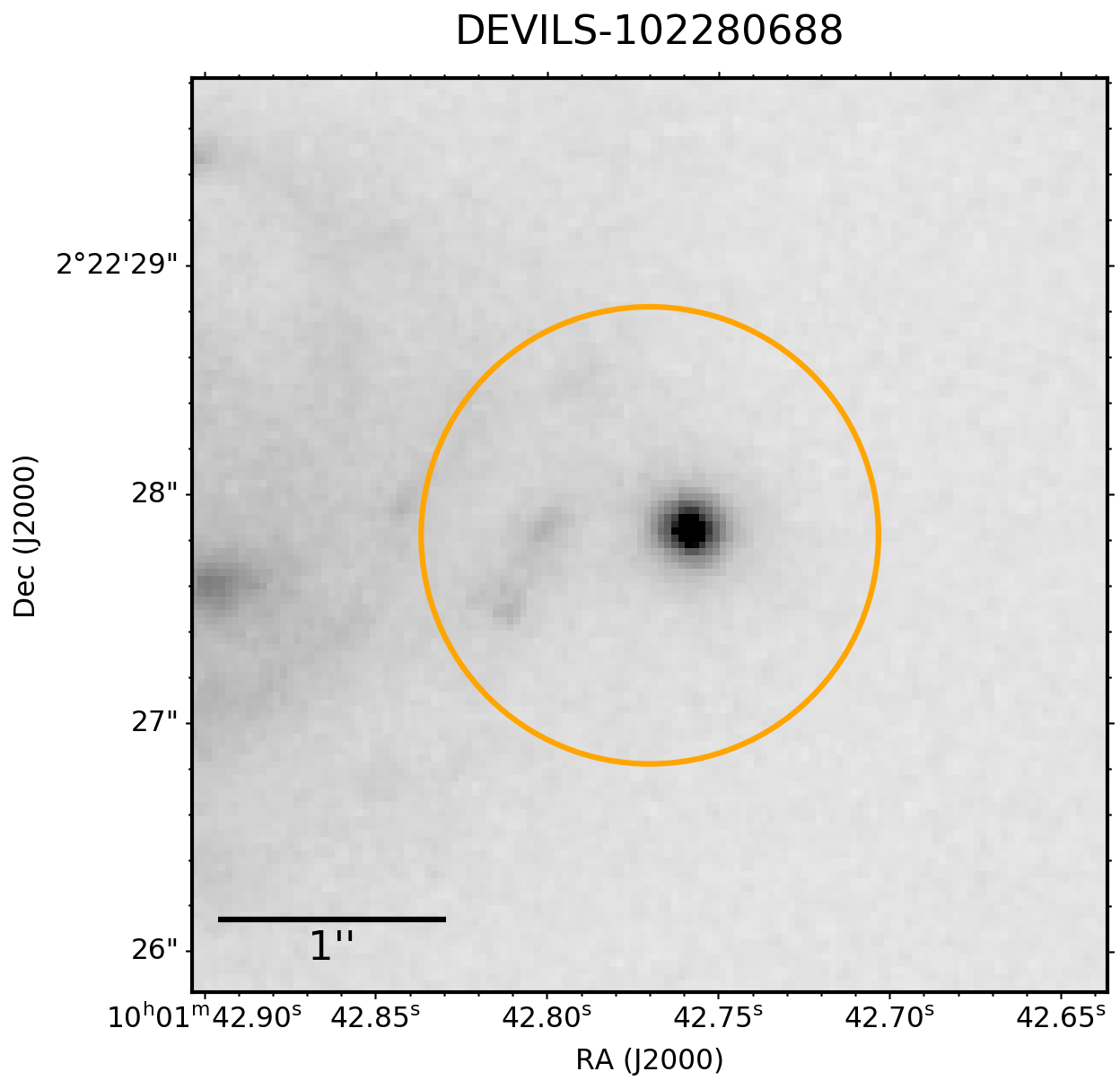}
\includegraphics[width=0.2\textwidth]{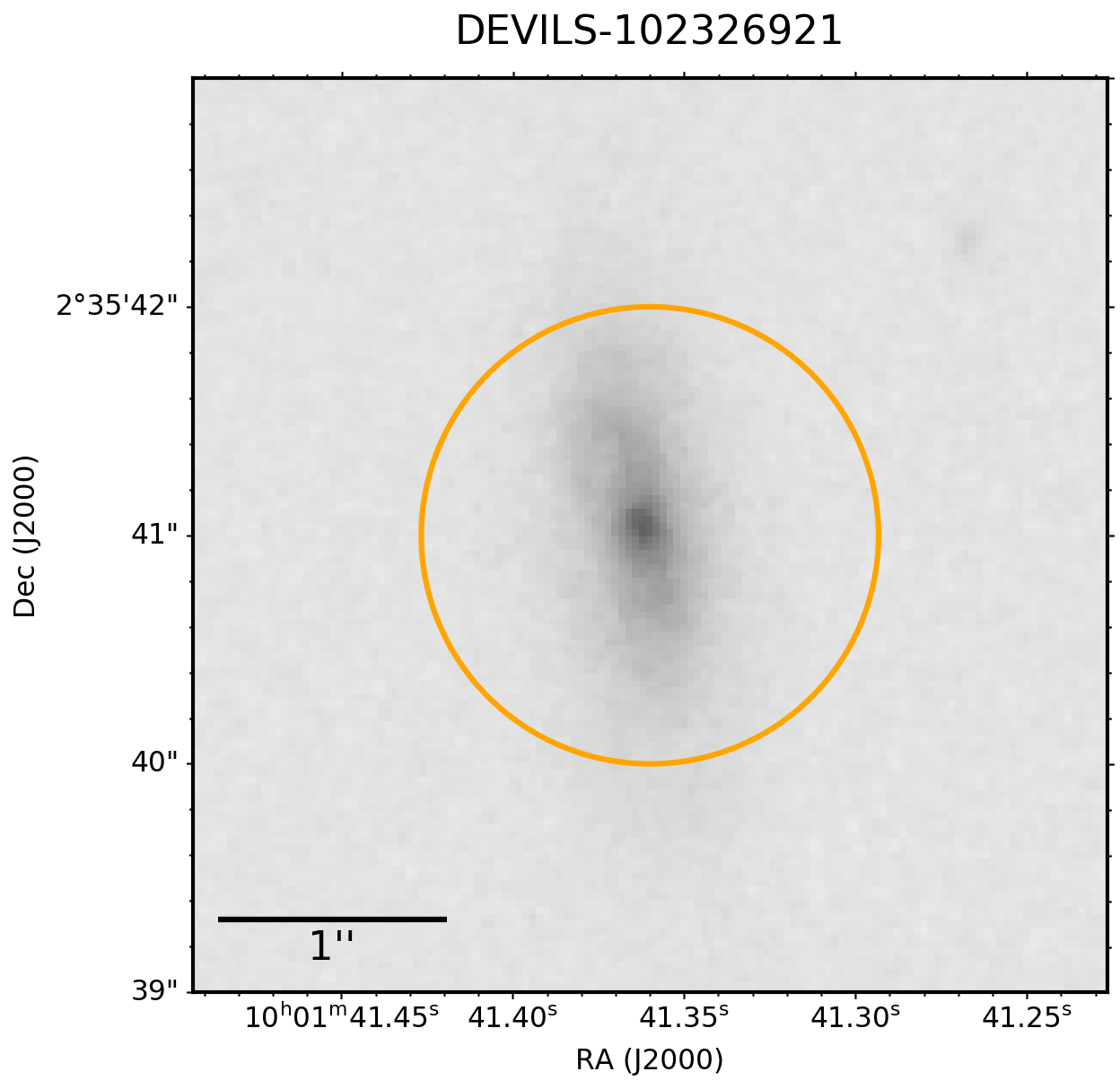}
\includegraphics[width=0.2\textwidth]{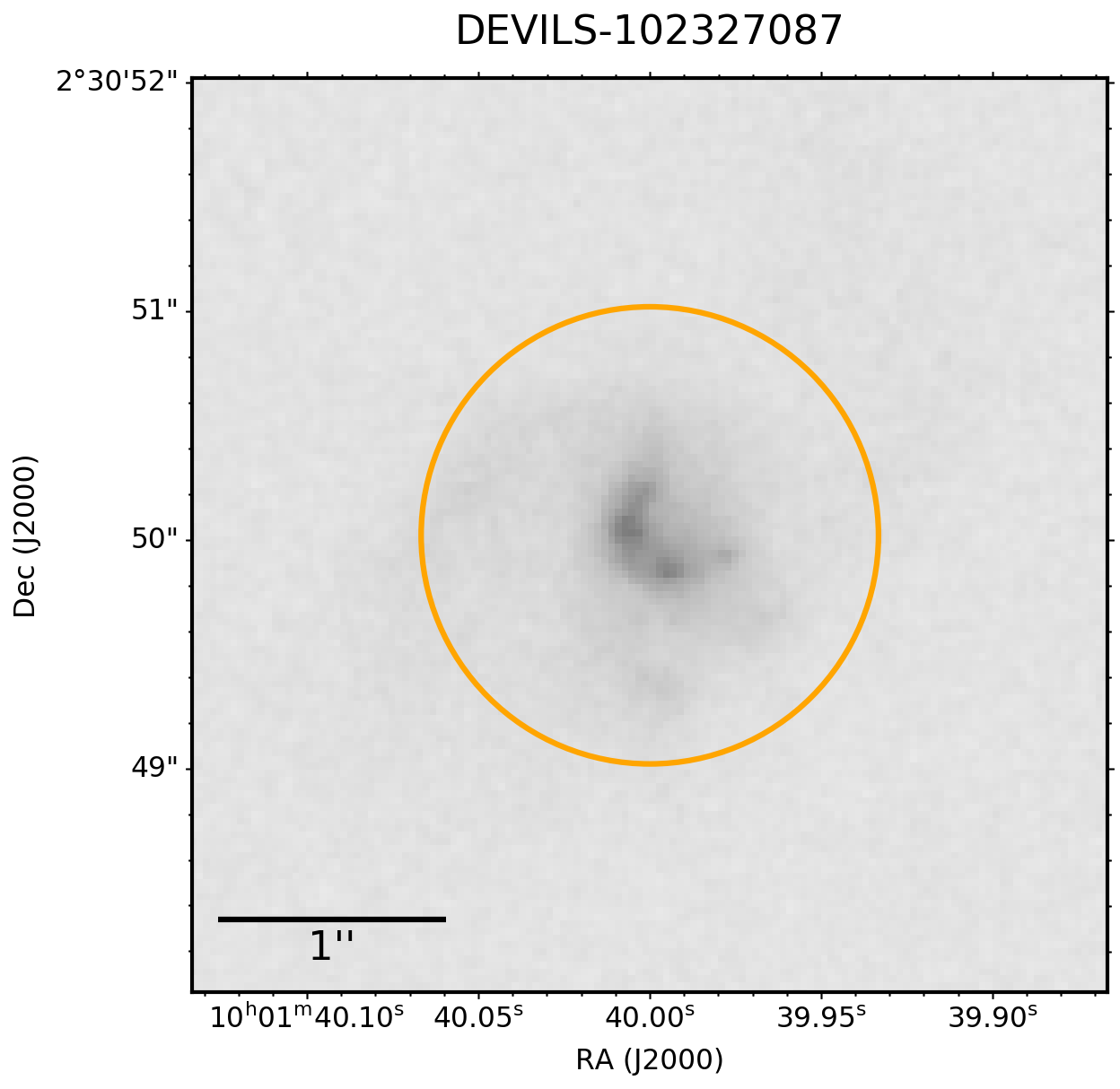}
\includegraphics[width=0.2\textwidth]{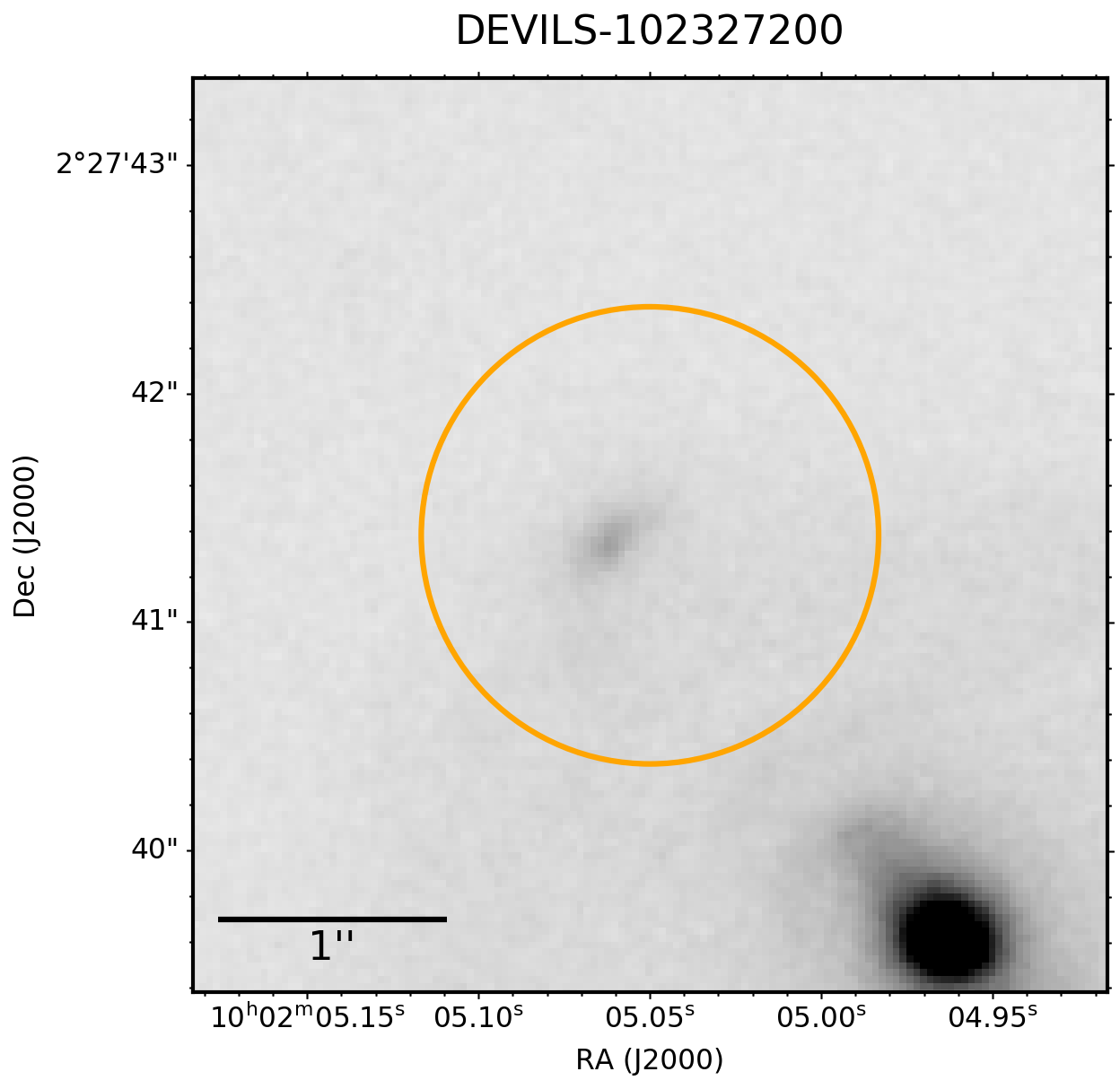}
\includegraphics[width=0.2\textwidth]{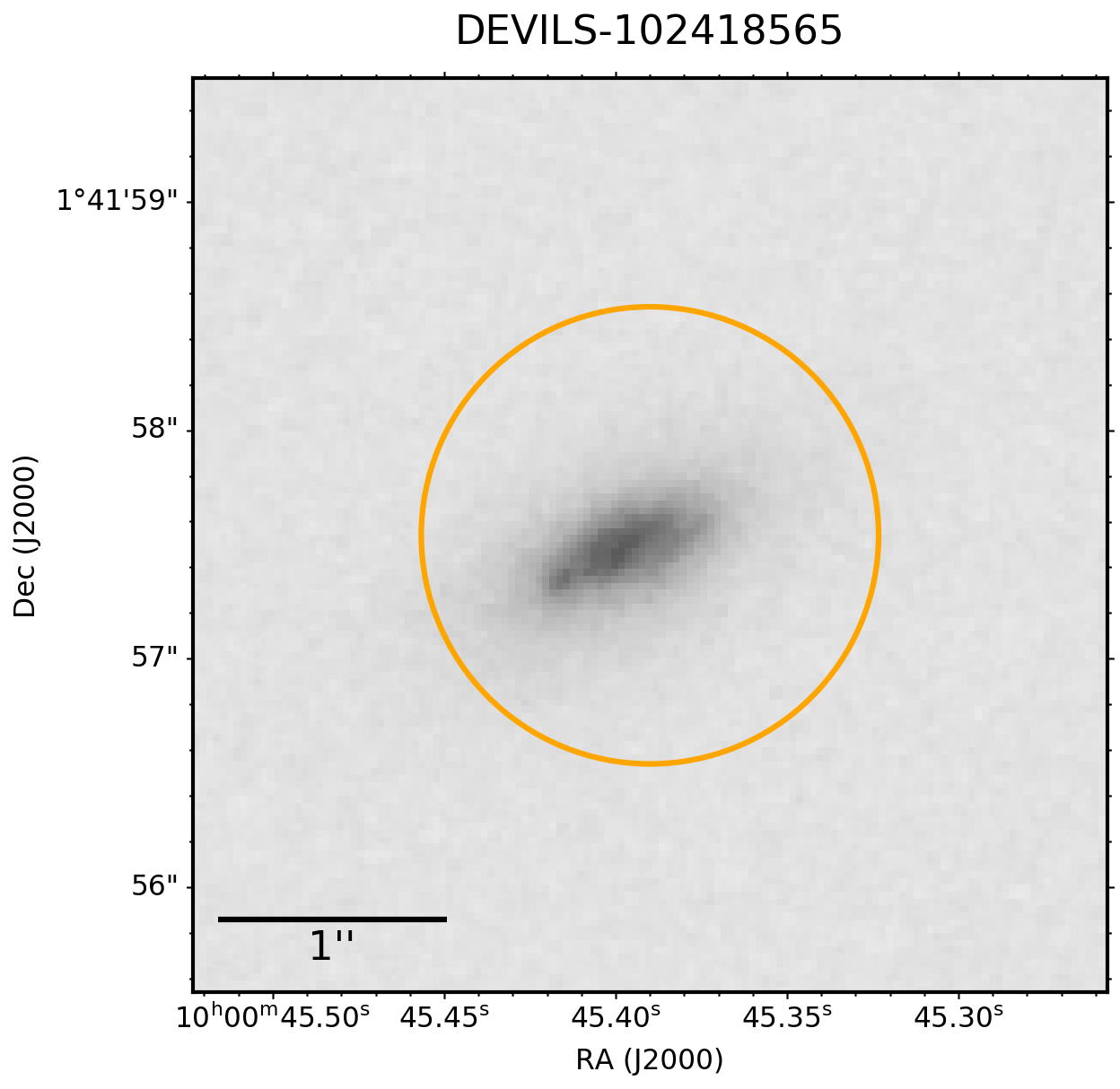}
\includegraphics[width=0.2\textwidth]{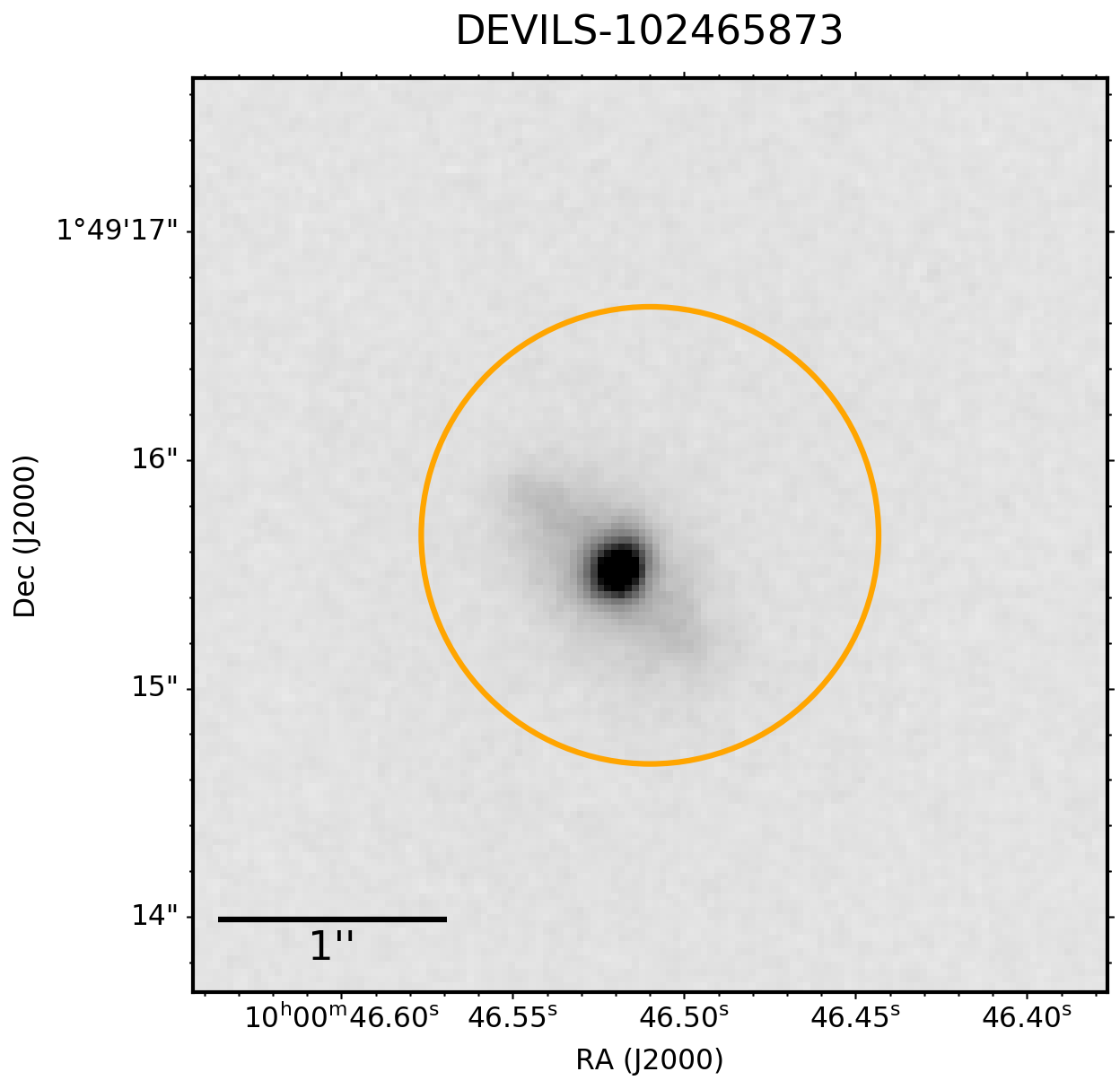}
\includegraphics[width=0.2\textwidth]{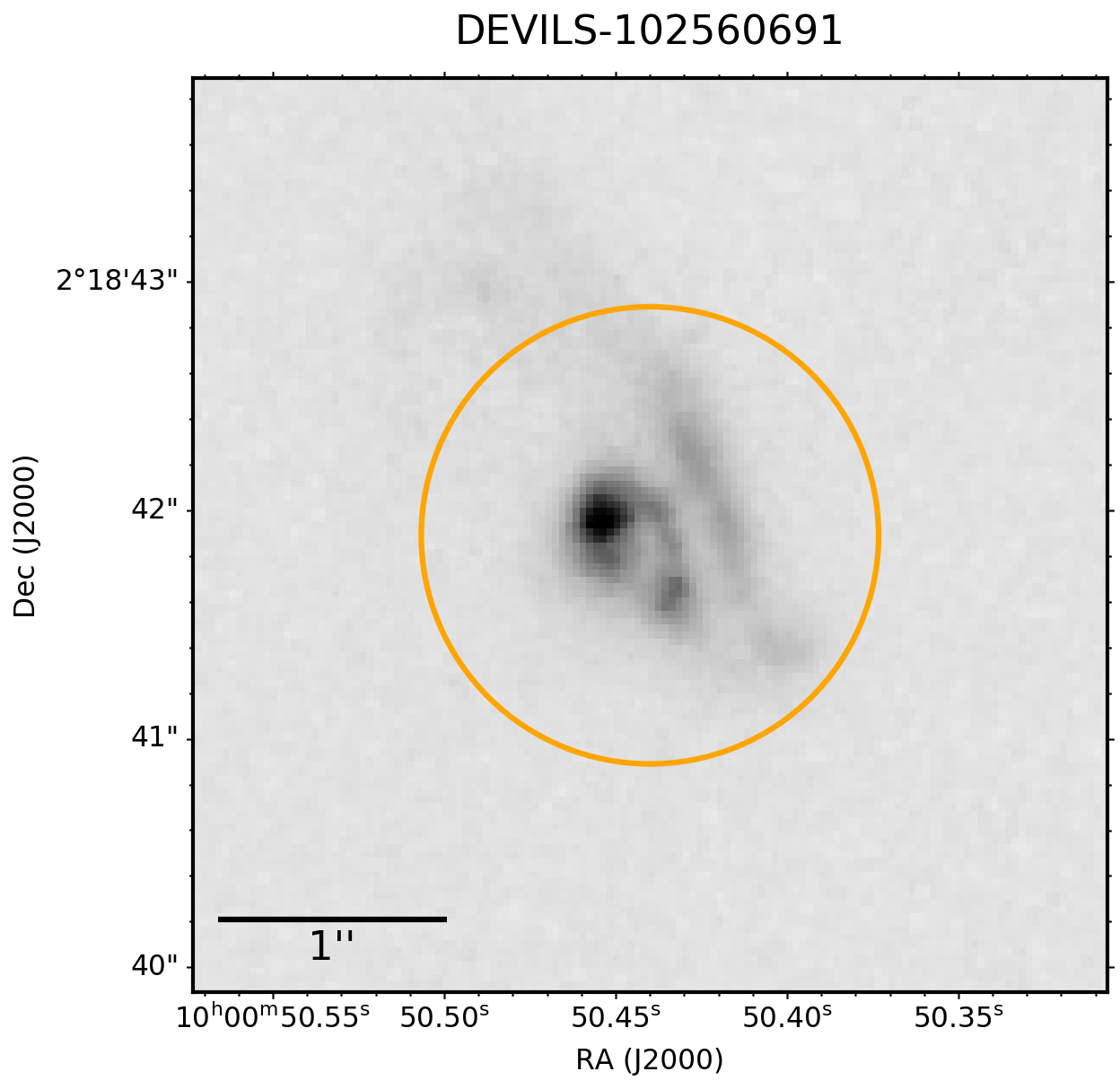}
\includegraphics[width=0.2\textwidth]{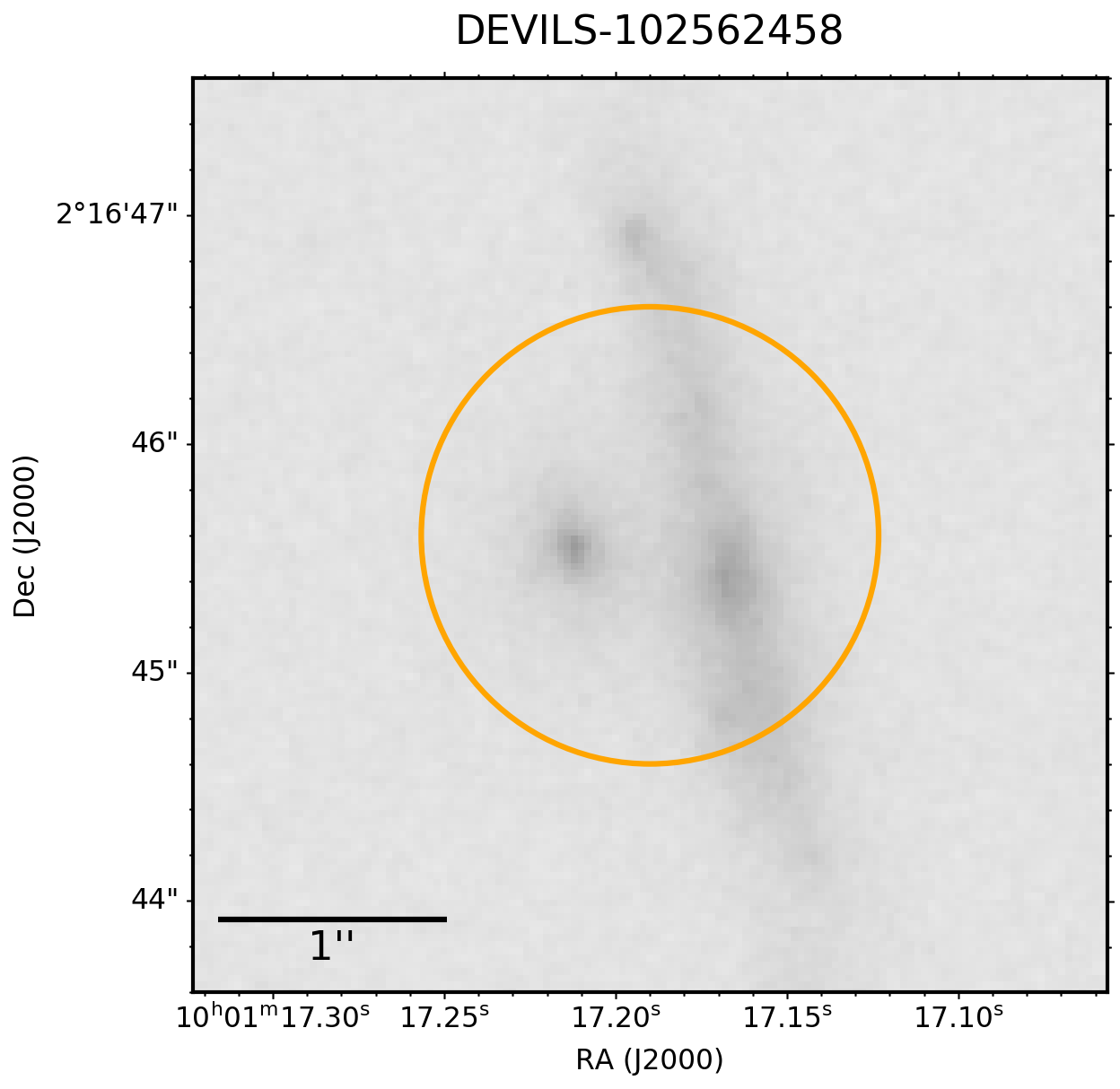}
\includegraphics[width=0.2\textwidth]{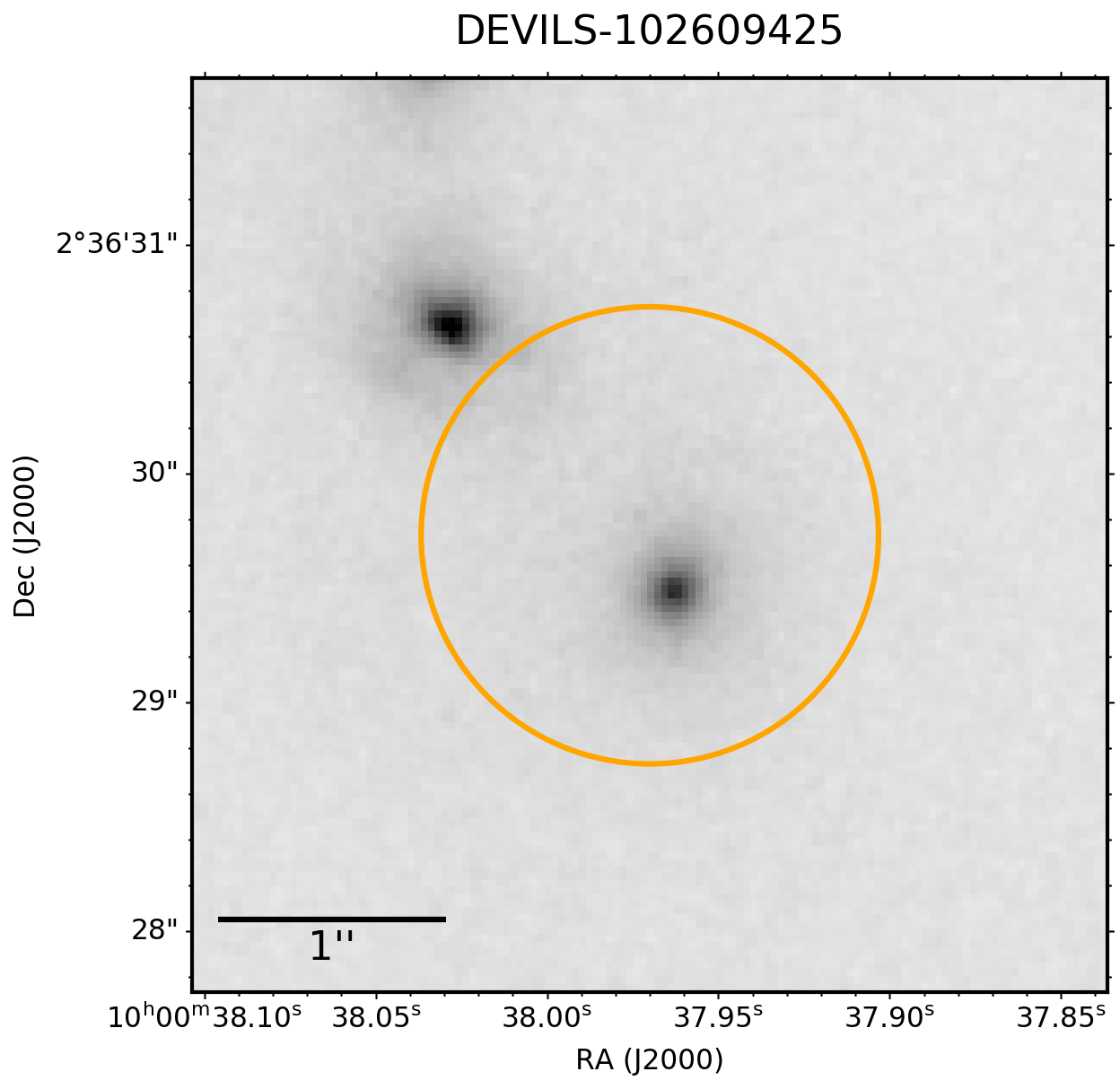}
\includegraphics[width=0.2\textwidth]{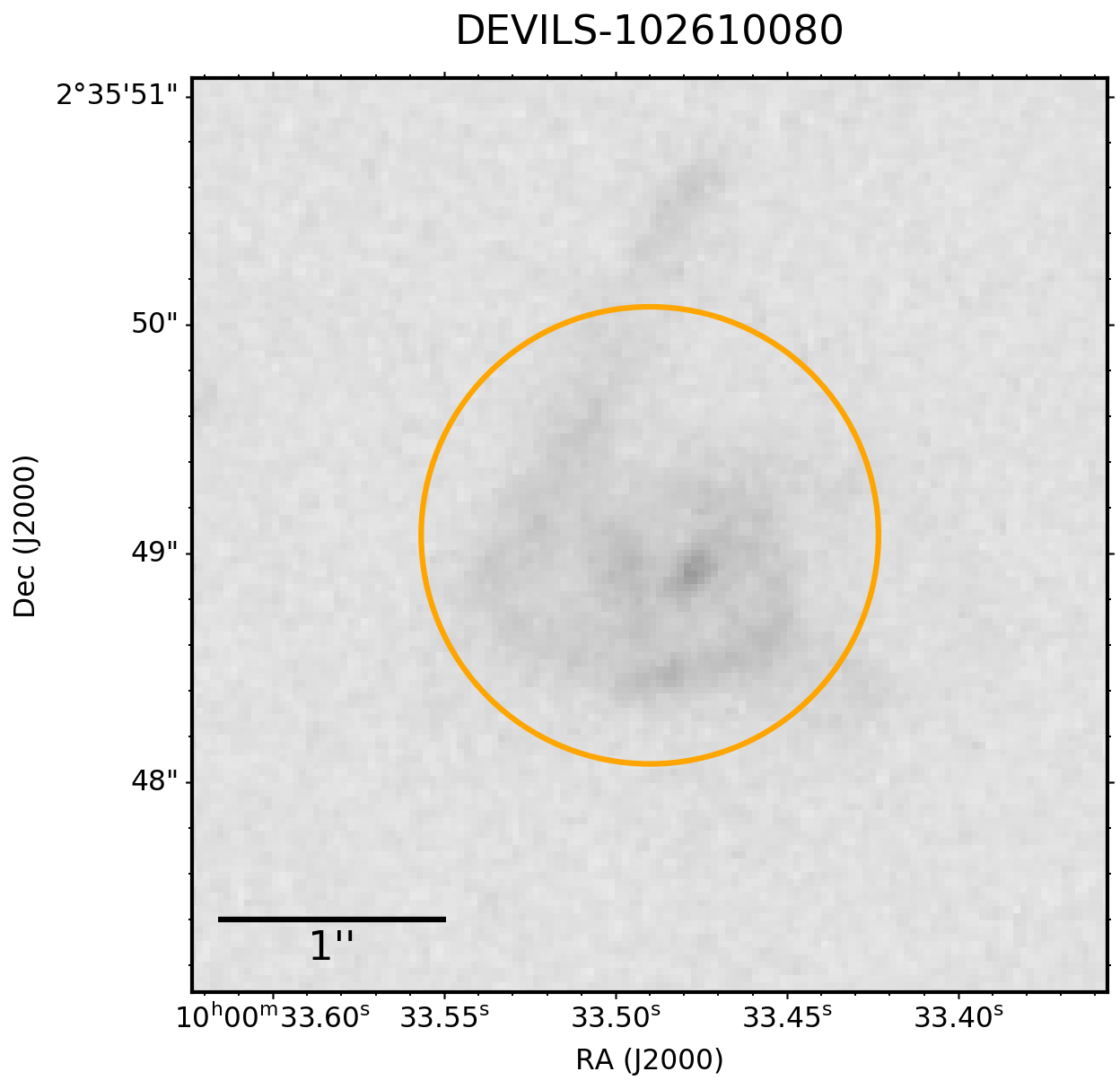}
\includegraphics[width=0.2\textwidth]{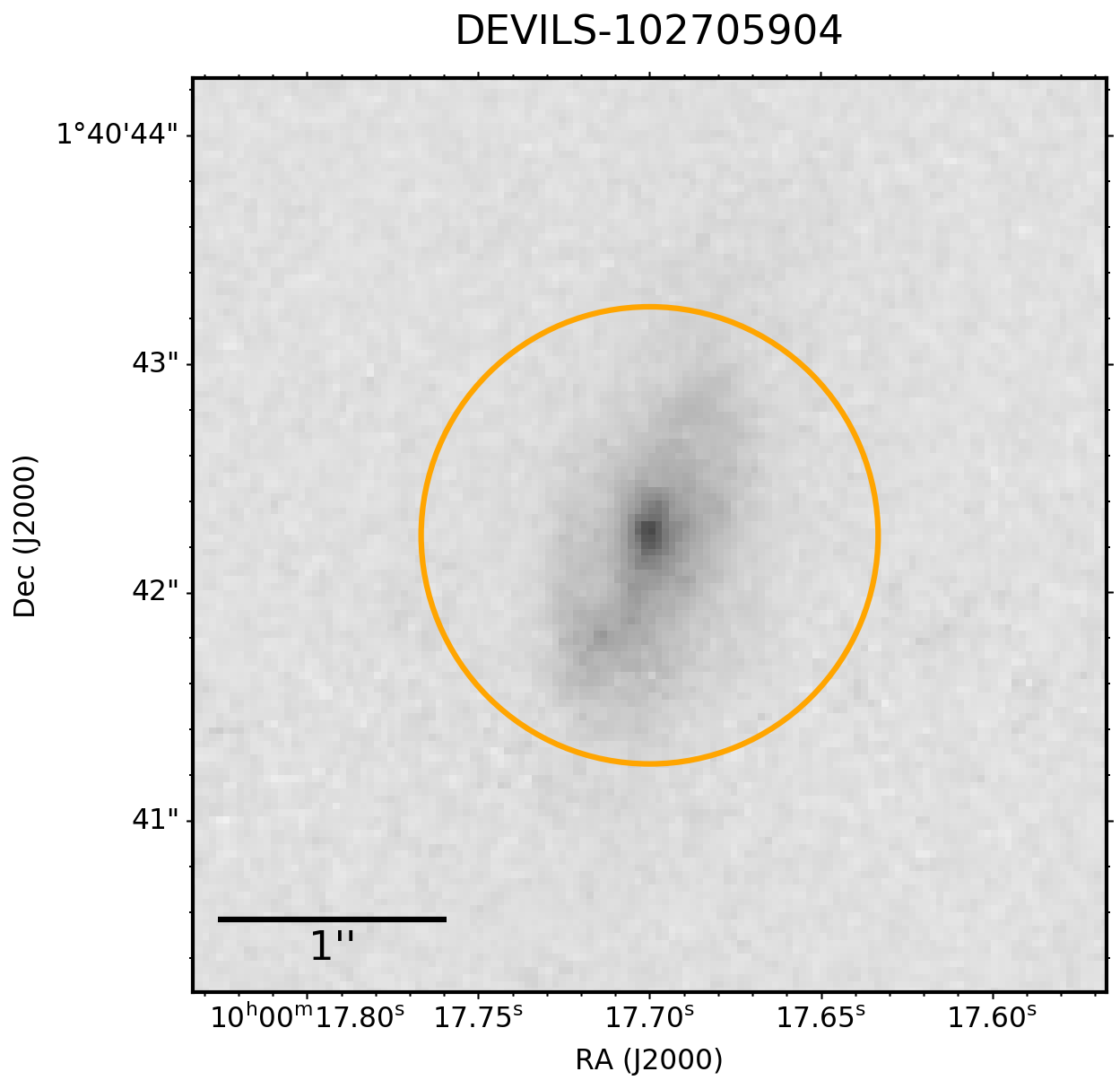}
\includegraphics[width=0.2\textwidth]{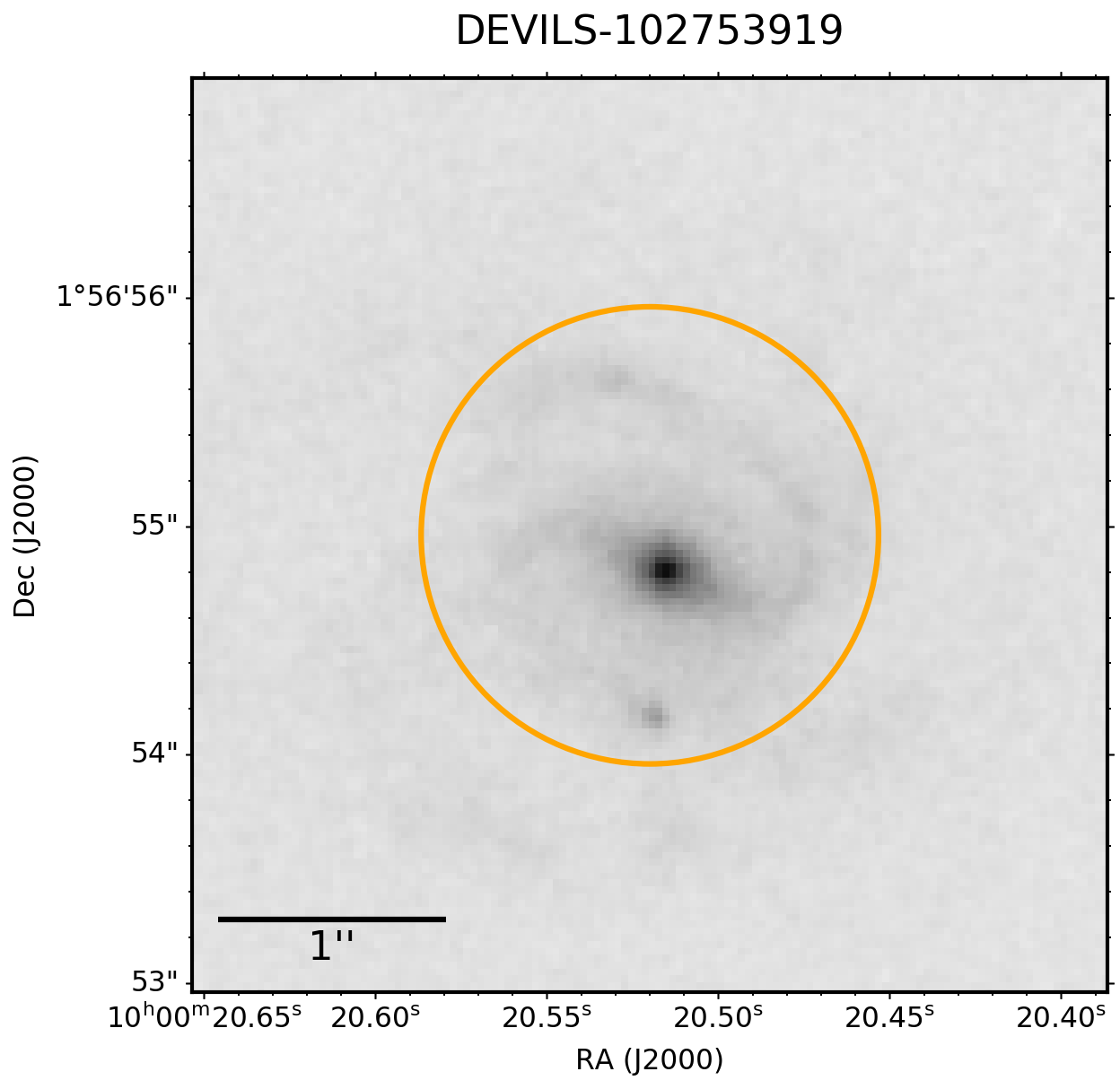}
\includegraphics[width=0.2\textwidth]{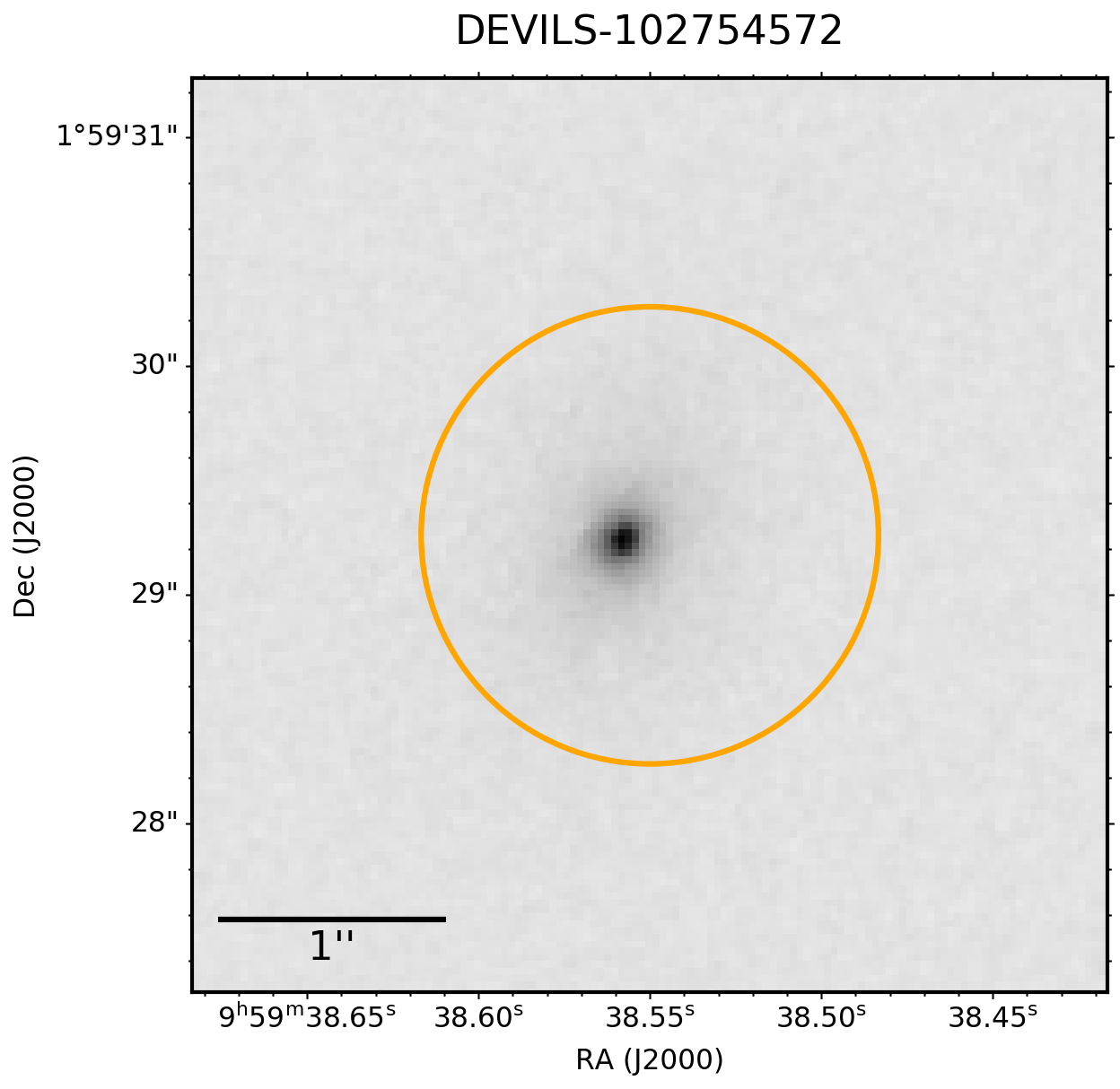}
\includegraphics[width=0.2\textwidth]{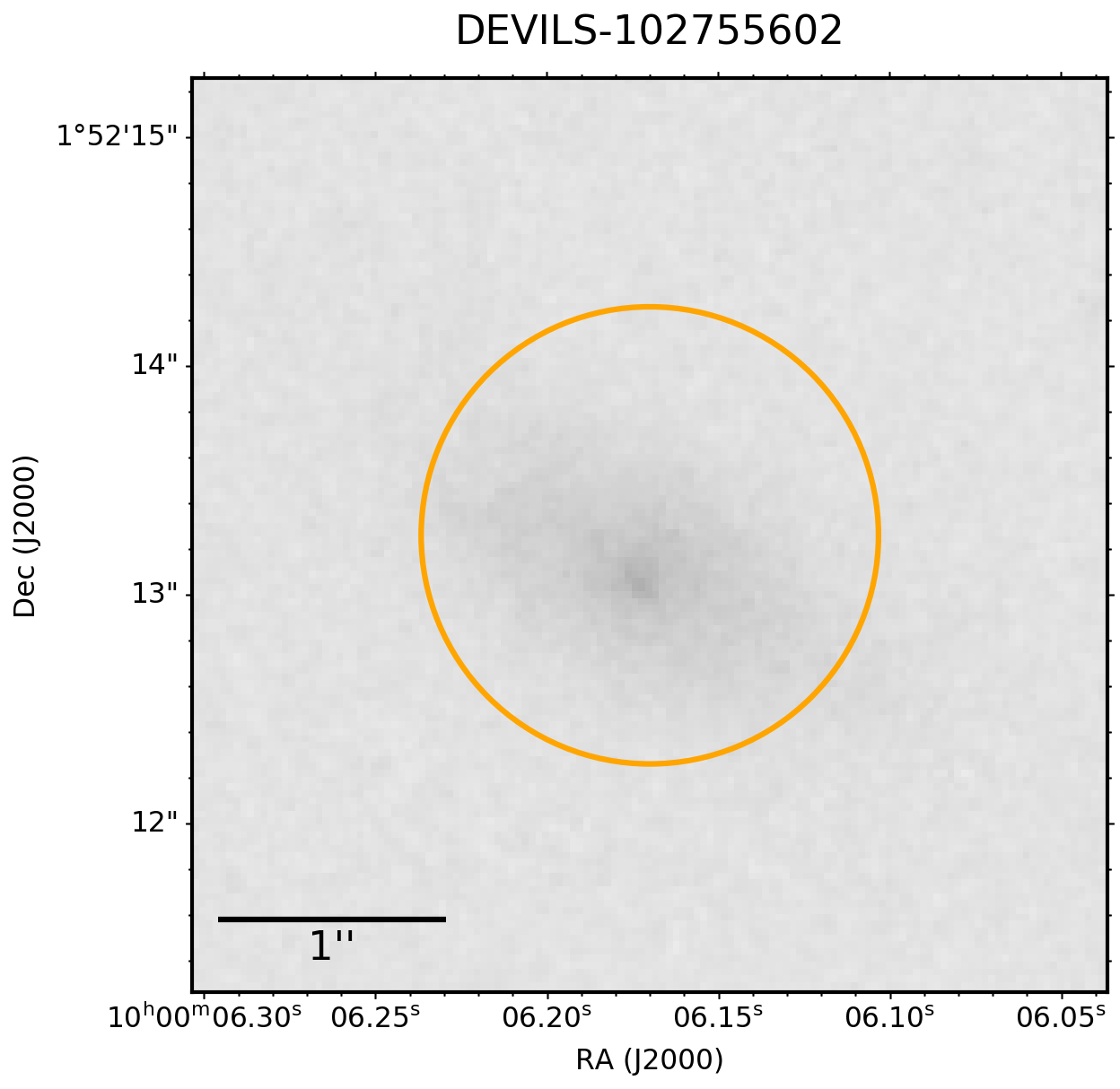}
\includegraphics[width=0.2\textwidth]{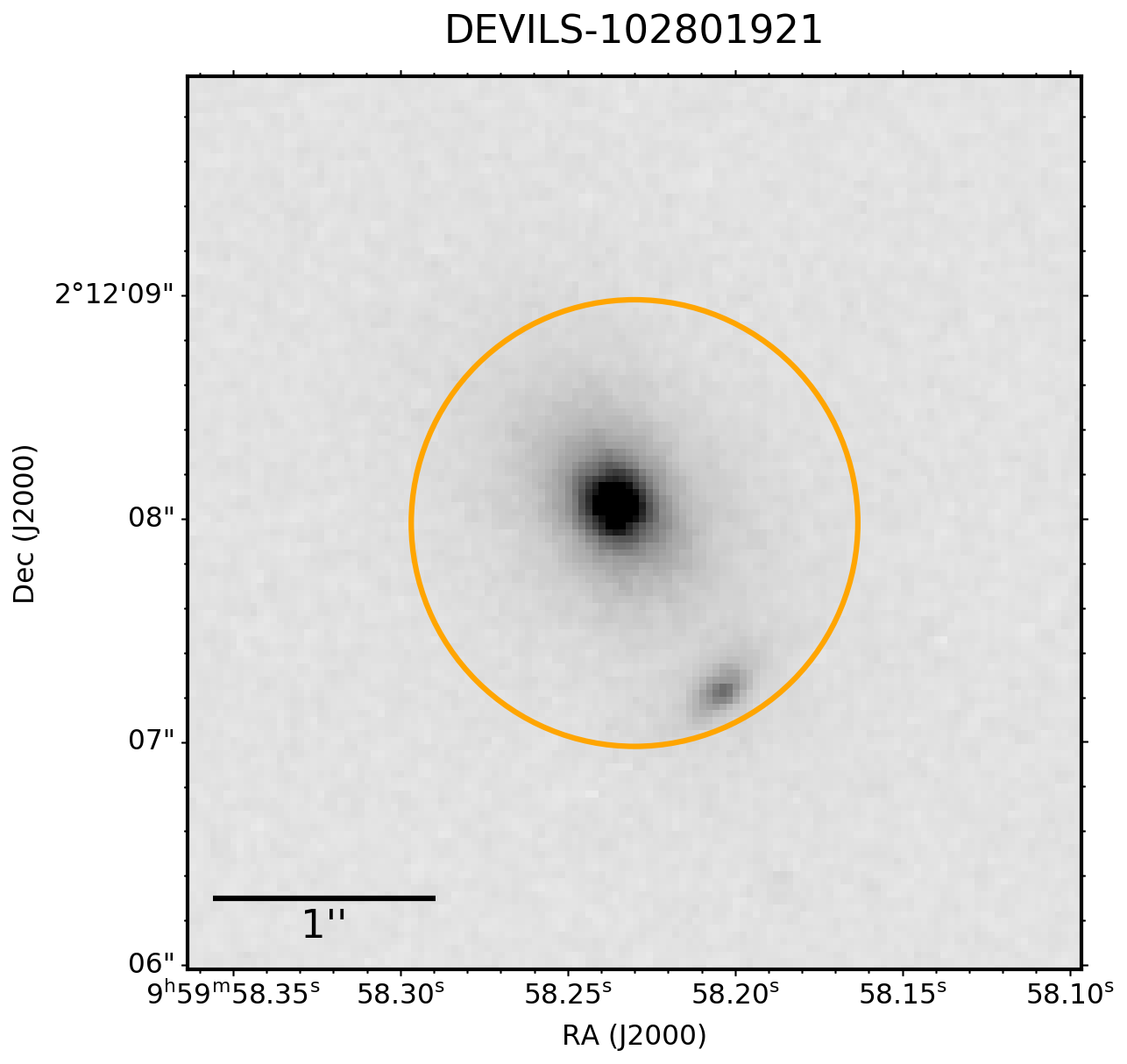}

    \caption{The cutouts from the HST/ACS COSMOS mosaic of all the blended spectra identified in DEVILS DR1. Pairs of galaxies along the line of sight are the most common blended spectra with passive elliptical systems that may be lensing a more distant source in the minority. Position of the DEVILS spectroscopic aperture marked with an orange circle. Their spectra are shown are Figure \ref{f:spectra:appendix} in the Appendix.}
    \label{f:ACS:cutouts}
\end{figure*}

\setcounter{figure}{7}

\begin{figure*}
    \centering
    \includegraphics[width=0.2\textwidth]{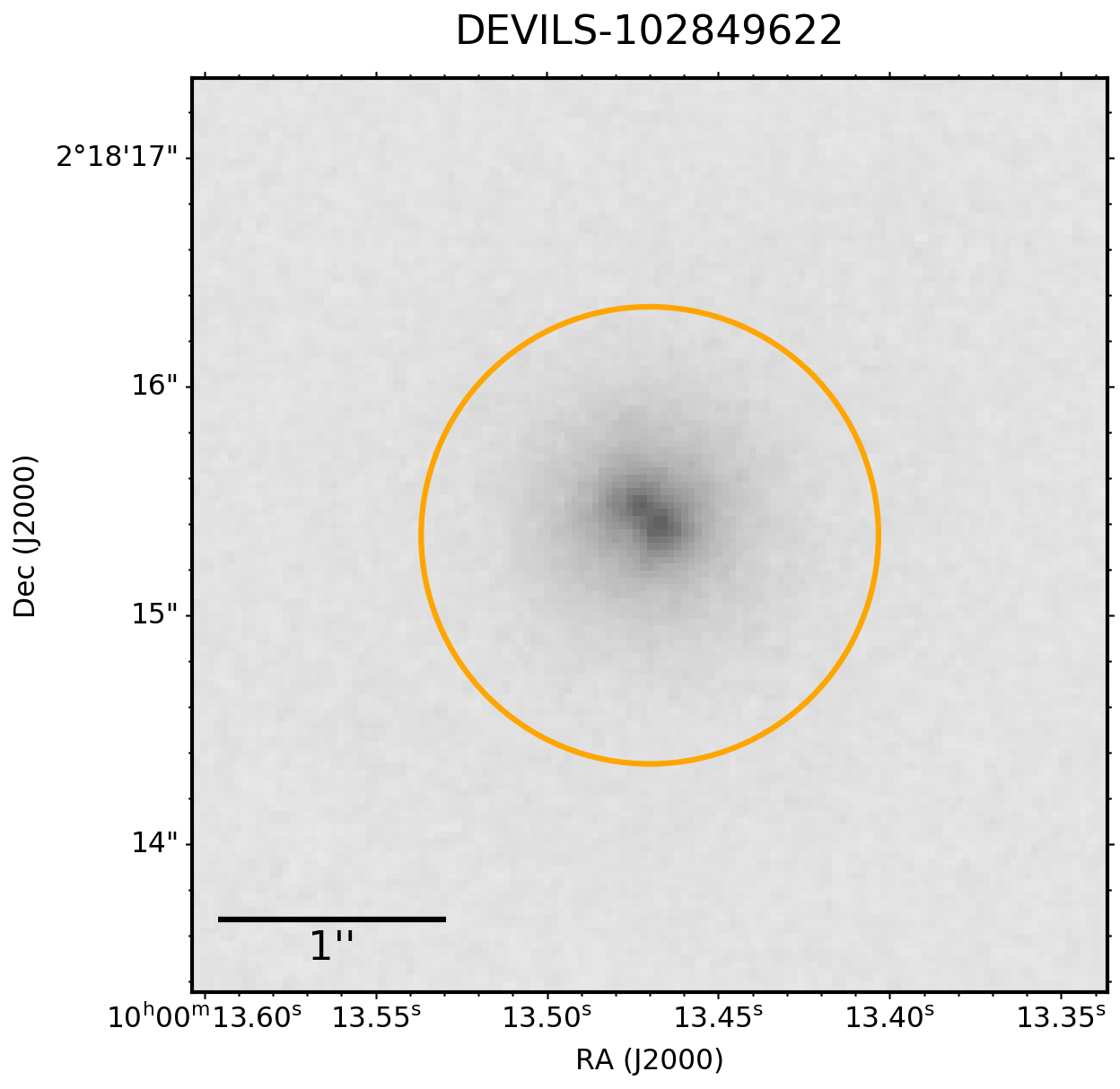}
\includegraphics[width=0.2\textwidth]{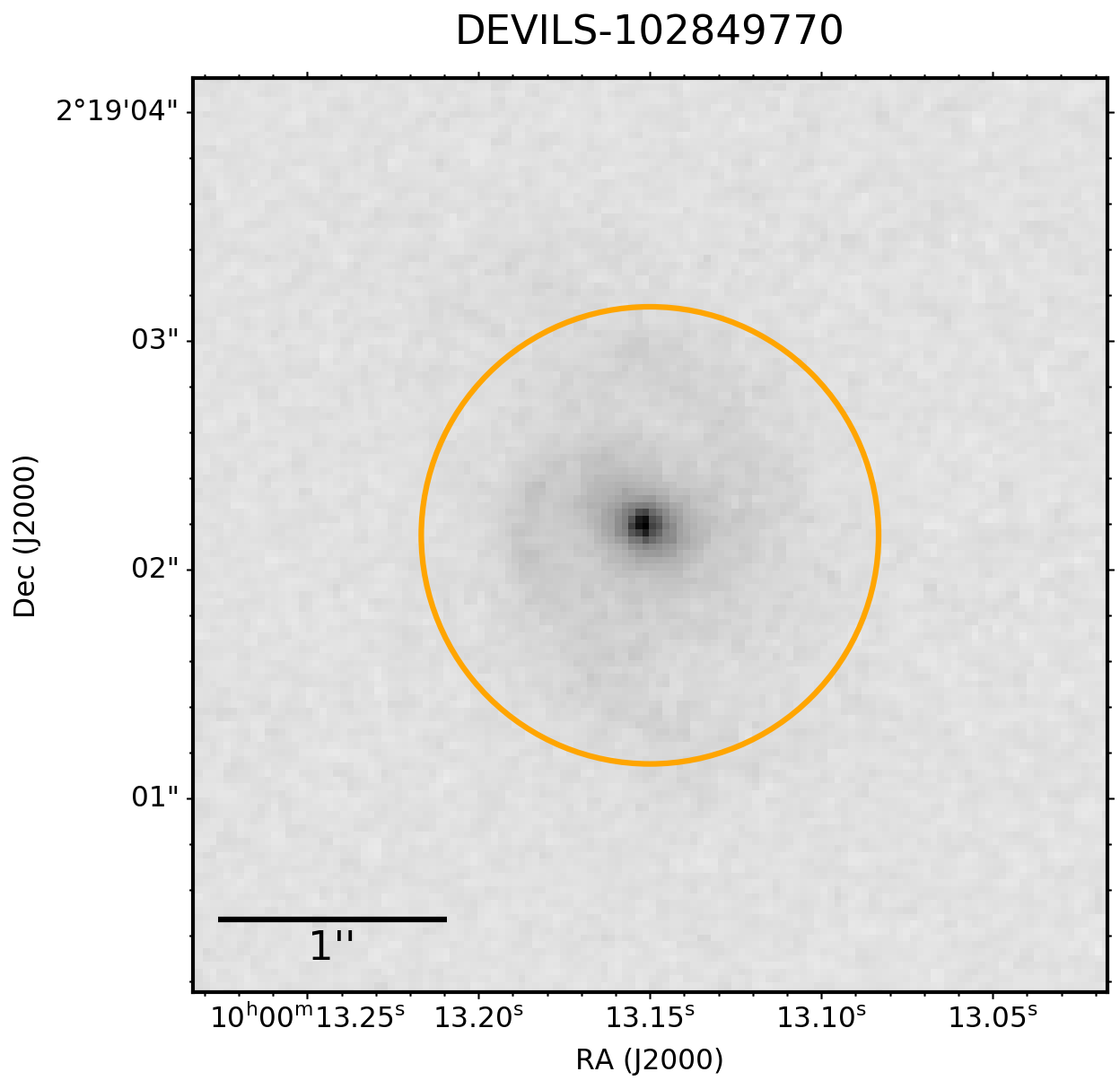}
\includegraphics[width=0.2\textwidth]{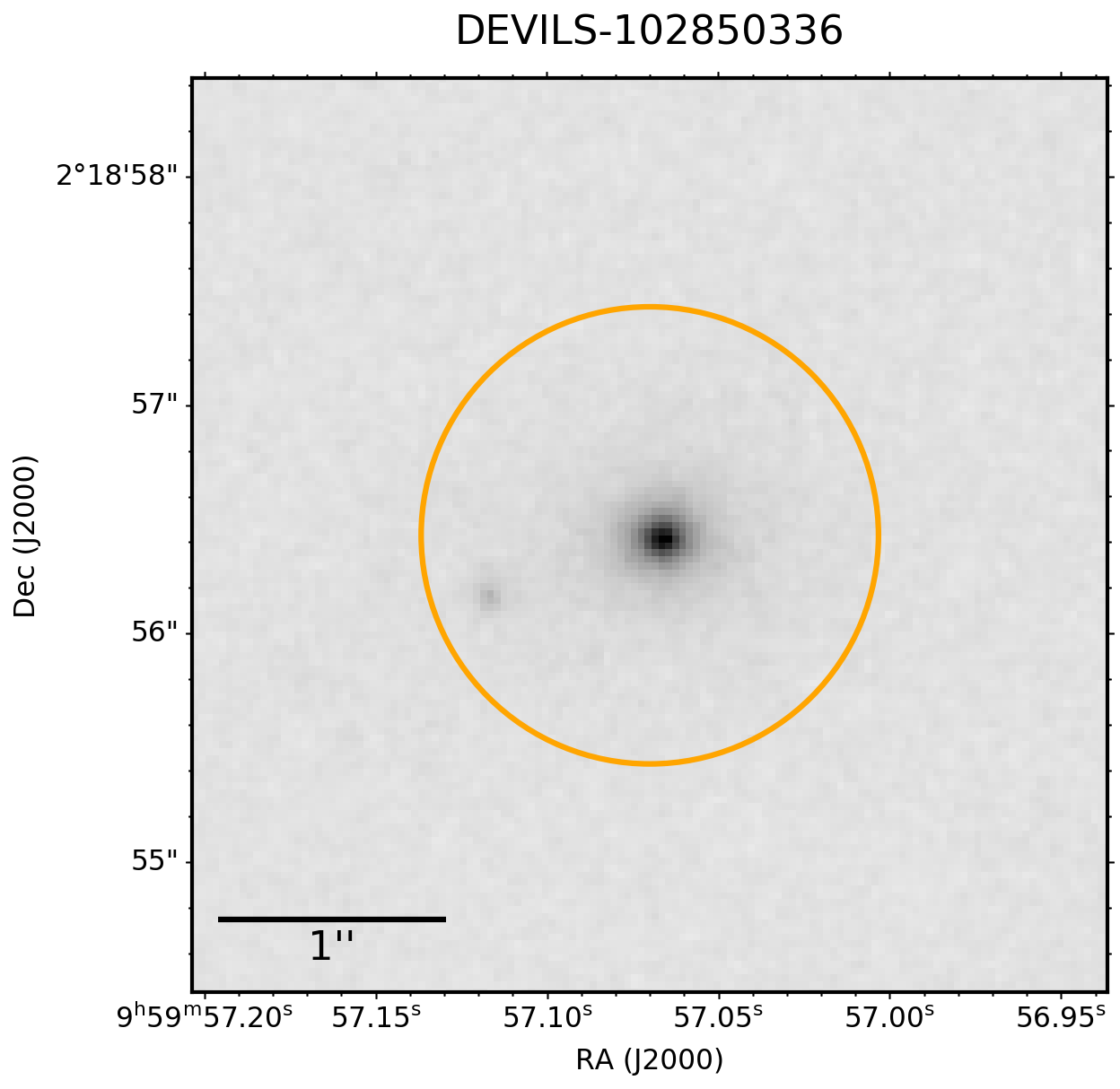}
\includegraphics[width=0.2\textwidth]{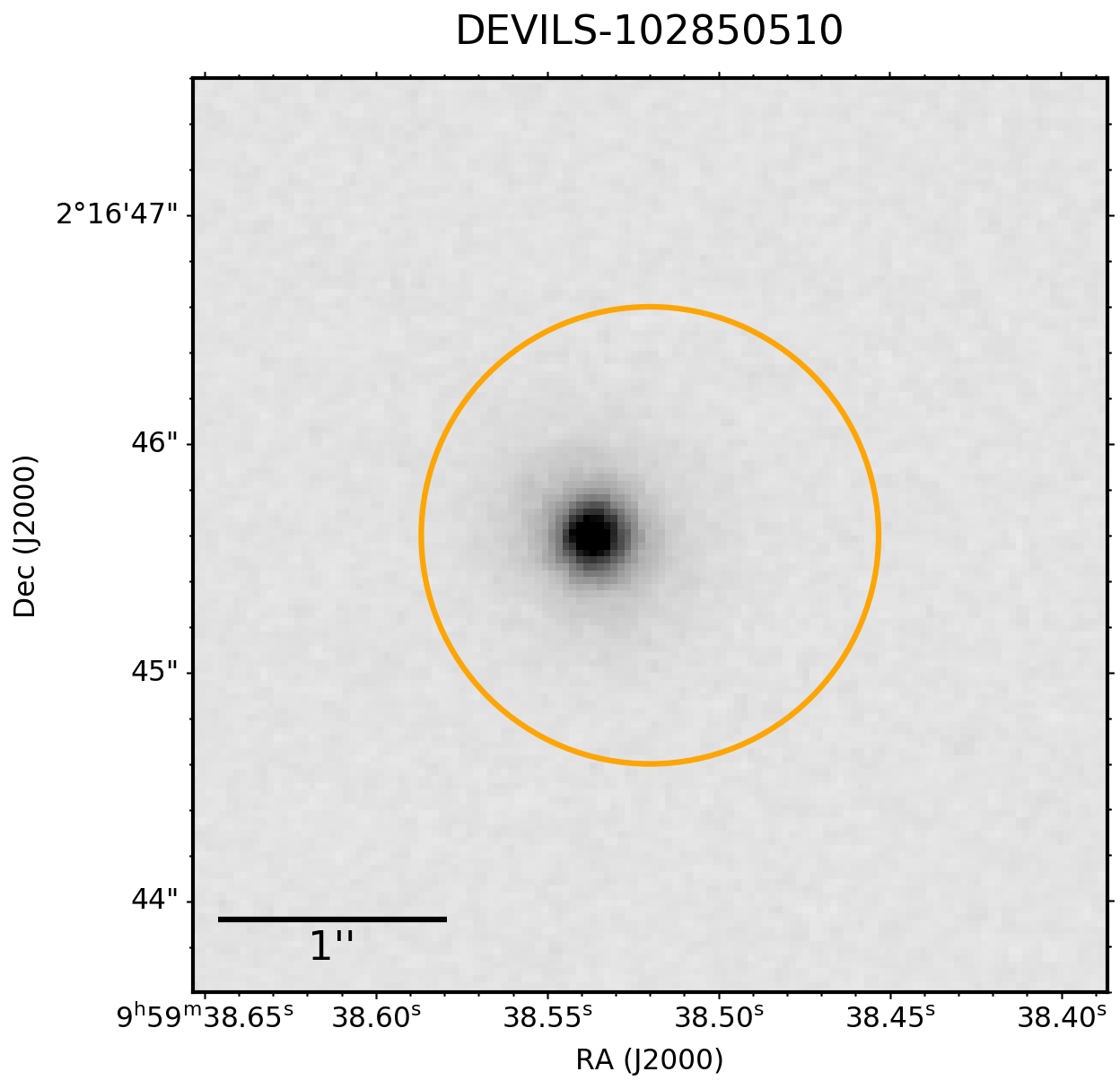}
\includegraphics[width=0.2\textwidth]{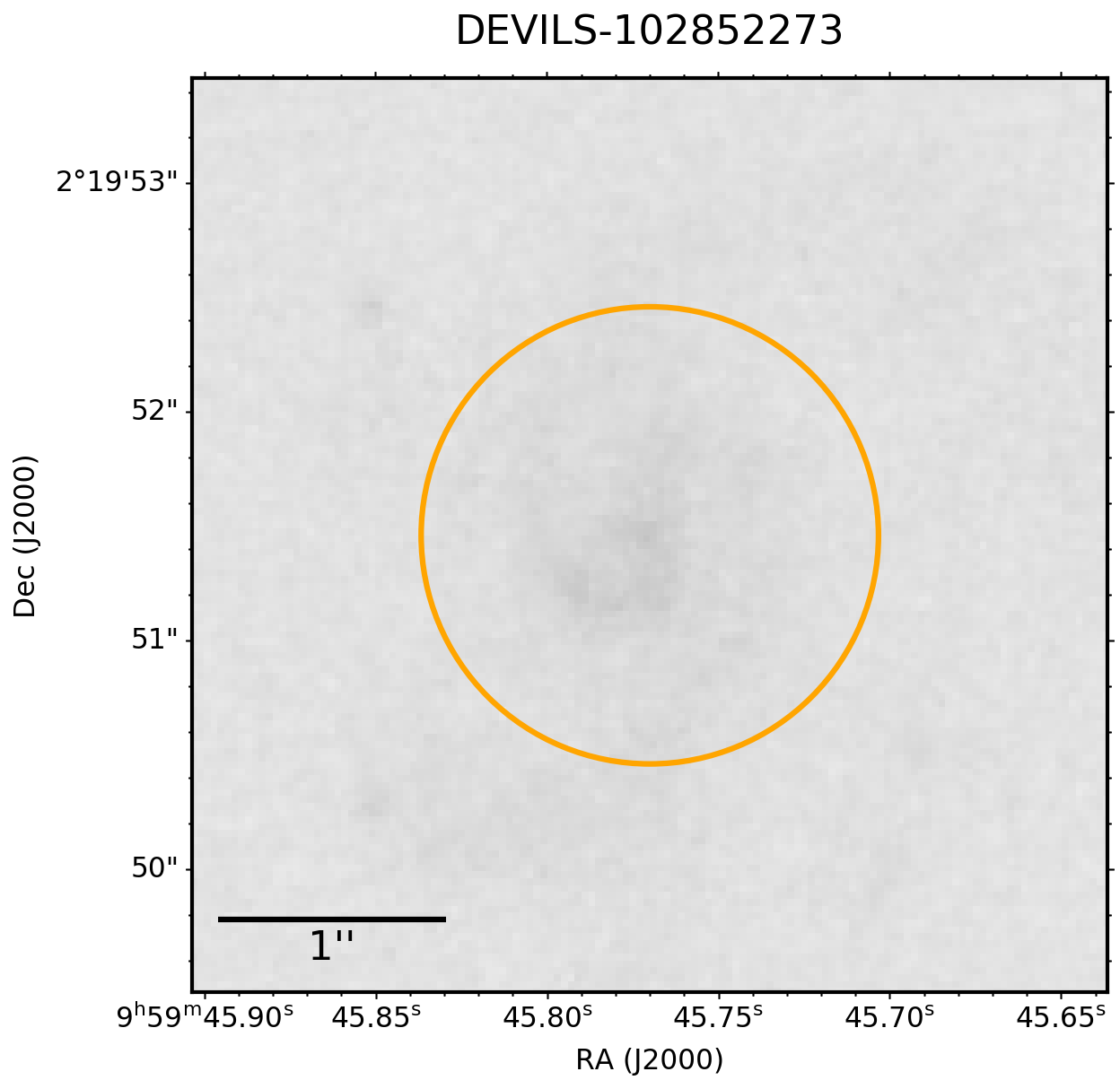}
\includegraphics[width=0.2\textwidth]{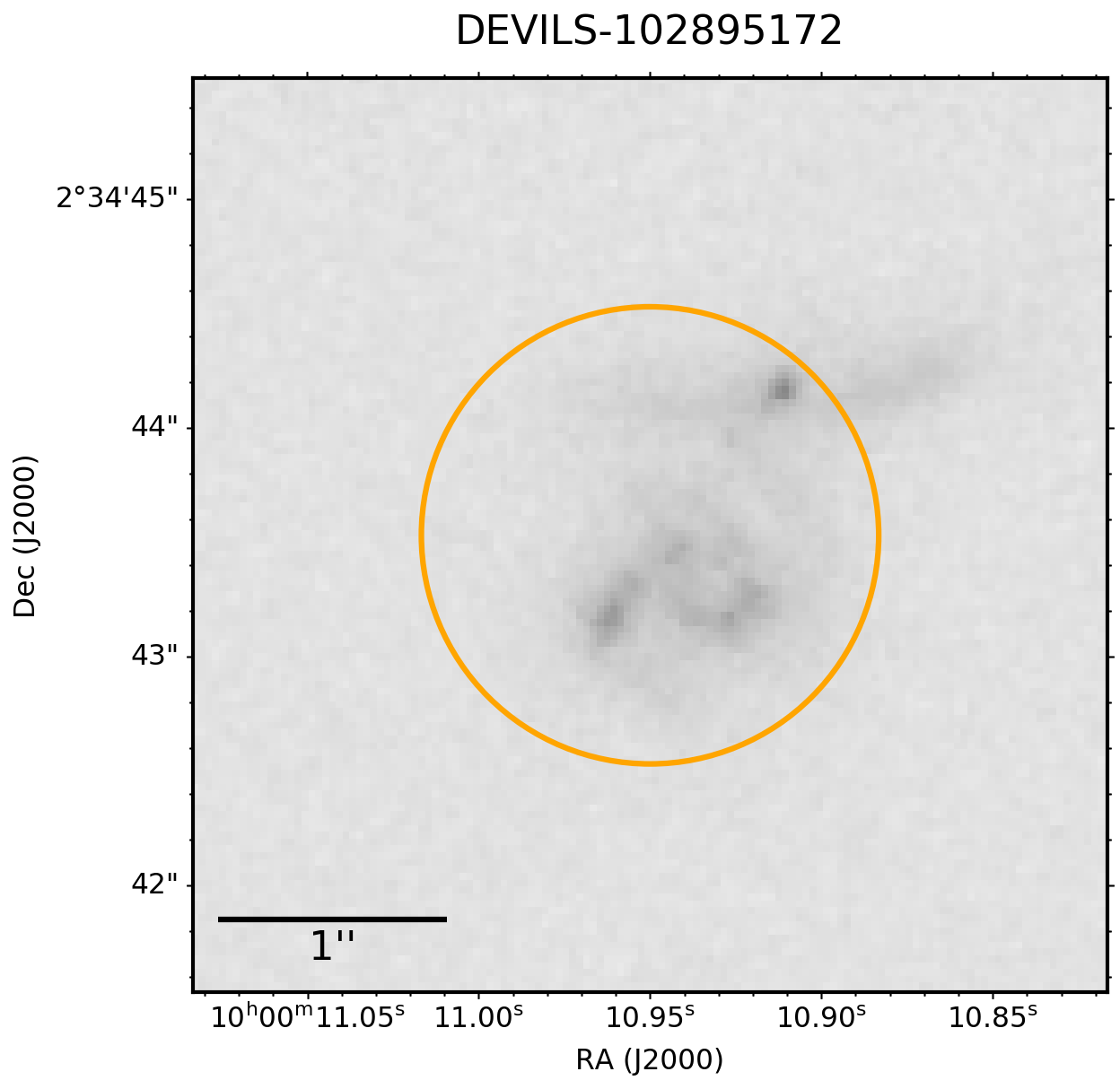}
\includegraphics[width=0.2\textwidth]{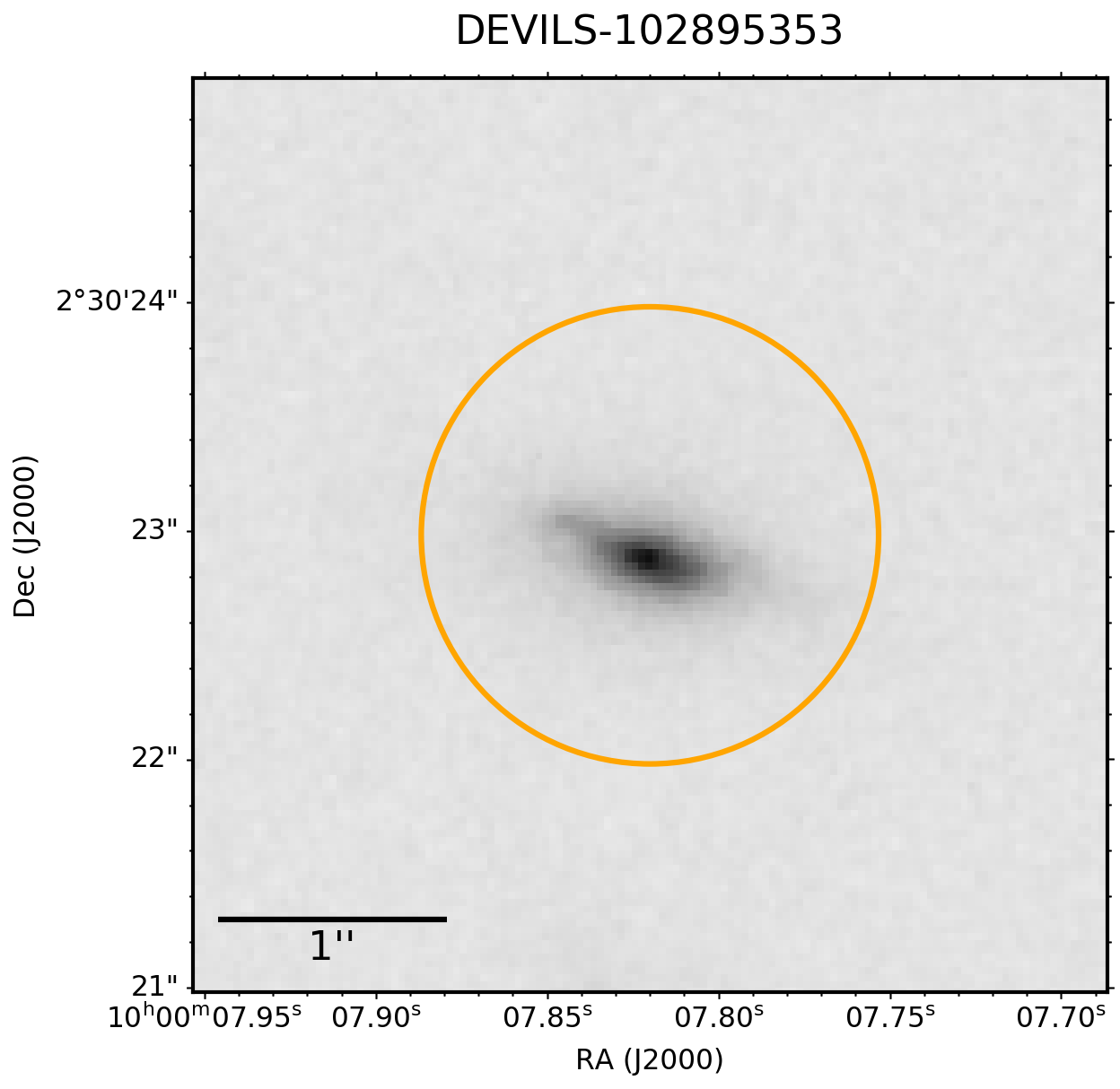}
\includegraphics[width=0.2\textwidth]{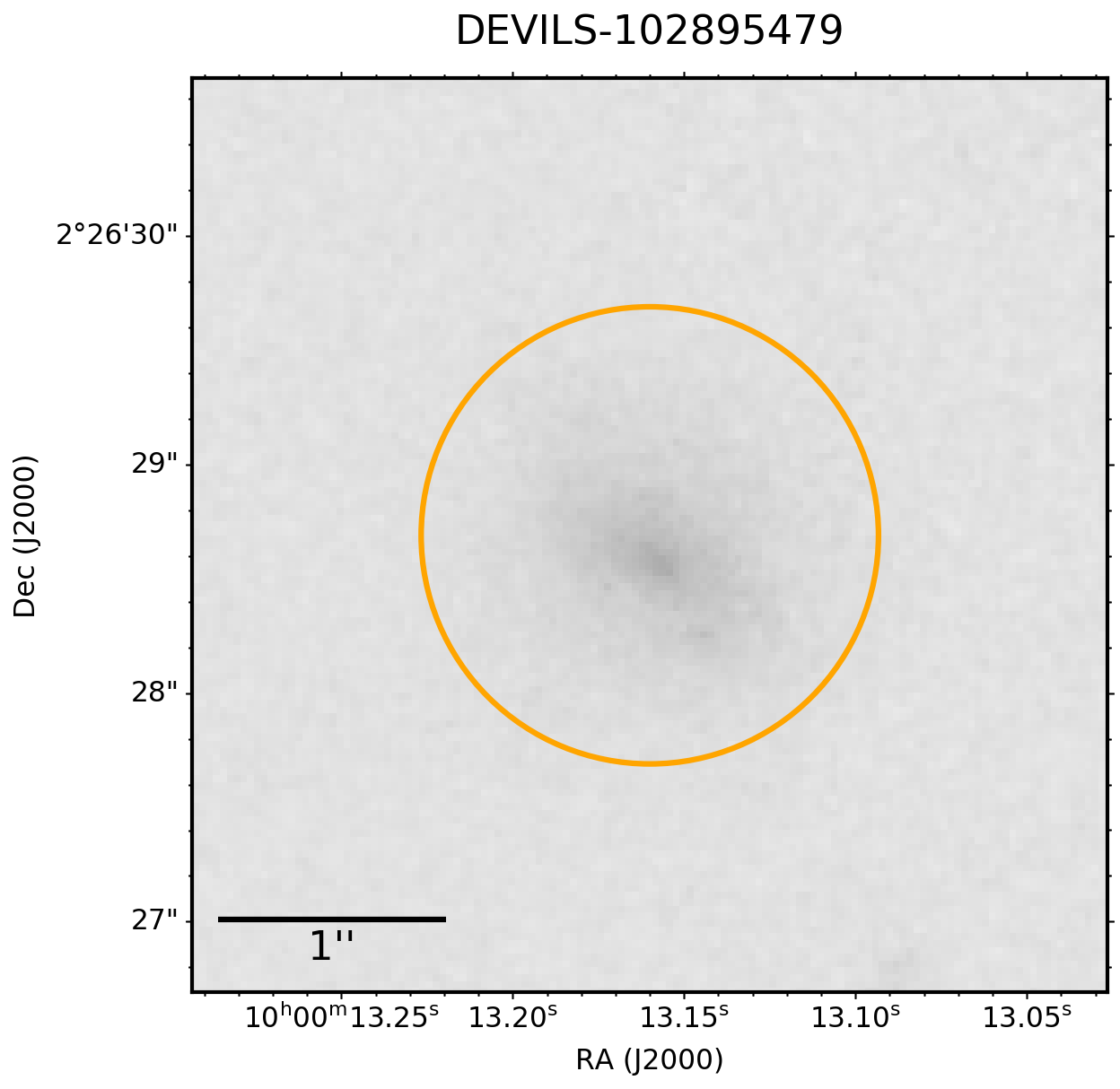}
\includegraphics[width=0.2\textwidth]{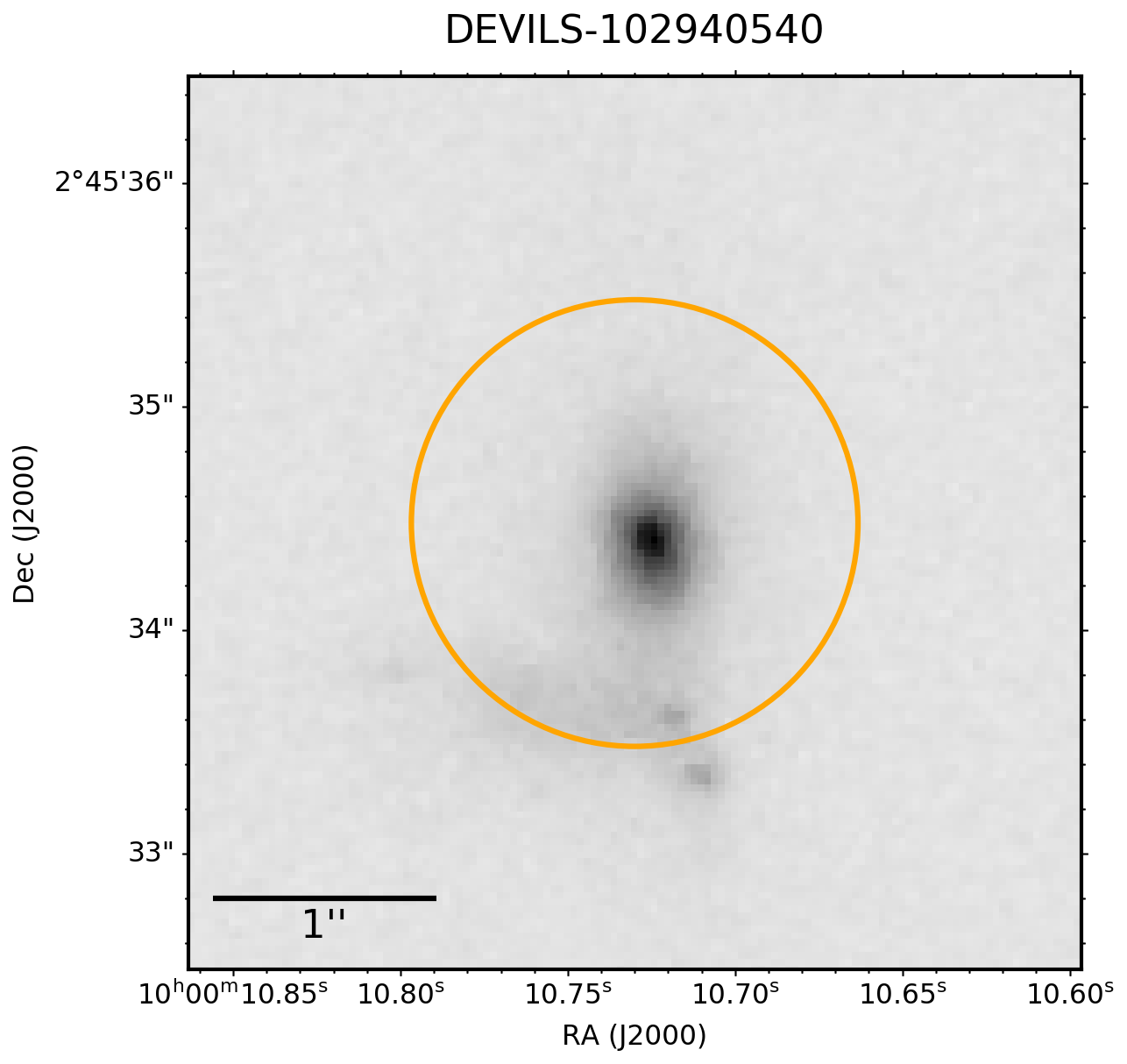}
\includegraphics[width=0.2\textwidth]{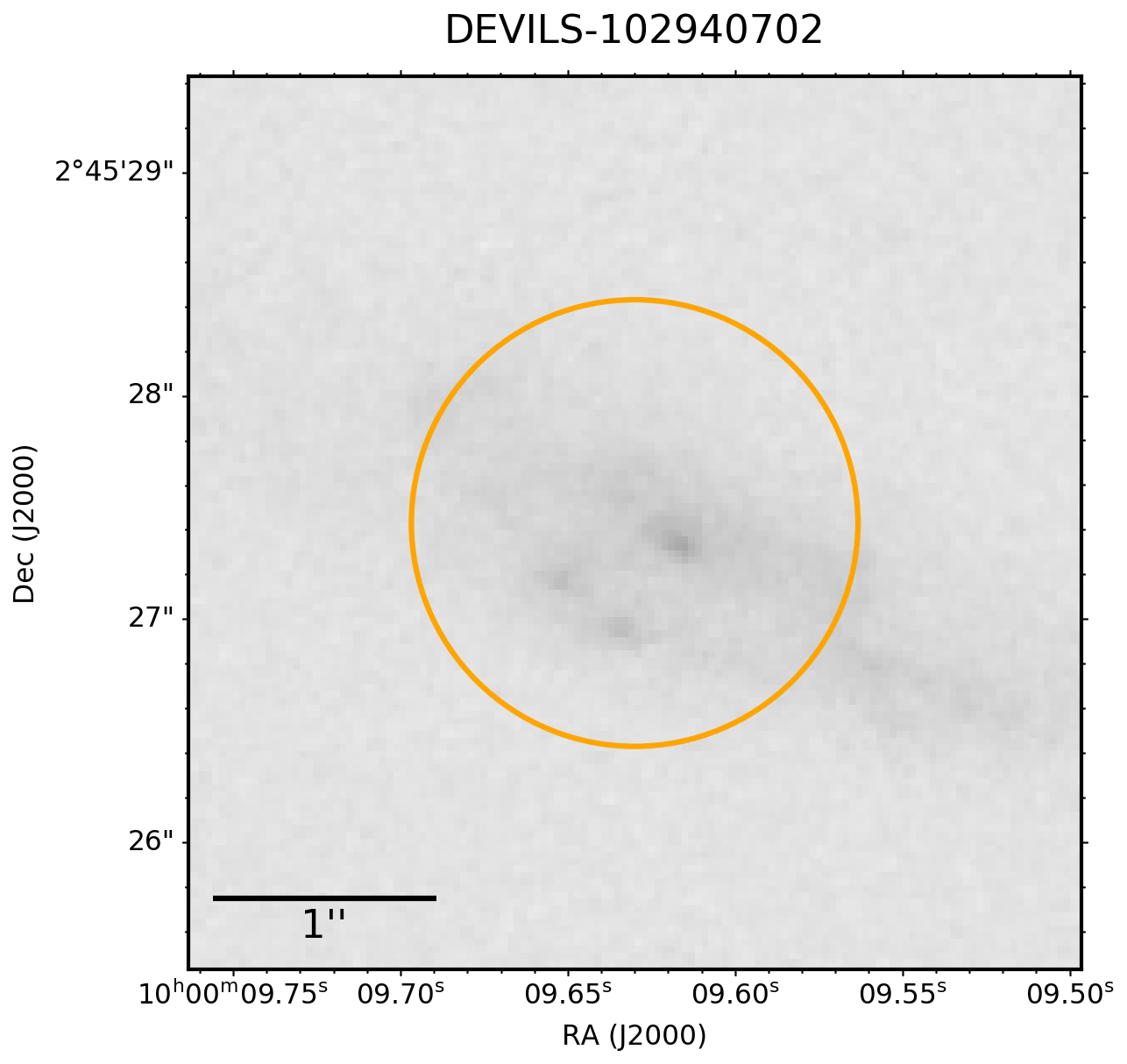}
\includegraphics[width=0.2\textwidth]{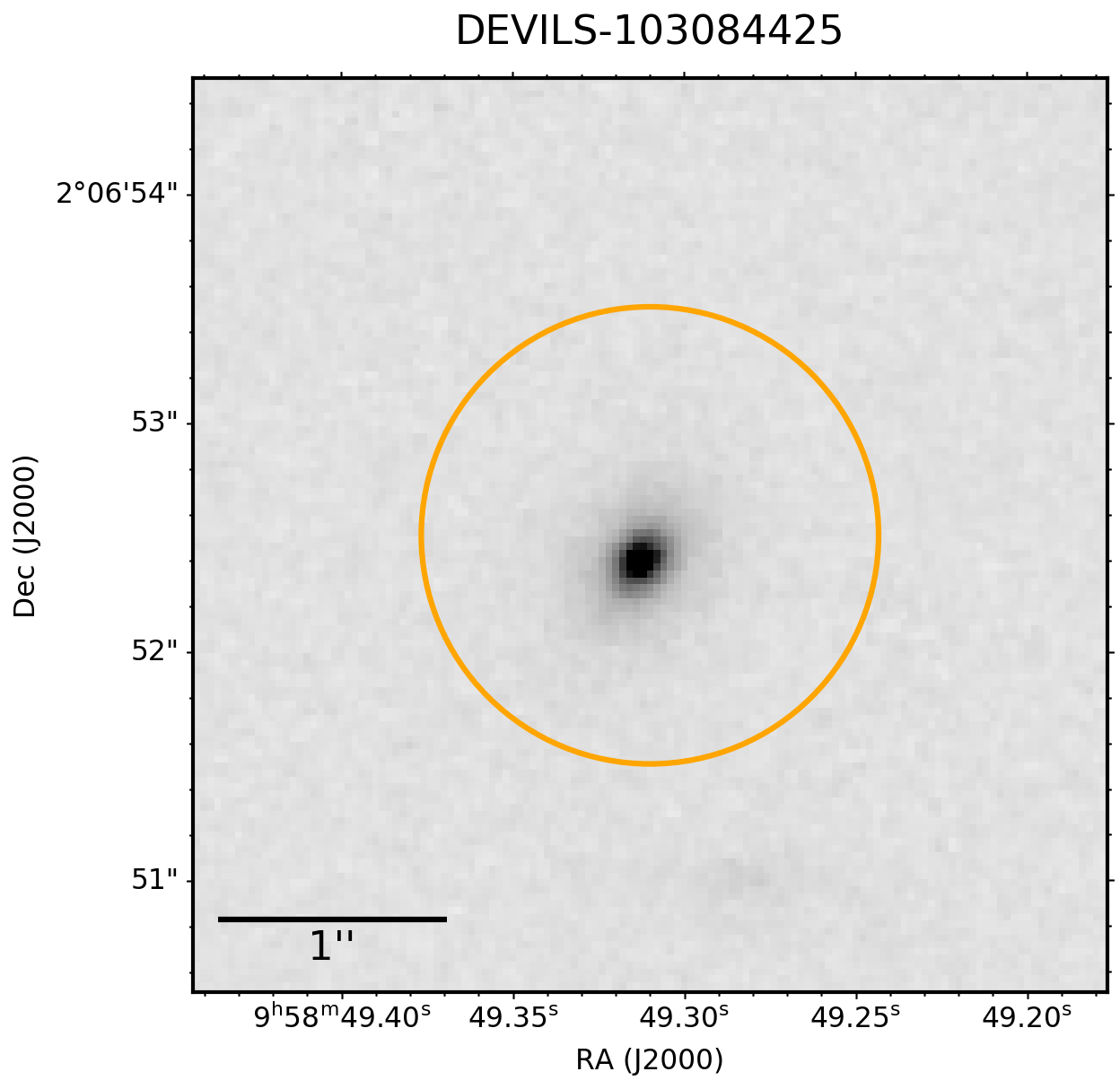}
\includegraphics[width=0.2\textwidth]{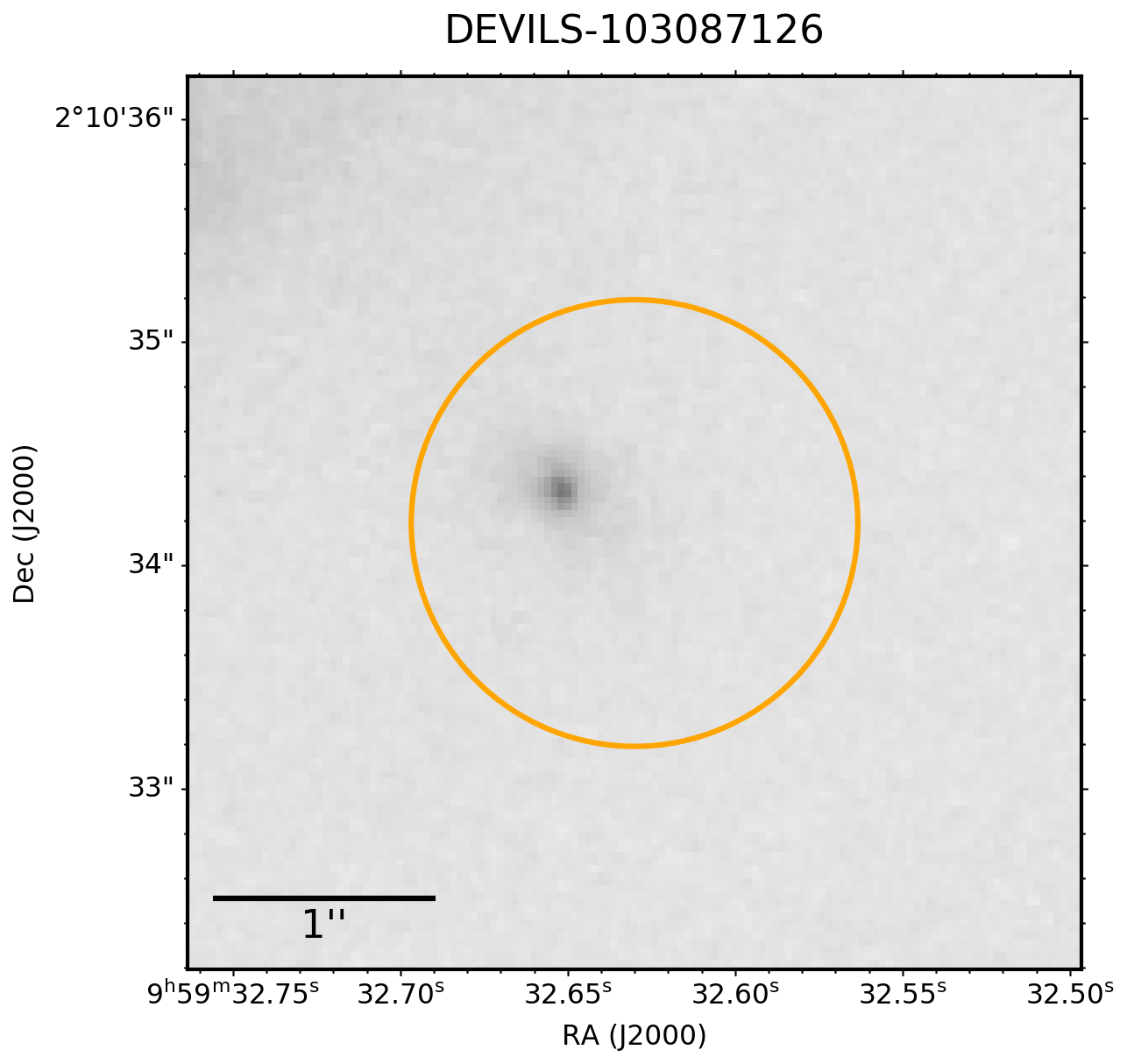}
\includegraphics[width=0.2\textwidth]{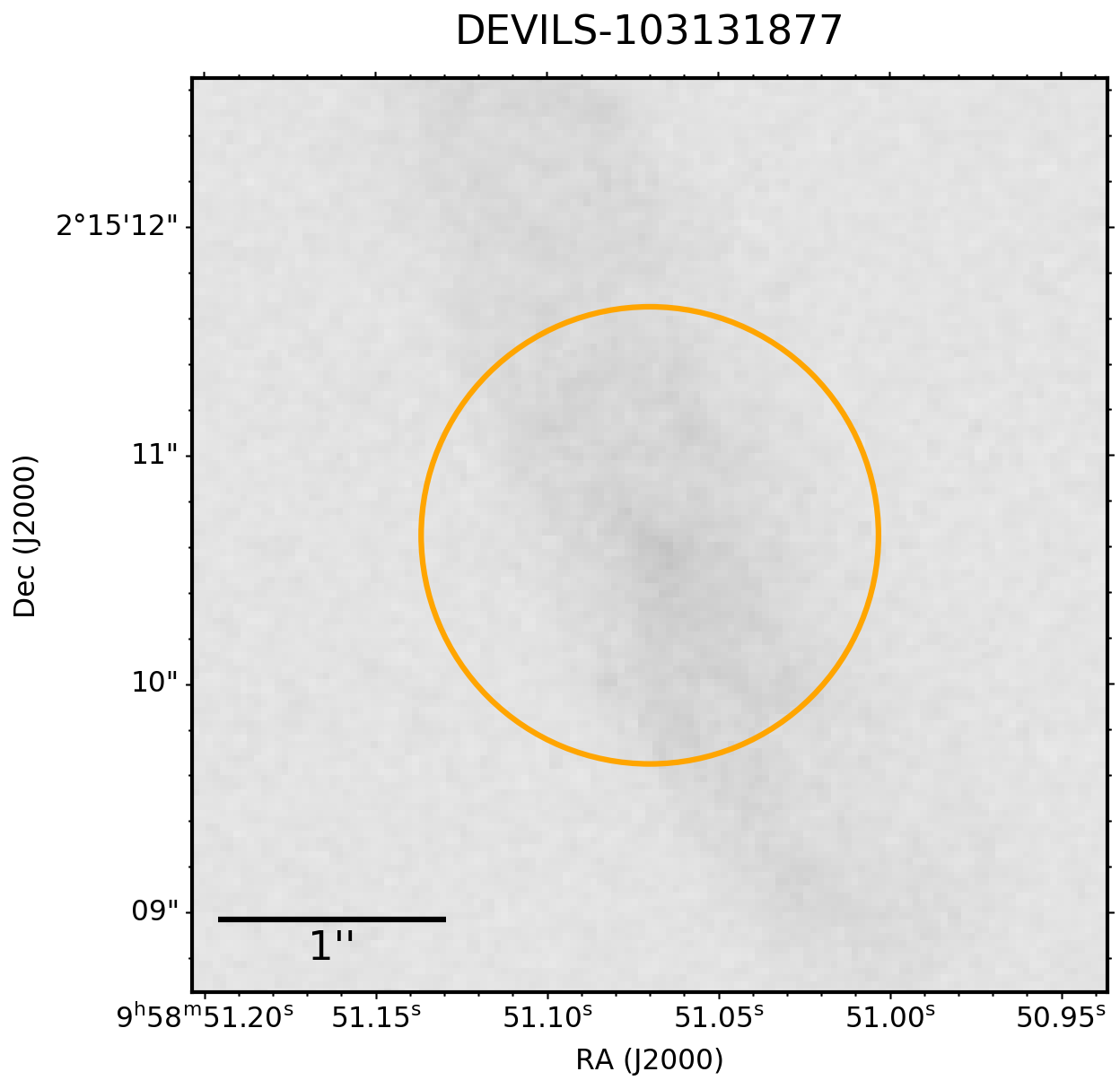}
\includegraphics[width=0.2\textwidth]{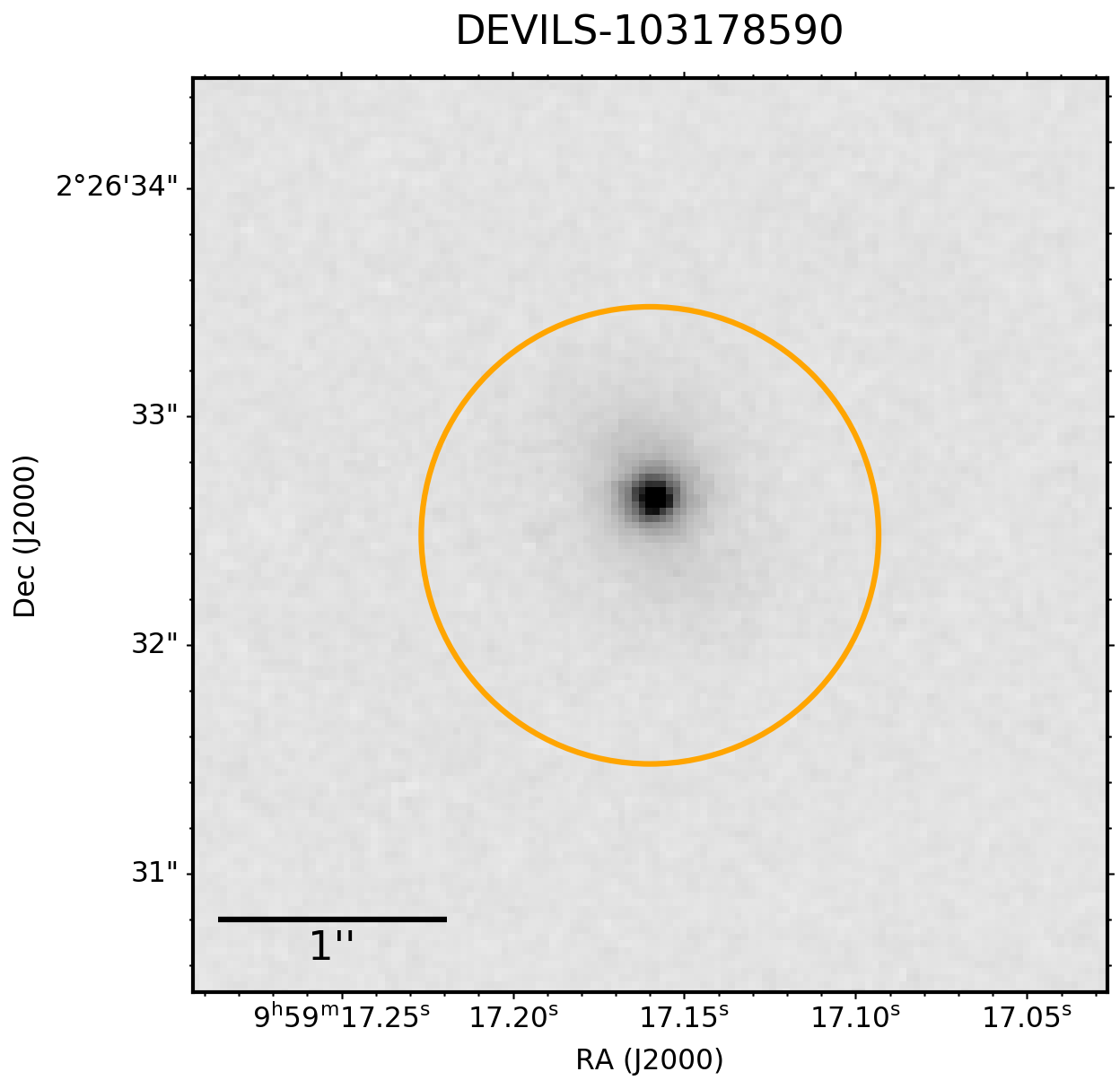}
\includegraphics[width=0.2\textwidth]{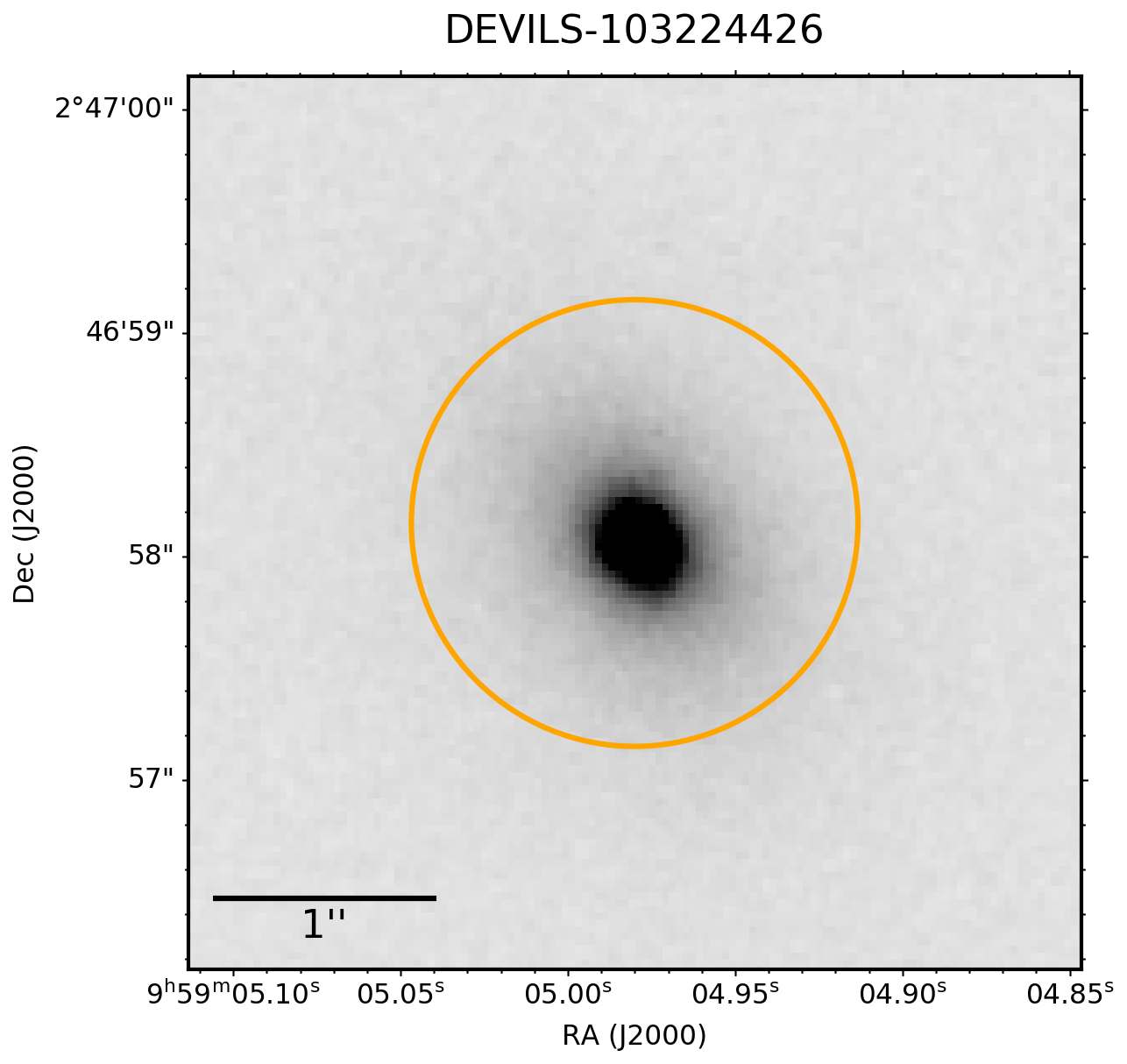}
\includegraphics[width=0.2\textwidth]{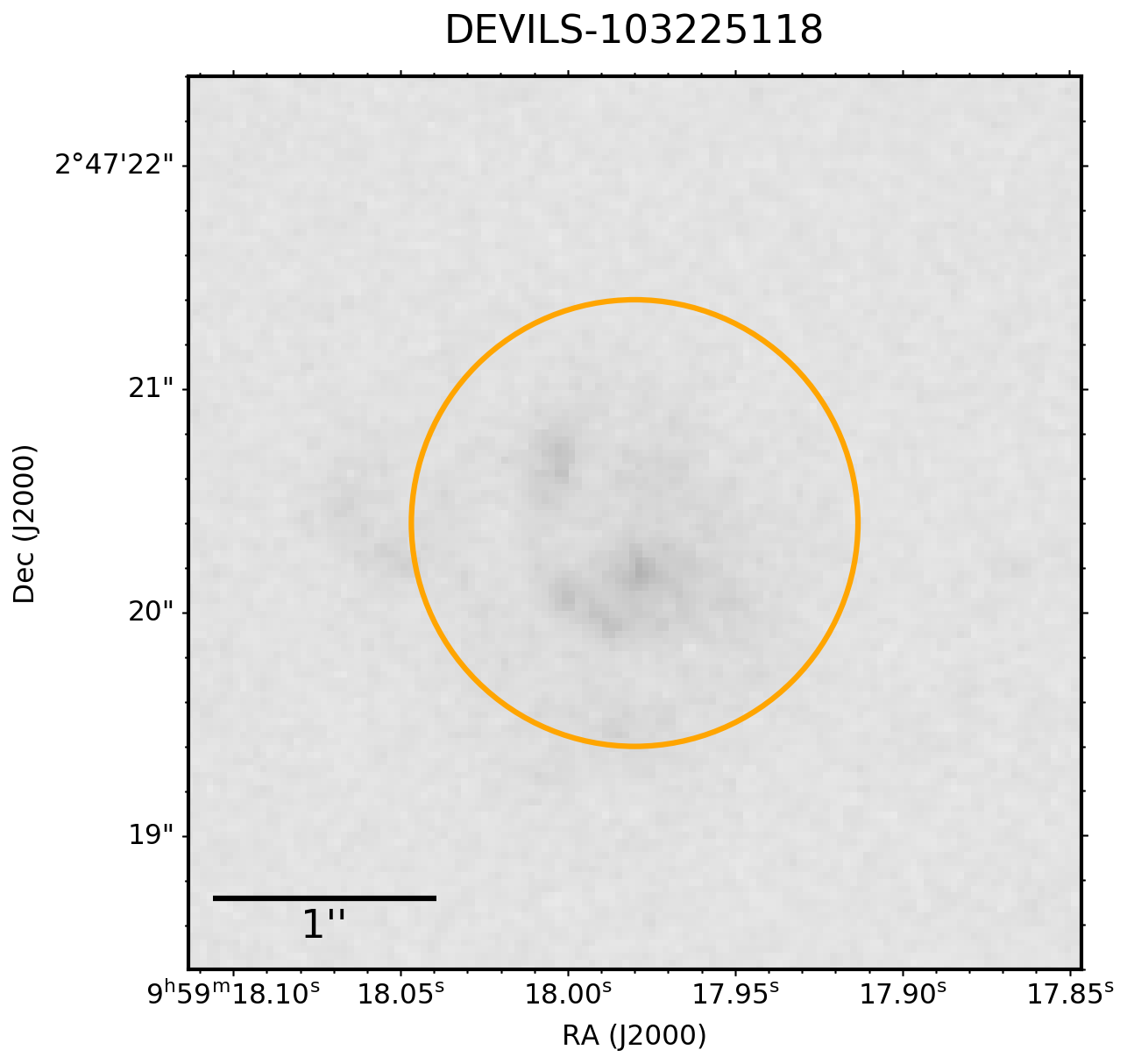}
\includegraphics[width=0.2\textwidth]{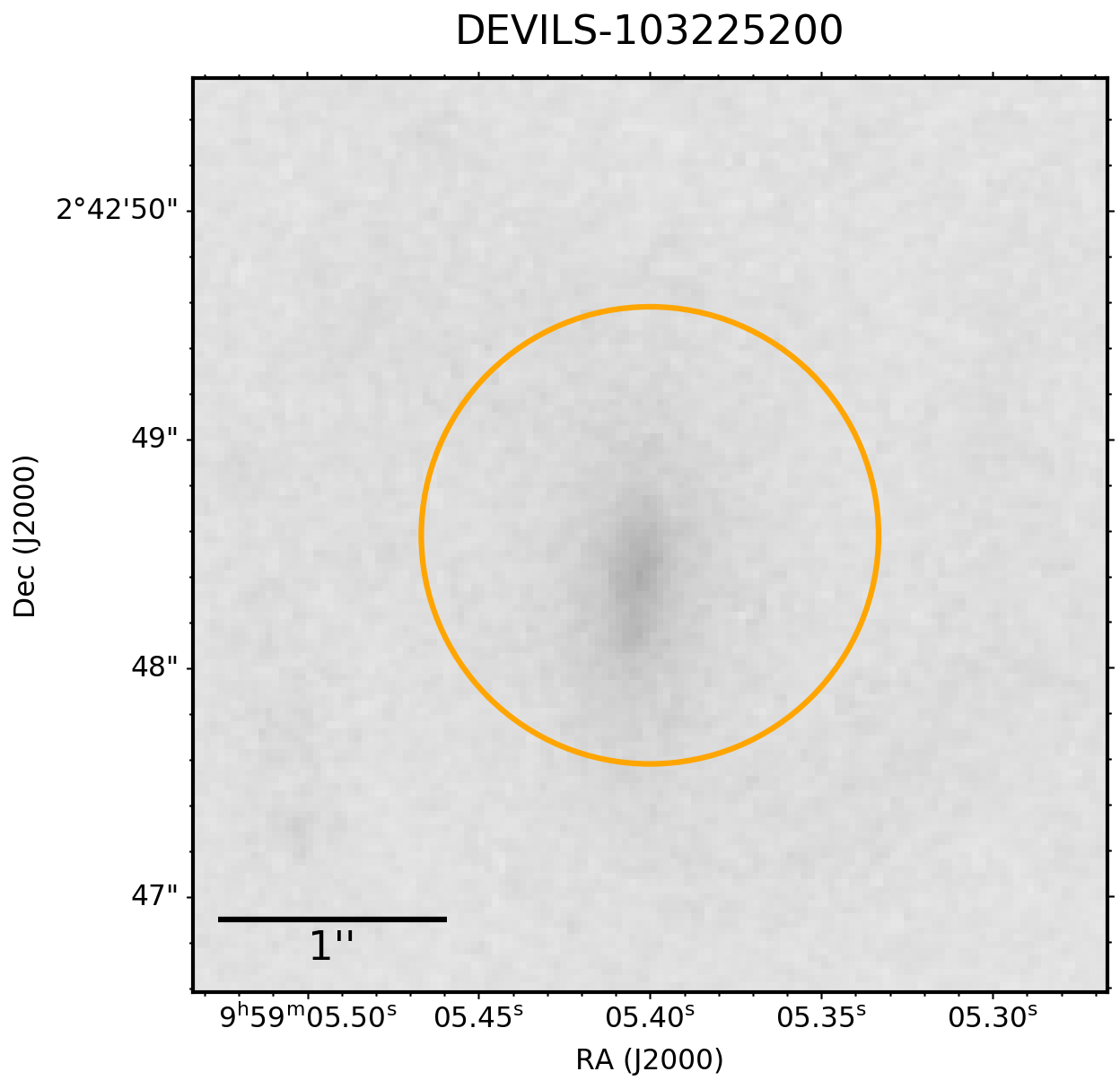}
\includegraphics[width=0.2\textwidth]{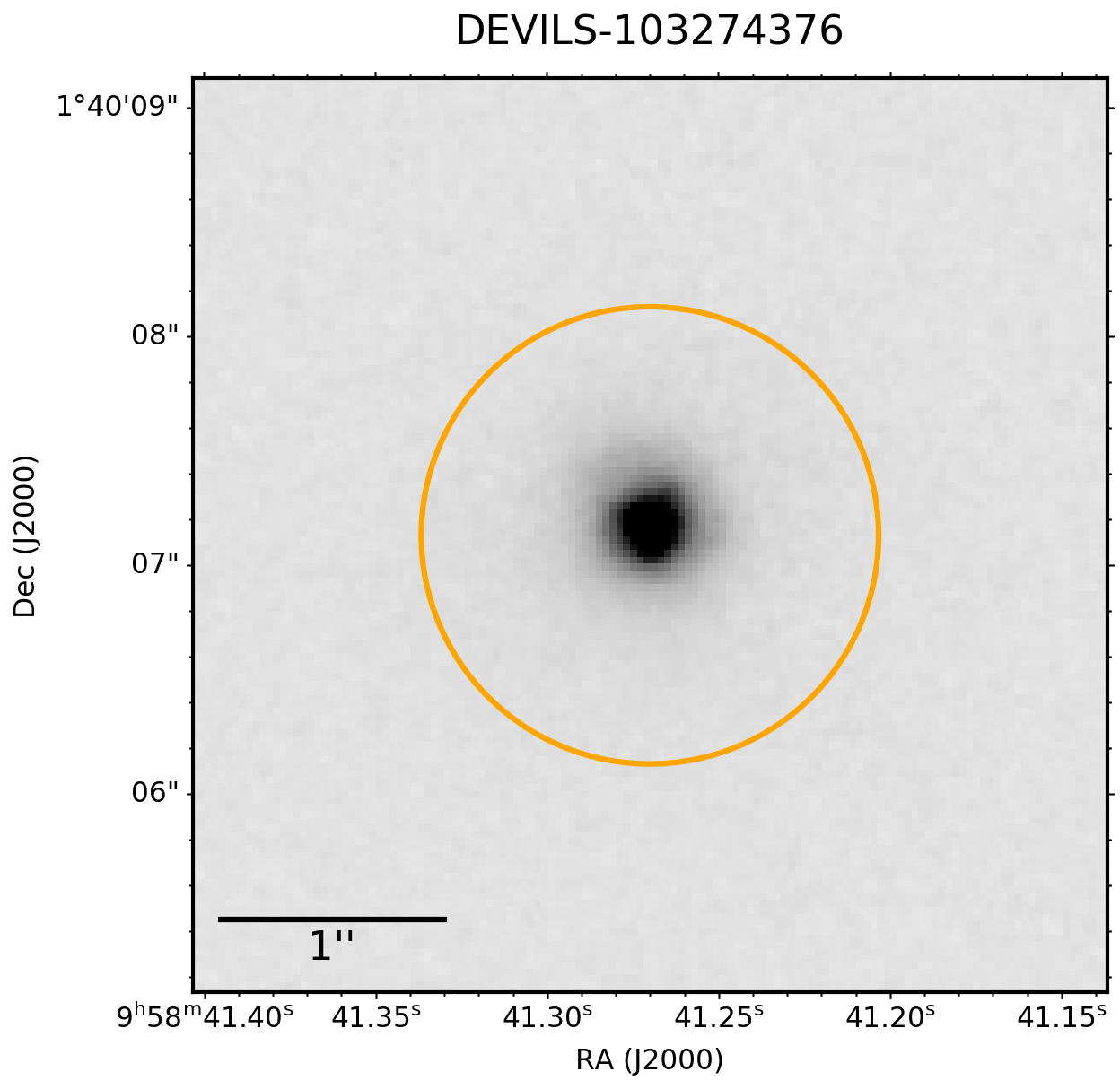}
\includegraphics[width=0.2\textwidth]{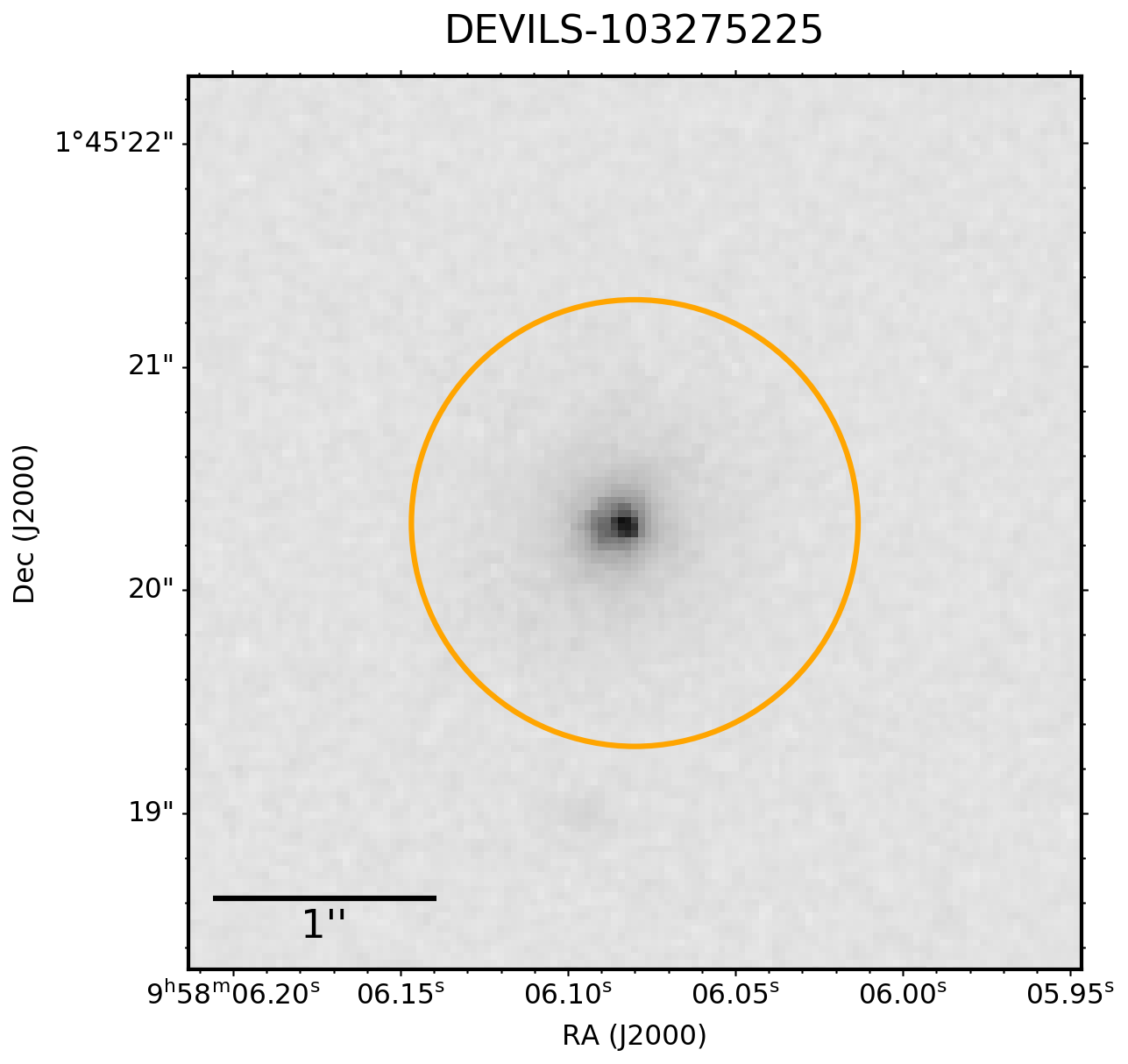}
\includegraphics[width=0.2\textwidth]{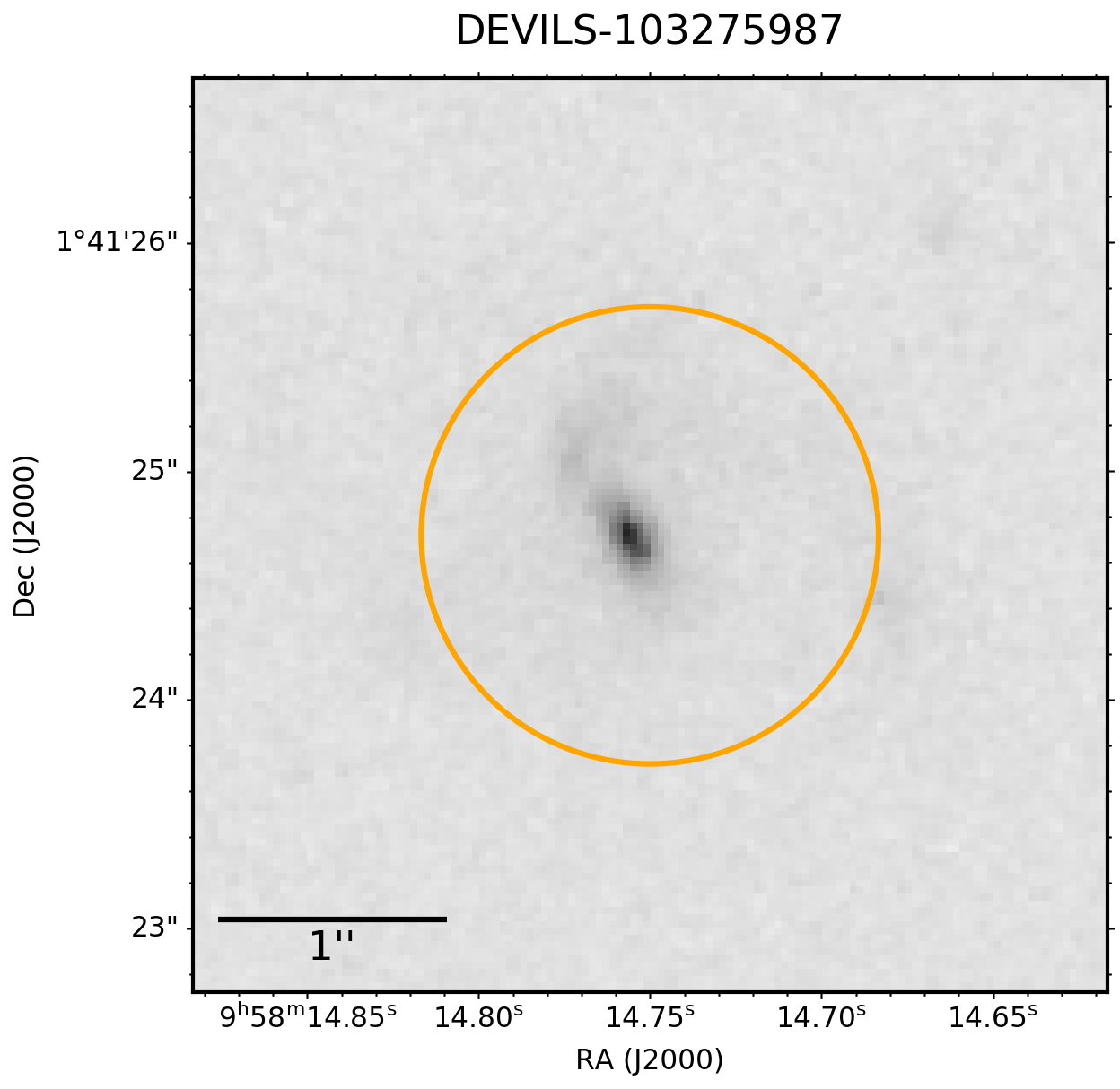}
\includegraphics[width=0.2\textwidth]{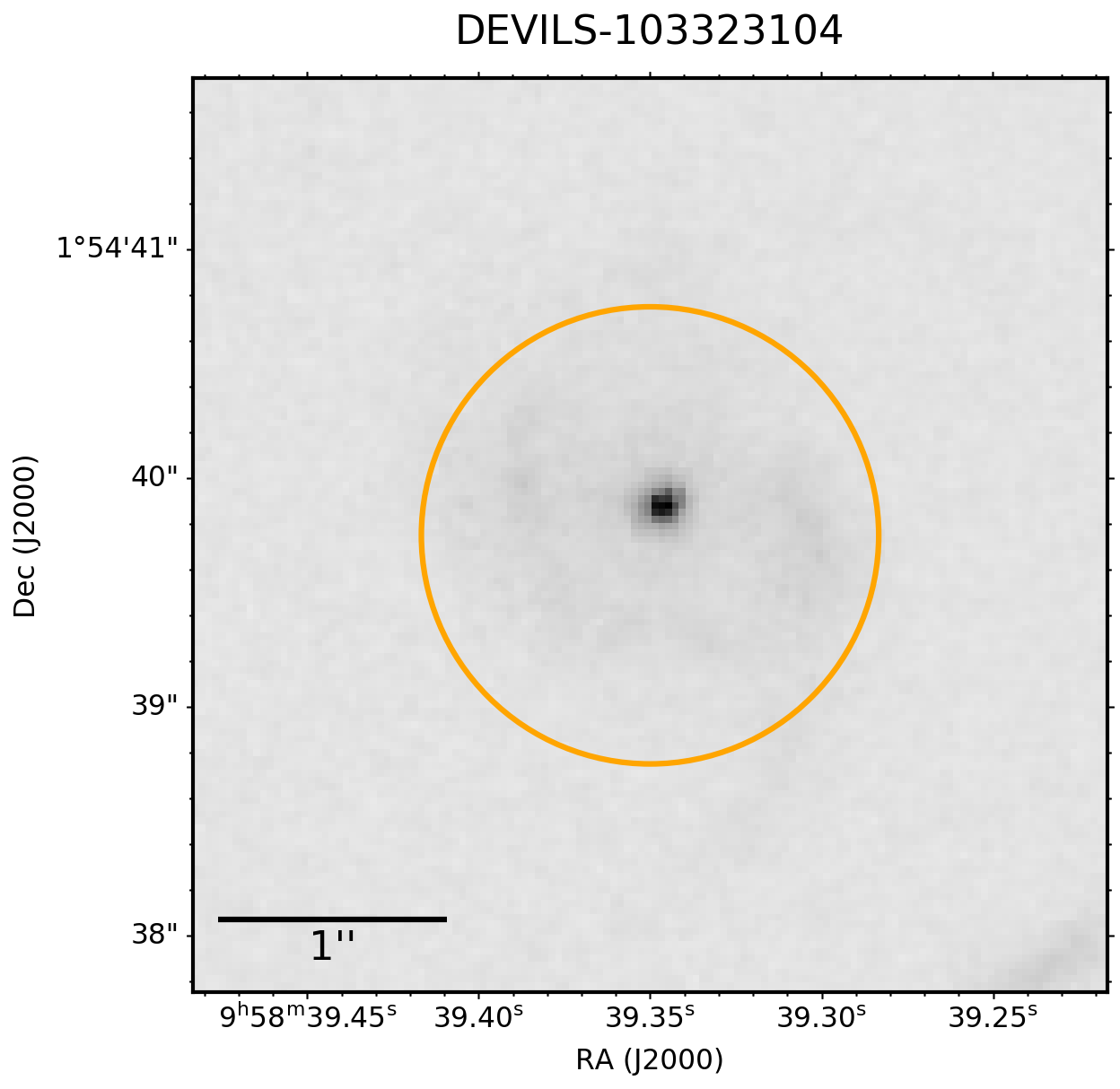}
\includegraphics[width=0.2\textwidth]{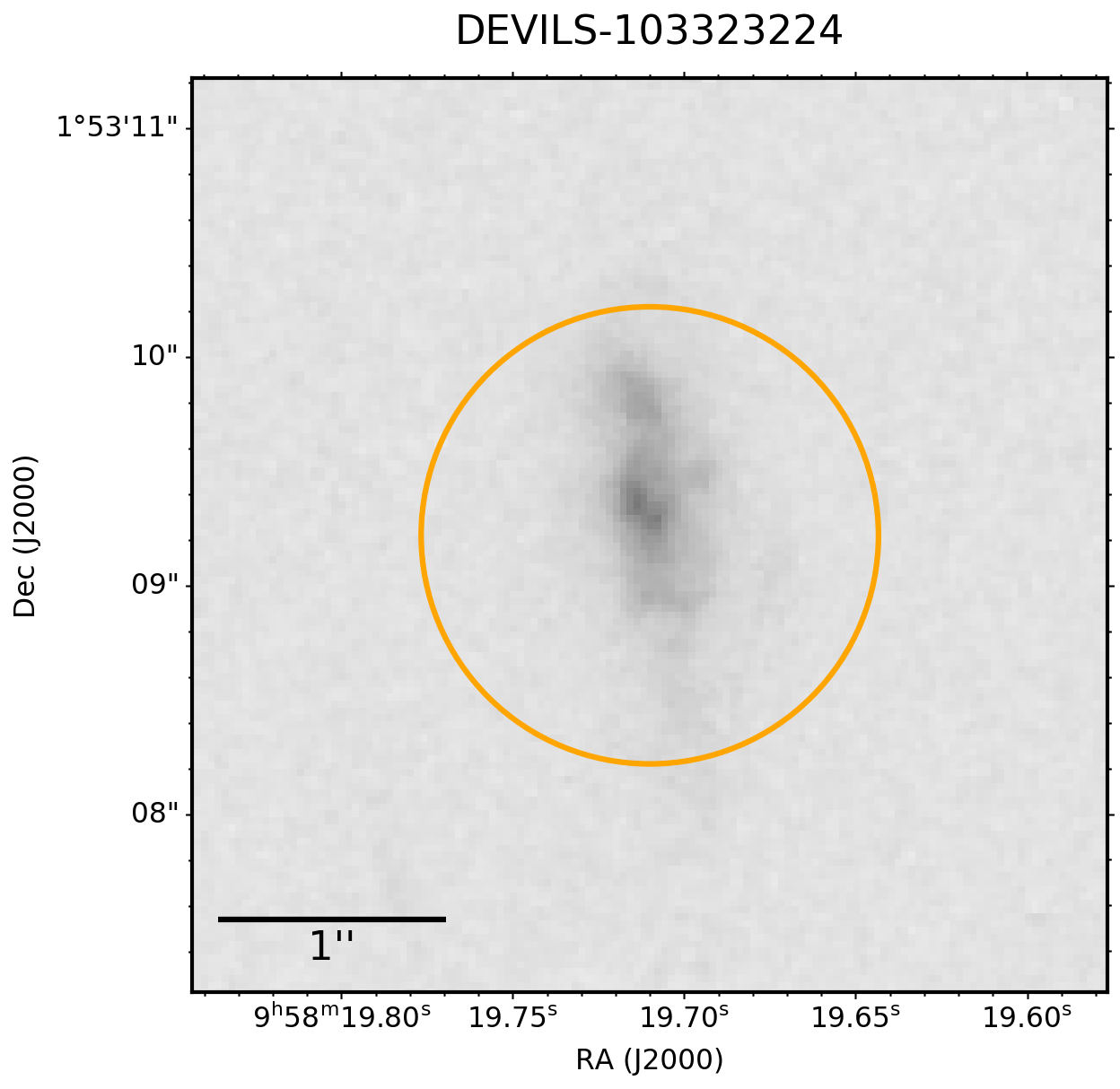}
\includegraphics[width=0.2\textwidth]{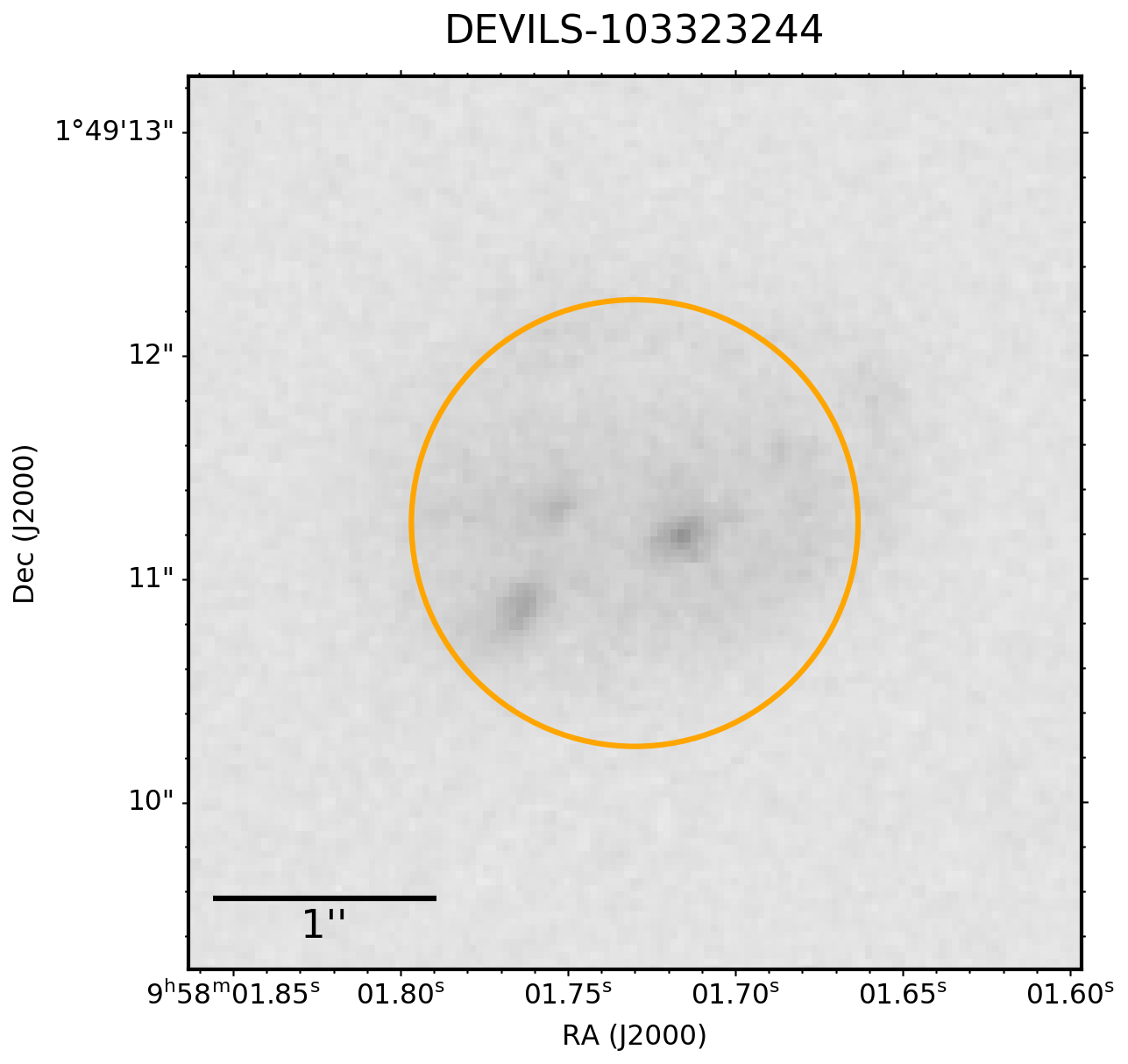}
\includegraphics[width=0.2\textwidth]{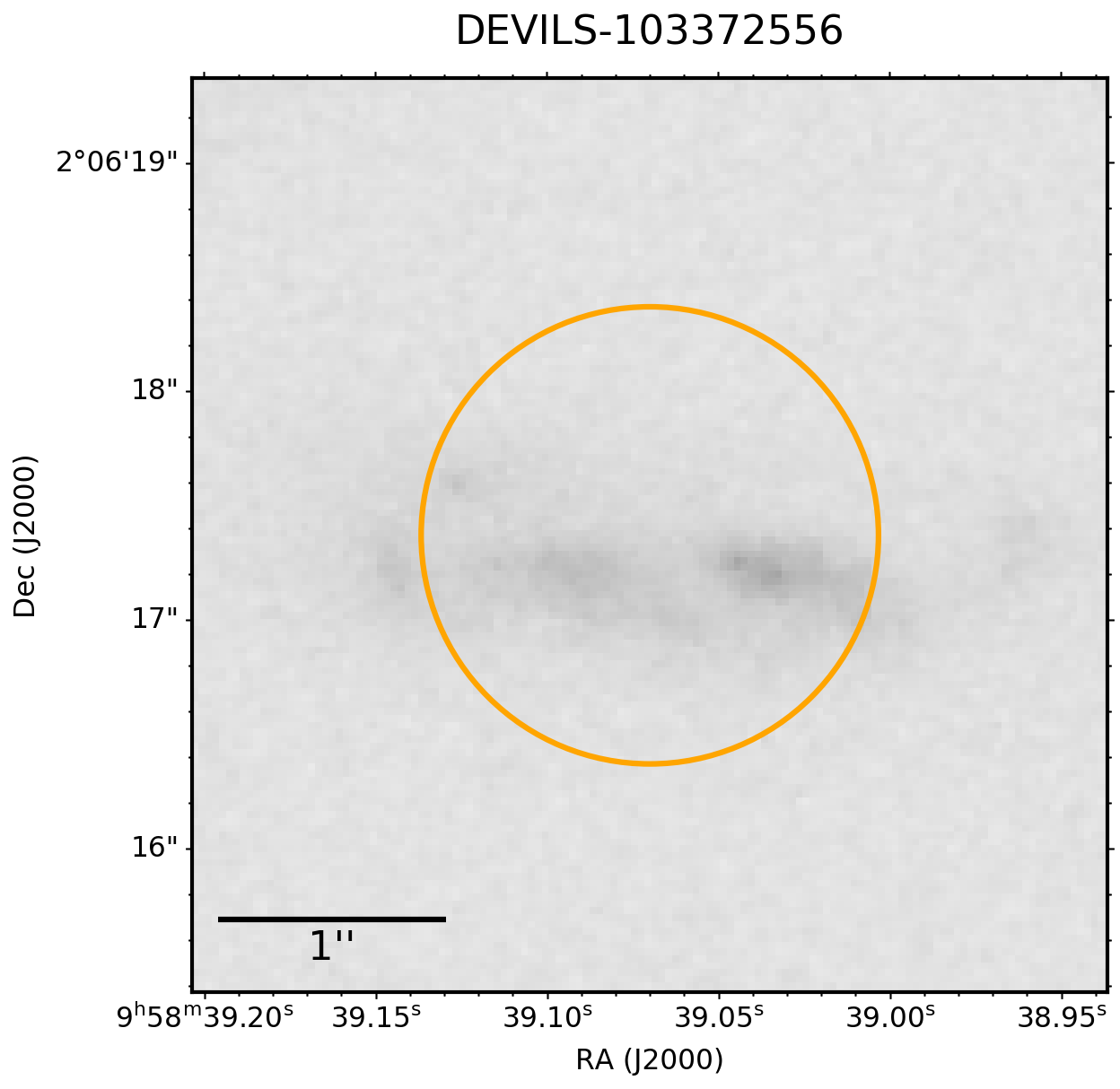}
\includegraphics[width=0.2\textwidth]{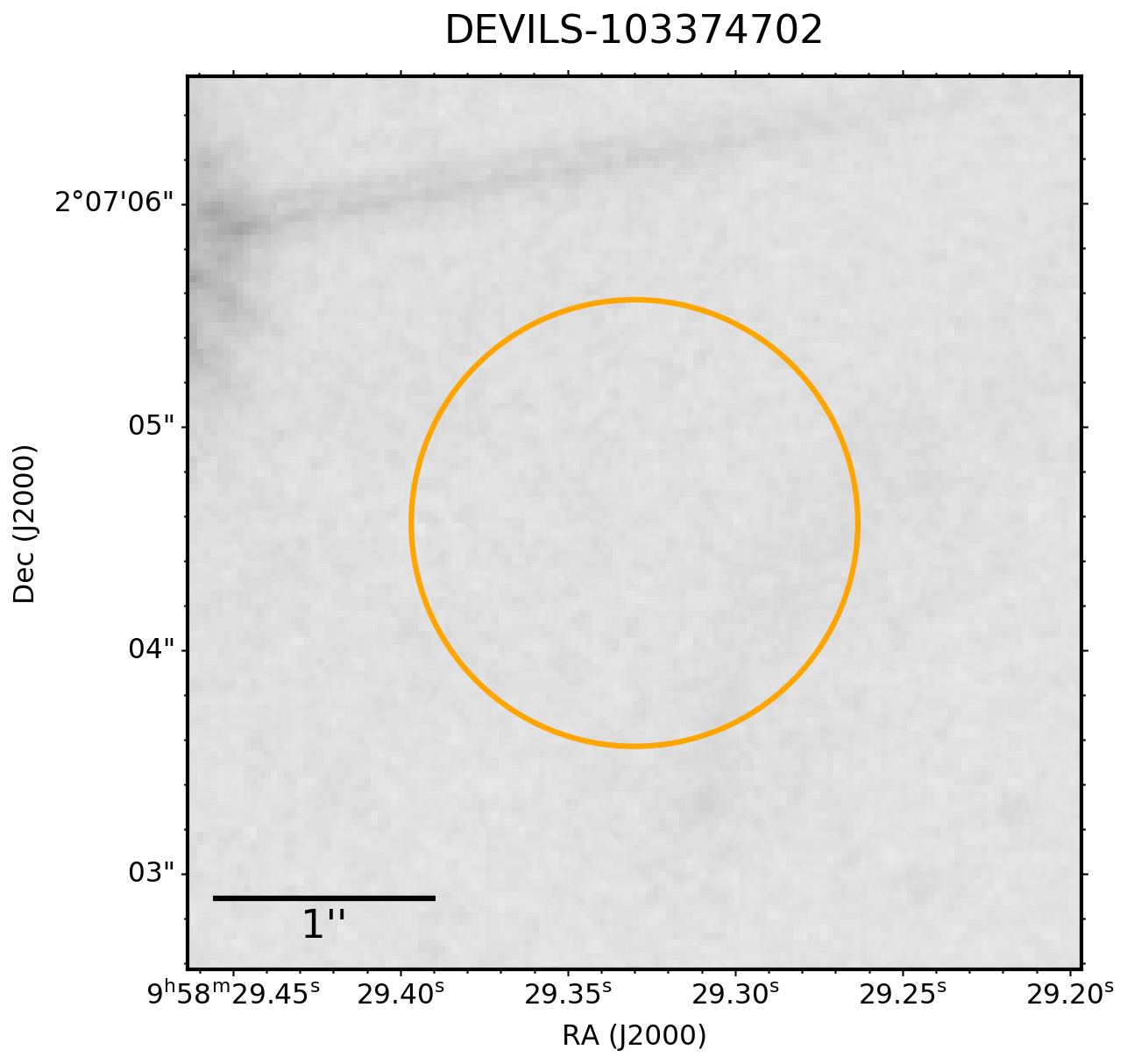}
\includegraphics[width=0.2\textwidth]{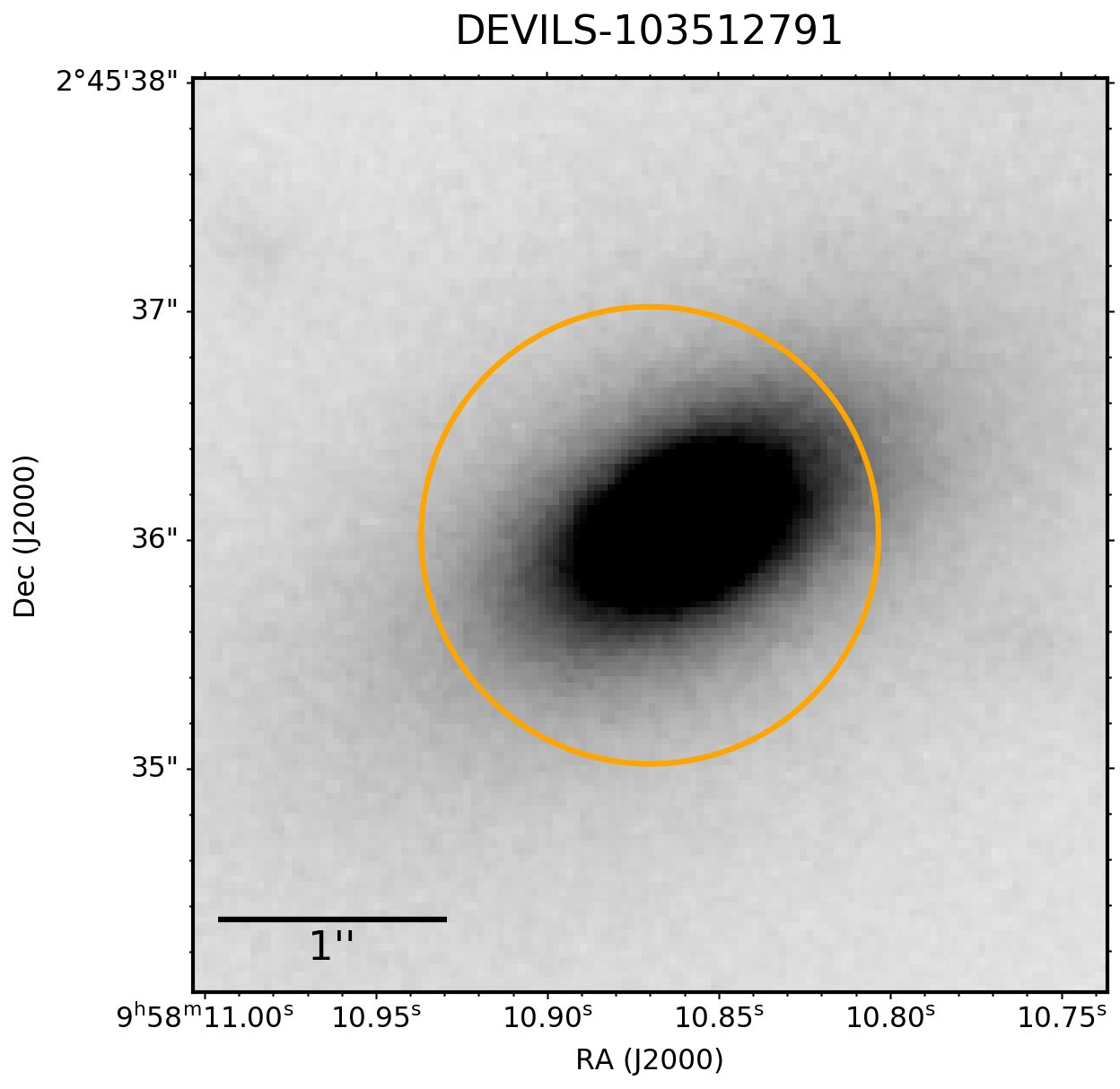}
\includegraphics[width=0.2\textwidth]{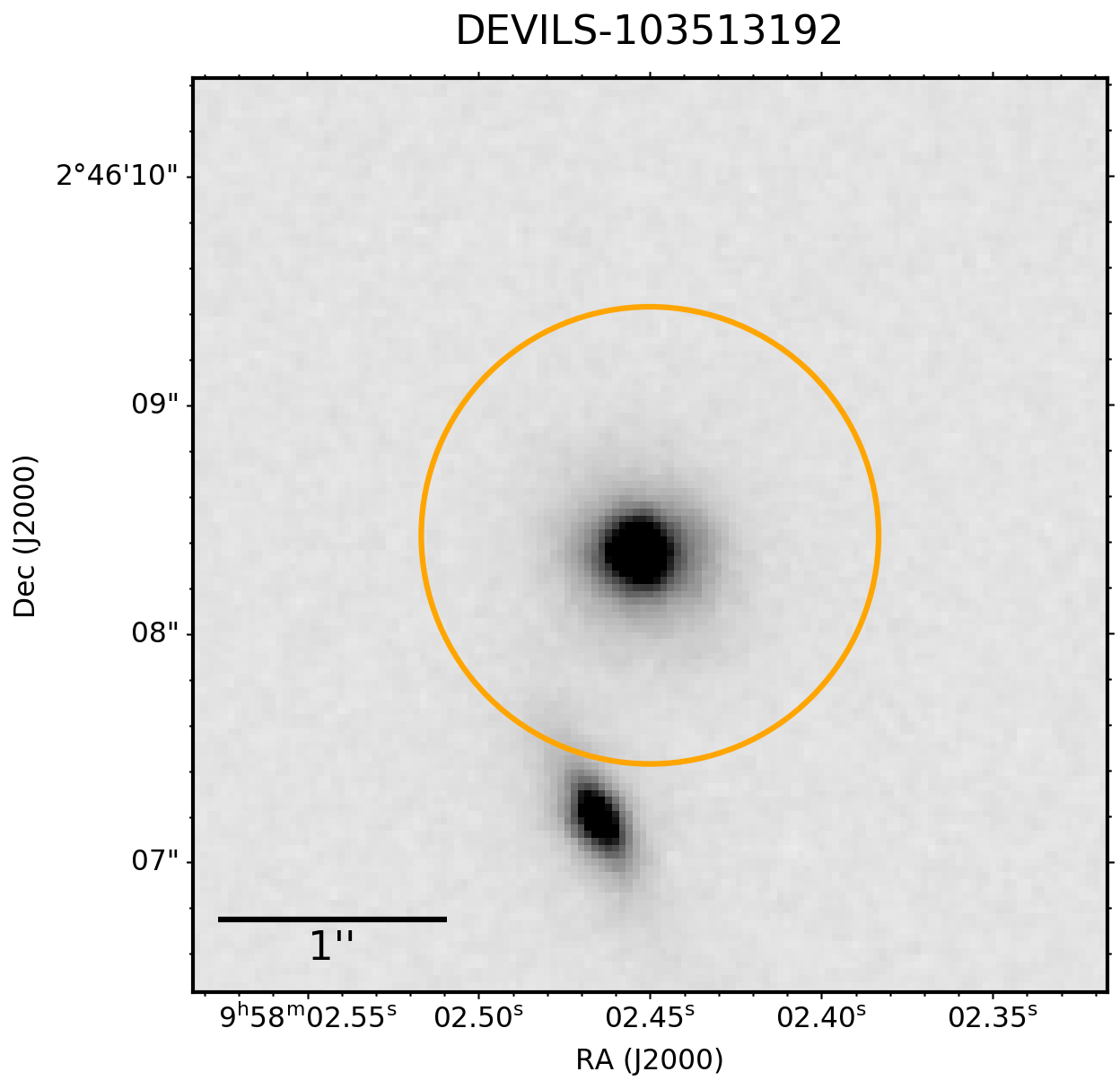}
\includegraphics[width=0.2\textwidth]{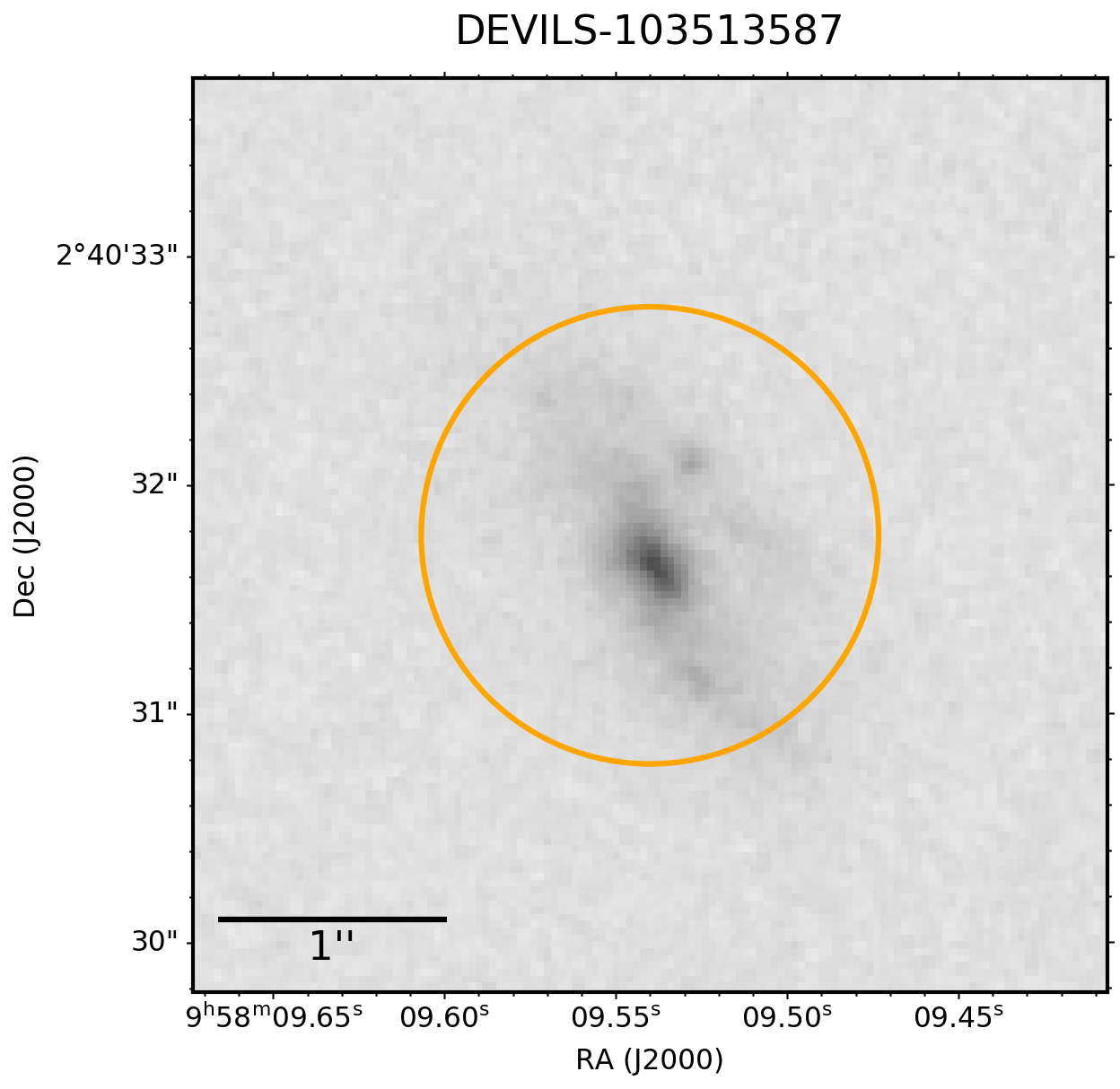}
    \caption{-- \textit{continued}.}
\end{figure*}


    
\begin{figure*}
    \centering
    \includegraphics[width=0.32\textwidth]{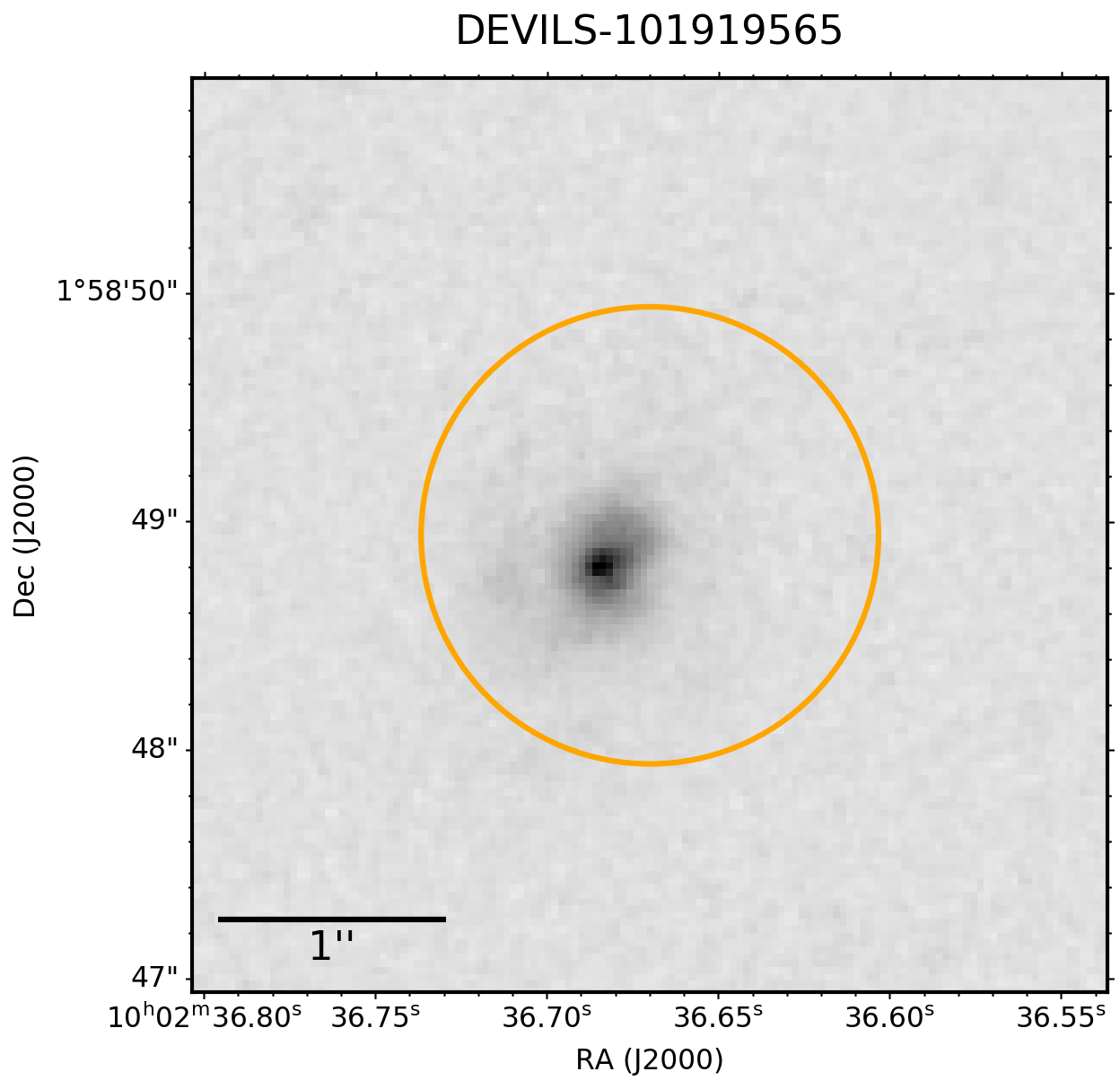}
    \includegraphics[width=0.32\textwidth]{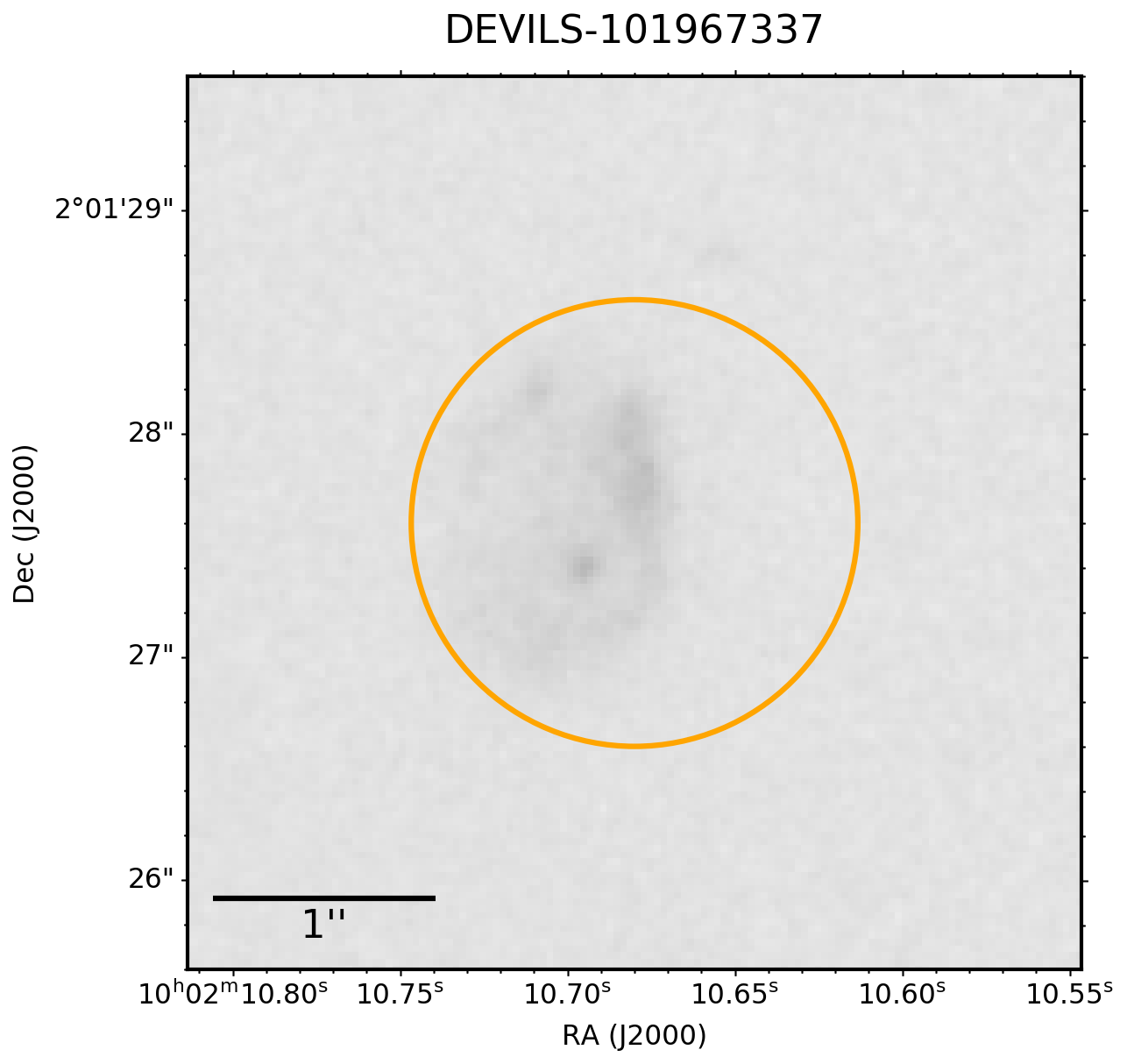}
    \includegraphics[width=0.32\textwidth]{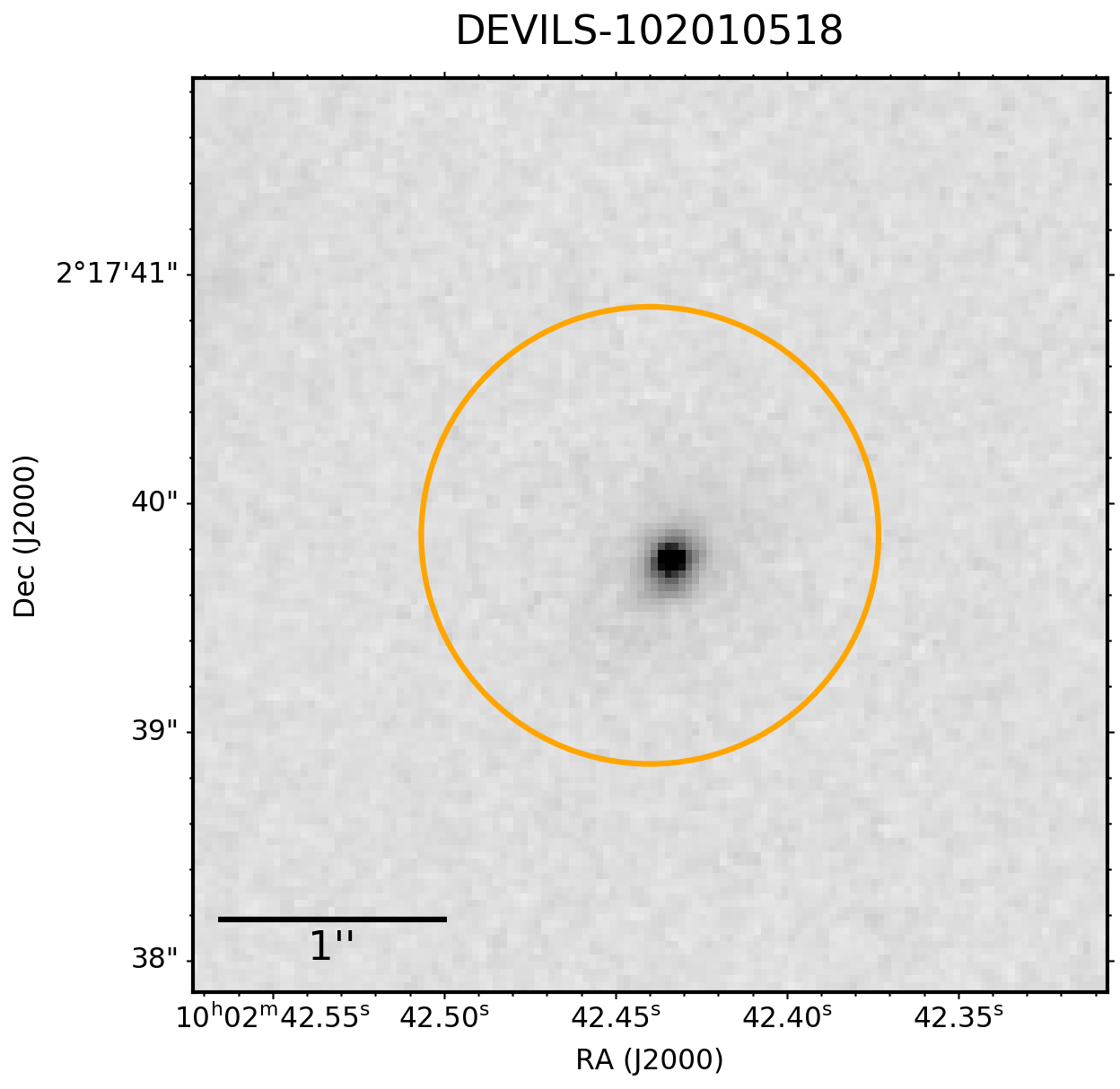}\\
    \includegraphics[width=0.32\textwidth]{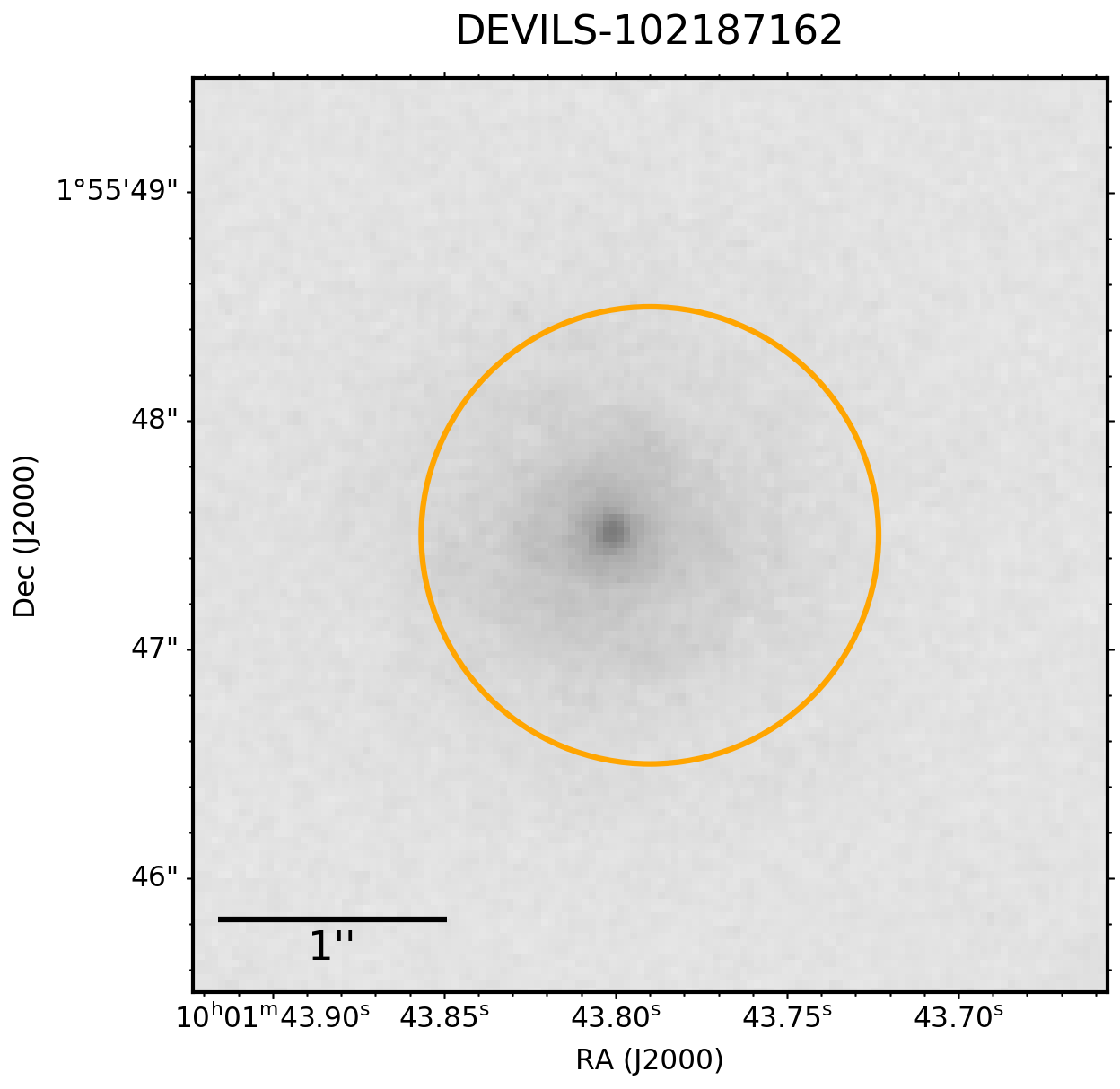}
    \includegraphics[width=0.32\textwidth]{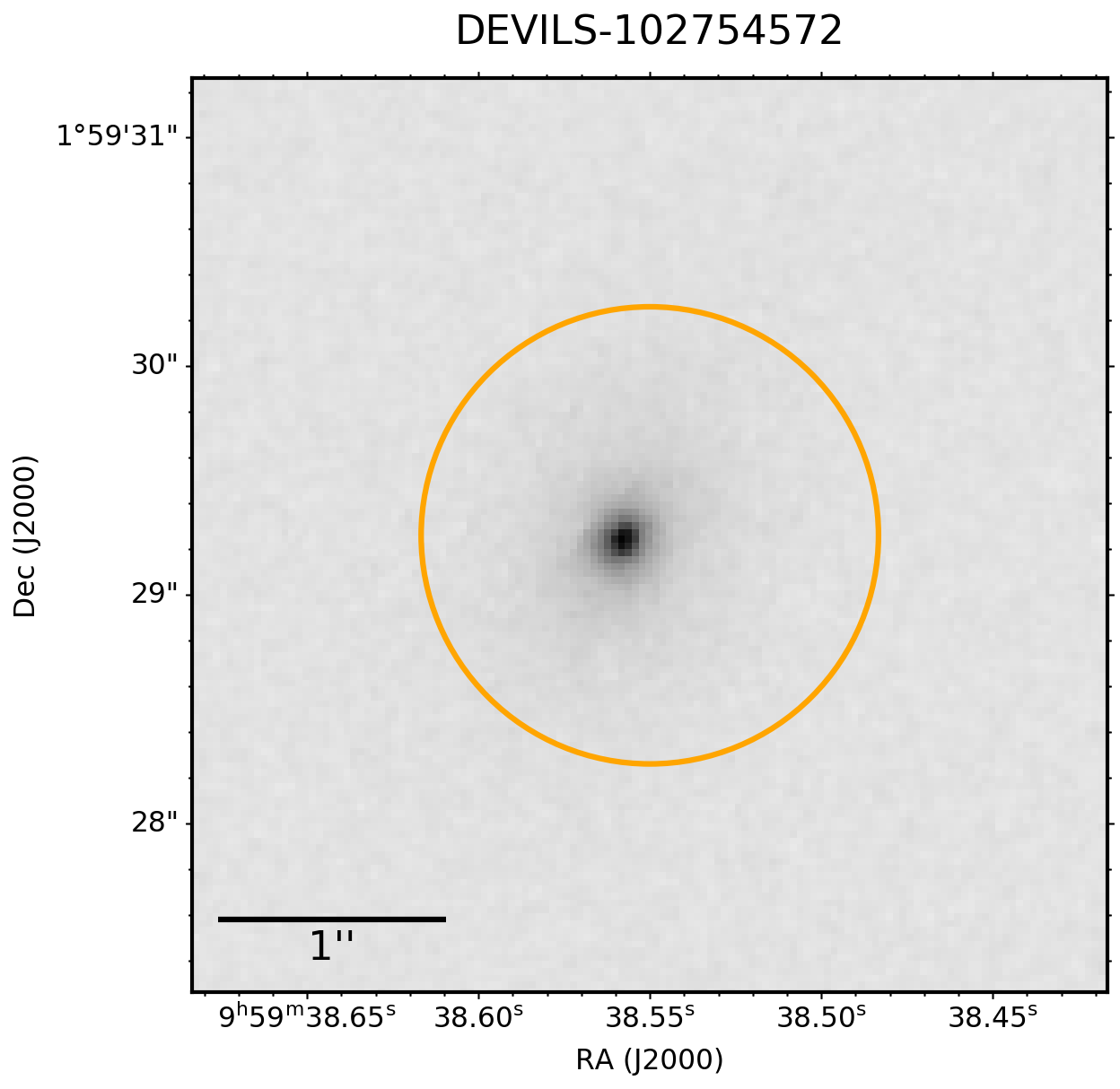}
    \includegraphics[width=0.32\textwidth]{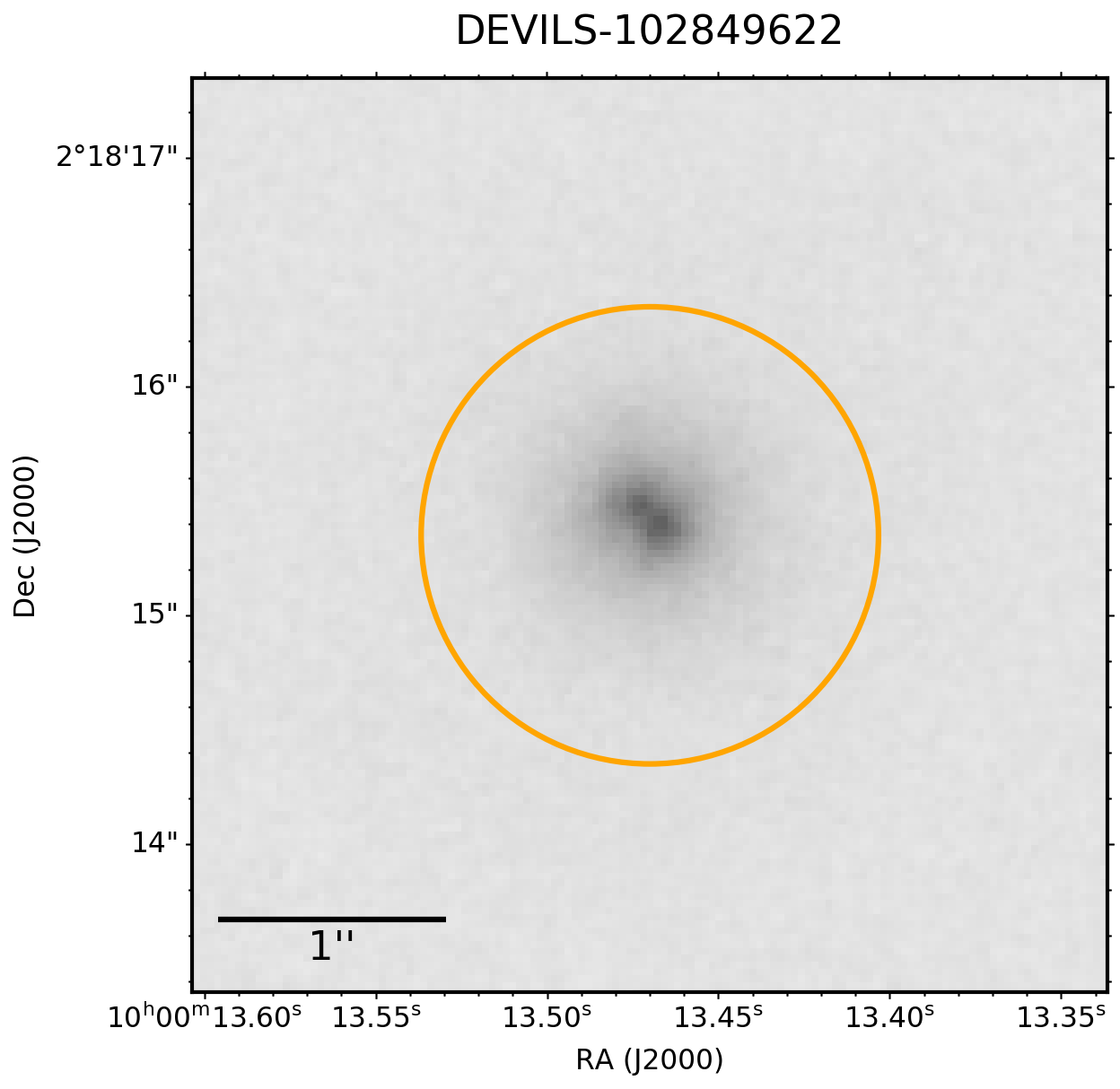}\\
    \includegraphics[width=0.32\textwidth]{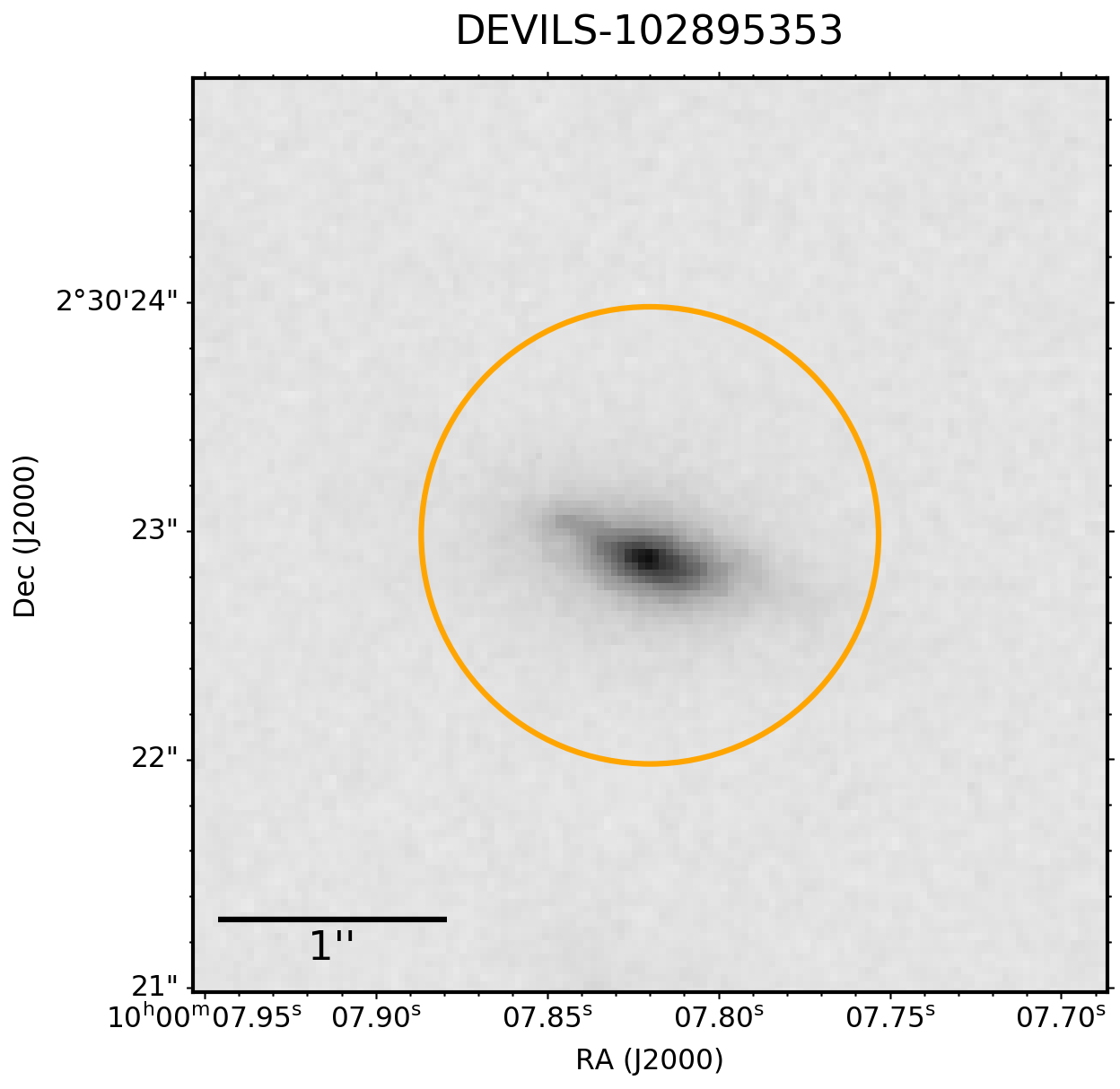}
    \includegraphics[width=0.32\textwidth]{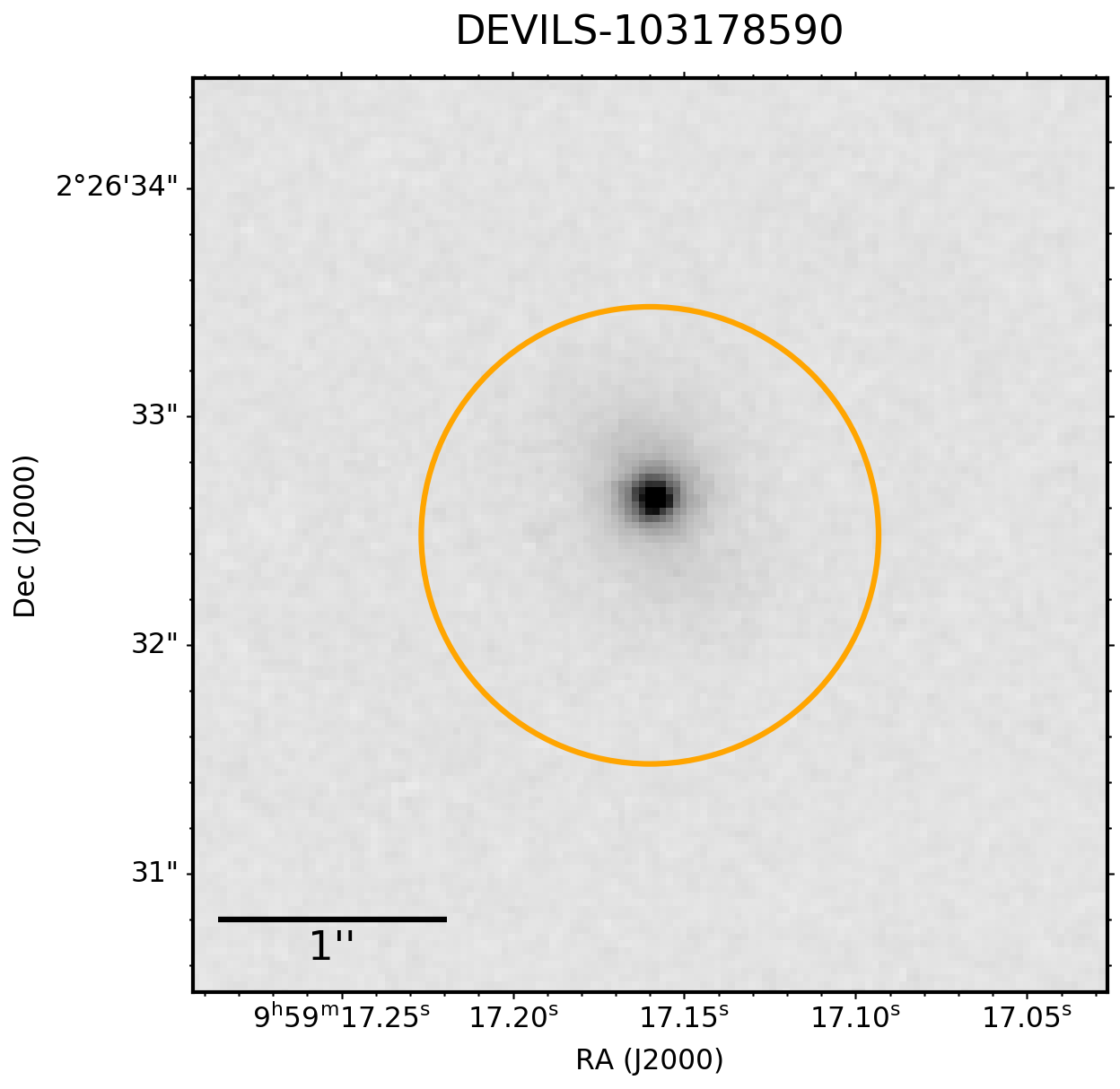}
    \includegraphics[width=0.32\textwidth]{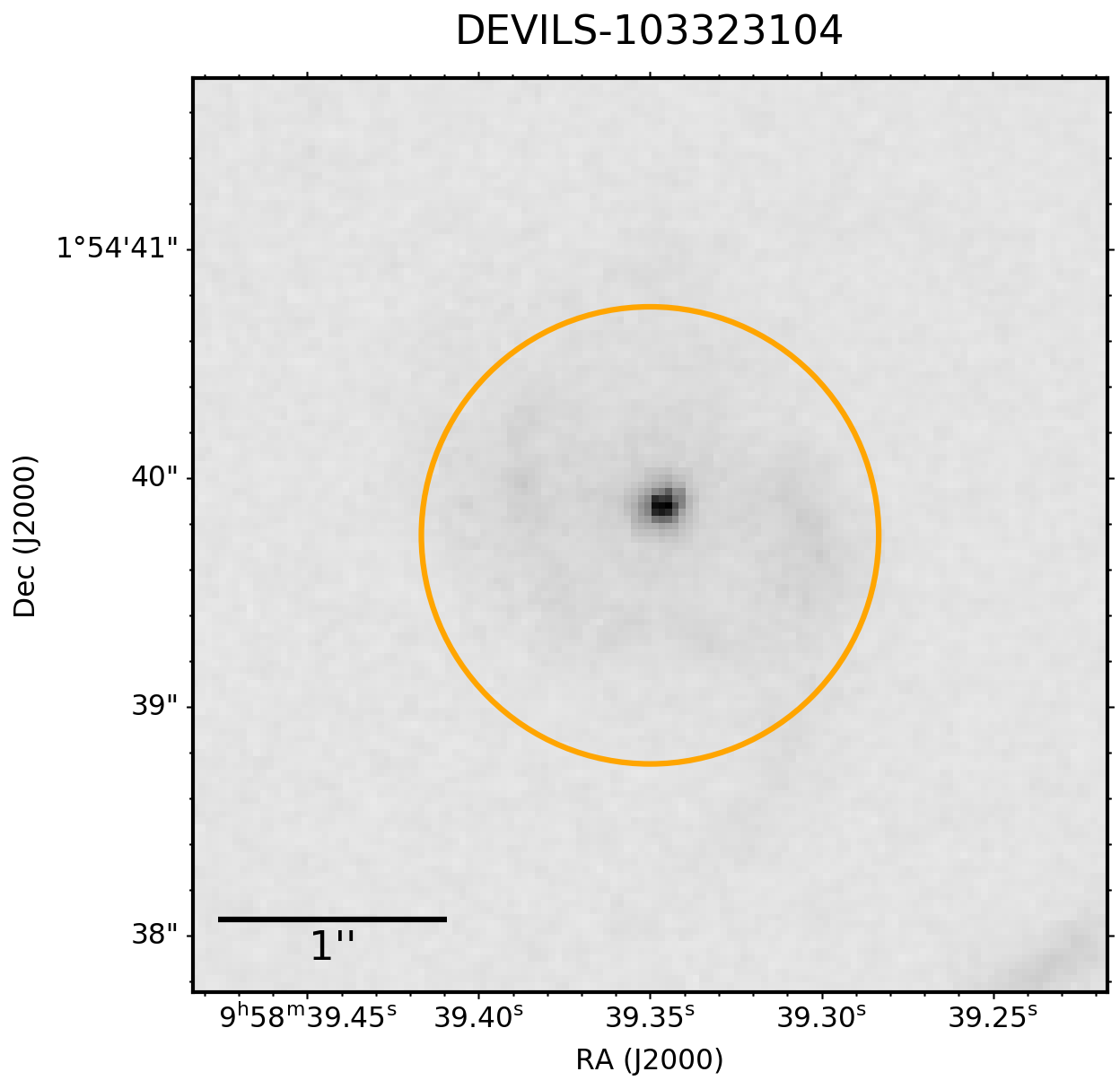}
    \caption{The candidate classical strong gravitational lenses in the DEVILS D10 sample, a massive elliptical with a more distant star-forming dwarf galaxy. In these COSMOS F814W images, it is difficult to identify source flux as it probes stellar continuum, not the star-formation (e.g. restframe H$\alpha$) of the source galaxy. }
    \label{f:ACS:lenscandidates}
\end{figure*}

\begin{table*}
    \centering
    \begin{tabular}{l l l l l l l l l}
    DEVILS ID & RA & DEC & $z_1$ & template-1 & $z_2$ & template-2 & Classification & Notes \\
    \hline
    \hline
  101919565 &  150.65279167  &   1.98026111 & 0.97 &   43 & 0.12 &   45 & lens  & \\ 
  101919822 &  150.61966667  &   1.91290278 & 0.31 &   44 & 0.38 &   45 & S-S   & \\ 
  101967337 &  150.54450000  &   2.02433333 & 0.06 &   46 & 0.87 &   45 & lens  & \\ 
  102010304 &  150.66491667  &   2.24220278 & 0.12 &   44 & 0.98 &   45 & S-E   & \\ 
  102010518 &  150.67683333  &   2.29440556 & 0.95 &   42 & 1.02 &   43 & lens  & \\ 
  102186670 &  150.42608333  &   1.97702778 & 0.32 &   45 & 0.25 &   44 & S     & single galaxy?\\ 
  102187162 &  150.43245833  &   1.92986111 & 0.25 &   47 & 0.68 &   47 & lens  & \\ 
  102187389 &  150.37729167  &   1.93966389 & 0.72 &   46 & 0.57 &   42 & E     & \\ 
  102233089 &  150.46775000  &   2.06401667 & 0.55 &   42 & 0.43 &   45 & S     & edgeon lens?  \\ 
  102233254 &  150.44391667  &   2.18347778 & 0.30 &   47 & 0.33 &   44 & S-S   & distant edge-on\\ 
  102233747 &  150.42983333  &   2.05285833 & 0.70 &   47 & 0.27 &   47 & S-E   & tidal arc? \\ 
  102234408 &  150.46829167  &   2.07676667 & 0.27 &   44 & 0.24 &   47 & S-S   & big spiral in front.  \\ 
  102280576 &  150.39179167  &   2.36102778 & 0.71 &   42 & 0.19 &   47 & S-S   & distant edge-on spiral  \\ 
  102280688 &  150.42820833  &   2.37439444 & 0.27 &   45 & 0.22 &   47 & S-E   & ocg  \\ 
  102326921 &  150.42233333  &   2.59472222 & 0.77 &   47 & 0.00 &   46 & S     & single\\ 
  102327087 &  150.41666667  &   2.51389444 & 0.89 &   46 & 0.07 &   46 & S-S   & small background galaxy\\ 
  102327200 &  150.52104167  &   2.46149444 & 0.48 &   42 & 0.25 &   46 & S-S   & ocg  \\ 
  102418565 &  150.18912500  &   1.69931667 & 0.23 &   46 & 0.20 &   46 & S     & HII or very distant galaxy?  \\ 
  102465873 &  150.19379167  &   1.82101944 & 0.89 &   42 & 0.83 &   42 & E-S   & ring galaxy?  \\ 
  102560691 &  150.21016667  &   2.31163611 & 0.75 &   46 & 0.30 &   47 & S-S   & triple system interaction.\\ 
  102562458 &  150.32162500  &   2.27933333 & 0.25 &   45 & 0.89 &   42 & S-E   & edge-on. OCG \\ 
  102609425 &  150.15820833  &   2.60825833 & 0.89 &   42 & 0.74 &   45 & E-E   & quadruple+ system \\ 
  102610080 &  150.13954167  &   2.59696667 & 0.72 &   45 & 0.28 &   47 & S-S   & distant edge-on ocg \\ 
  102705904 &  150.07375000  &   1.67840278 & 0.32 &   45 & 0.29 &   47 & S-?   & HII or distant galaxy  \\ 
  102753919 &  150.08550000  &   1.94860000 & 0.67 &   47 & 0.29 &   45 & S-?   & big face-on spiral, no clear background  \\ 
  102754572 &  149.91062500  &   1.99146111 & 0.85 &   42 & 0.79 &   42 & lens  &    \\ 
  102755602 &  150.02570833  &   1.87035000 & 0.12 &   45 & 0.98 &   45 & S     & no bg galaxy  \\ 
  102801921 &  149.99262500  &   2.20221667 & 0.72 &   45 & 0.28 &   47 & S-E   & ocg  \\ 
  102849622 &  150.05612500  &   2.30426389 & 0.73 &   45 & 0.29 &   47 & E/lens & double nucleus? arc?  \\ 
  102849770 &  150.05479167  &   2.31726389 & -0.00 &   14 & 1.06 &   42 & S    & face-on, no bg galaxy\\ 
  102850336 &  149.98779167  &   2.31567500 & 0.93 &   40 & 0.82 &   42 & S-E  & small bg galaxy  \\ 
  102850510 &  149.91050000  &   2.27933333 & 0.73 &   47 & 0.29 &   47 & E & no clear bg galaxy  \\ 
  102852273 &  149.94070833  &   2.33096111 & 0.73 &   45 & 0.29 &   47 & S & large face-on spiral, no clear bg  \\ 
  102895172 &  150.04562500  &   2.57875833 & 0.74 &   46 & 0.30 &   47 & S-S & ocg \\ 
  102895353 &  150.03258333  &   2.50638333 & 0.29 &   46 & 0.73 &   45 & lens & faint arc?  \\ 
  102895479 &  150.05483333  &   2.44130278 & 0.73 &   45 & 0.29 &   46 & S & faint bg galaxy?  \\ 
  102940540 &  150.04470833  &   2.75957778 & 0.33 &   46 & 0.13 &   47 & S-E & interaction/ocg  \\ 
  102940702 &  150.04012500  &   2.75761944 & 0.63 &   46 & 0.21 &   47 & S-S   & substructure or bg galaxy?  \\ 
  103084425 &  149.70545833  &   2.11458611 & 0.10 &   46 & 0.48 &   45 & E & no sign of bg source  \\ 
  103087126 &  149.88595833  &   2.17616389 & 0.44 &   45 & 0.07 &   47 & S-E & ocg, big face-on spiral  \\ 
  103131877 &  149.71279167  &   2.25295833 & 0.70 &   47 & 0.89 &   41 & S & substructure or bg galaxy?  \\ 
  103178590 &  149.82150000  &   2.44235556 & 0.23 &   47 & 0.65 &   47 & E/lens & LSB ring?  \\ 
  103224426 &  149.77075000  &   2.78281944 & 0.51 &   40 & 0.28 &   46 & E & no bg source  \\ 
  103225118 &  149.82491667  &   2.78900000 & 1.00 &   46 & 0.14 &   46 & S-S & multicomponent system  \\ 
  103225200 &  149.77250000  &   2.71349444 & 0.66 &   45 & 0.23 &   47 & S & big face-on spiral, bg source?  \\ 
  103274376 &  149.67195833  &   1.66864722 & 0.42 &   47 & 0.06 &   47 & E & bg source close to nucleus?  \\ 
  103275225 &  149.52533333  &   1.75563889 & 0.75 &   47 & 0.30 &   47 & E & faint nearby companion  \\ 
  103275987 &  149.56145833  &   1.69020000 & 0.65 &   47 & 0.23 &   47 & S-S & multicomponent system.  \\ 
  103323104 &  149.66395833  &   1.91104167 & 0.92 &   45 & 0.09 &   45 & lens & face-on ring spiral or lens?  \\ 
  103323224 &  149.58212500  &   1.88589444 & 0.63 &   47 & 0.21 &   47 & S & disturbed spiral  \\ 
  103323244 &  149.50720833  &   1.81979167 & 0.59 &   43 & 0.19 &   44 & S-E &   \\ 
  103372556 &  149.66279167  &   2.10482500 & 0.63 &   47 & 0.22 &   47 & S & bg or substructure  \\ 
  103374702 &  149.62220833  &   2.11793611 & 0.00 &   09 & 0.65 &   42 & S & LSB milky way star contamination  \\ 
  103512791 &  149.54529167  &   2.76000556 & 0.27 &   42 & 0.30 &   44 & E-S & big elliptical, bg spirals.  \\ 
  103513192 &  149.51020833  &   2.76900833 & 0.41 &   45 & 0.05 &   47 & S-E & ocg  \\ 
  103513587 &  149.53975000  &   2.67549444 & 0.43 &   47 & 0.92 &   45 & S-E &  \\ 
  103708094 &  149.45433333  &   2.44372500 & 0.61 &   46 & 0.31 &   45 & -  \\     
  \hline\hline
  \end{tabular}
    \caption{The DEVILS blended spectra in COSMOS (D10). ``ocg'' means an occulting galaxy pair, pairs good for measuring disk transparency at higher redshifts.The   }
    \label{t:cosmos:candidates}
\end{table*}

\section{Concluding Remarks}
\label{s:conluding}

In this paper we report the blended spectra in DEVILS DR1. For the COSMOS (D10) field, nine classical candidate lenses (elliptical galaxies as the potential lens) are identified with DEVILS promising to deliver more distant strong gravitational lenses and any previous survey. 
We stress that these are candidates to be confirmed still with other methods (AO imaging, an IFU map, or lensed \hi{} at the source galaxy). 
At the distance out to which DEVILS can identify strong lenses, resolving the characteristic arc may become challenging for even the most high resolution imaging. IFU and AO may help disentangle the signal from lens and source, and the detection of \hi{} line at the source redshift is considered strong confirmation. We note that for \hi{} lensing events, the lensing galaxy need not be an elliptical. Change alignment would allow the \hi{} signal to be magnified by any galaxy mass along the line of sight as lens galaxies are functionally transparent to the \hi{} signal. 

Spectroscopic identification of lensing galaxies thus far have had a high rate of confirmation, e.g. \cite{Chan16} reported a $>$70\% confirmation rate for a subset of 14 of the \cite{Holwerda15} lens candidates. Similar numbers have been reported for SLACS \citep{slacs1,slacs2,slacs3,slacs4}. However, confirmation that these are strongly lensing galaxies may have to come from other means for these higher redshift candidate lenses. The resolution needed to resolve the characteristic arc of a gravitational lens in these cases is even better than HST (Figure \ref{f:RE:hist}) but may be possible for ALMA or future observatories. 

In practice, confirmation is more likely to come from using 21cm emission line of atomic hydrogen (\hi). The lens galaxy is effectively transparent for the \hi{} signal from the source and it can be readily identified as it is separated in velocity space and the DEVILS spectroscopy provides a source redshift. 

\cite{Lipnicky18a} reported the first strongly lensed \hi{} detection, demonstrating this is a feasible approach. The DEVILS fields are all being targeted by intermediate and deep \hi{} surveys. COSMOS is the focus of CHILES with the JVLA \citep{Dodson15,Jones16a,Fernandez16,Hess19,Blue-Bird20}.  
XMM-LSS is targeted with L-band observations for MIGHTEE \citep{Jarvis16,Delhaize21,Maddox21} and ECDF-S is the target for the L- and UHF-band observations for LADUMA \citep{Holwerda11seeon,Blyth15a,Blyth16,Baker18}. 
The \hi{} disk of the source galaxy is much larger than the star-forming one, increasing the odds that a magnified signal can be identified \citep[see][]{Deane15}. Starting from candidate selection using DEVILS and similar spectroscopy, the odds of detecting strongly lensed and therefore magnified \hi{} from distant (i.e. $z>0.1$ for \hi{}) sources is set to increase significantly. Reporting the position and redshifts of potential lensed \hi{} sources for early \hi{} follow-up was the main motivation for this paper. 

The scientific use of such \hi{} detections is to constrain the \hi{} mass function \citep[\hi MF][]{Zwaan05} at higher redshift. At present, the \hi{} content at higher redshifts is mostly measured from stacked \hi{} spectra \citep{Fabello11b,Fabello12,Delhaize13,Maddox13,Gereb15,Bera19,Guo20,Chowdhury20,Chowdhury21}.  With strongly lensed \hi{} detections, and a reasonable prior for the magnification distribution, a wholly separate constraint can be given on the low-mass end of the \hi MF. 

In addition to the \hi MF science, these gravitational lens candidates can be a useful cosmological probe. 
Increased numbers of strong gravitational lenses to monitor for supernovae in the source galaxy open the tantalizing possibility of time-delay cosmography \citep{Treu16, Oguri19}. The possibility of the detection of a strongly lensed supernova has long been considered to be a realistic prospect in statistical samples with modern survey cadence and sensitivity.  \citep{Wang00,Porciani00,Goobar02a,Goobar02b,Holz01,Kostrzewa-Rutkowska13}. 
A few strongly lensed supernova in galaxy-galaxy lenses have already been reported in iPTF, a precursor to the Zwicky Transient Factory \citep{Goobar17}, and in Pan-STARRS \citep{Quimby14}. 
The most viable observational approach is to monitor a great number of suspected galaxy-galaxy lenses using  near-future observatories, e.g. the James Webb Space Telescope \citep[JWST,][]{Petrushevska18}, the Vera Rubin Observatory \citep{Liao18,Tu19}, and Zwicky Transient Survey \citep[ZTF,][]{Shu18,Goldstein19}.

A few well-characterized lenses can make a significant difference in cosmography to determine the expansion rate of the Universe. For example, the H0LICOW\footnote{H0 Lenses in COSMOGRAIL’s Wellspring \citep{Suyu17}.} project has monitored \textit{four} strong gravitational lenses with a quasar behind it for some time \citep[cf.][]{Suyu17, Birrer19} and constrained the expansion rate independently of the other two measurements, Supernovae Type Ia distances  \citep[$H_0 = ~ 73 ~ \rm km/s/Mpc$,][]{Beaton16, Riess16, Riess18, Riess19} and the cosmic microwave background \citep[$H_0 = ~ 68 ~ \rm km/s/Mpc$,][]{Planck-Collaboration20}. The H0LICOW result has not yet broken the current tension between the two values for $H_0$ but this approach shows great promise.  Quasars do change over time but much less predictably than supernovae. With a few lensed sources and more reliable timing, lensed supernovae offer a very attractive alternative for cosmography. Strongly lensing candidates at higher redshifts, even without resolved imaging, may offer additional constraints. 

Many of the new DEVILS candidate lenses are at greater distances offering a longer expected time-delay, which is a function of the lens and source distances. Presently, time-delay supernovae are limited to lower redshift \cite[e.g. $z<0.1$,][]{Quimby14,Goobar17}.   
The longer expected time-delay offers the option to measure the Hubble constant ($H_0$) from the multiple imaged supernova more accurately. The multiple images need not be resolved but each lightcurve does need to be resolved in the time-series. With a time-delay of order of weeks, a supernova lightcurve could easily be modeled for each event, allowing for a greater relative accuracy of the delay and hence inferred Hubble expansion rate. Each DEVILS target field is also a Vera Rubin Observatory deep drilling field and the LADUMA target field (CDFS) is in the THUNDERKAT \citep{Fender17} transient search, in conjunction with the MeerLicht optical transient searches \citep{Woudt21,Hosenie21}. 
The emphasis on transient searches in these fields, makes  candidate strong gravitational lenses in them especially attractive for the topic of cosmography.

The next generation of spectroscopic surveys is being planned with the 4MOST instrument \citep{de-Jong12a,Depagne15}. Planned extra-galactic surveys are the two-tiered Wide Area Vista Extragalactic Survey \citep[WAVES,][]{Driver19}, the Optical, Radio Continuum and HI Deep Spectroscopic Survey (ORCHIDSS, Duncan et al. \textit{in prep.}), and a cosmological low-s/n wide-area survey \citep[CRS,][]{Richard19}. These 4MOST surveys are expected to achieve high completeness in their target fields and yield a boon of spectroscopically confirmed strong lensing systems, especially those with more distant lenses and lower masses.

Strong gravitational lenses will continue to be identified in future spectroscopic surveys, e.g. WAVES and ORCIDSS using the 4MOST instrument. Confirmation of these lenses may come in the form of magnified \hi{} rather \citep[cf][]{Deane15, Blecher19} than resolved imaging of the lensing arcs or high resolution imaging even with the Roman or Euclid space telescopes or targeted follow-up with JWST. Future data-releases from DEVILS promise to improve the numbers of strong lenses known at $z\sim1$ in the immediate future.  

\section{Data Availability Statement}

DEVILS DR1 will be made available with a separate publication via the project website\footnote{\url{https://devilsurvey.org/wp/}}. Relevant blended spectrum information will be part of the full DR1 as well as listed in the tables of this manuscript.

\section*{Acknowledgements}

This work has been supported by the European Union's Horizon 2020 Research and Innovation Programme under the Maria Sklodowska-Curie grant agreement (No. 754510), the Polish National Science Centre (UMO-2016/23/N/ST9/02963), the Spanish Ministry of Science and Innovation through the Juan de la Cierva-formacion programme (FJC2018-038792-I)

This research made use of APLpy, an open-source plotting package for Python hosted at \url{http://aplpy.github.com}
This research made use of Astropy, a community-developed core Python package for Astronomy \citep{Astropy-Collaboration13,Astropy-Collaboration18}.


\appendix

\section{Blended Spectra in DEVILS}

In this Appendix, we list the DEVILS object flagged as candidate blended spectra, listing their positions and redshifts in two additional tables, one for XMM-LSS (D02), Table \label{t:xmm:candidates} and one for ECDF-S (D03), Table \label{t:cdfs:candidates}. We show the spectra for the candidates in the COSMOS (D10) field listed in Table \label{t:cosmos:candidates} with the spectra shifted to the resframe wavelength for both redshift solutions in Figure \ref{f:spectra}. 

\begin{table*}
    \centering
    \begin{tabular}{l l l l l l l }
    DEVILS ID & RA & DEC & $z_1$ & template-1 & $z_2$ & template-2  \\
    \hline
    \hline
23945817 &   35.50616667  &  -4.54419722 & 0.95 &   43 & 1.02 &   43 \\
  23995608 &   35.48341667  &  -4.33065000 & 0.76 &   47 & 0.00 &   45 \\
  23996308 &   35.46912500  &  -4.21832500 & 0.70 &   45 & 0.27 &   46 \\
  24041251 &   35.37587500  &  -5.07945833 & 0.38 &   47 & 0.27 &   44 \\
  24142011 &   35.24208333  &  -4.51141389 & 0.79 &   46 & 0.01 &   46 \\
  24143221 &   35.21520833  &  -4.59569444 & 0.27 &   46 & 0.30 &   46 \\
  24193078 &   35.24100000  &  -4.38532500 & 0.46 &   45 & 0.08 &   47 \\
  24241442 &   34.98895833  &  -5.11926667 & 0.62 &   47 & 0.20 &   47 \\
  24241567 &   34.96800000  &  -5.15127222 & 0.25 &   44 & 0.22 &   47 \\
  24242039 &   35.03679167  &  -4.99350833 & 0.14 &   45 & 1.00 &   45 \\
  24242215 &   35.06750000  &  -5.16637500 & 0.63 &   46 & 0.21 &   47 \\
  24293199 &   35.08458333  &  -4.94779167 & 0.27 &   44 & 0.30 &   47 \\
  24293268 &   35.01629167  &  -4.71430556 & 0.25 &   44 & 0.26 &   47 \\
  24293310 &   35.12950000  &  -4.70440556 & 0.74 &   45 & 0.29 &   46 \\
  24293967 &   34.98733333  &  -4.90861111 & 0.82 &   45 & 0.03 &   47 \\
  24294352 &   35.15616667  &  -4.87894167 & 0.75 &   47 & 0.30 &   47 \\
  24345222 &   35.02795833  &  -4.61946944 & 0.63 &   41 & 0.24 &   47 \\
  24345466 &   35.02870833  &  -4.63459444 & 0.77 &   45 & 0.00 &   45 \\
  24345663 &   35.17904167  &  -4.66834444 & 0.95 &   43 & 1.02 &   43 \\
  24394824 &   35.09225000  &  -4.27165833 & 0.60 &   42 & -4.71 &   12 \\
  24395213 &   35.18966667  &  -4.27158889 & 0.48 &   45 & 0.10 &   47 \\
  24440571 &   34.84237500  &  -5.08841389 & 0.31 &   47 & 0.28 &   47 \\
  24442735 &   34.88283333  &  -5.10223056 & 0.71 &   47 & 0.27 &   47 \\
  24491257 &   34.95154167  &  -4.90585833 & 0.76 &   47 & 0.31 &   47 \\
  24491480 &   34.87079167  &  -4.86123611 & 0.31 &   44 & 0.27 &   47 \\
  24491527 &   34.93812500  &  -4.86090556 & 0.28 &   46 & 0.72 &   47 \\
  24491585 &   34.92358333  &  -4.74284444 & 0.75 &   46 & 0.30 &   47 \\
  24491650 &   34.90016667  &  -4.80743333 & 0.65 &   45 & 0.23 &   47 \\
  24491864 &   34.76579167  &  -4.79886389 & 0.33 &   45 & 0.30 &   47 \\
  24491889 &   34.85179167  &  -4.84960833 & 0.71 &   47 & 0.27 &   47 \\
  24492071 &   34.95025000  &  -4.85865833 & 0.82 &   42 & 0.04 &   45 \\
  24492650 &   34.91337500  &  -4.89559167 & 0.21 &   46 & 0.62 &   45 \\
  24492762 &   34.73716667  &  -4.75255278 & 0.89 &   46 & 0.07 &   46 \\
  24544023 &   34.79745833  &  -4.53279167 & 0.03 &   46 & 0.81 &   43 \\
  24544177 &   34.88254167  &  -4.46093611 & 0.33 &   47 & 0.33 &   47 \\
  24596487 &   34.86800000  &  -4.31477778 & 0.75 &   41 & 0.04 &   47 \\
  24596709 &   34.83625000  &  -4.26286667 & 0.44 &   47 & 0.07 &   47 \\
  24596900 &   34.96254167  &  -4.21835000 & 0.69 &   47 & 0.26 &   47 \\
  24640004 &   34.60112500  &  -4.99404167 & 0.71 &   47 & 0.81 &   41 \\
  24640069 &   34.63387500  &  -5.01778889 & 0.65 &   47 & 0.87 &   40 \\
  24690693 &   34.49441667  &  -4.81551111 & 0.71 &   45 & 0.28 &   47 \\
  24690952 &   34.51208333  &  -4.89002778 & 0.75 &   45 & 0.30 &   47 \\
  24691156 &   34.63895833  &  -4.75832500 & 0.71 &   47 & 0.27 &   47 \\
  24692379 &   34.62654167  &  -4.79908333 & 0.72 &   45 & 0.28 &   47 \\
  24740073 &   34.65158333  &  -4.45387778 & 0.79 &   47 & 0.02 &   44 \\
  24837980 &   34.28087500  &  -5.15114167 & 0.20 &   45 & 1.12 &   45 \\
  24838186 &   34.37887500  &  -5.02776389 & 0.80 &   47 & 0.02 &   44 \\
  24838207 &   34.24262500  &  -5.06420000 & 0.25 &   47 & 0.68 &   45 \\
  24838544 &   34.46829167  &  -5.09759167 & 0.62 &   40 & 0.53 &   44 \\
  24942871 &   34.37150000  &  -4.63441111 & 0.71 &   47 & 0.27 &   47 \\
  24943476 &   34.28445833  &  -4.58350833 & 0.21 &   45 & 0.18 &   47 \\
  24992849 &   34.46066667  &  -4.28522778 & 0.74 &   47 & 0.45 &   45 \\
  24992987 &   34.38341667  &  -4.40945278 & 0.43 &   47 & 0.07 &   47 \\
  24993238 &   34.29933333  &  -4.28556389 & 0.15 &   45 & 0.16 &   47 \\
  24993260 &   34.33441667  &  -4.38700833 & 0.83 &   47 & 0.04 &   47 \\
  24993840 &   34.33962500  &  -4.40013611 & 0.25 &   47 & 0.68 &   45 \\
  25039757 &   34.22391667  &  -5.11365278 & 0.39 &   47 & 0.71 &   46 \\
  25039908 &   34.21112500  &  -5.13786667 & 0.67 &   47 & 0.25 &   45 \\
  25040091 &   34.03766667  &  -4.97813611 & 0.48 &   45 & 0.10 &   47 \\
  25040892 &   34.09312500  &  -5.06527778 & 0.14 &   45 & 1.01 &   45 \\
  25108384 &   34.00550000  &  -4.81034167 & 0.64 &   45 & 0.38 &   46 \\
  25108806 &   34.16625000  &  -4.73569444 & 0.71 &   45 & 0.27 &   46 \\
  25108841 &   34.21095833  &  -4.82175556 & 0.36 &   47 & 0.01 &   47 \\
  25109419 &   34.17883333  &  -4.69961667 & 0.30 &   47 & 0.74 &   45 \\
  \hline\hline
\end{tabular}
    \caption{The DEVILS blended spectra in XMM-LSS (D02). }
    \label{t:xmm:candidates}
\end{table*}

\setcounter{table}{2}

\begin{table*}
    \centering
    \begin{tabular}{l l l l l l l }
    DEVILS ID & RA & DEC & $z_1$ & template-1 & $z_2$ & template-2  \\
    \hline
    \hline
  25109652 &   34.11029167  &  -4.92928611 & 0.46 &   46 & 0.08 &   47 \\
  25175636 &   34.11670833  &  -4.46486944 & 0.72 &   43 & 0.69 &   40 \\
  25176091 &   34.01475000  &  -4.55083056 & 0.50 &   47 & 0.12 &   47 \\
  25176779 &   34.17937500  &  -4.53499722 & 0.75 &   47 & 0.30 &   46 \\
  25241942 &   34.19083333  &  -4.35923611 & 0.42 &   42 & 0.25 &   47 \\
  25242440 &   34.18587500  &  -4.26509167 & 0.35 &   40 & 9.11 &   12 \\
  25242758 &   34.06904167  &  -4.29557778 & 0.72 &   47 & 0.28 &   47 \\
  25242789 &   34.19308333  &  -4.23933056 & 0.43 &   47 & 0.09 &   46 \\
  25243097 &   34.08058333  &  -4.40158056 & 0.14 &   47 & 0.49 &   45 \\
  25243428 &   34.17683333  &  -4.32388333 & 0.45 &   47 & 0.08 &   47 \\
  25243492 &   34.07950000  &  -4.28745000 & 0.45 &   47 & 0.07 &   47 \\
  \hline\hline
\end{tabular}
    \caption{-- \textit{continued} }
\end{table*}

\begin{table*}
    \centering
    \begin{tabular}{l l l l l l l }
    DEVILS ID & RA & DEC & $z_1$ & template-1 & $z_2$ & template-2  \\
    \hline
    \hline
31364972 &   53.89120833  & -27.76997222 & 0.95 &   42 & 1.02 &   43 \\
  31366150 &   53.77891667  & -27.95961944 & 0.63 &   45 & 0.00 &   22 \\
  31420864 &   53.85745833  & -27.67635556 & 0.20 &   44 & 0.00 &   47 \\
  31470431 &   53.53220833  & -28.35941944 & 0.31 &   47 & -2.40 &   09 \\
  31470803 &   53.55166667  & -28.44153889 & 0.08 &   47 & 0.65 &   42 \\
  31472266 &   53.58391667  & -28.37151944 & 0.71 &   47 & 0.27 &   47 \\
  31580189 &   53.54012500  & -27.90944444 & 0.10 &   44 & 0.41 &   42 \\
  31580674 &   53.71579167  & -27.82735556 & 0.61 &   47 & 0.19 &   47 \\
  31580973 &   53.63183333  & -27.77484722 & 0.47 &   47 & 0.92 &   45 \\
  31581136 &   53.60870833  & -27.83990833 & 0.74 &   47 & 0.30 &   47 \\
  31581150 &   53.62883333  & -27.90129722 & 0.40 &   46 & 0.88 &   45 \\
  31581359 &   53.56654167  & -27.78446389 & 0.67 &   45 & 0.24 &   46 \\
  31585572 &   53.43470833  & -27.87049167 & 0.73 &   45 & 0.29 &   47 \\
  31635177 &   53.49095833  & -27.60029722 & 0.77 &   45 & 0.01 &   44 \\
  31635948 &   53.57295833  & -27.57112222 & 0.78 &   46 & 0.01 &   43 \\
  31737748 &   53.30095833  & -28.02351111 & 0.29 &   47 & 0.73 &   45 \\
  31738799 &   53.14987500  & -28.05283611 & 0.21 &   46 & 0.63 &   47 \\
  31739322 &   53.19358333  & -28.00018889 & 0.28 &   47 & 0.73 &   45 \\
  31897468 &   52.86245833  & -28.30308056 & 0.28 &   47 & 0.73 &   47 \\
  31897890 &   52.93775000  & -28.32748333 & 0.71 &   47 & 0.28 &   47 \\
  31898084 &   52.84445833  & -28.40419167 & 0.86 &   40 & 0.79 &   47 \\
  31898601 &   53.08812500  & -28.25800556 & 0.77 &   45 & 0.01 &   45 \\
  31898627 &   52.97508333  & -28.30872500 & 0.64 &   47 & 0.22 &   47 \\
  31898958 &   53.05341667  & -28.34696111 & 0.71 &   47 & 0.28 &   47 \\
  31899017 &   52.92825000  & -28.41990278 & 0.28 &   47 & 0.72 &   45 \\
  31953454 &   52.84104167  & -28.15541389 & 0.62 &   45 & 0.21 &   47 \\
  31953908 &   52.88858333  & -28.09198611 & 0.00 &   13 & 0.36 &   47 \\
  31954005 &   53.05487500  & -28.20985556 & 1.03 &   43 & 0.15 &   45 \\
  32008128 &   52.80141667  & -27.83991111 & 0.28 &   44 & 0.25 &   46 \\
  32064222 &   52.86333333  & -27.70248333 & 0.57 &   46 & 0.48 &   41 \\
  32064590 &   53.00425000  & -27.55768333 & 0.73 &   47 & 0.29 &   47 \\
  32064684 &   53.06500000  & -27.51943889 & 0.58 &   45 & 0.18 &   47 \\
  32064950 &   52.79958333  & -27.66902222 & 0.31 &   44 & 0.28 &   46 \\
  32065013 &   53.00483333  & -27.55390556 & 0.73 &   47 & 0.29 &   47 \\
  32115739 &   52.64220833  & -28.26730833 & 0.32 &   45 & 0.77 &   45 \\
  32115855 &   52.71337500  & -28.28865278 & 0.75 &   47 & 0.30 &   46 \\
  32116509 &   52.79425000  & -28.32276667 & 0.68 &   47 & 0.25 &   44 \\
  32170820 &   52.74045833  & -28.09380000 & 0.49 &   47 & 0.11 &   46 \\
  32171195 &   52.76537500  & -28.03151389 & 0.66 &   41 & 0.30 &   44 \\
  32172424 &   52.66054167  & -28.11606111 & 0.74 &   45 & 0.29 &   47 \\
  32172833 &   52.76887500  & -28.09402222 & 0.93 &   45 & 0.10 &   43 \\
  32227404 &   52.48862500  & -27.77536667 & 0.74 &   45 & 0.30 &   47 \\
  32227786 &   52.58745833  & -27.82791389 & 0.28 &   47 & 0.73 &   45 \\
  32227924 &   52.68816667  & -27.78745833 & 0.79 &   45 & 0.01 &   47 \\
  32228358 &   52.62358333  & -27.84935556 & 0.30 &   46 & 0.74 &   46 \\
  32283863 &   52.72366667  & -27.60104722 & 0.65 &   46 & 0.23 &   47 \\
  32283912 &   52.63470833  & -27.64496944 & -0.00 &   44 & 0.76 &   45 \\
  32283973 &   52.79820833  & -27.52830556 & 0.12 &   46 & 0.23 &   46 \\
  32377963 &   52.40470833  & -28.24047222 & 0.76 &   40 & 0.28 &   44 \\
  \hline\hline
\end{tabular}
    \caption{The DEVILS blended spectra in ECDF-S (D03). }
    \label{t:cdfs:candidates}
\end{table*}


\clearpage
\newpage

\begin{figure*}
    \centering
\includegraphics[width=0.49\textwidth]{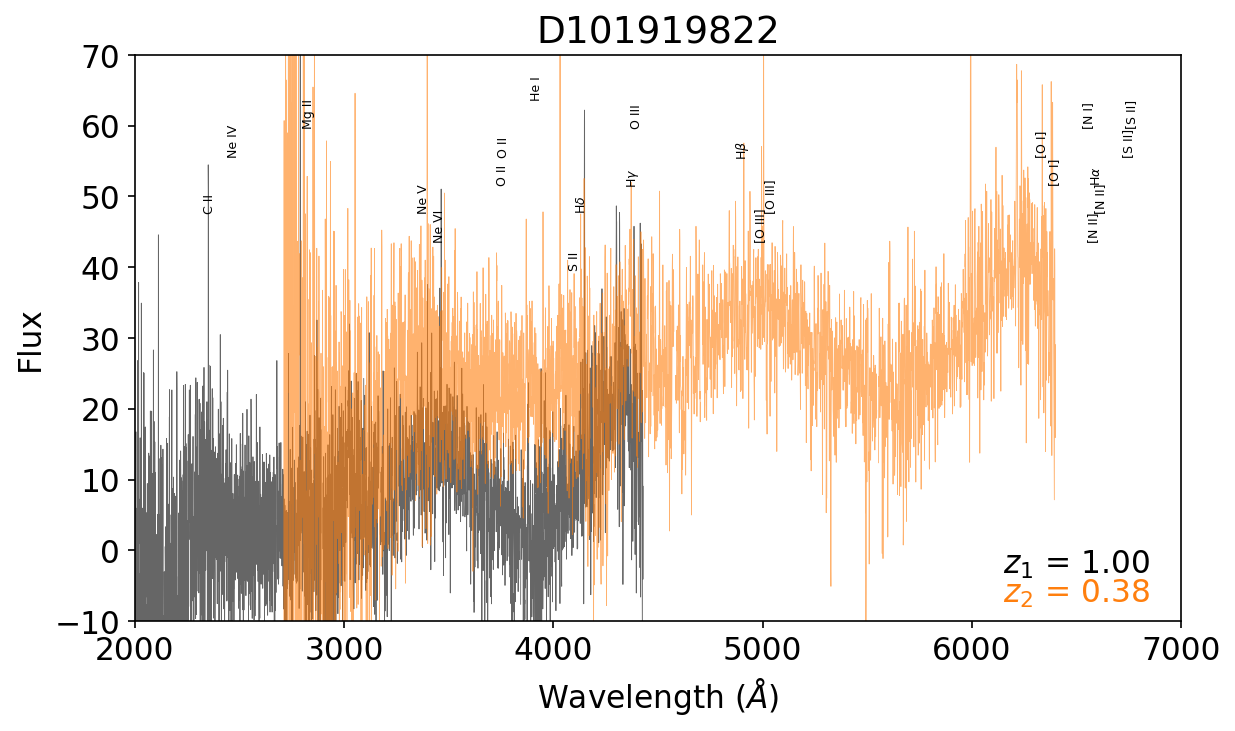}
\includegraphics[width=0.49\textwidth]{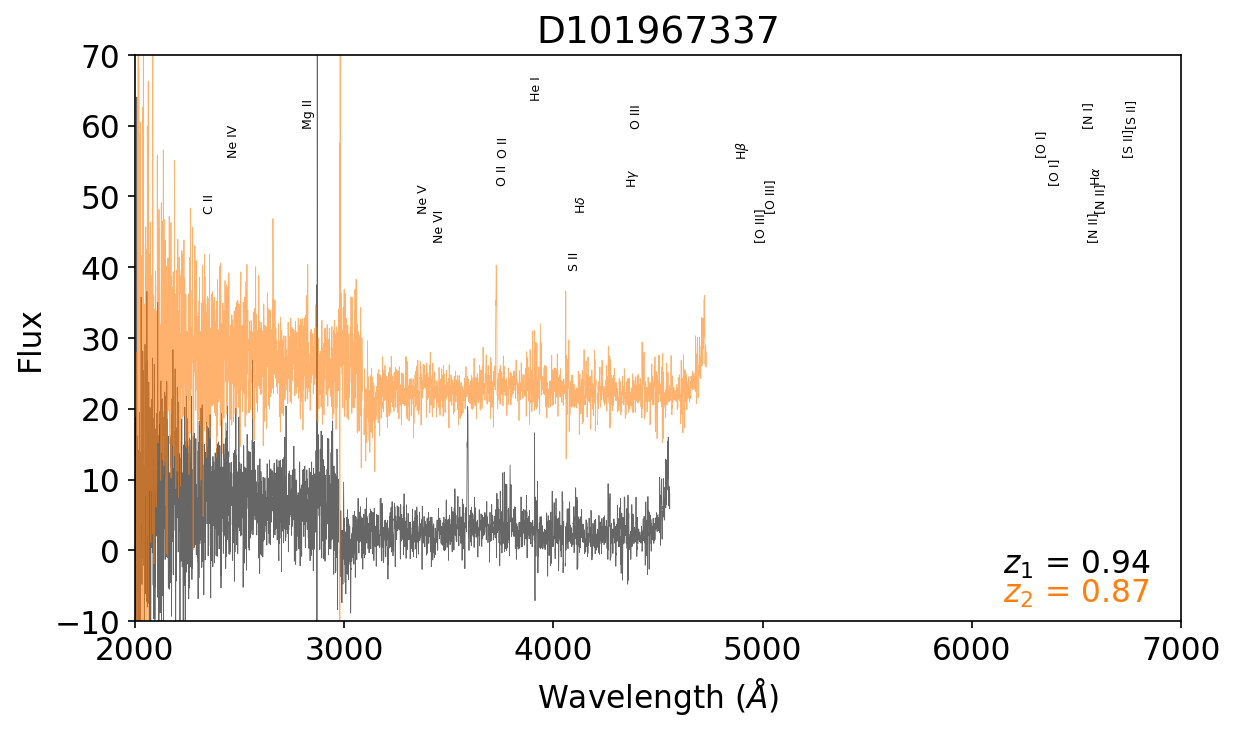}
\includegraphics[width=0.49\textwidth]{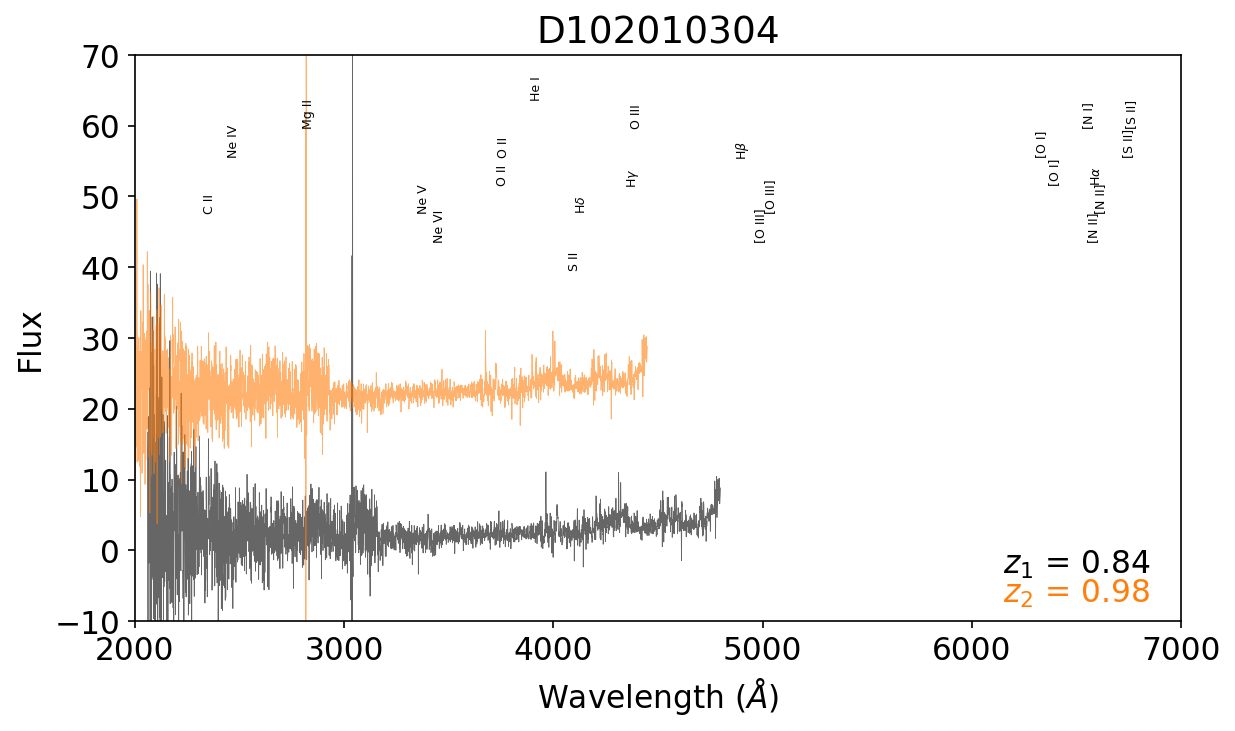}
\includegraphics[width=0.49\textwidth]{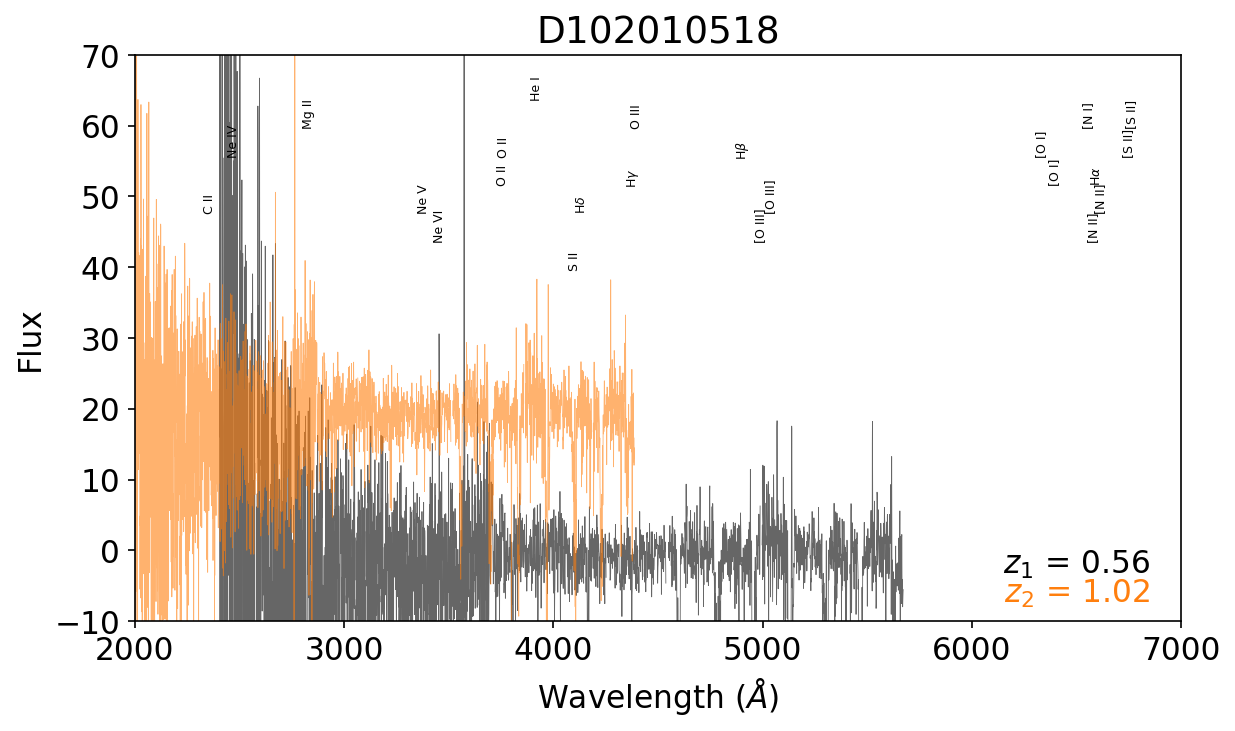}
\includegraphics[width=0.49\textwidth]{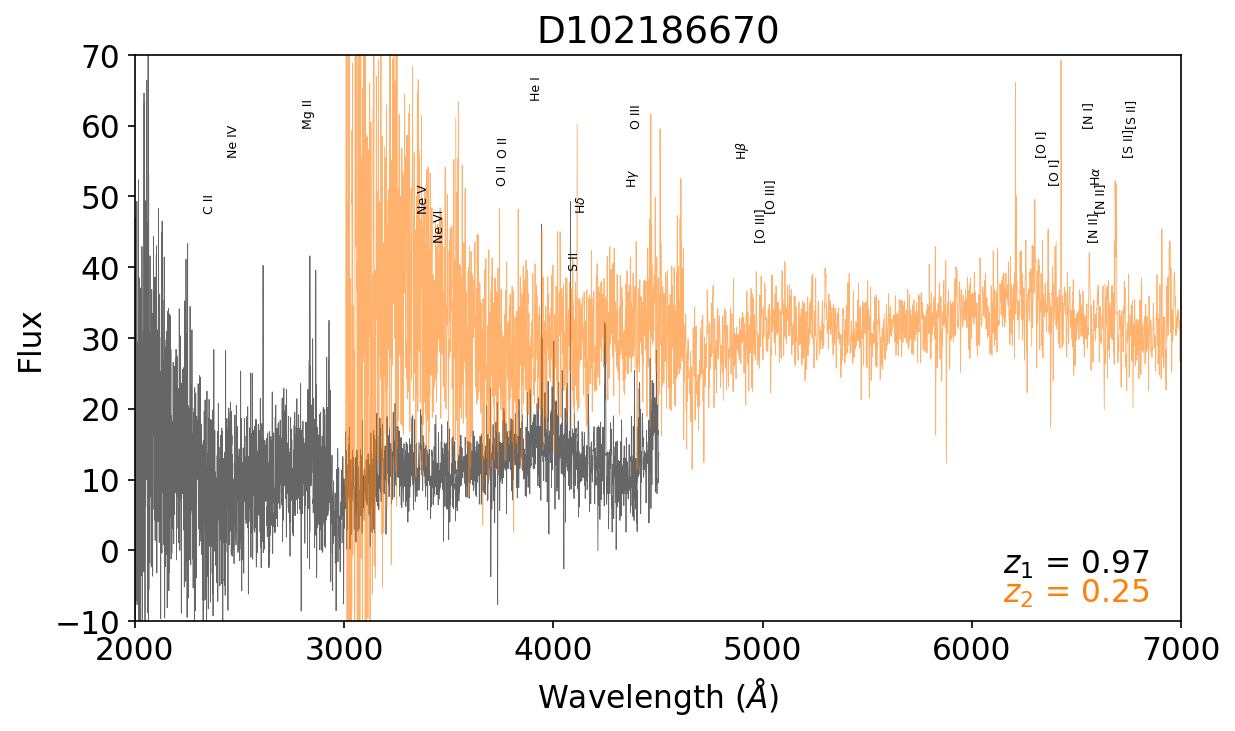}
\includegraphics[width=0.49\textwidth]{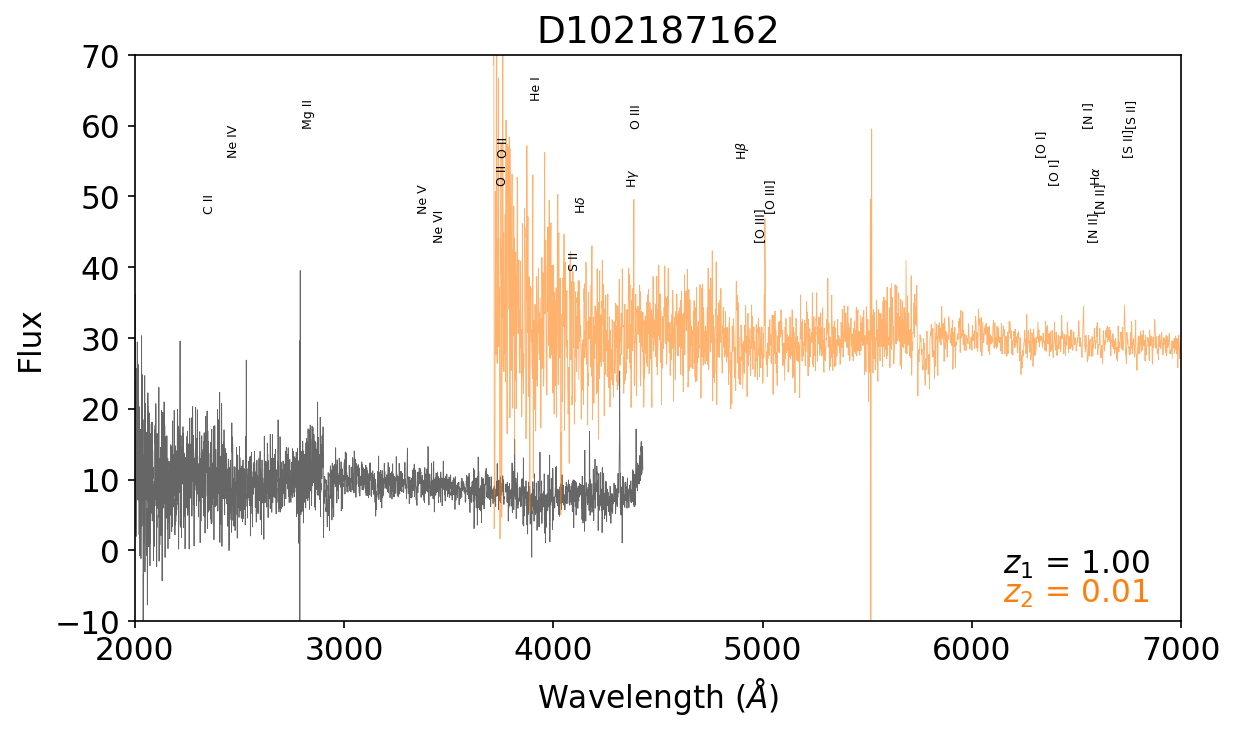}
\includegraphics[width=0.49\textwidth]{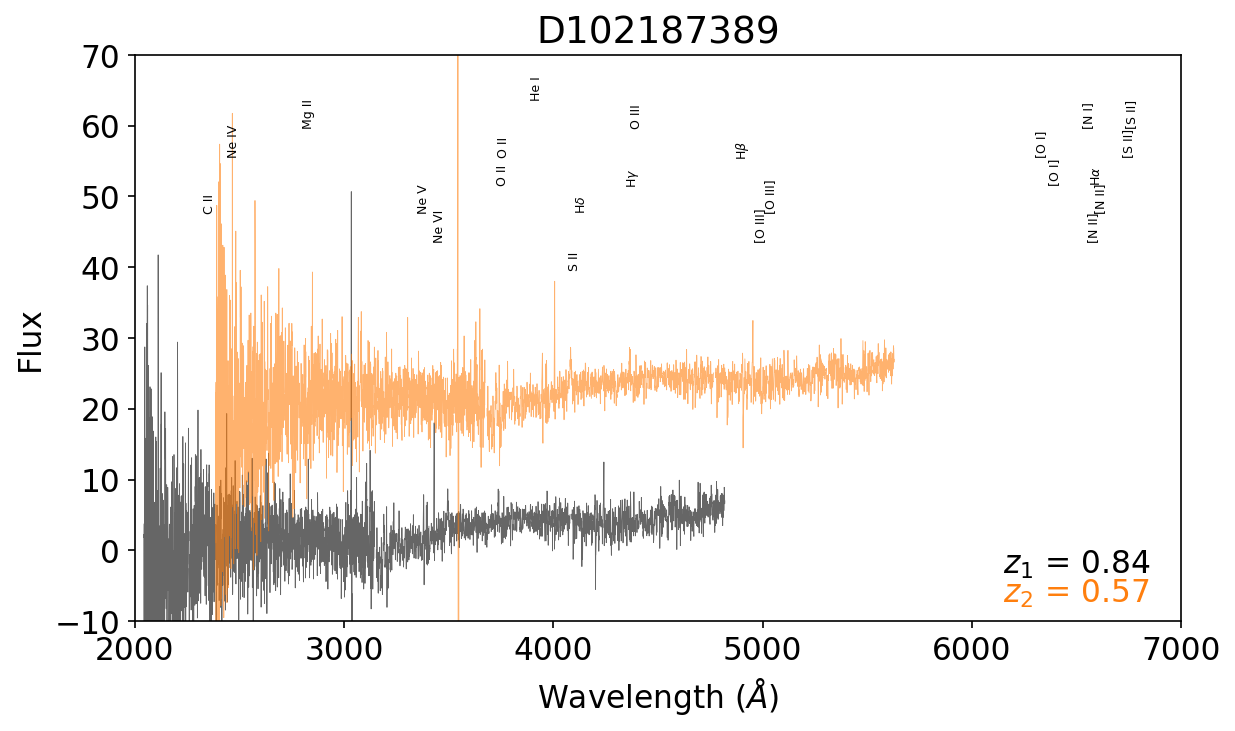}
\includegraphics[width=0.49\textwidth]{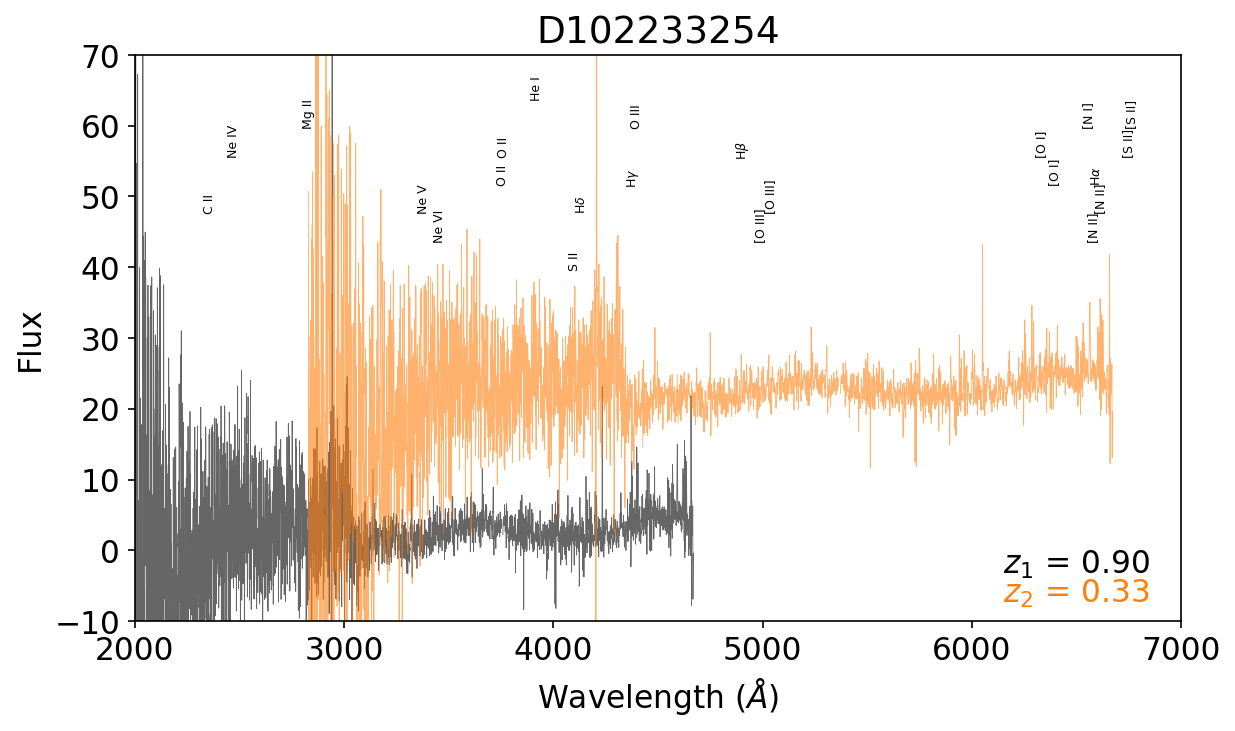}
   \caption{\label{f:spectra:appendix} DEVILS spectra identified as blends by the crosscorrelation parameters from {\sc autoz}. Emission and absorption line features that can be expected are marked in black and red for the first and second highest crosscorrelation peak respectively.}
\end{figure*}

\setcounter{figure}{9}

\begin{figure*}
    \centering
\includegraphics[width=0.49\textwidth]{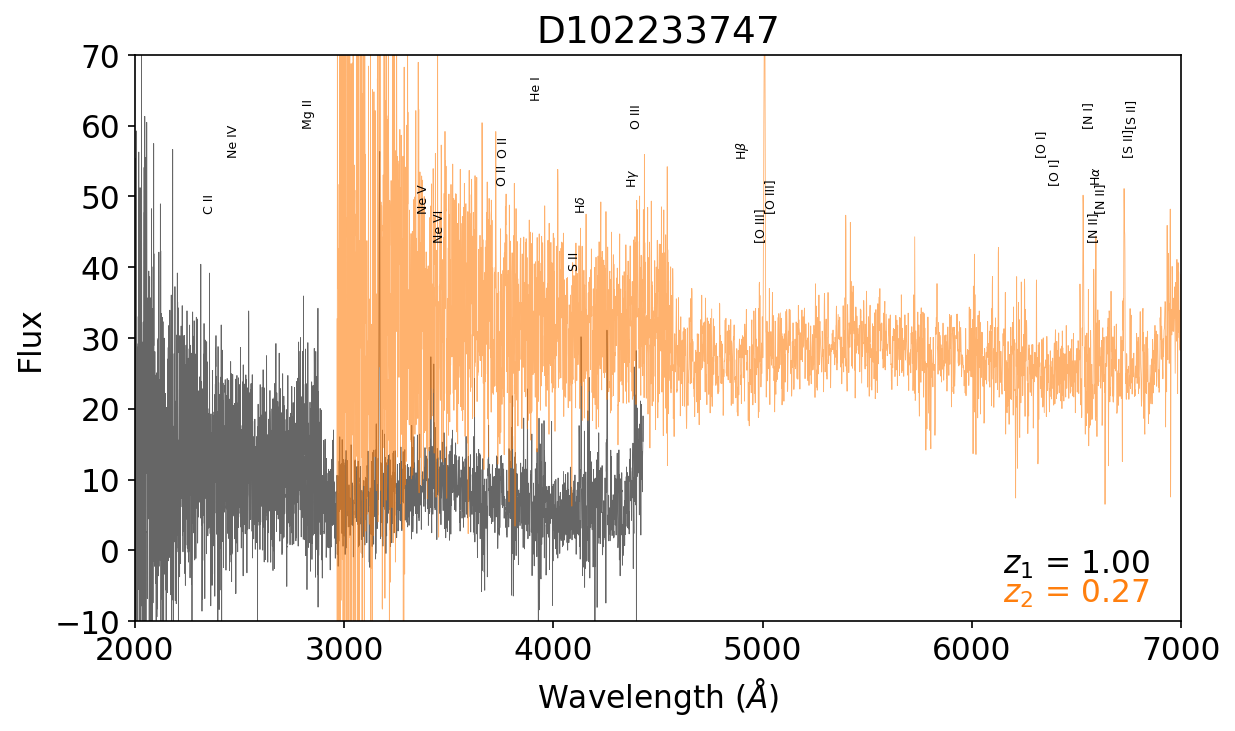}
\includegraphics[width=0.49\textwidth]{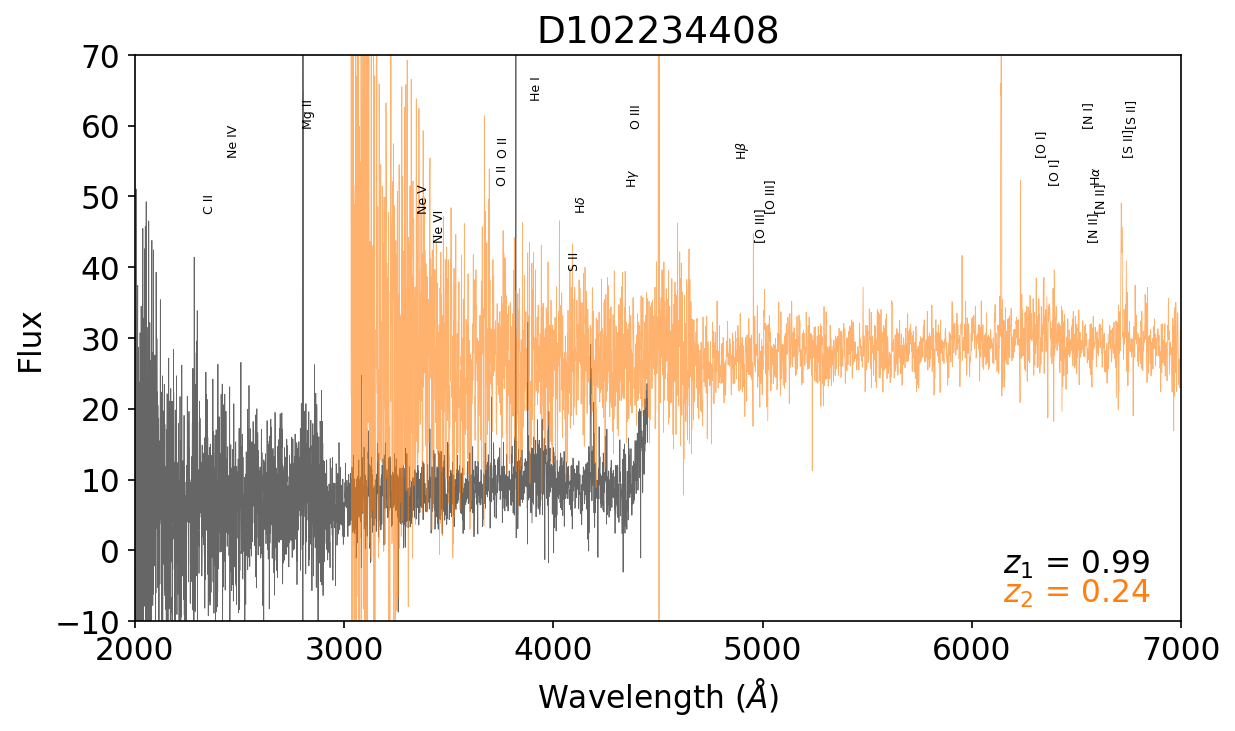}
\includegraphics[width=0.49\textwidth]{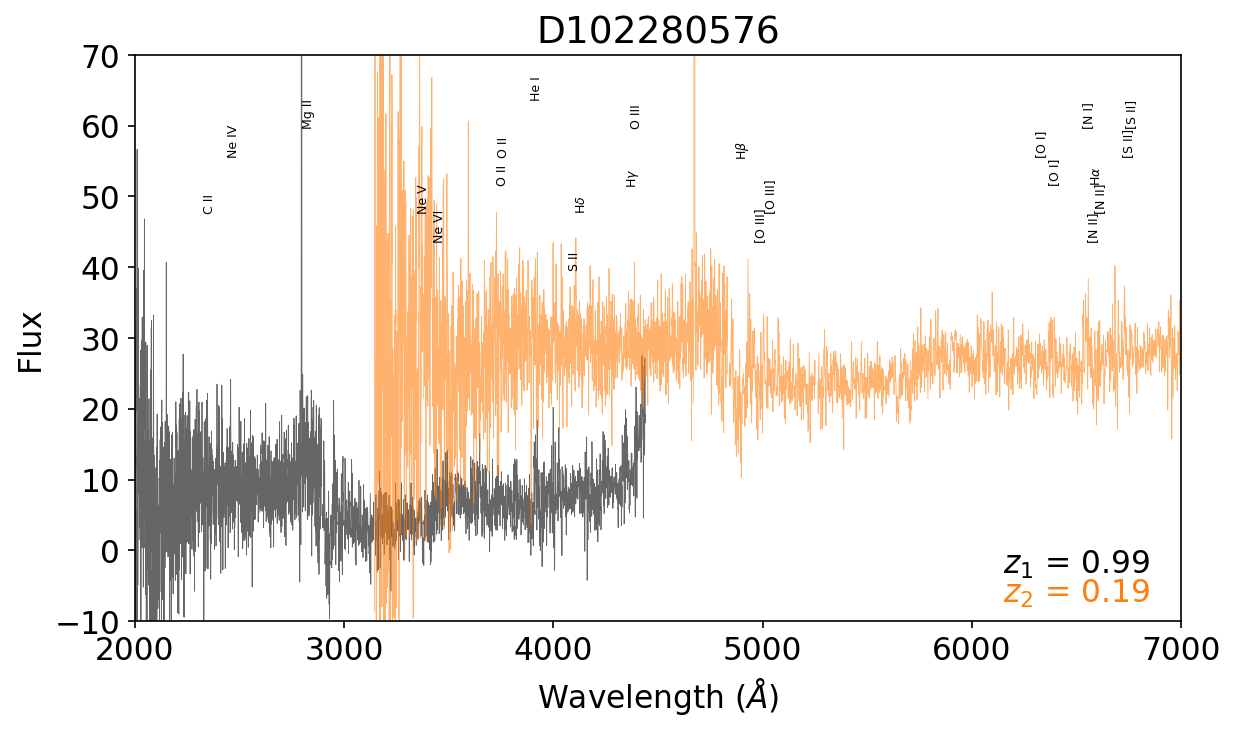}
\includegraphics[width=0.49\textwidth]{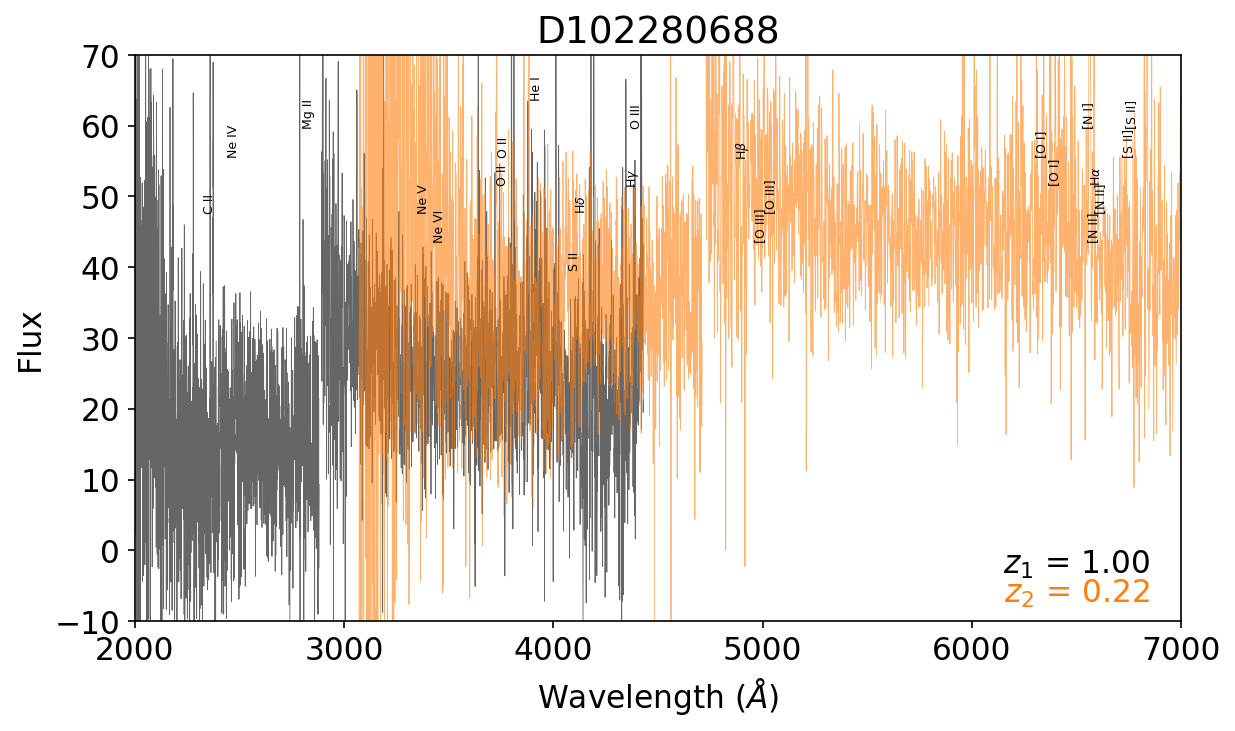}
\includegraphics[width=0.49\textwidth]{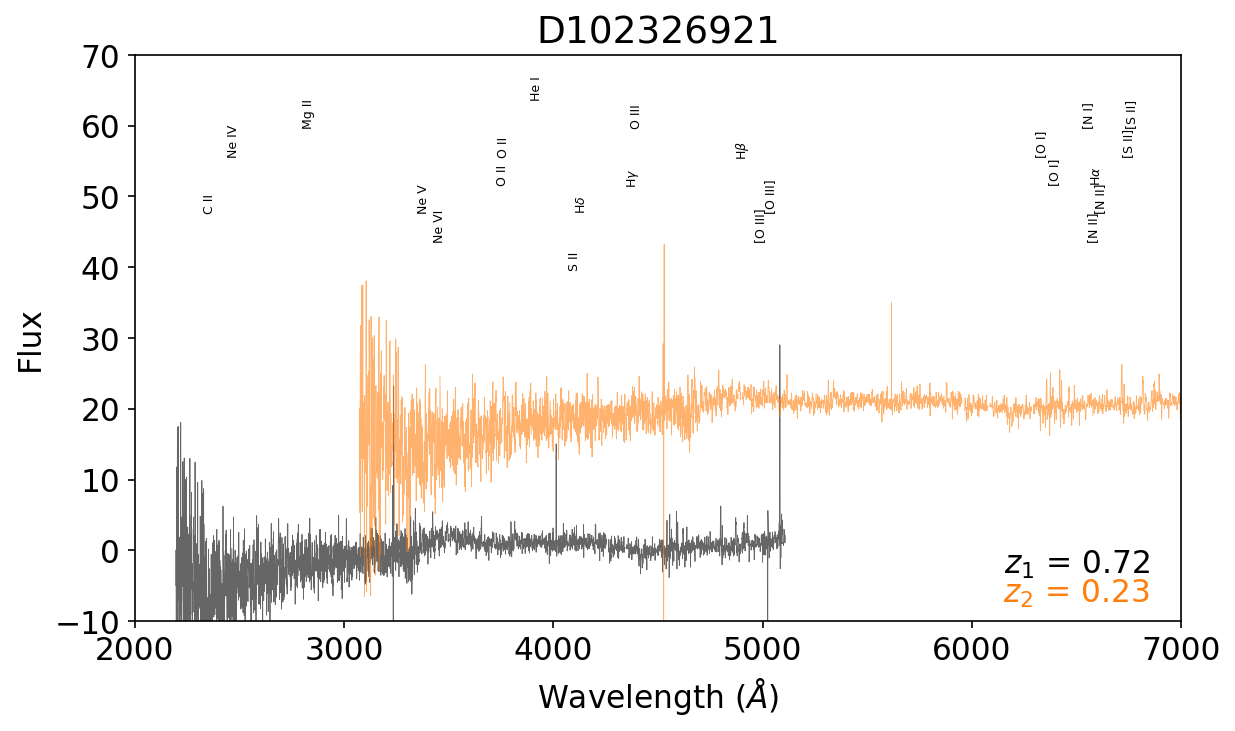}
\includegraphics[width=0.49\textwidth]{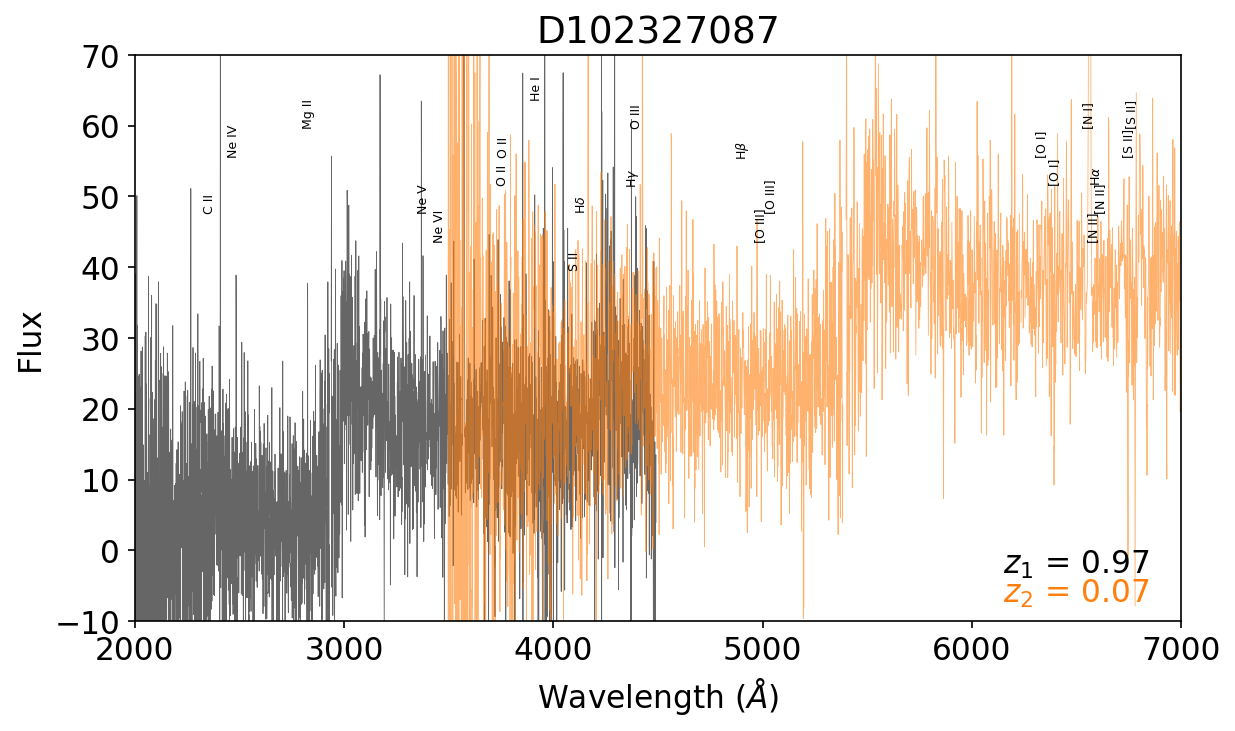}
\includegraphics[width=0.49\textwidth]{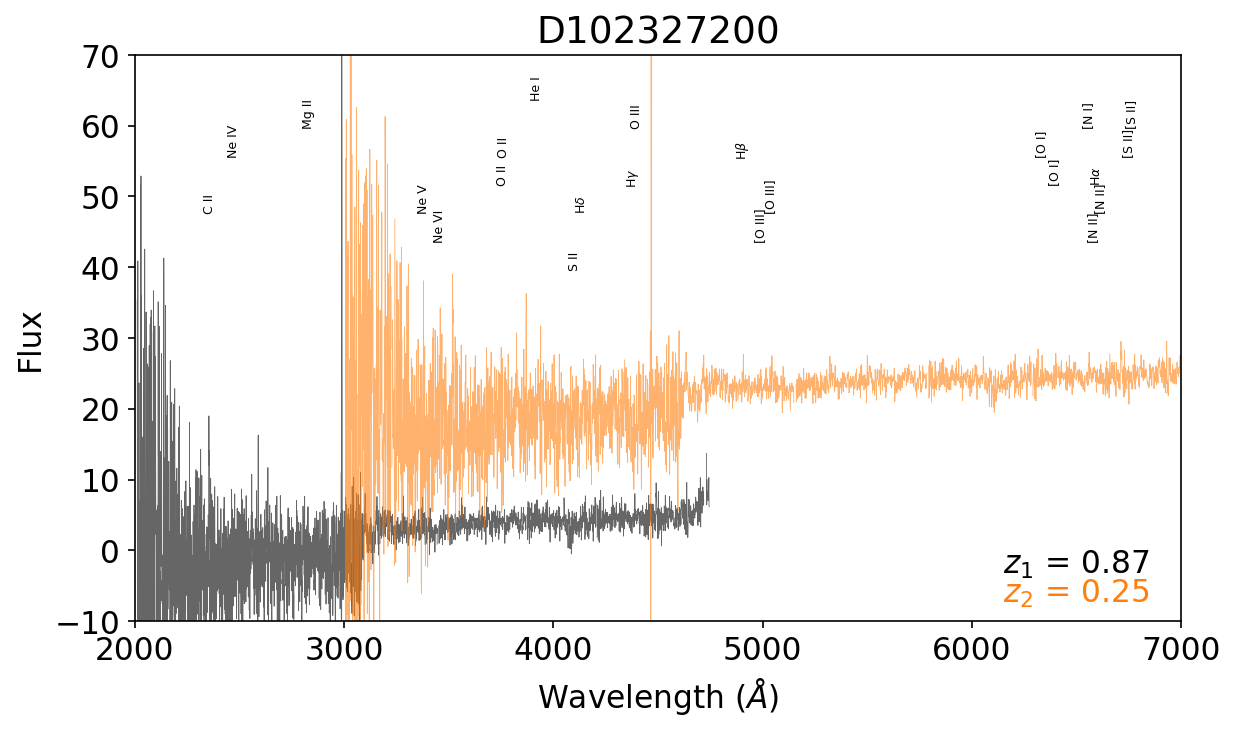}
\includegraphics[width=0.49\textwidth]{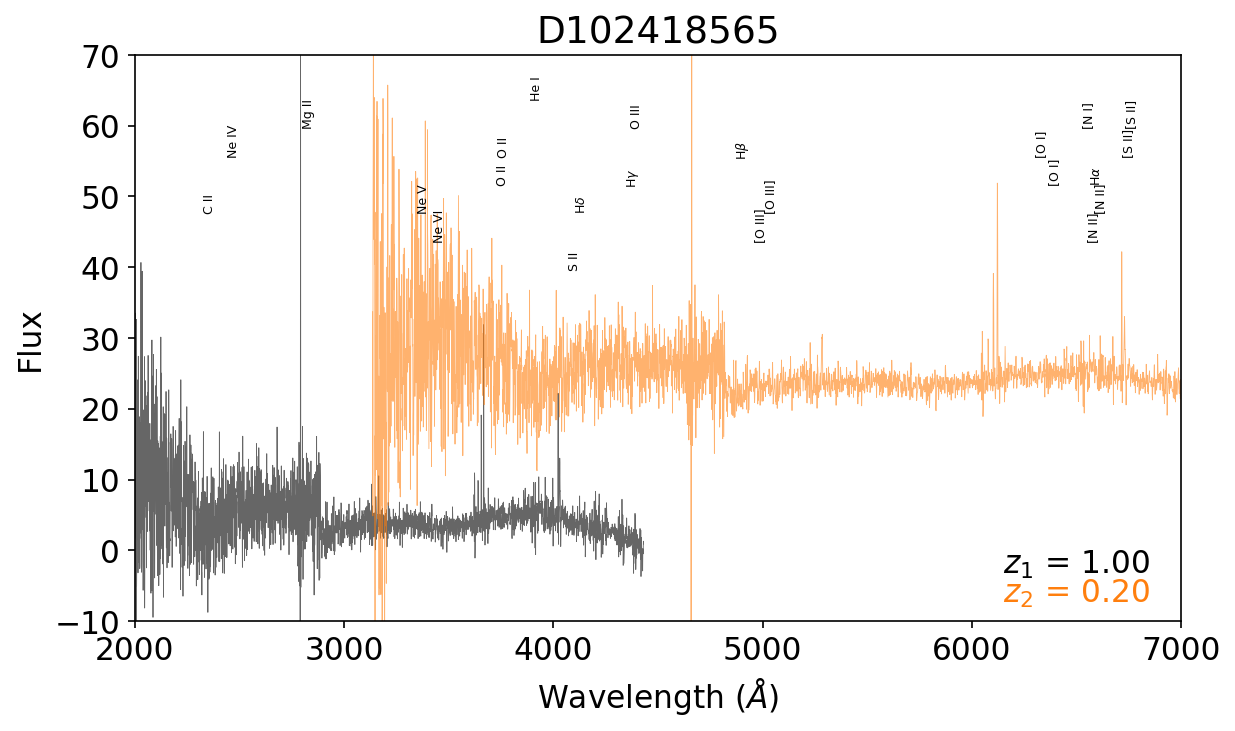}
\caption{-- \textit{continued}.}
\end{figure*}

\setcounter{figure}{0}

\begin{figure*}
    \centering
\includegraphics[width=0.49\textwidth]{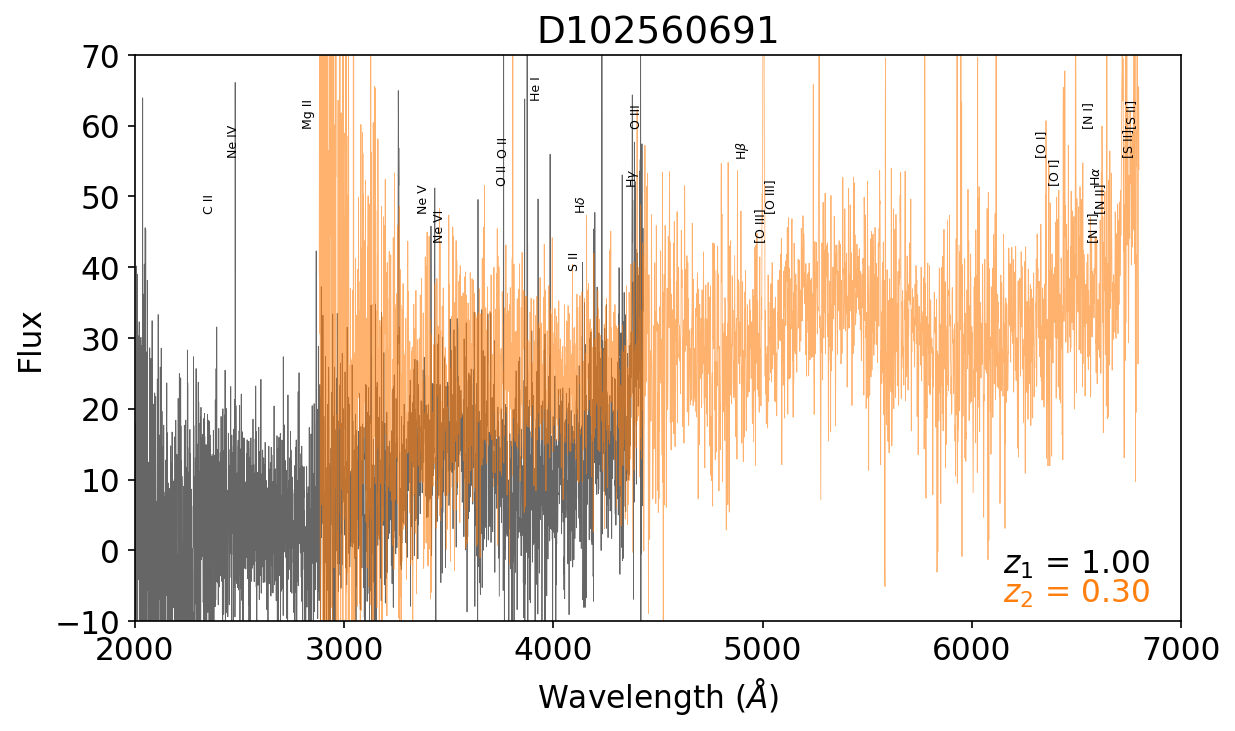}
\includegraphics[width=0.49\textwidth]{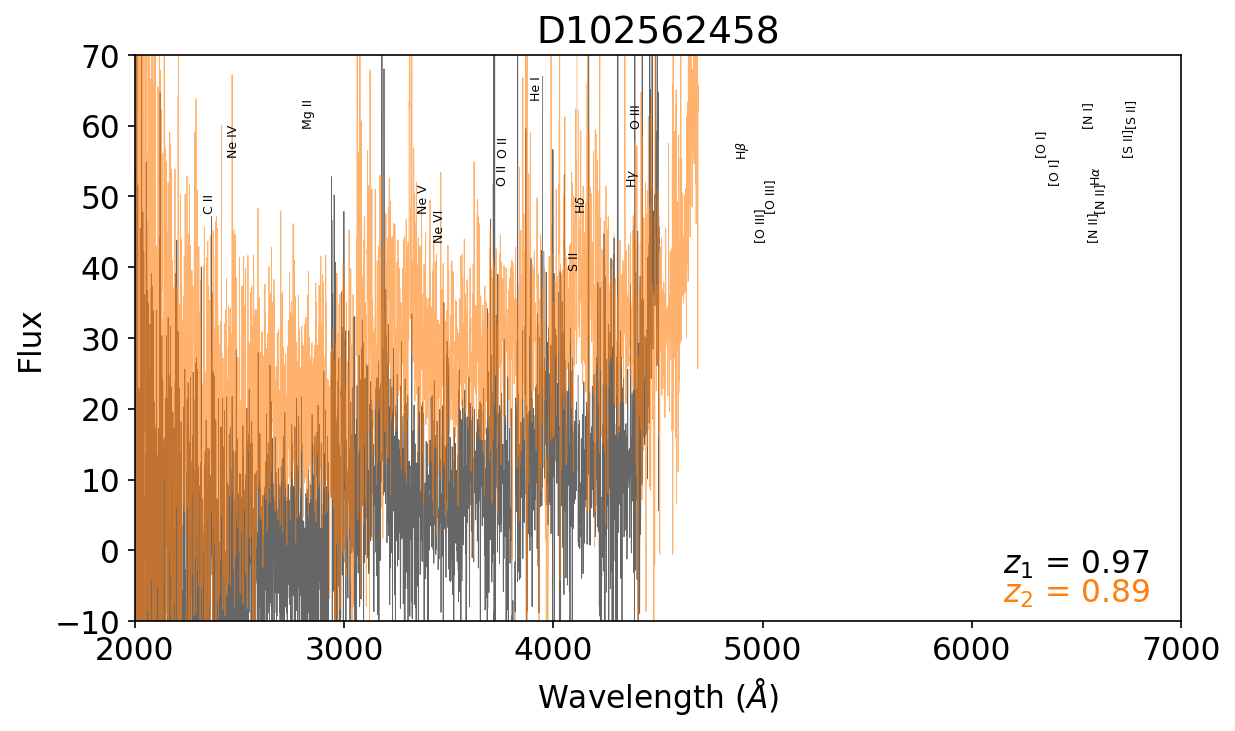}
\includegraphics[width=0.49\textwidth]{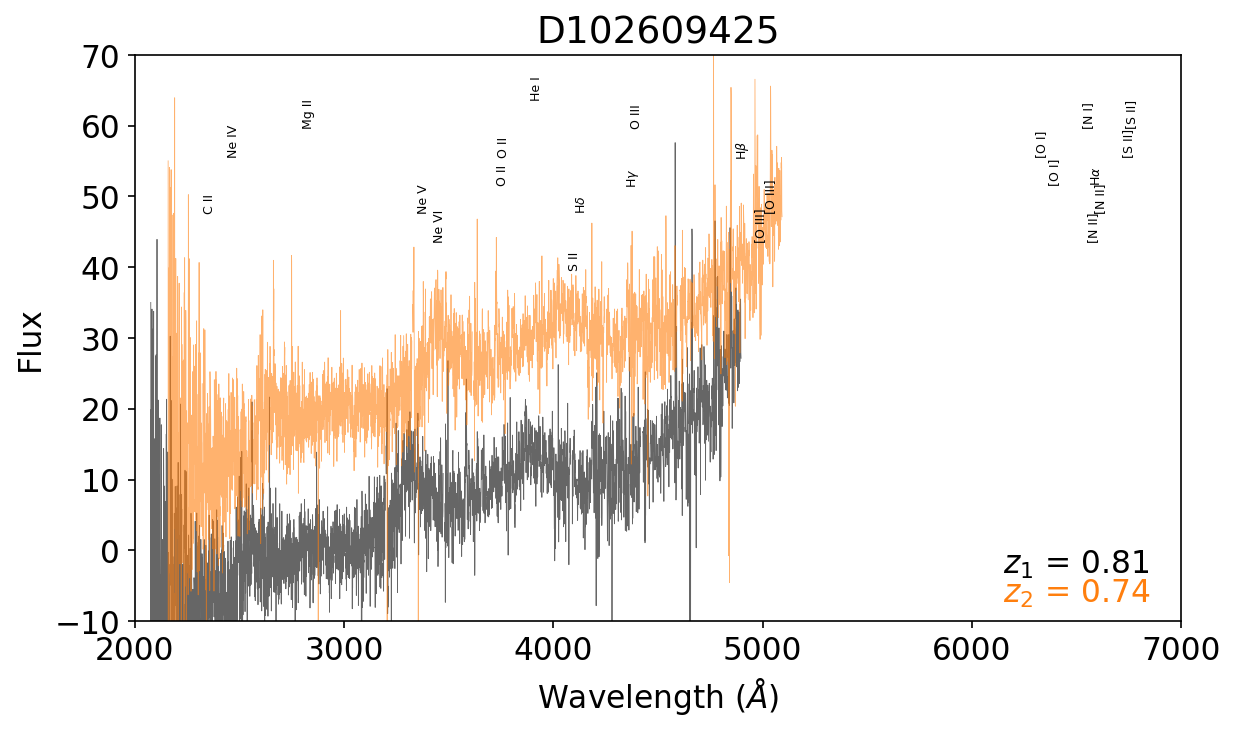}
\includegraphics[width=0.49\textwidth]{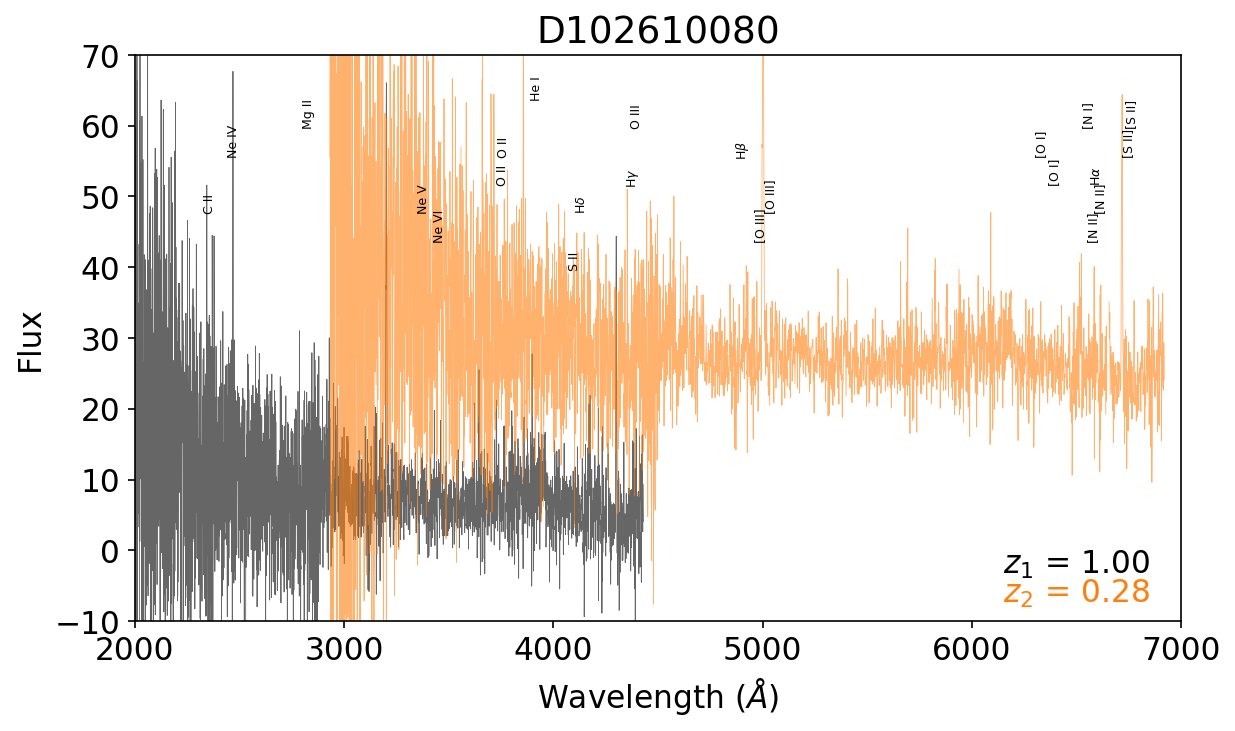}
\includegraphics[width=0.49\textwidth]{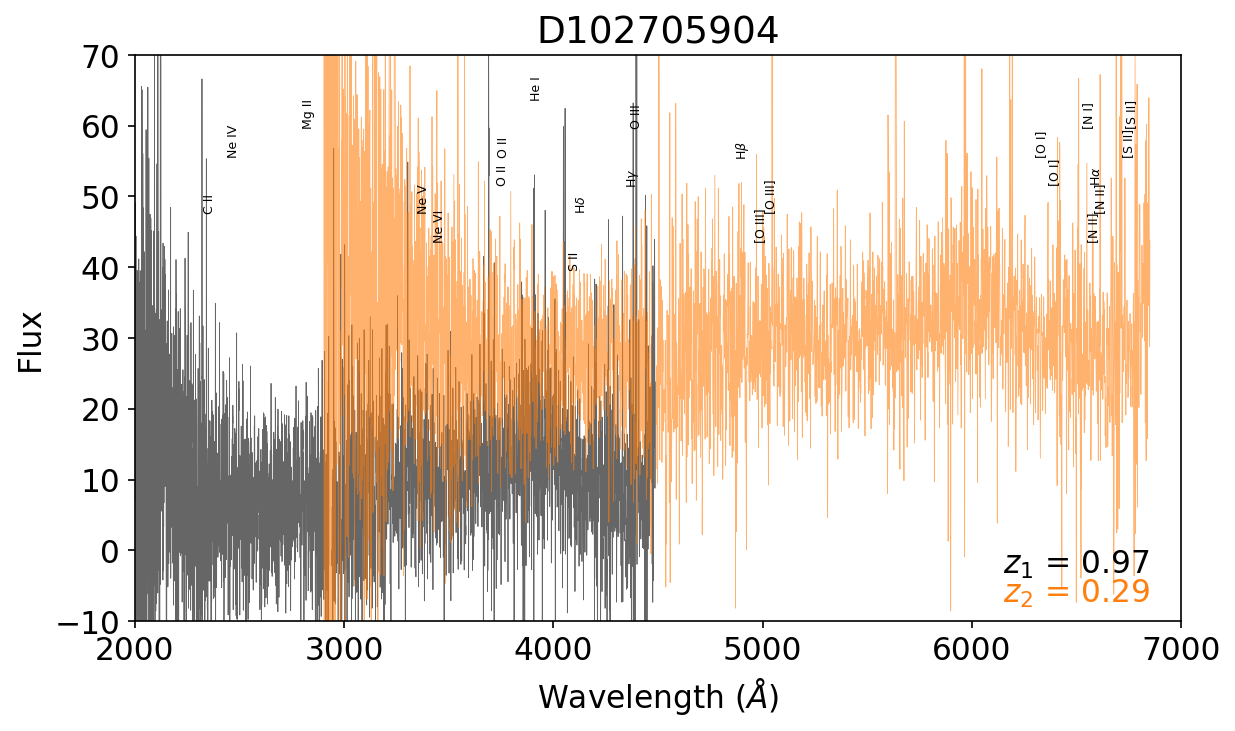}
\includegraphics[width=0.49\textwidth]{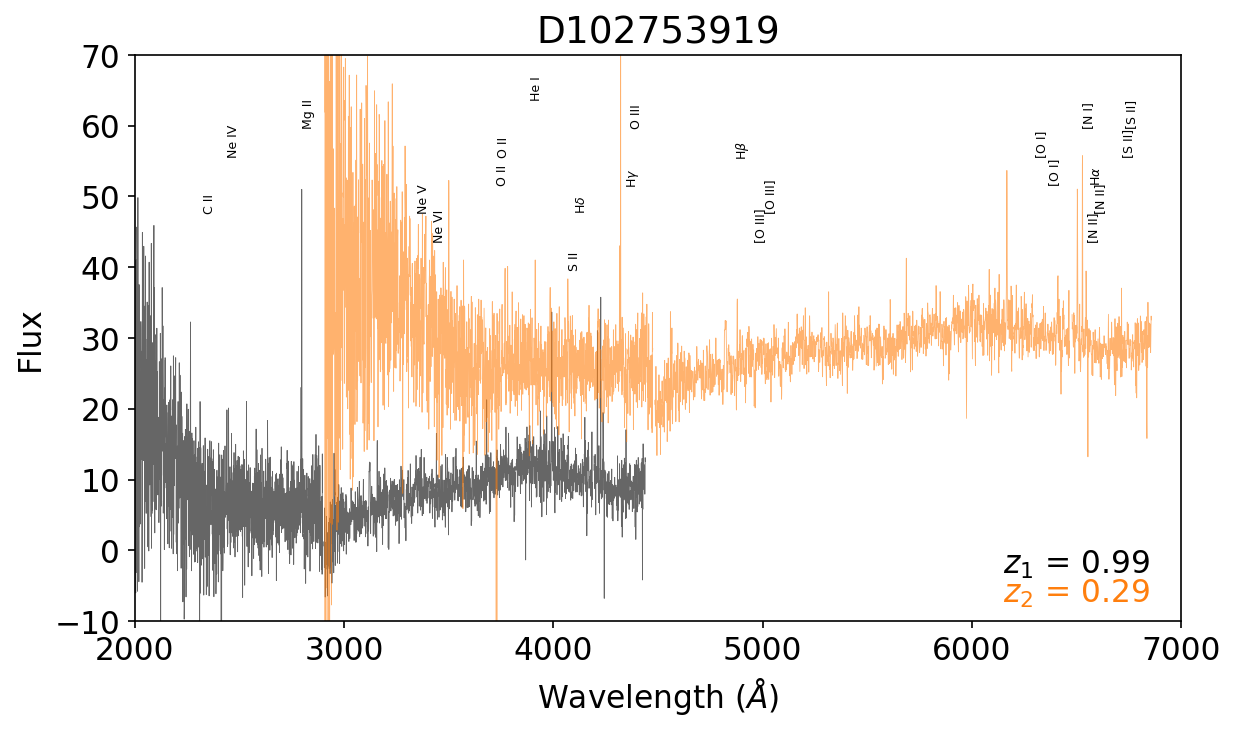}
\includegraphics[width=0.49\textwidth]{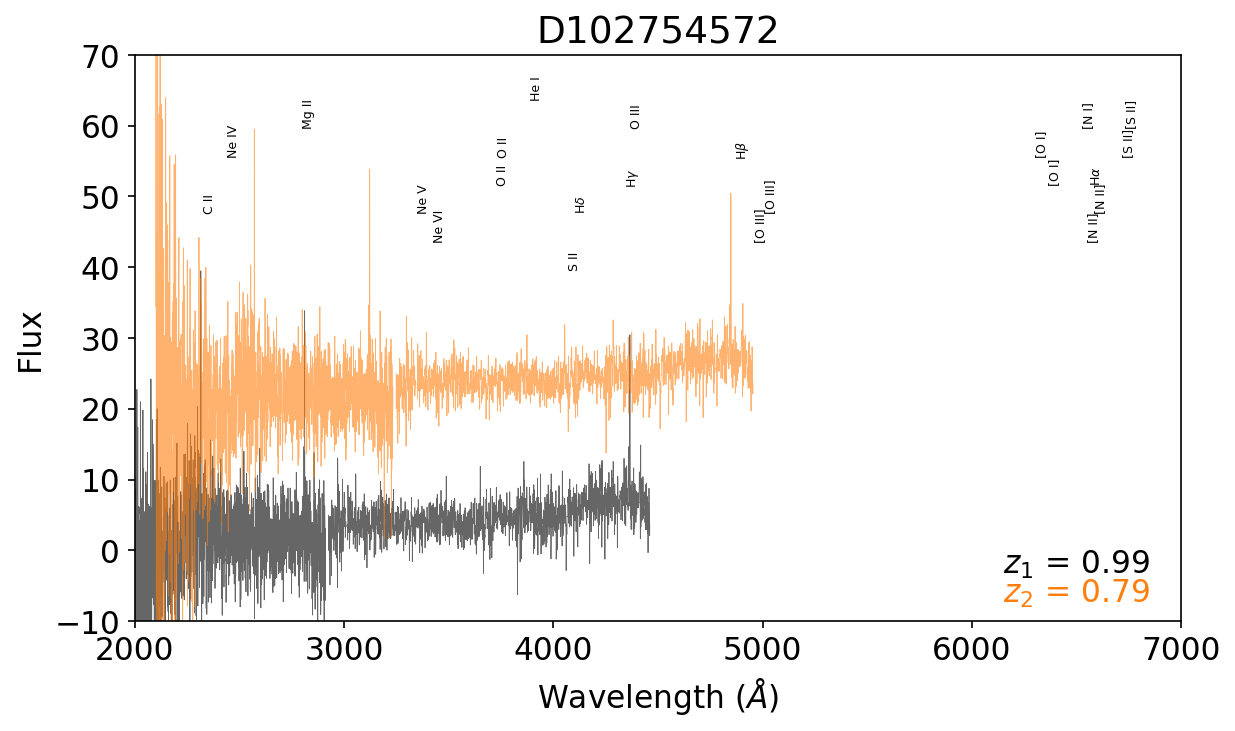}
\includegraphics[width=0.49\textwidth]{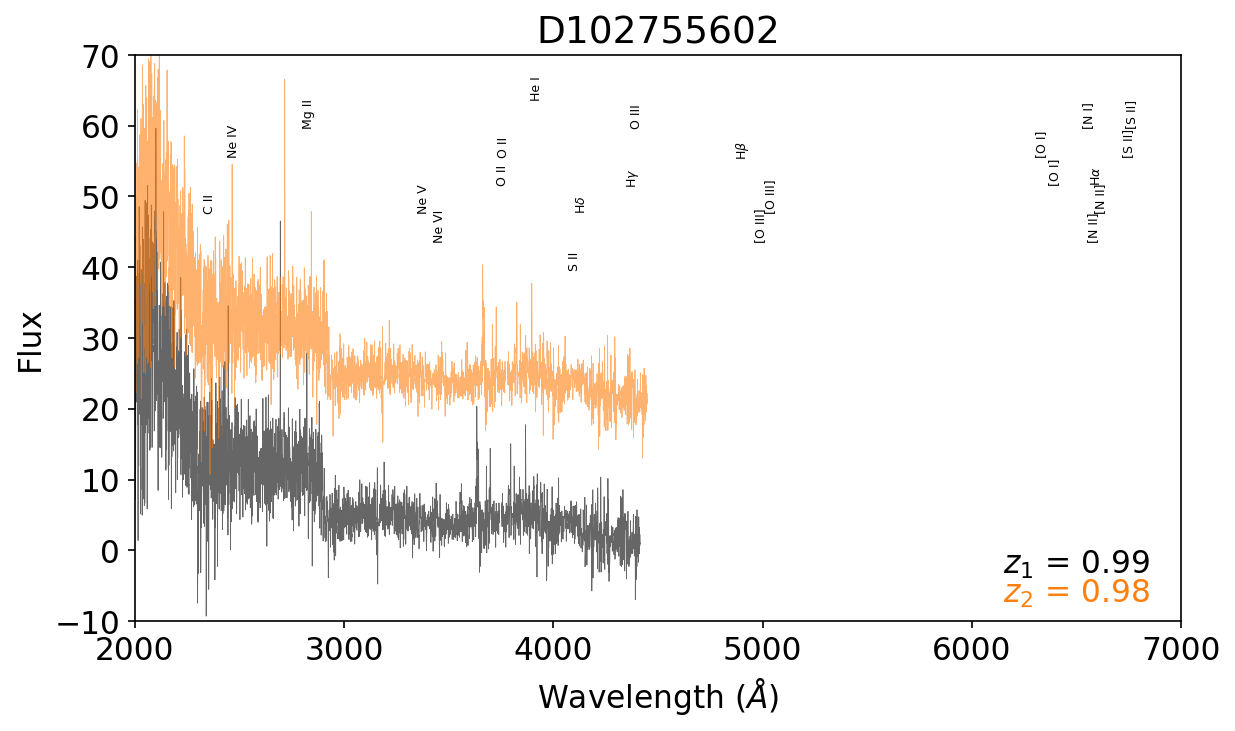}
\caption{-- \textit{continued}.}
\end{figure*}

\setcounter{figure}{9}

\begin{figure*}
    \centering
\includegraphics[width=0.49\textwidth]{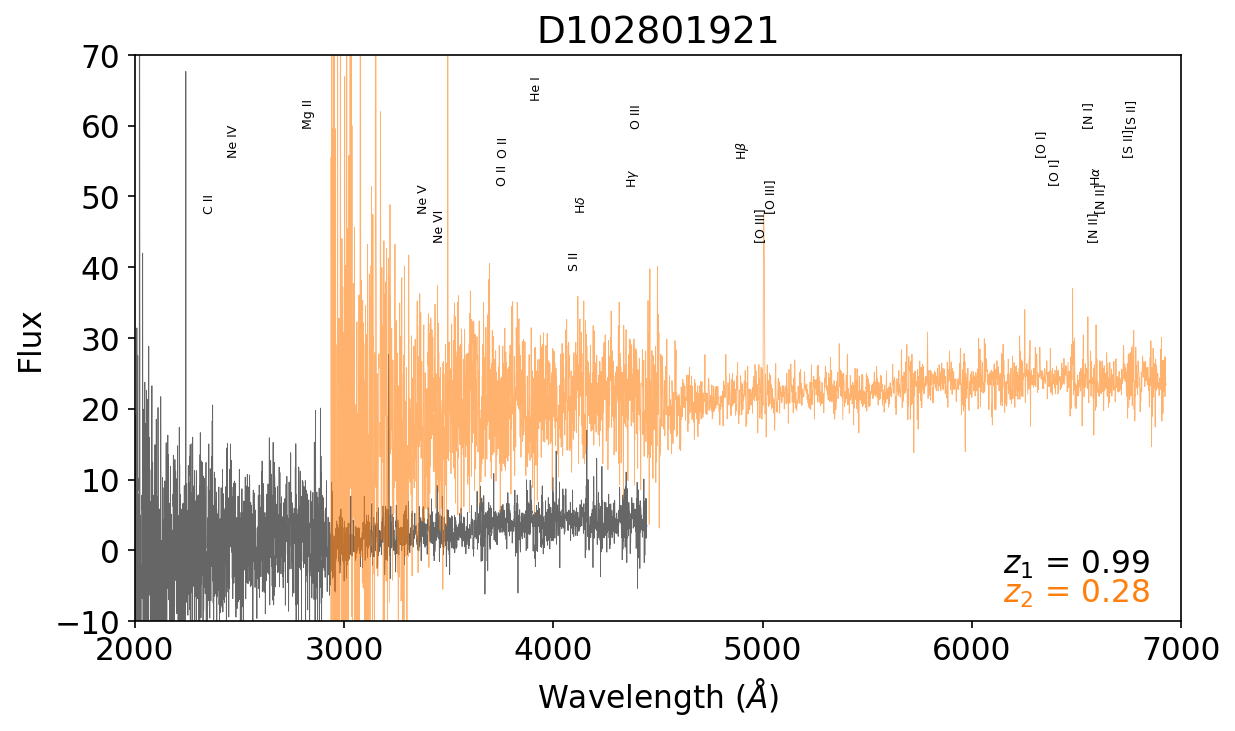}
\includegraphics[width=0.49\textwidth]{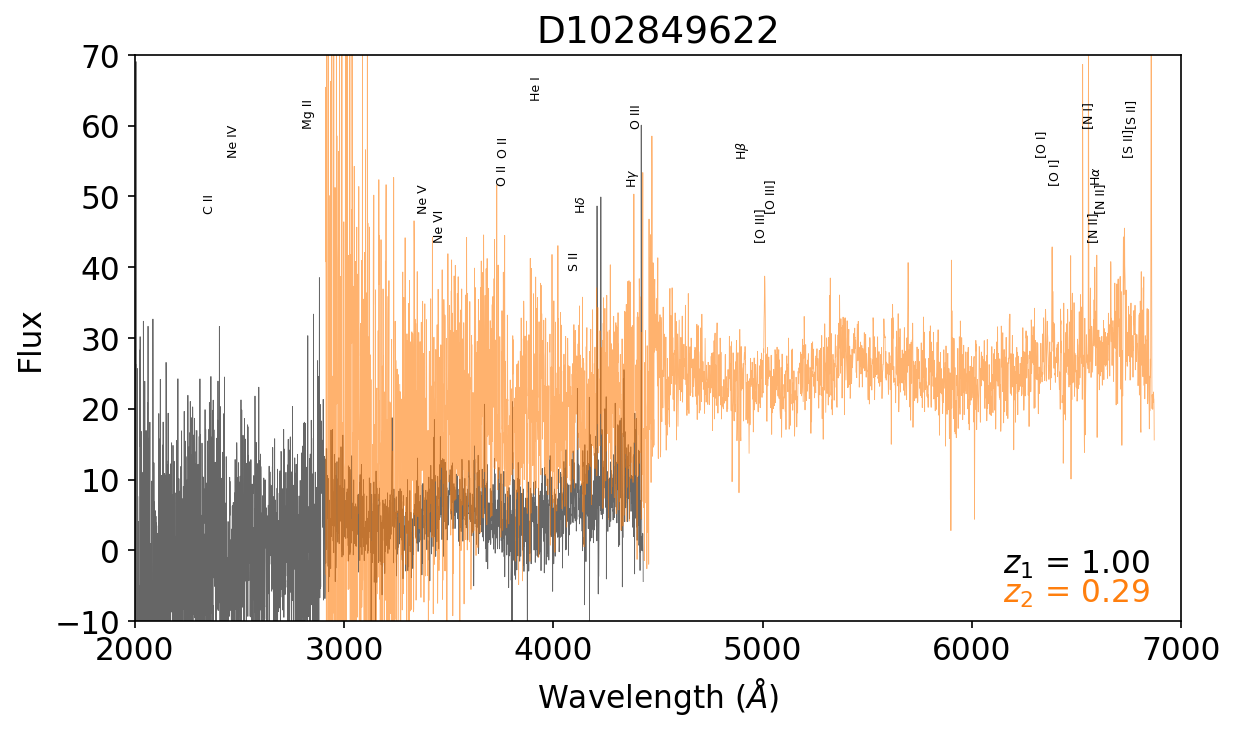}
\includegraphics[width=0.49\textwidth]{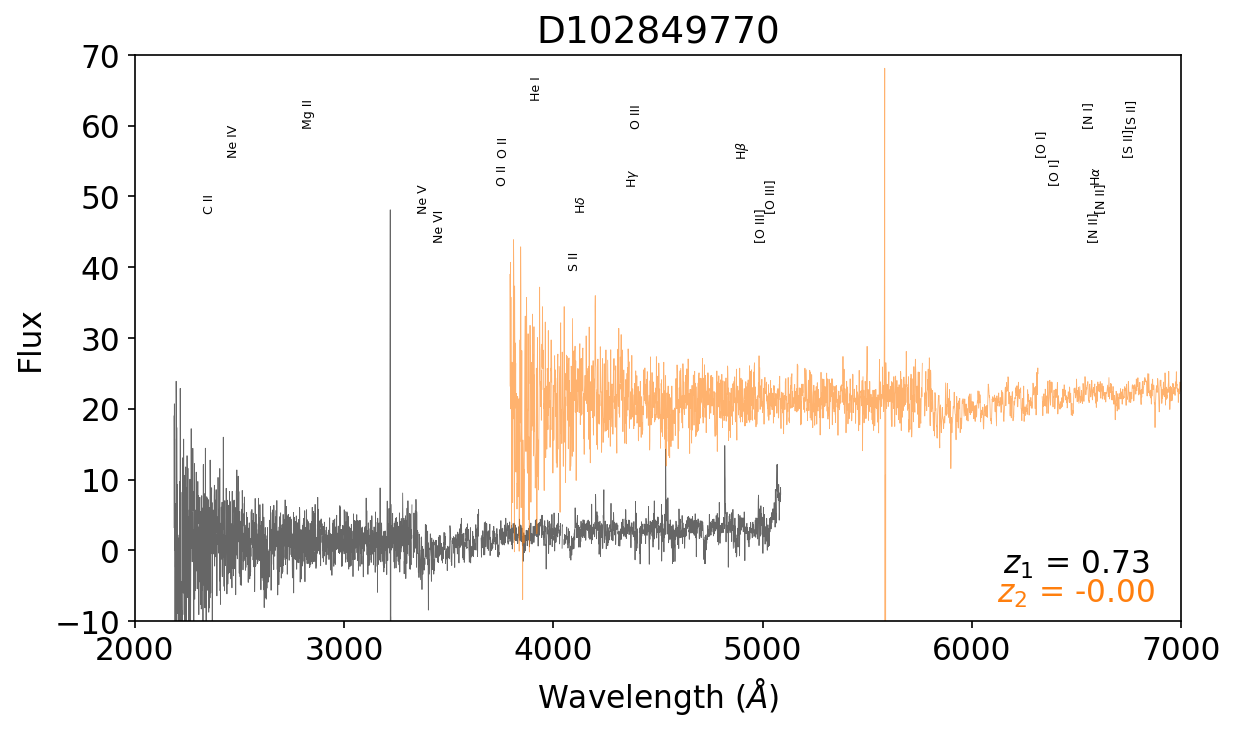}
\includegraphics[width=0.49\textwidth]{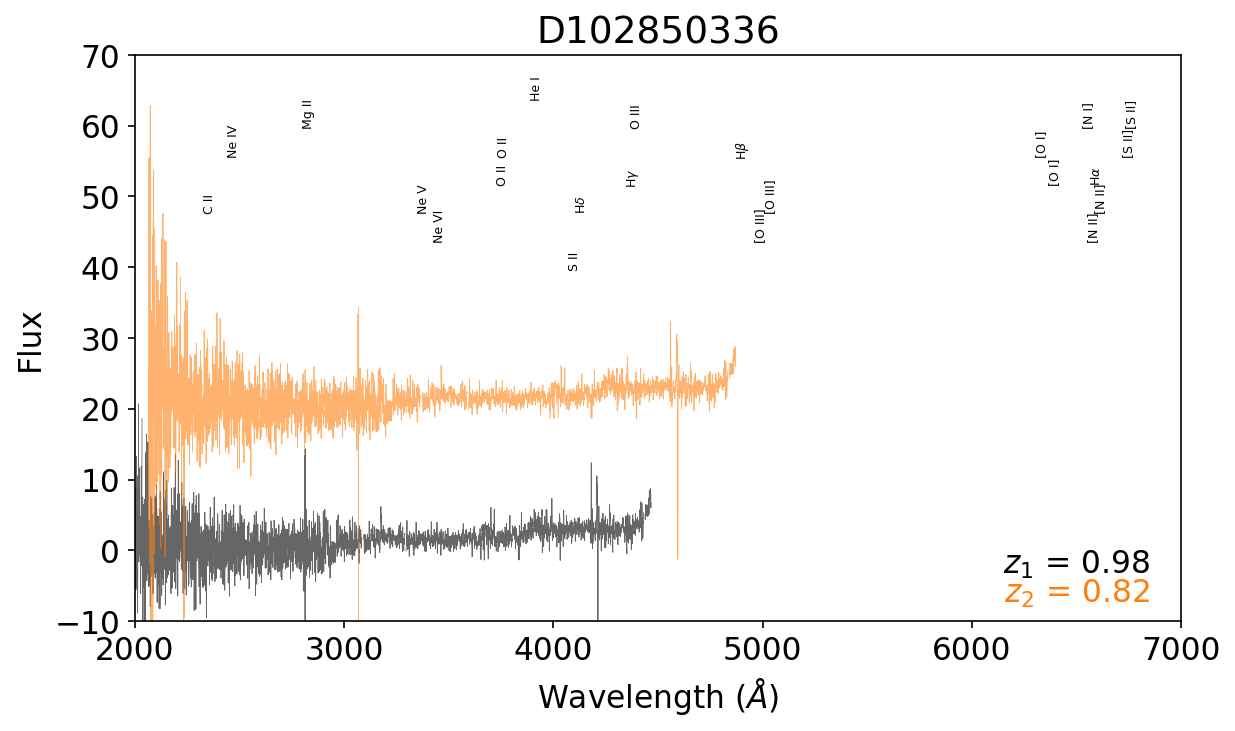}
\includegraphics[width=0.49\textwidth]{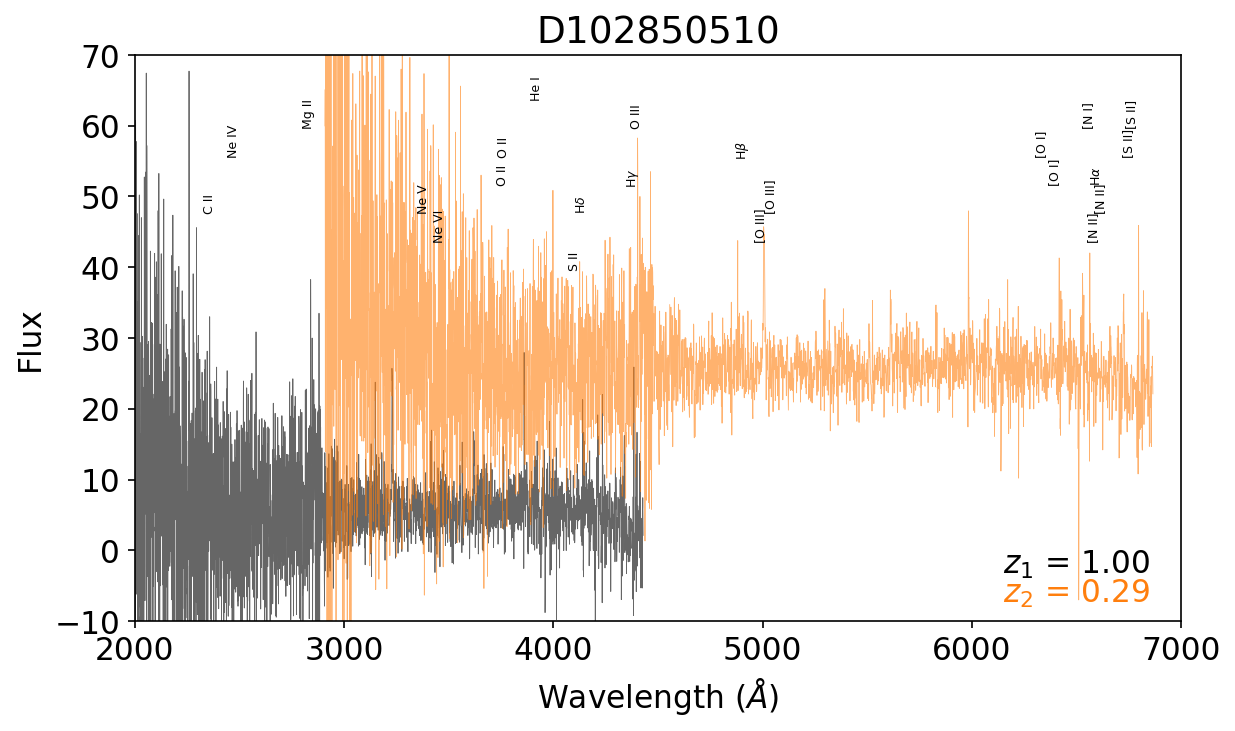}
\includegraphics[width=0.49\textwidth]{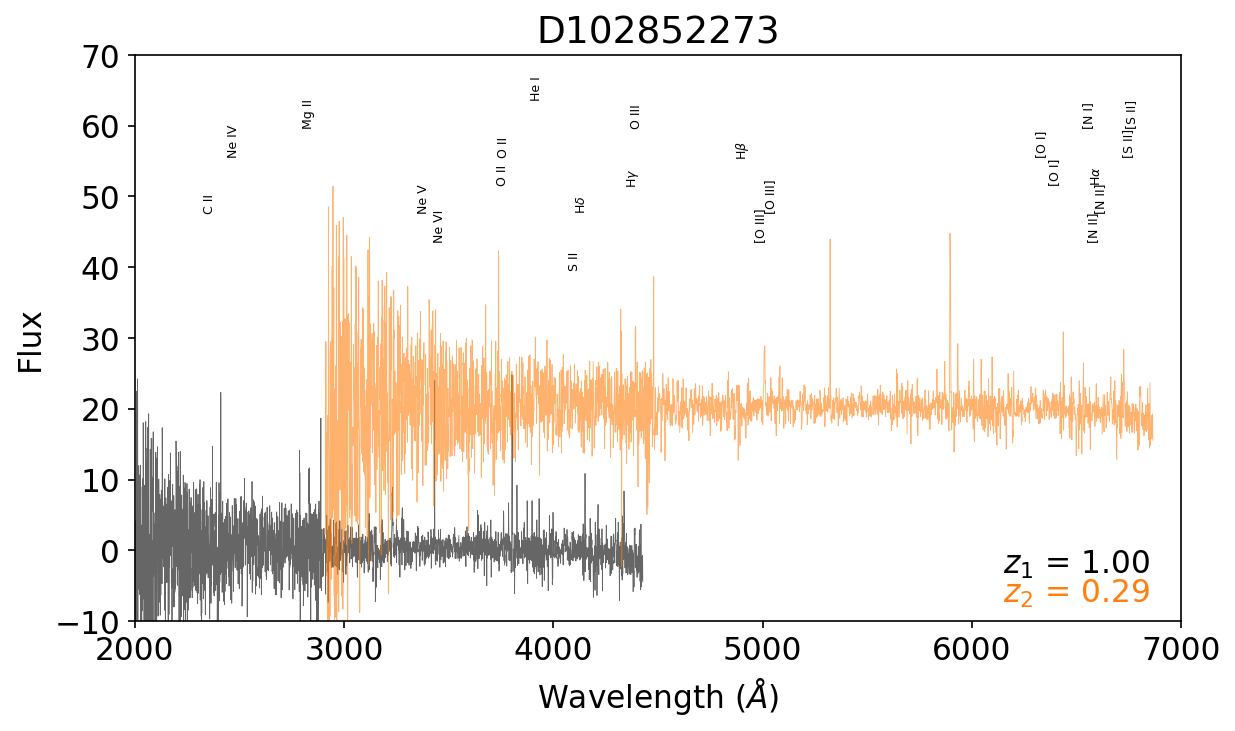}
\includegraphics[width=0.49\textwidth]{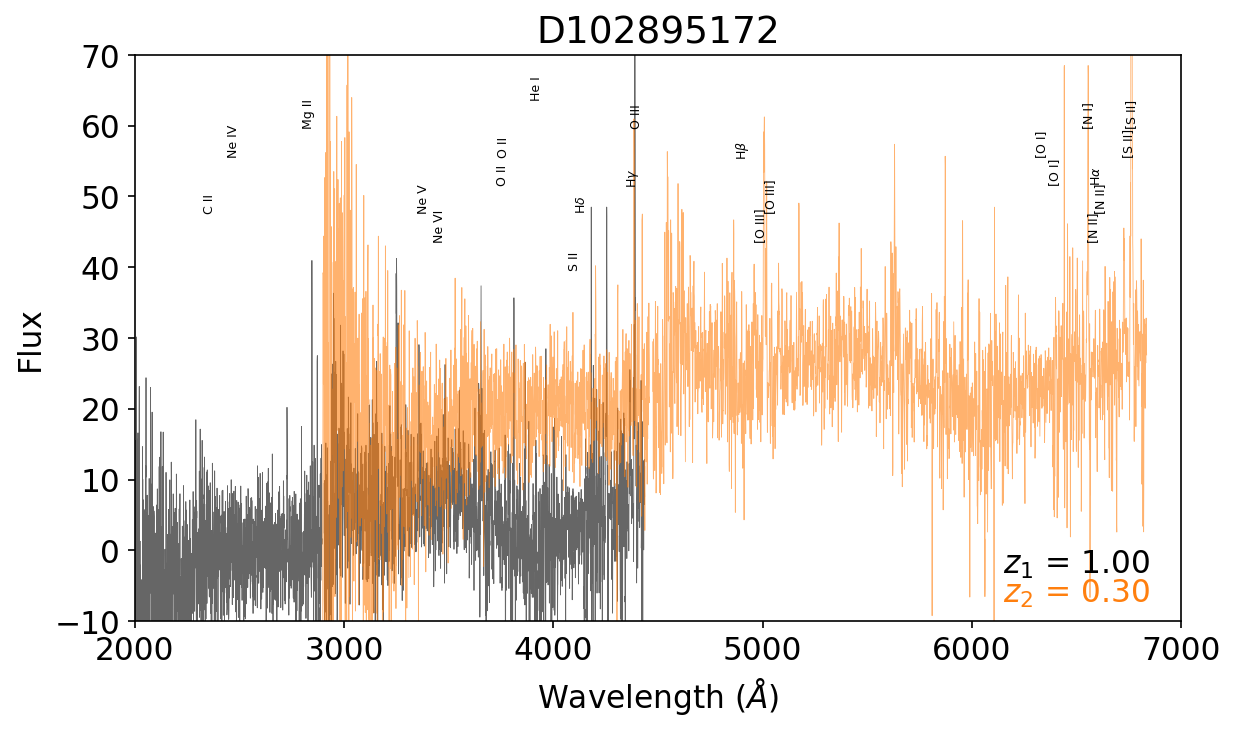}
\includegraphics[width=0.49\textwidth]{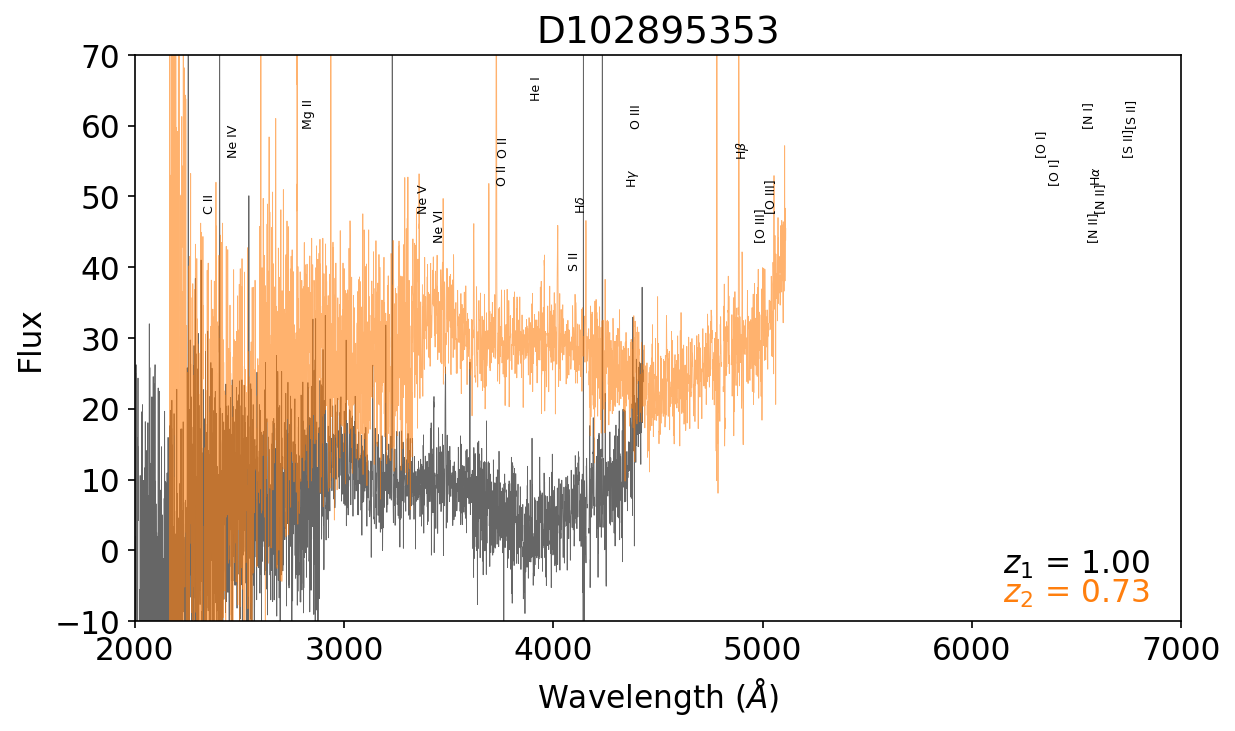}
    \caption{-- \textit{continued}.}
\end{figure*}

\setcounter{figure}{0}

\begin{figure*}
    \centering
\includegraphics[width=0.49\textwidth]{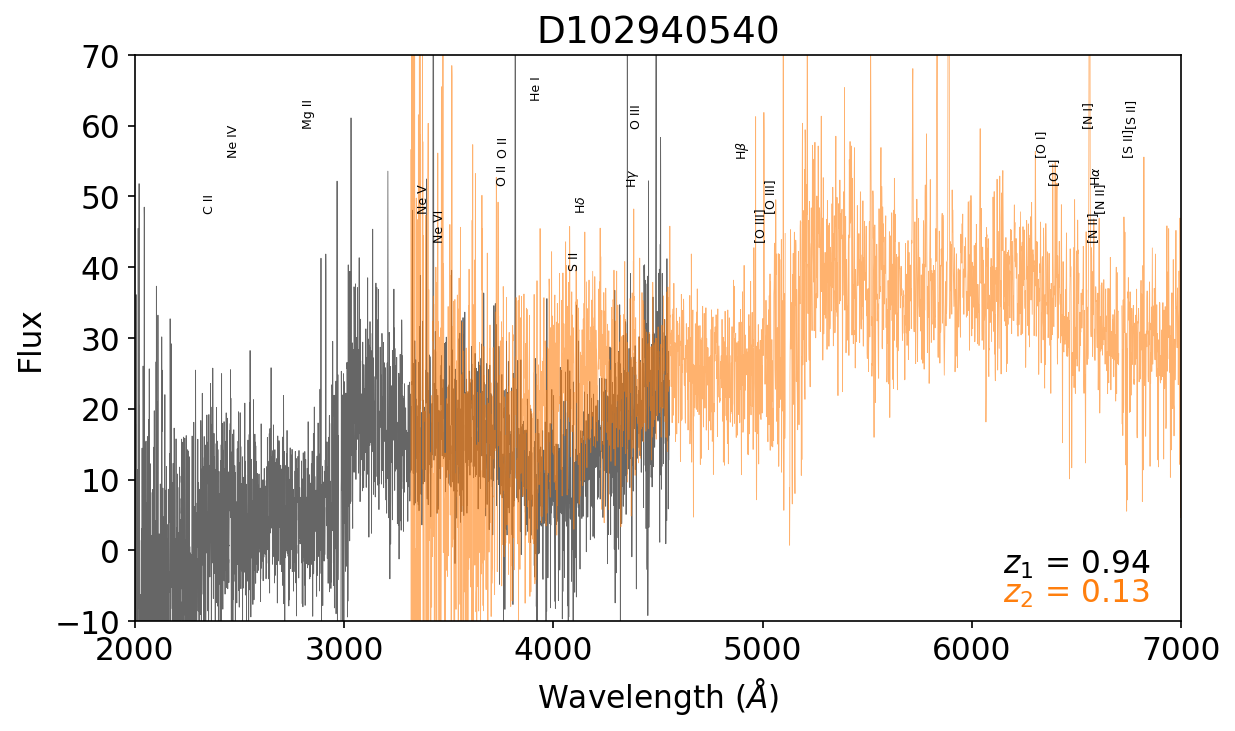}
\includegraphics[width=0.49\textwidth]{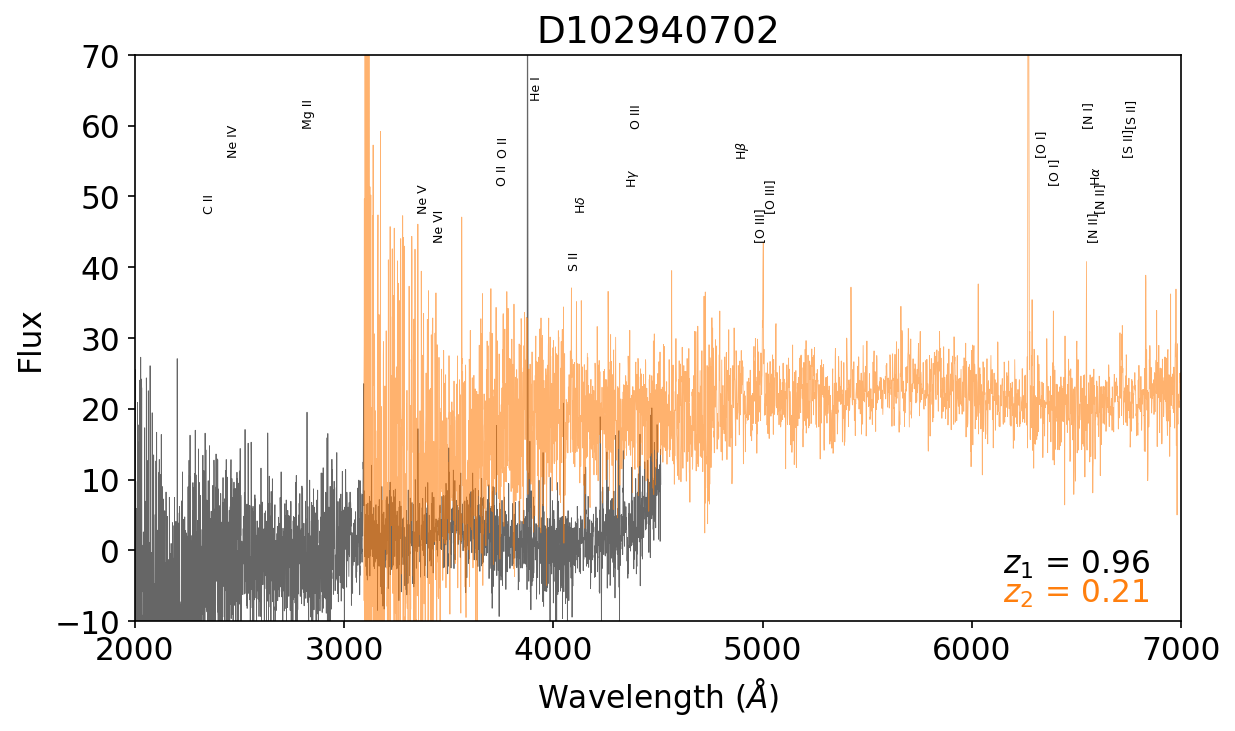}
\includegraphics[width=0.49\textwidth]{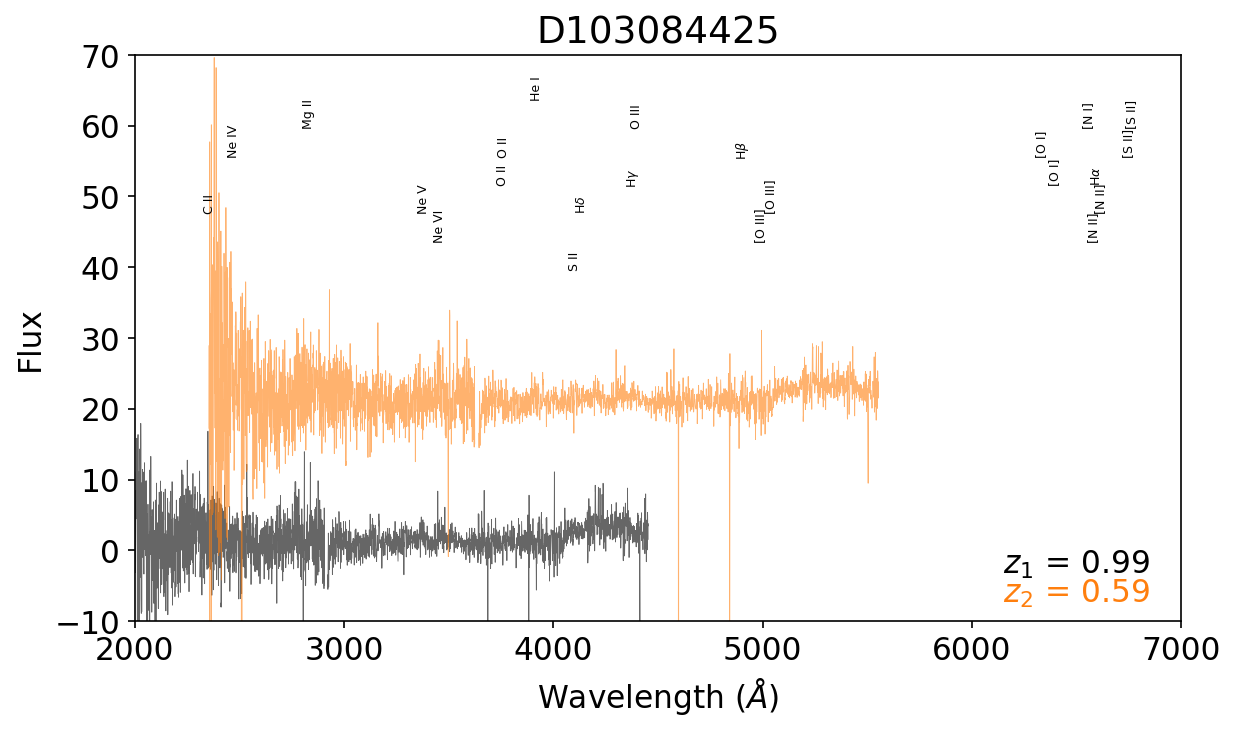}
\includegraphics[width=0.49\textwidth]{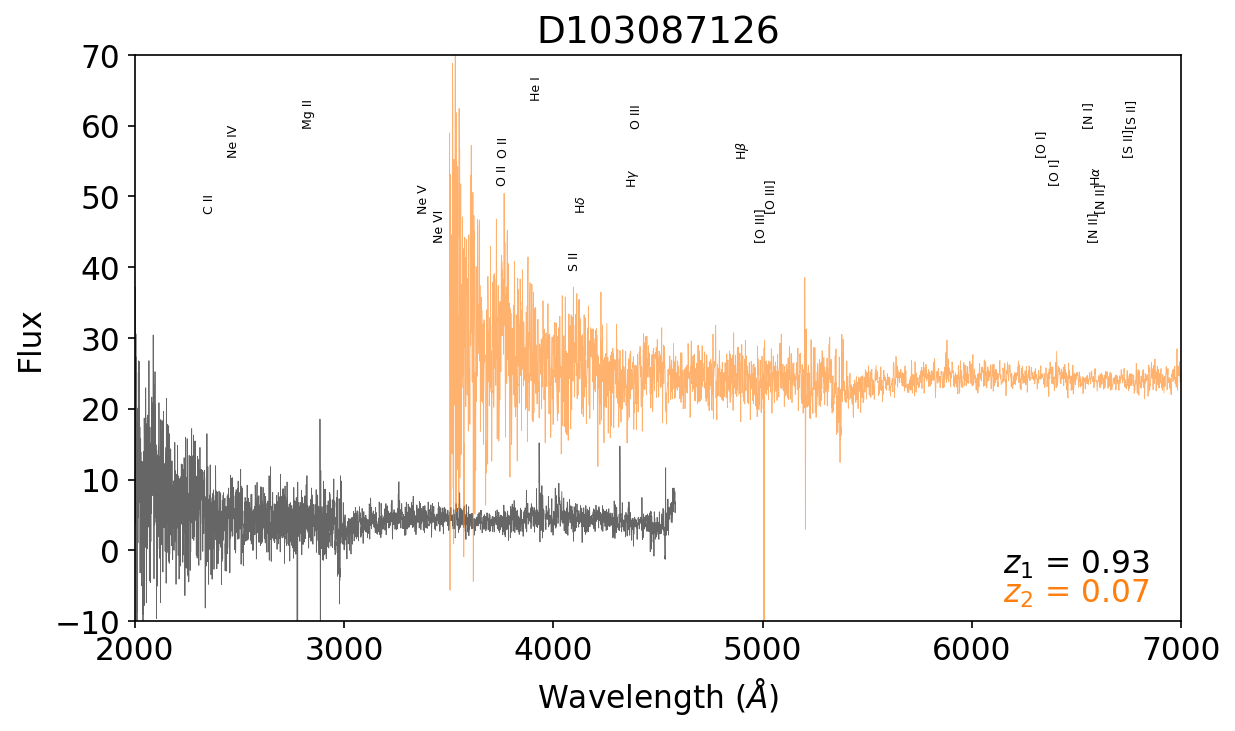}
\includegraphics[width=0.49\textwidth]{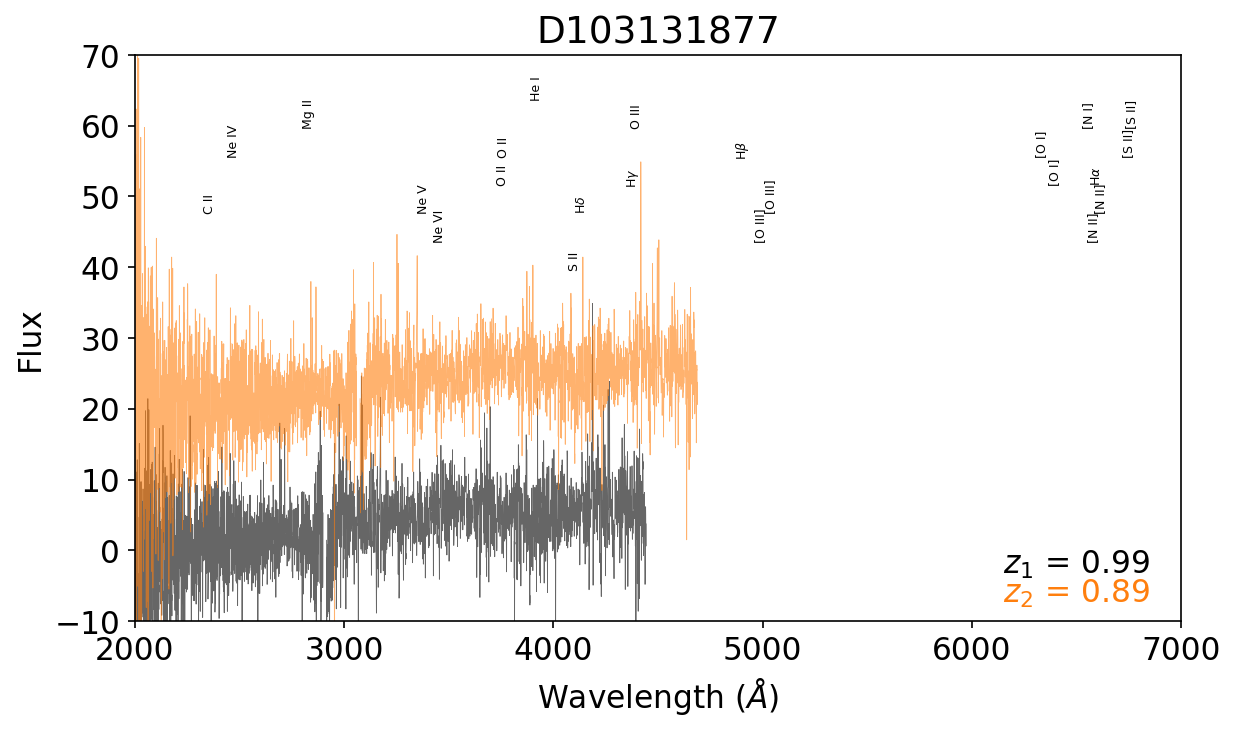}
\includegraphics[width=0.49\textwidth]{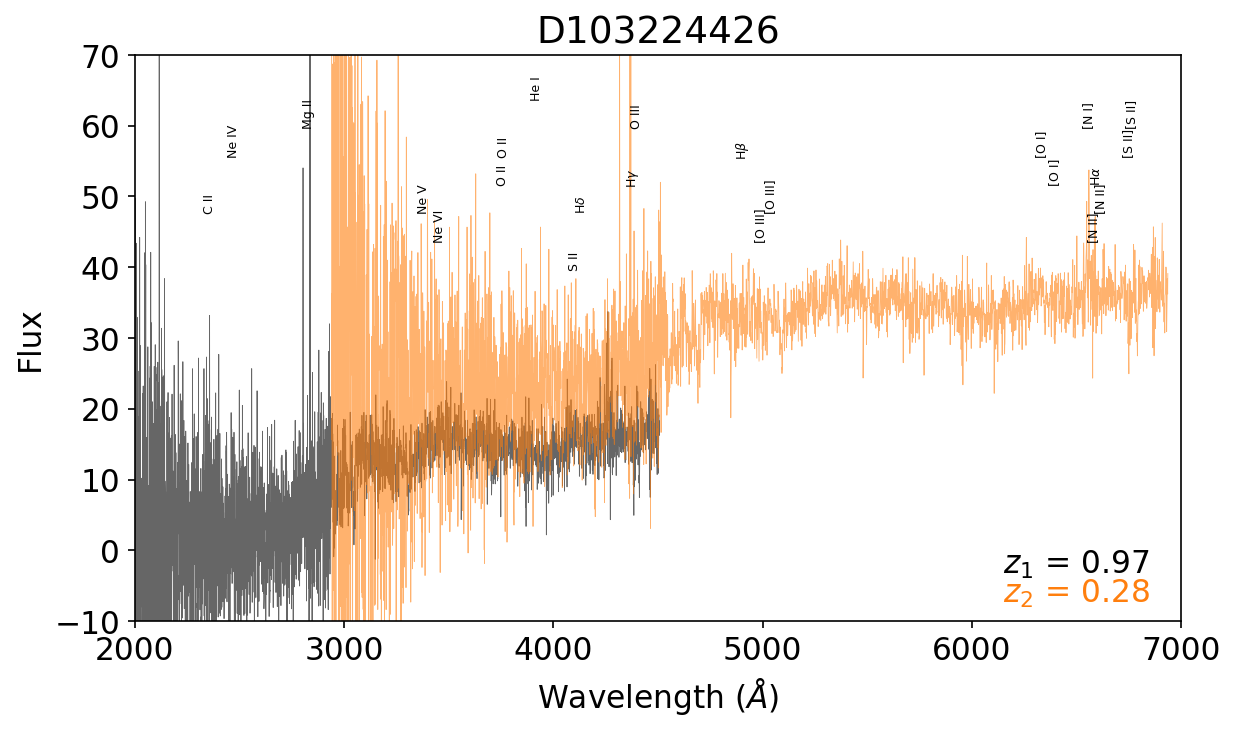}
\includegraphics[width=0.49\textwidth]{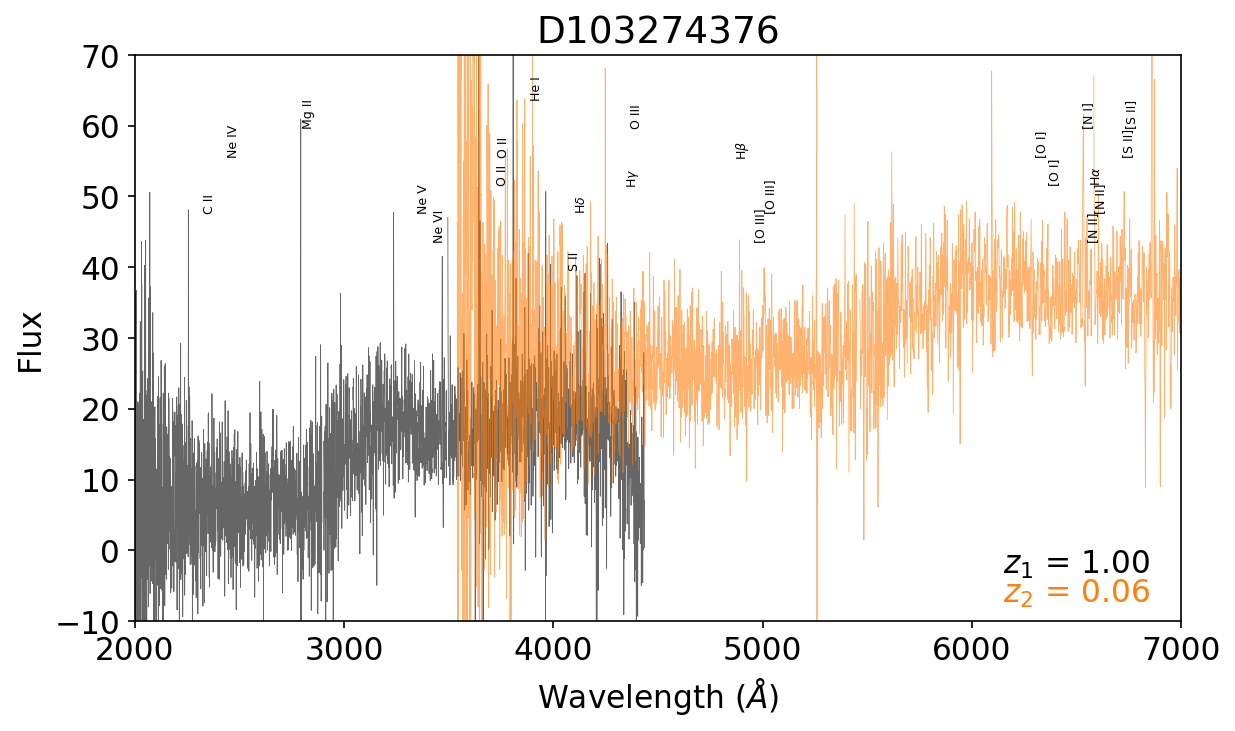}
\includegraphics[width=0.49\textwidth]{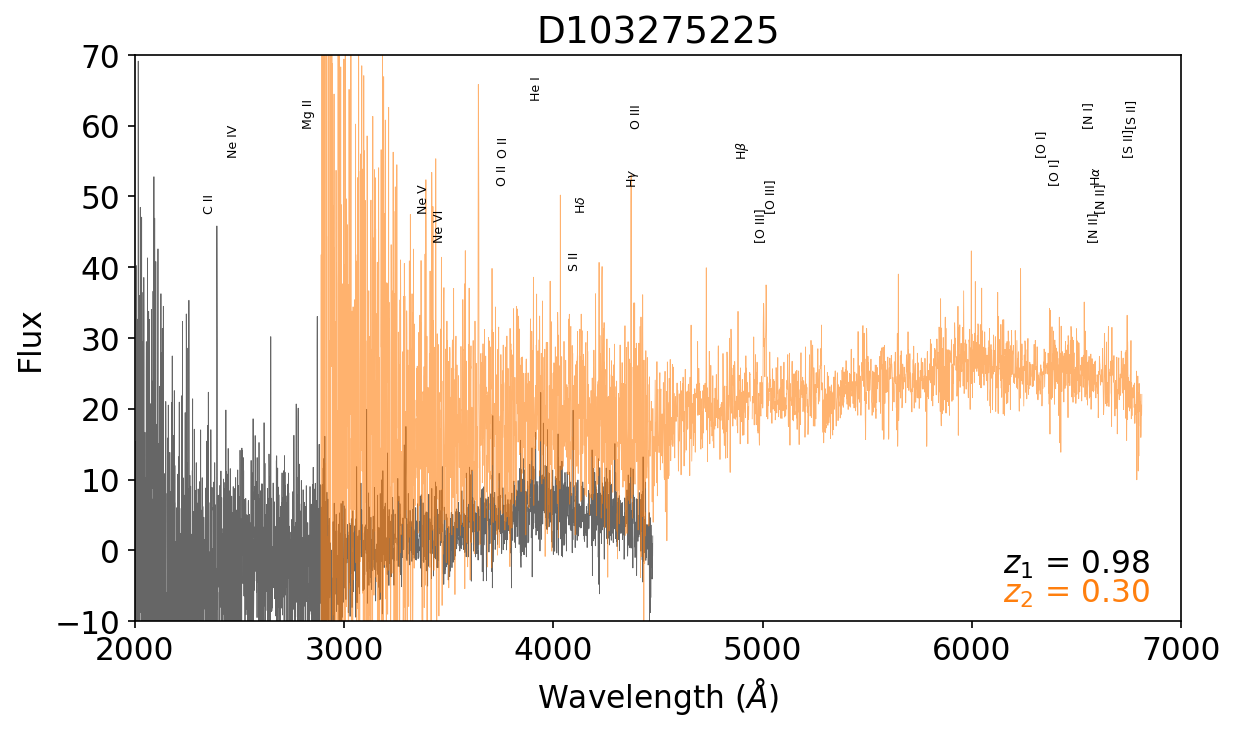}
    \caption{-- \textit{continued}.}
\end{figure*}

\setcounter{figure}{0}

\begin{figure*}
    \centering
\includegraphics[width=0.49\textwidth]{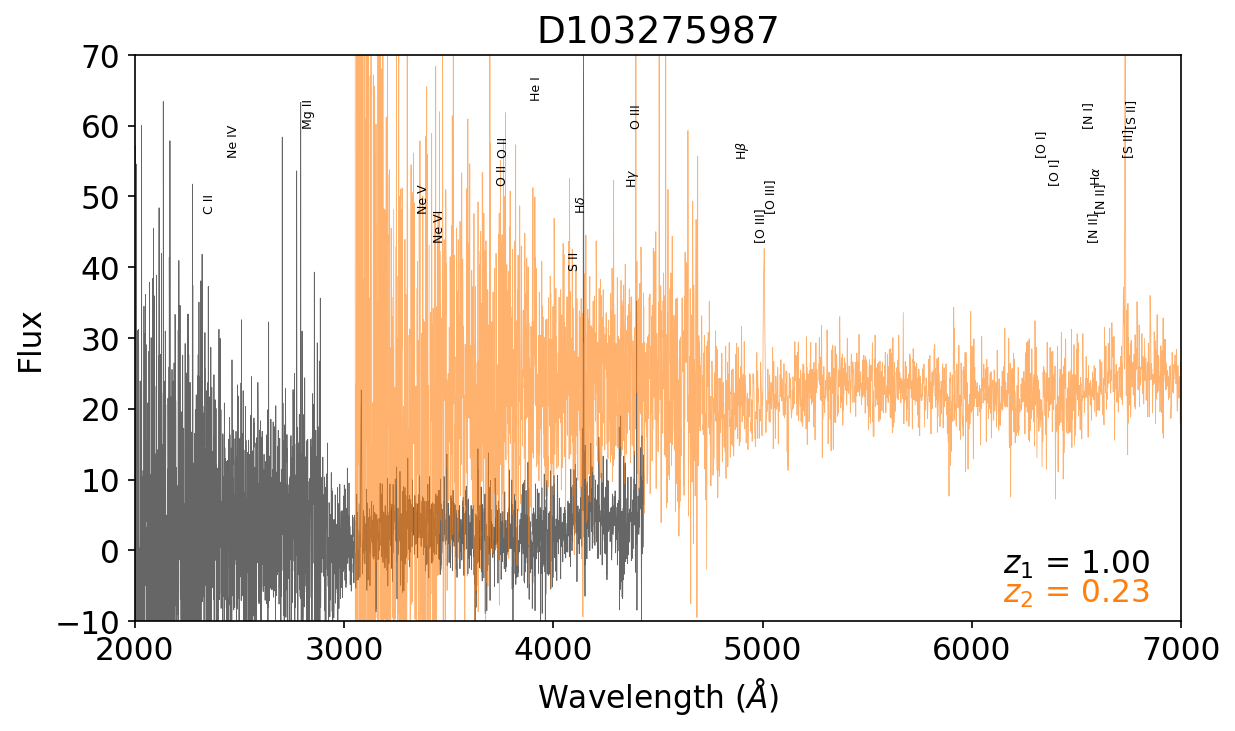}
\includegraphics[width=0.49\textwidth]{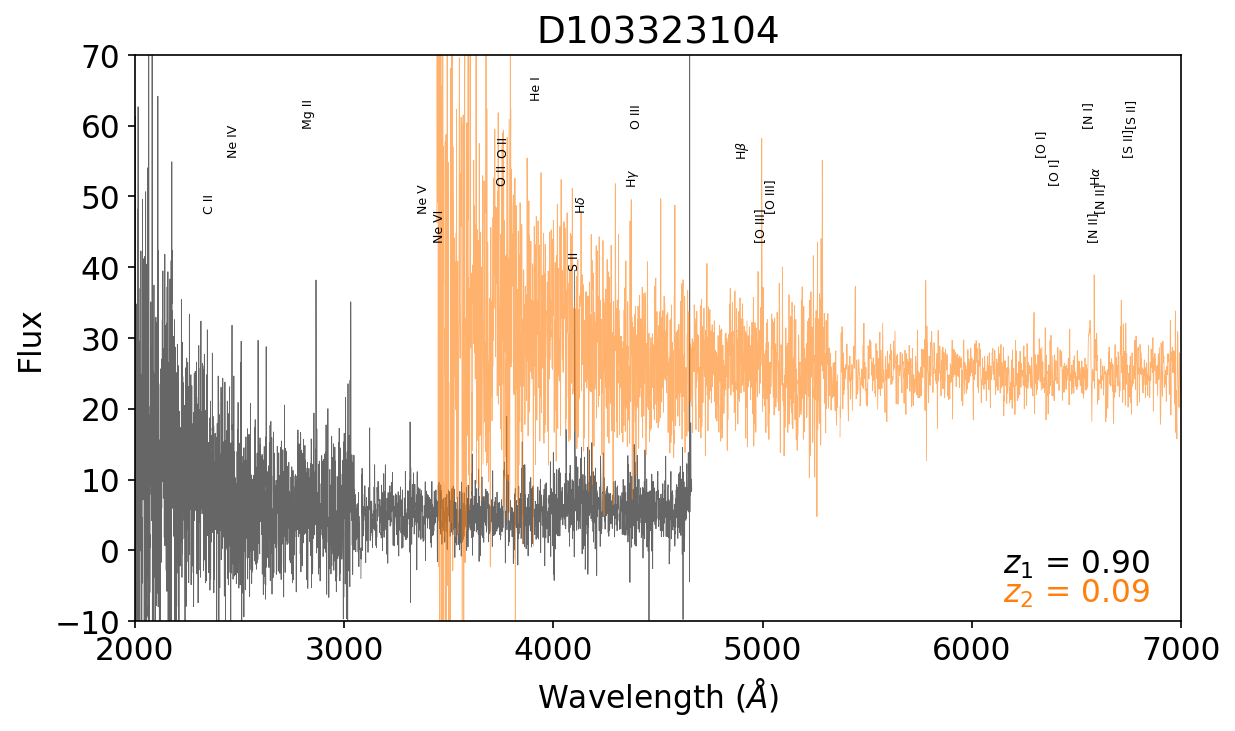}
\includegraphics[width=0.49\textwidth]{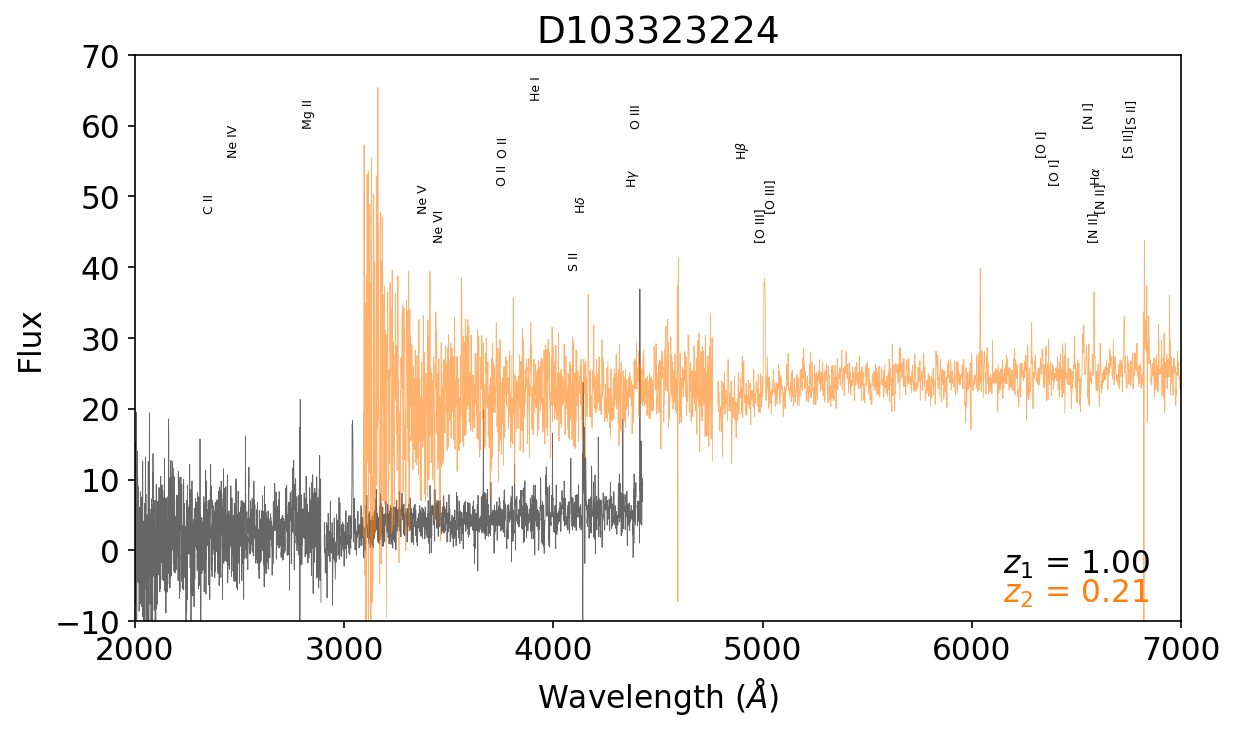}
\includegraphics[width=0.49\textwidth]{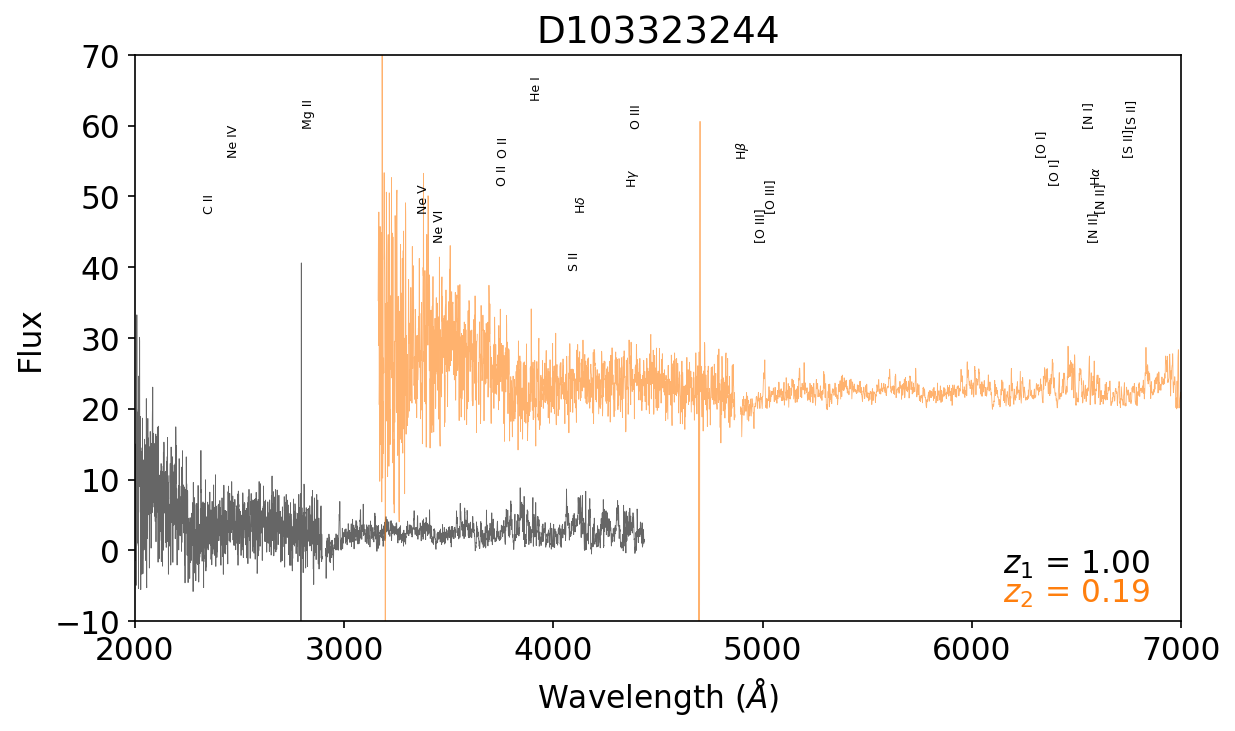}
\includegraphics[width=0.49\textwidth]{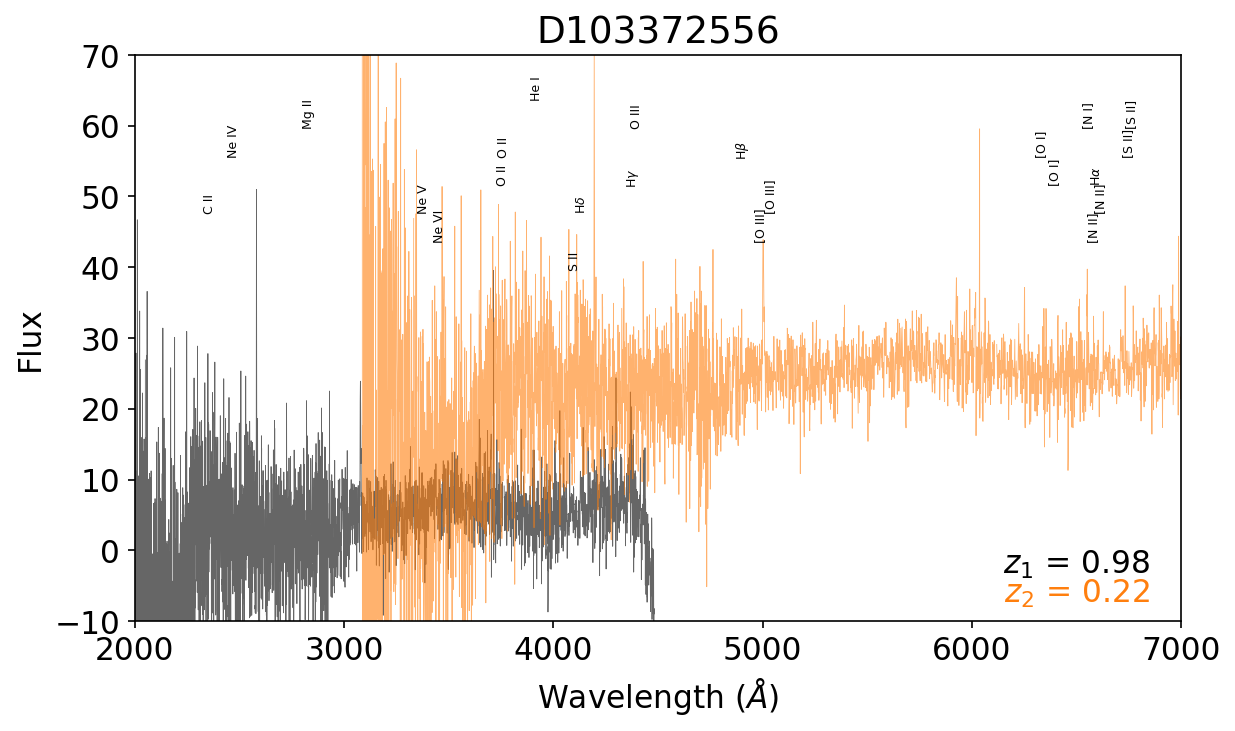}
\includegraphics[width=0.49\textwidth]{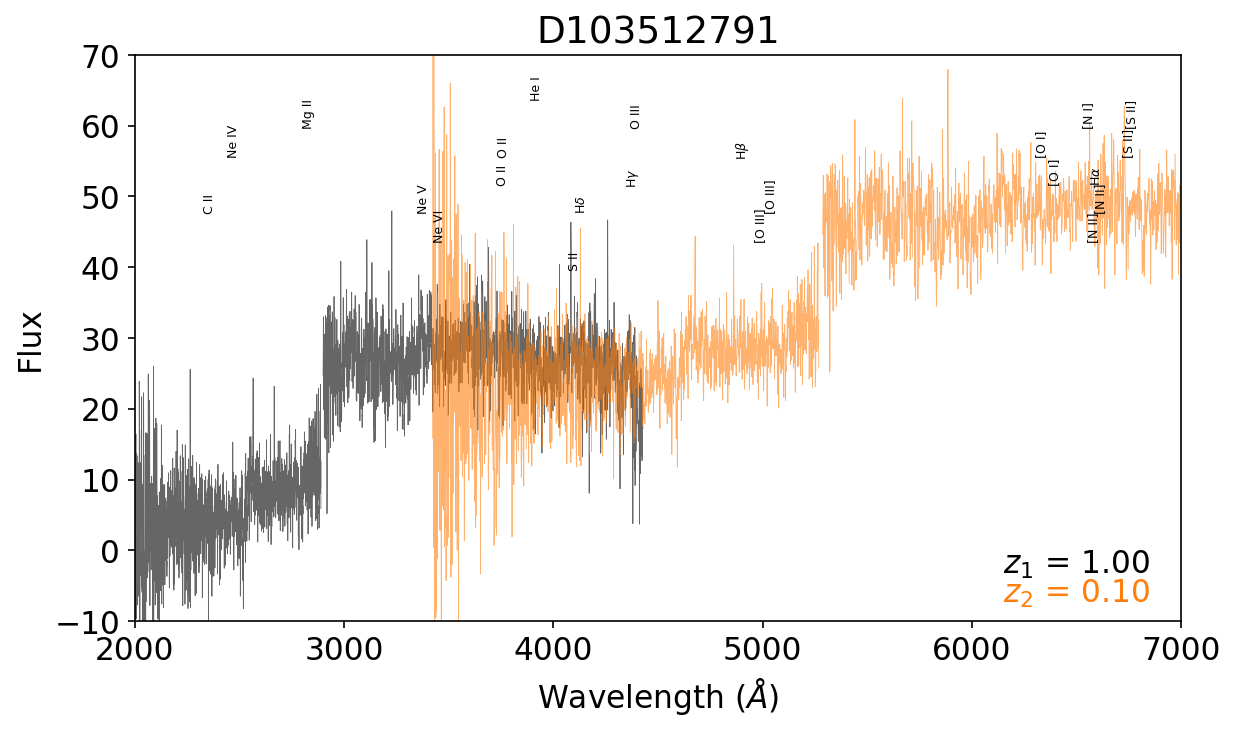}
\includegraphics[width=0.49\textwidth]{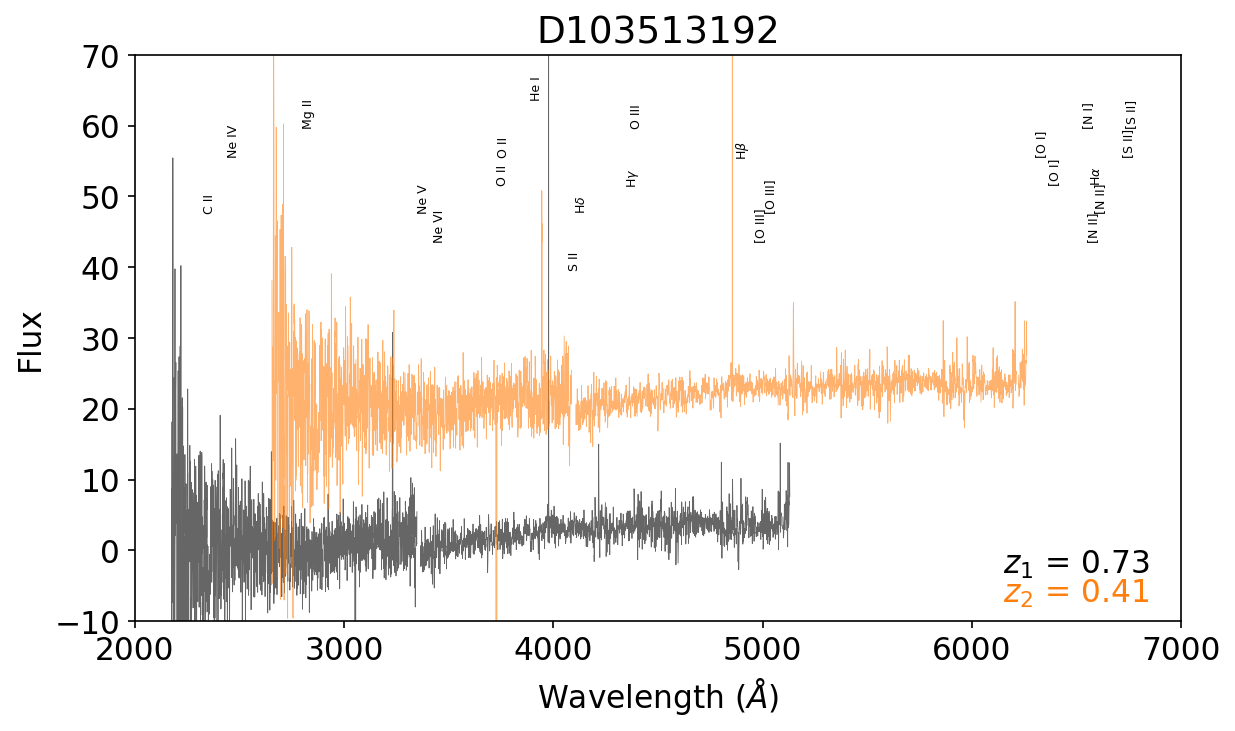}
\includegraphics[width=0.49\textwidth]{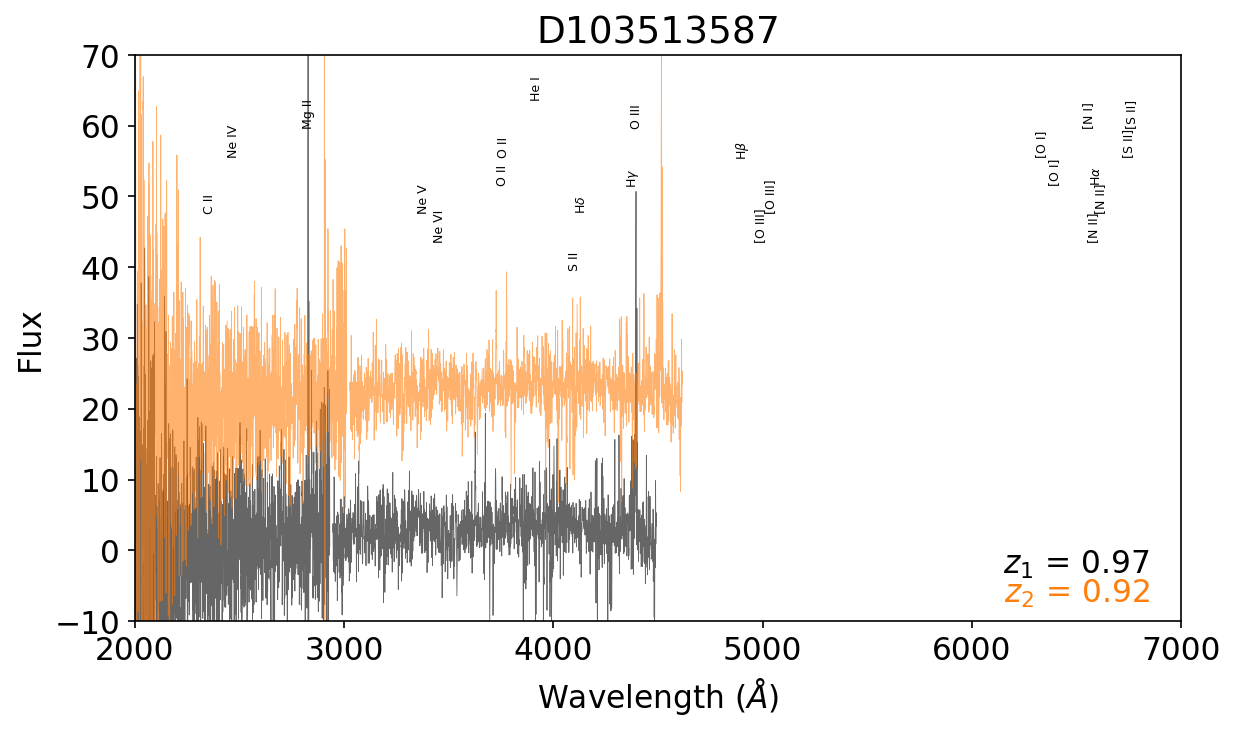}
   \caption{-- \textit{continued}.}
\end{figure*}

\setcounter{figure}{0}

\begin{figure*}
    \centering
\includegraphics[width=0.49\textwidth]{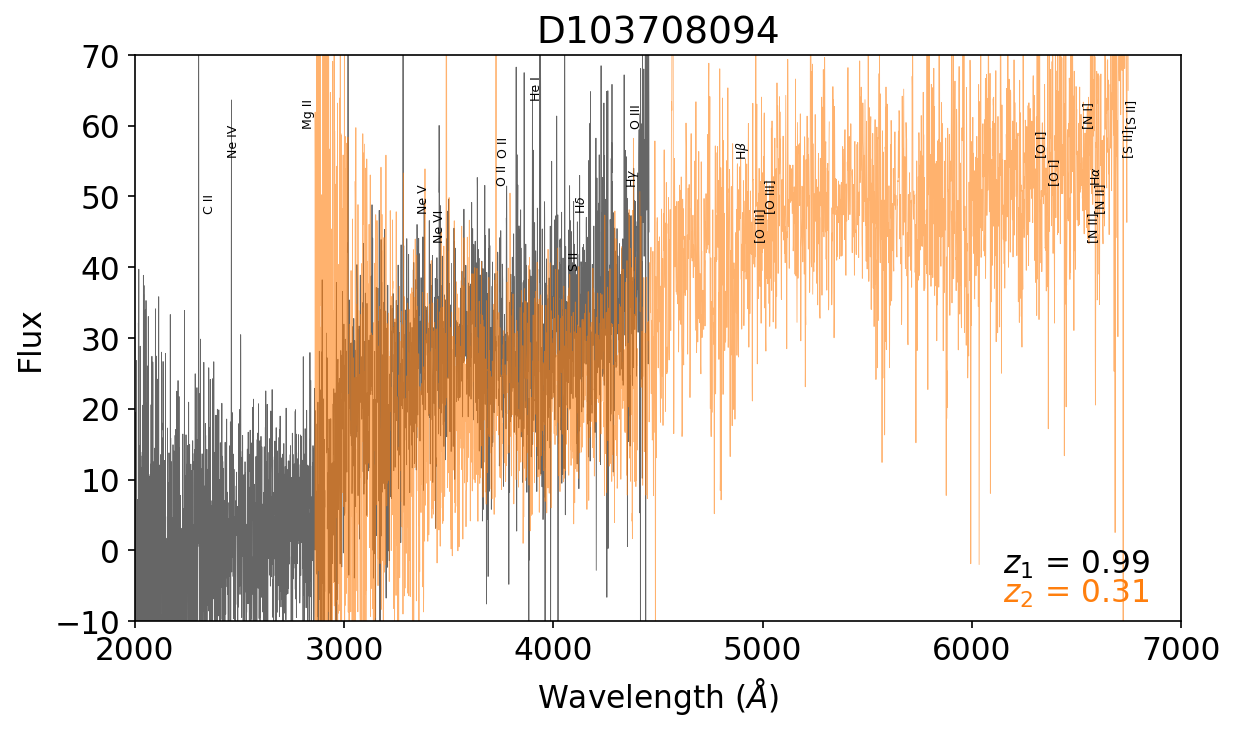} 
\includegraphics[width=0.49\textwidth]{Holwerda21_A1_D103513587.png}
\includegraphics[width=0.49\textwidth]{Holwerda21_A1_D103708094.png} 
   \caption{-- \textit{continued}.}
\end{figure*}

\bsp	
\label{lastpage}
\end{document}